\documentclass[a4paper,11pt]{article}
\usepackage{graphicx,color}
\usepackage{amsmath}
\usepackage{amssymb}
\usepackage{multirow}
\usepackage{url}
\usepackage{subfigure}
\usepackage{bm}
\usepackage{authblk}
\usepackage{fullpage}
\usepackage{lineno}
\usepackage{xifthen}
\usepackage{comment}
\usepackage{listings}
\usepackage{gensymb}
\usepackage{wasysym}
\makeindex
\graphicspath{{figures/}}

\def\draftmodename{\detokenize{T2K-II-beamline-TDR-2017-draft}}
\newif\ifdraftcompile
\ifthenelse{\equal{\jobname}{\draftmodename}}{\draftcompiletrue}{\draftcompilefalse}
\newcommand{\MODCOLOR}{black}
\newcommand{\MODCOLORB}{black}

\begin{document}
\title{J-PARC Neutrino Beamline Upgrade Technical Design Report}
\author{T2K Collaboration and J-PARC Neutrino Facility group}
\date{\today}
\maketitle
\abstract{%
In this document, technical details of the upgrade plan of the
J-PARC neutrino beamline for the extension of the T2K experiment are
described~\cite{T2K2proposal}. T2K has proposed to accumulate data
corresponding to $2\times{}10^{22}$ protons-on-target in the next decade,
aiming at an initial observation of CP violation with $3\sigma$ or higher
significance in the case of maximal CP violation. Methods to
increase the neutrino beam intensity, which are necessary to achieve the
proposed data increase, are described. %
}

\clearpage
\noindent{\Large The T2K Collaboration}\newline
\noindent
K.\,Abe$^{\,52}$,
H.\,Aihara$^{\,51,26}$,
A.\,Ajmi$^{\,29}$,
C.\,Alt$^{\,9}$,
C.\,Andreopoulos$^{\,50,31}$,
M.\,Antonova$^{\,24}$,
S.\,Aoki$^{\,28}$,
Y.\,Asada$^{\,65}$,
Y.\,Ashida$^{\,29}$,
A.\,Atherton$^{\,50}$
% ...

\newline\newline\noindent
% ...
$^{9}$~ETH Zurich, Institute for Particle Physics, Zurich, Switzerland\\
% ...
$^{24}$~Institute for Nuclear Research of the Russian Academy of Sciences, Moscow, Russia\\
% ...
$^{26}$~Kavli Institute for the Physics and Mathematics of the Universe (WPI), The University of Tokyo Institutes for Advanced Study, University of Tokyo, Kashiwa, Chiba, Japan\\
% ...
$^{28}$~Kobe University, Kobe, Japan\\
$^{29}$~Kyoto University, Department of Physics, Kyoto, Japan\\
% ...
$^{31}$~University of Liverpool, Department of Physics, Liverpool, United Kingdom\\
% ...
$^{50}$~STFC, Rutherford Appleton Laboratory, Harwell Oxford,  and  Daresbury Laboratory, Warrington, United Kingdom\\
$^{51}$~University of Tokyo, Department of Physics, Tokyo, Japan\\
$^{52}$~University of Tokyo, Institute for Cosmic Ray Research, Kamioka Observatory, Kamioka, Japan\\
% ...
$^{65}$~Yokohama National University, Faculty of Engineering, Yokohama, Japan\\
% ...

\clearpage
\tableofcontents
\clearpage

\section*{Executive summary}

T2K, Tokai to Kamioka long-baseline neutrino oscillation experiment, is an accelerator-based
neutrino experiment using high intense muon neutrino beam from J-PARC accelerator and
Super-Kamiokande as a far detector, located 295~km away from J-PARC. T2K revealed the existence
of $\nu_{\mu} \to \nu_e$ oscillation phenomenon in 2013~\cite{Abe:2013hdq}. After the discovery, 
CP violation in neutrino sector has been explored,
increasing the statistics o both $\nu_{\mu} \to \nu_e$ and $\bar{\nu}_{\mu} \to \bar{\nu}_e$ oscillations.
In 2017, T2K reported the results of 2$\sigma$ significance to exclude the CP conservation with
data up to 2.2$\times 10^{21}$ protons on target. T2K has proposed to extend the run from 
7.8$\times 10^{21}$ POT to 20$\times 10^{21}$ POT to improve the sensitivity of
neutrino CP violation with significantly improved statistics. 
To increase the statistics, J-PARC accelerator and T2K neutrino beamline groups proposed
beam power improvement toward 1.3~MW and neutrino beamline upgrades to be achieved
by 2026. In this document, technical details of the neutrino beamline upgrades are described.

\clearpage
\section{Introduction}

\subsection{Overview}
The Tokai to Kamioka Long Baseline Neutrino Oscillation Experiment (T2K)
uses the \(\nu_\mu\) (muon neutrino) beam produced by the J-PARC neutrino beamline
to measure \(\nu_\mu\) to \(\nu_e\) (electron neutrino) and \(\nu_\mu\) to \(\nu_\mu\) 
neutrino oscillations after the neutrinos propagate 295~km to the Super-Kamiokande
far detector.  Since the neutrino cross section is small, the neutrino flux should
be as large as possible to maximize the number of interactions in the far
detector.  Various upgrades to the J-PARC accelerator are under way in order to increase 
the J-PARC neutrino flux for T2K and subsequent experiments, as discussed in this document.

\begin{figure}[h] 
\centering 
\includegraphics[width=0.8\textwidth] {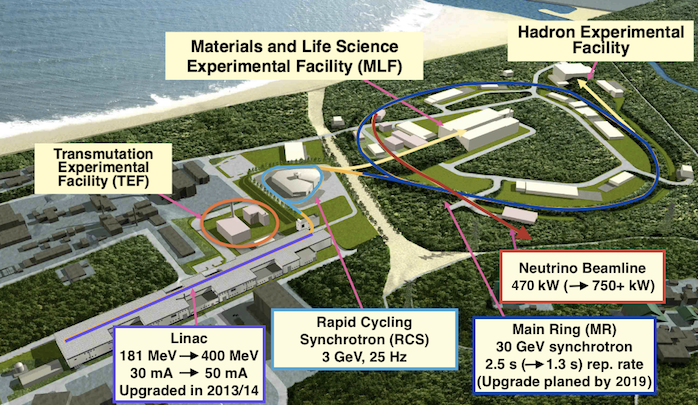}
\caption{\small The J-PARC accelerator complex.}
\label{fig:jparc} 
\end{figure}

The J-PARC complex is shown in Fig.~\ref{fig:jparc}.  
The neutrino beam for the T2K experiment is produced using 30~GeV protons from the
J-PARC Main Ring accelerator (MR).  Protons extracted from 
the MR are bent towards the Super-Kamiokande direction and focused onto the neutrino production target in
the neutrino primary beamline, which is described in Sec.~\ref{sec:primary}.  Neutrino parent hadrons 
outgoing from the target are focused and allowed to decay, producing a neutrino beam, 
in the neutrino secondary beamline, which is described in Sec.~\ref{sec:secondary}.

\subsection{The requirement for the neutrino beam for extended running of T2K}
\label{sec:beamrequirements}
After the discovery of $\nu_{\mu} \to \nu_{e}$  oscillation by T2K,
neutrino CP violation is being explored increasing the statistics of both
$\nu_{\mu} \to \nu_{e}$  and $\bar{\nu}_{\mu} \to \bar{\nu}_{e}$  oscillations.
In 2017, T2K has shown the results of 2$\sigma$ significance to exclude the CP conservation
with data up to $22.3 \times 10^{20}$ protons on the target (POT), which is 30\% of
the original T2K target POT. 
The next generation projects such as Hyper-Kamiokande~\cite{HK} and DUNE~\cite{DUNE} aim to 
achieve greater than 3$\sigma$ sensitivity to CPV across a wide range of $\delta_{CP}$
on the time scale of 2026 and beyond. 

The T2K collaboration proposes to extend the run from $7.8\times 10^{21}$ POT to 
$20 \times 10^{21}$ POT in a five or six year period after the currently approved running
to explore CP violation with sensitivity greater than $3\sigma$ if $\delta_{CP} \sim -\frac{\pi}{2}$
and the mass hierarchy is normal. 
\color{\MODCOLOR} We can also improve a precision of neutrino oscillation parameters, 
$\theta_{23}$ and $\Delta m^2_{32}$, to 1.7$^\circ$ or better and 1\%, respectively, 
with the proposed POT. 
\color{black}
We refer to this extended running as ``T2K-II''~\cite{T2K2proposal}. 

\begin{figure}[h] 
\centering 
\includegraphics[width=0.8\textwidth] {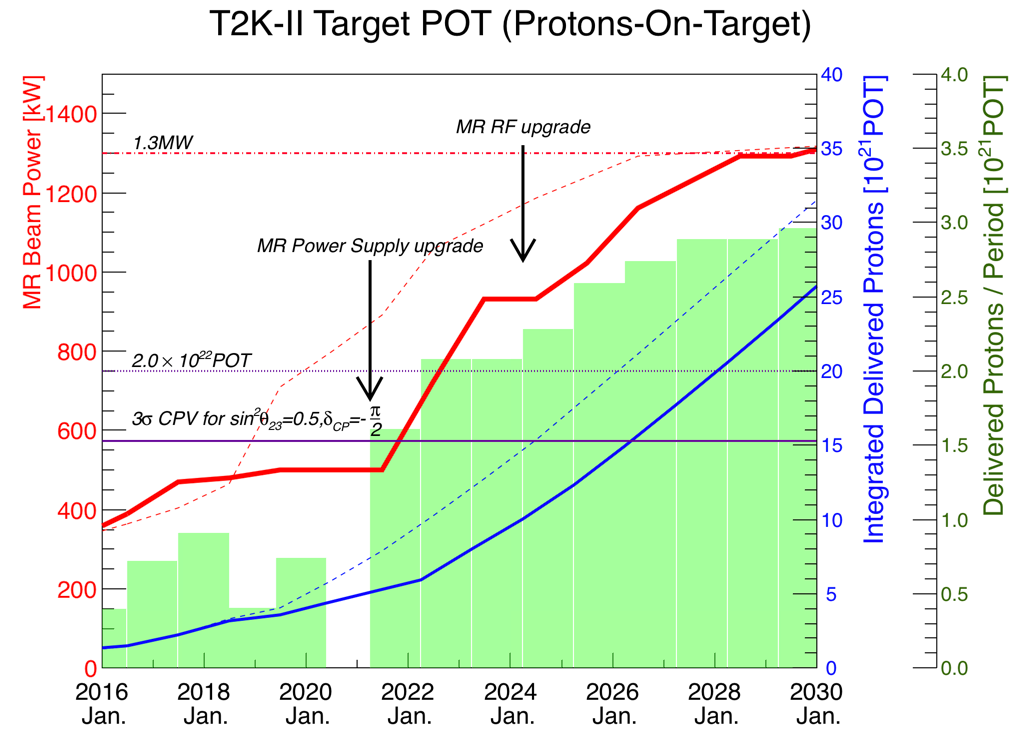}
\caption{\small \color{\MODCOLOR}
Our target MR beam power and accumulated POT as a function of Japanese Fiscal Year (JFY). 
The solid lines are the our target MR beam power (red) and accumulated POT (blue) 
where six months of the MR operation with the fast-extraction mode each year and 
the running time efficiency of 90\% are assumed. It is also assumed that 
the MR beam power upgrade is performed earlier than the schedule shown in 
IAC2018\cite{IAC2018} as described in the text. 
In this case, the MR main power supply installation is assumed to be performed 
from July 2020 to September 2021. 
The dashed lines are the target MR beam power (red) and accumulated POT (blue) shown in 
T2K-II proposal\cite{T2K2proposal} where the MR main power supply installation was scheduled in 2019.%
\color{black}
}
\label{fig:expectedpot} 
\end{figure}

In this proposal, T2K requires the increase of the MR beam power up to 1.3~MW. 
\color{\MODCOLOR}
After the T2K-II proposal was submitted, the schedule of 
the MR beam power improvement was revised based on the funding situation
and then the installation of the MR main power supply, which is one of 
major upgrade items described in Sec.~\ref{sec:accelerator}, will be performed 
in 2021 instead of 2019\cite{IAC2018,Koseki:2018ccn}. 
Figure~\ref{fig:expectedpot} shows our target data accumulation scenario 
where six months of the MR operation with the fast-extraction mode each year
and the running time efficiency of 90\% are assumed.
In this plot, it is also assumed that 
the MR beam power upgrade is performed earlier than the schedule shown in 
IAC2018\cite{IAC2018} based on the discussion with the MR group, 
where it is considered that half a year ahead schedule of the MR main power supply and 
one year ahead schedule of the MR RF upgrade assuming allocation of the supplemental budget. 
The CPV sensitivity greater than 3$\sigma$ is expected to achieve 
around 2026 in the case of $\delta_{CP} = -\frac{\pi}{2}$ and the $\sin^2\theta_{23} = 0.5$  
where these values are close to the T2K 2018 summer results\cite{Neutrino2018T2K}. 
The $20 \times 10^{21}$ POT required in the T2K-II proposal is also expected to achieve 
by the end of the Japanese fiscal year of 2027. 
\color{black}

Moreover, further improvements are planned to increase the effective statistics and the sensitivity.
Increasing the electromagnetic horn current from the present 250~kA to the designed 320~kA will
result in 10\% greater neutrino flux while reducing the wrong sign components.
Improvement of neutrino flux prediction uncertainty can make benefits on better understanding 
of the neutrino interaction models and neutrino oscillation analysis.

In the following sections, the detailed plan how to realize those requirements on the 
neutrino beam for T2K-II will be discussed.
Those requirements are also common for the future projects, Hyper-Kamiokande,
since the same beamline is utilized.

\subsection{The upgrade plan of J-PARC accelerator and  the prospects of proton beam intensity of J-PARC MR}
\label{sec:accelerator}

The J-PARC accelerator consists of a normal-conducting LINAC as an injection system, a Rapid Cycling Synchrotron (RCS), and
a Main Ring synchrotron (MR). H$^-$ ion beams are accelerated to 400~MeV by the LINAC and then injected to the RCS ring,
where conversion into a proton beam is achieved by charge-stripping foils.
The RCS accumulates and accelerates two proton beam bunches up to 3~GeV at a repetition rate of 25~Hz.
Four beam pulses are injected from the RCS to the MR at 40~ms intervals to form eight bunches in a cycle, and accelerated
up to 30~GeV. The circulating proton beam bunches are extracted within a single turn into the neutrino primary beamline
by a kicker/septum magnet system in fast extraction (FX) mode operation. 

The MR beam power has been increased gradually since the start of the user operation in 2010.
The 485~kW proton beam with 2.5$\times 10^{14}$ protons per pulse (ppp) at 2.48 s cycle was successfully provided to the neutrino beamline as of April 2018. 
The current plan to achieve the design beam power of 750~kW is to reduce the repetition cycle to 1.3~s.
With the currently achieved beam intensity, the MR beam power can be increased to 900~kW at 1.3~s cycle.
In a high intensity trial, a single-shot beam with 2.6$\times 10^{14}$ ppp was successfully extracted to the neutrino beamline.
Diligent studies to increase the beam intensity have been performed continuously in these years.
Thanks to the recent progress in the high intensity studies, the long-term MR power upgrade plan was revised to aim for 1.3~MW with 3.2$\times 10^{14}$ ppp
at 1.16~s cycle by 2026. Beam operation parameters for the achieved and proposed beam power are summarized in Tab.~\ref{tab:MR_power_parameters}.
\begin{table}
\centering
\small
\caption{\small Summary of the MR operation parameters for the current and proposed beam power.}
\label{tab:MR_power_parameters}
\begin{tabular}{lcccc}
\hline\hline
Beam power             &  485~kW              & 511~kW              & 750~kW              &  1.3~MW             \\
                       & (achieved)           & (demonstrated)      & (proposed)          & (proposed)          \\\hline
Beam energy            & 30~GeV               & 30~GeV              & 30~GeV              & 30~GeV              \\
Beam intensity (ppp)   & 2.5$\times 10^{14}$  & 2.6$\times 10^{14}$ & 2.0$\times 10^{14}$ & 3.2$\times 10^{14}$ \\ 
Repetition cycle       & 2.48~s               & 1~shot              & 1.32~s              & 1.16~s              \\\hline\hline
\end{tabular}
\end{table}

\color{\MODCOLOR}
There are five upgrade items to be planned for reducing the repetition cycle for 750~kW as following.
\color{black}
\begin{itemize}
\item Upgrade of all the power supplies for the MR magnets. 
\item Upgrade of MR RF cavities to achieve a higher gradient acceleration.
\item Upgrade of injection and extraction devices for the MR.
\item Increase of MR ring collimator capacity from 2 to 3.5~kW. 
\item \color{\MODCOLOR} Upgrade of one family of the MR quadrupole magnet to enlarge its aperture. \color{black}
\end{itemize}

In the MR power supply upgrade all the power supplies for the MR magnets are replaced with new ones
where energy storage capacitors and symmetric circuit are adopted to satisfy
both faster repetition rate and low current ripple.
One of the new power supplies, which is a middle-scale one used for a quadrupole magnet in straight section,
has already been produced and operated stably since 2016. It has successfully reduced the output current ripple by 
an order of magnitude. One of the large-scale power supplies has also been produced in 2017 and is now 
under commissioning. Three new buildings for the new power supplies have already been constructed in FY2016 and
FY2017. Production of the rest of the new power supplies are currently planned to be done in 2$\sim$3 years
and their installation to be done with one-year-long shutdown.  
After the power supply installation, the commissioning of the MR operation with the new power supplies
is supposed to be started and then user operation with 1.3~s cycle is expected soon after the commissioning.

Regarding the RF cavity upgrade, higher acceleration gradient is required to satisfy the 1.3~s cycle.
To meet the requirement, a newly developed core material, FINEMET$^{\tiny{\textregistered}}$ core FT3L, 
which had twice higher impedance than the old one FT3M, was adopted to the new RF cavity.
The old FT3M cavities have been already replaced with the new FT3L ones by FY2016. In the current operation,
seven cavities are used as fundamental harmonic RF cavities with nominal RF voltage of 300$\sim$390~kV.
The rest of two cavities are used as second harmonic cavities in order to stretch the beam bunch structure
longitudinally and to reduce the peak beam intensity to avoid beam instability.
Higher RF voltage of 510~kV is required for the 1.3~s operation, and all the nine cavities must be used 
as the fundamental ones. Two additional cavities should be installed for the second RF ones,
and the old FT3M ones are planned to be refurbished.

For the upgrade toward 1.3~MW, the following upgrade items are being planned.
\begin{itemize}
\item Upgrade of the MR RF anode power supply.
\item Installation of two additional fundamental RF cavities.
\item Upgrade of the MR beam position monitors for higher precision measurement.
\item Upgrade of the MR FX kicker magnet to lower impedance one.
\end{itemize}

The beam intensity and repetition cycle are further improved to 3.2$\times 10^{14}$ ppp and
1.16~s, respectively. To satisfy the requirements further upgrade of the MR RF system is planned.
For the higher beam intensity RF voltage of 600~kV is required for the fundamental RF cavities.
Two additional FT3L cavities will be installed for the fundamental RF ones (i.e., 11 cavities in total).
In addition, it is also necessary to increase peak anode current in the RF cavities.
The anode power supply for the RF system will also be upgraded by adding four more inverter units (i.e.,
15 $\to$ 19 inverter units in total). Configuration of the MR RF system is summarized in 
Tab.~\ref{tab:MRRF_config}.
\begin{table}
\centering
\small 
\caption{\small Summary of MR RF configuration for the current and proposed beam power.}
\label{tab:MRRF_config}
\begin{tabular}{lcccc}
\hline\hline
Beam power             &  485~kW              & 511~kW              & 750~kW              &  1.3~MW             \\
                       & (achieved)           & (demonstrated)      & (proposed)          & (proposed)          \\\hline
\# of fundamental RF   & 7 (FT3L)             & 7 (FT3L)            & 9 (FT3L)            & 11 (FT3L)           \\
\# of second RF        & 2 (FT3L)             & 2 (FT3L)            & 2 (FT3M)            & 2  (FT3M)           \\
Fundamental RF voltage & 300$\sim$390~kV      & 300$\sim$390~kV     & 510~kV              & 600~kV              \\
Second RF voltage      & 110~kV               & 110~kV              & 120~kV              & 120~kV              \\
RF anode power supply  & 15 inverter units    & 15 inverter units   & 15 inverter units   & 19 inverter units   \\
\hline\hline
\end{tabular}
\end{table}

In addition to the hardware upgrades, it is really important to understand beam properties and to reduce beam loss
by intensive beam studies in order to achieve safe and reliable operation even with such a high intensity beam. 

\color{\MODCOLOR}
The MR upgrade plan is reviewed at the J-PARC Accelerator Technical Advisory Committee (A-TAC)\cite{atac}. 
The detailed information of these MR upgrade items are described in the technical design report of the MR upgrade 
which will be released in February 2019. 
\color{black}

\subsection{Beamline upgrade toward $>$1.3~MW operation}
\label{sec:summaryofbeamlineupgrade}
Toward the operation with 1.3MW and beyond beam power,
the neutrino beamline will be upgraded.
In order to operation the beam safely and stably at the new higher beam power,
the following subjects are necessary.
\begin{itemize}
\item Tolerable beam loss in the primary beamline 
\item Reliable beam parameter measurements
\item Protection of the beamline equipments against any unexpected
  beam injections (collimator, interlock)
\item Robustness against the direct proton beam exposure (target, beam window and
  profile monitors)
\item Cooling of the heat generated by beam
\item Remote maintenance capability for the high radio activated equipment
\item Handling of produced radioactive wastes
\end{itemize}

In the following sections, Sec.~\ref{sec:primary} for the primary beamline and
beam monitor, Sec.~\ref{sec:secondary} for the secondary beamline,
Sec.~\ref{sec:ctrldaq} for beamline DAQ and control, 
a summary of the beam operation up to now and
necessary improvements toward the high power beam operation 
will be discussed.
As discussed above, the improvement of systematic uncertainties is
also necessary to achieve higher sensitivity of neutrino CPV.
The plan of improvement of neutrino flux prediction will be discussed in
Sec.~\ref{sec:flux}.

\clearpage

\section{Primary Beamline and Beam Monitors}
\label{sec:primary}

\graphicspath{{figures/main_primary}}
\subsection{Introduction} 
The role of the primary beamline is to
stably deliver protons extracted from
the main ring to the neutrino production
target with proper beam position,
size and injection angle.
The extracted proton beam is transported
toward the T2K far detector, Super-Kamiokande,
with a off-axis angle of 2.5${}^{\circ}$. 
The beam transport has to be done with tolerable beam loss
through the beamline from the view point of
both the maintenance of the beamline equipment and
the radiation level at the boundary of the radiation
controlled area.

\begin{figure}[h] 
\centering 
\includegraphics[width=0.8\textwidth] {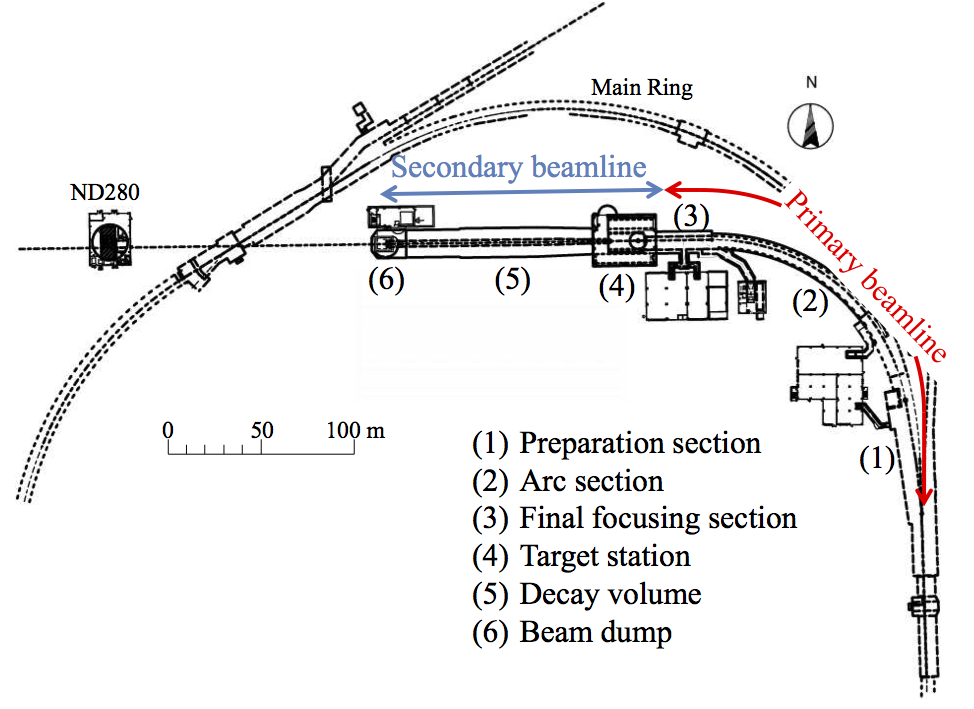}
\caption{Overview of the neutrino beamline.}
\label{fig:beamline}
\end{figure}

There are three sections of the primary beamline: 
preparation section, arc section and final focusing section
(Fig.~\ref{fig:beamline}).
The 30~GeV accelerated proton beam is extracted 
into the preparation section. The extracted beam 
is tuned the position and width with 
normal-conducting magnets in order to match
the beam optics at the arc section. 
The beam is then bent toward the direction of 
the Super-Kamiokande by 80.7$^\circ$ using 
super-conducting combined function magnets at the 
arc section.
After the arc section, the beam is tuned the position
and size and focused on the center of the target 
with normal-conducting magnets while directing the
beam downward by 3.647$^\circ$ at the final
focusing section. 
These sections consist of the following equipment:
\begin{itemize}
\item normal-conducting magnets and their power supplies,
\item super-conducting magnets and their power supply and
cryogenic system,
\item beam monitors,
\item beam plugs,
\item collimators,
\item vacuum system. 
\end{itemize}
The beam optics and the present design of the equipment except for the beam monitor 
is briefly described in Sec.~\ref{sec:primarymain} to clarify
which parts need modification toward higher beam power. 
The upgrade plan of the primary beamline is then discussed in Sec.~\ref{sec:primaryup}. 
The present design, operation status and upgrade plan of the proton beam monitors
is then discussed in Sec.~\ref{sec:protonmonitor}.
Sec.~\ref{sec:primarymainte} discusses the maintenance scenario of the primary beamline. 
The upgrade plan of  muon monitor, which is 
one of important beam monitors in T2K, is described in Sec.~\ref{sec:mumon}.

\subsection{Primary beamline} \label{sec:primarymain}
\subsubsection{Optics}
\begin{figure}[h] 
\centering
\includegraphics[width=10cm,clip] {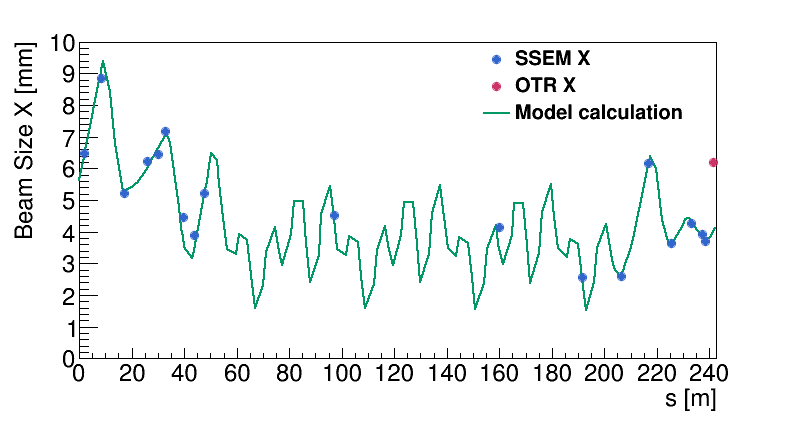}
\vspace{1ex}
\includegraphics[width=10cm,clip] {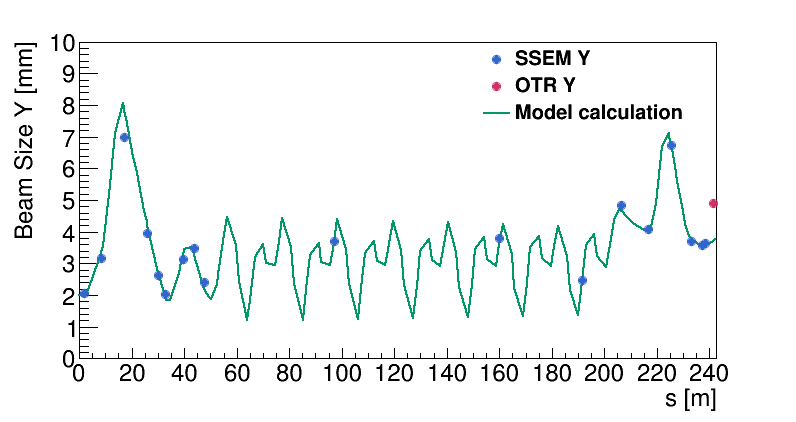}
\caption{Beam envelop of the present optics at 30~GeV. The beam size is measured 
by beam profile monitors (SSEM and OTR). Top (bottom) figure shows the horizontal
(vertical) envelop. 
\color{\MODCOLORB} $\Delta p/p = 0.001$ is assumed to calculate the beam size.\color{black} }
\label{fig:optics} 
\end{figure}

The primary beamline has been stably operated so far.
Stable operation with 485~kW beam power was achieved.
The beam emittance is expected to be 6$\sim$7$\pi$ mm mrad
(three sigma emittance) based on the beam width measurement
and the calculation of $\beta$ and momentum dispersion value.
The beam envelop of the present optics at 30~GeV is shown in
Fig.~\ref{fig:optics}.

An emittance of 10$\pi$ mm mrad was assumed for the extracted
beam from the MR in the design of primary beamline.
Normal-conducting magnets,
installed in the preparation section and in the final-focus section,
and their beam ducts were designed to accept an 81$\pi$ mm mrad
beam envelope both in horizontal and in vertical individually
with no momentum spread, and to have momentum acceptance of 2\%
for 10$\pi$ mm mrad emittance beam.
On the other hand, super-conducting magnets installed
in the arc section have an acceptance of 206$\pi$ mm mrad.

However, due to a last-minute optics re-calculation and magnet swap,
the aperture of some of the magnets may be too small for
the desired increase in beam power. 
The beam size is the largest at PQ1, 
the first quad after extraction, and at FQ2 and FQ3 for targeting.
Beam sizes at FQ2 and FQ3 are sensitive to the beam emittance,
and if the incoming beam emittance increases for higher beam power,
a larger aperture is required.

Beam optics tuning is regularly performed when the MR beam power or
the beam parameters is changed.
It is important to keep the neutrino beam direction stable
within 1~mrad because the neutrino energy peak can be shifted.
In order to keep the beam direction stable,
the beam position on the target is carefully tuned at the center
with checking the measured beam position at the muon monitor.
The beam direction is also confirmed by INGRID beam profile measurements. 
Beam size at the target surface is also tuned to keep 4mm of 
a gaussian standard deviation for both horizontal and vertical 
by adjusting the magnetic force of the four quadrupole magnets
at the final-focusing section in oder to keep the targeting
efficiency high enough while minimizing the thermal stress 
in the target. 
Understanding of the momentum dispersion is important. 
At the recent high power operation, $\sim 0.1$\% of the momentum
spreading is observed at the MR. On the other hand,
there is a finite dispersion on the horizontal direction. 
In order to keep fluctuation of the horizontal beam position and size small,
the horizontal momentum dispersion is required to keep less than 1m
during the beam optics tuning based on the optics model calculation.

\subsubsection{Beam loss}
\begin{figure}[h] 
\centering 
\includegraphics[width=10cm] {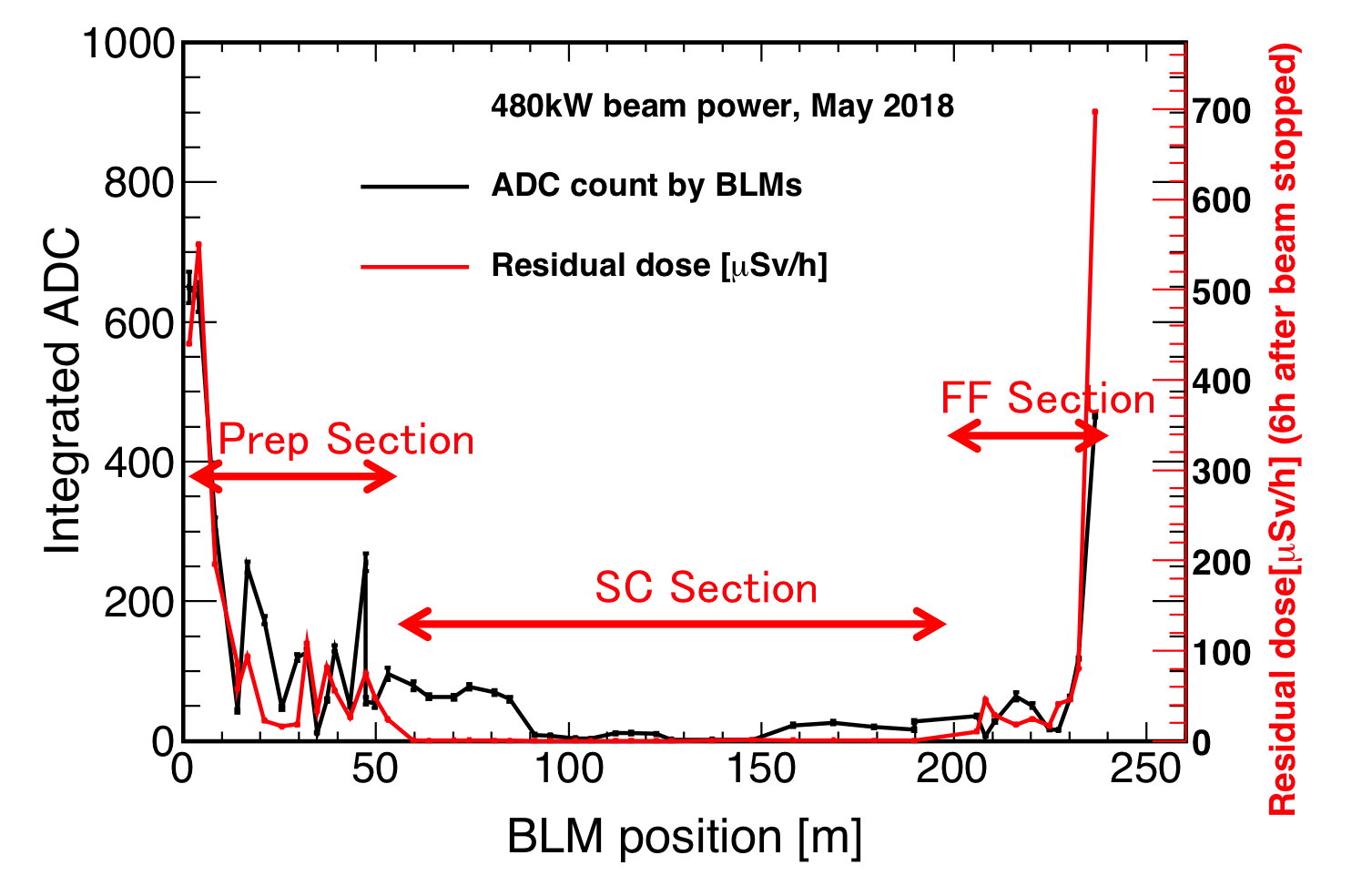}
  \caption{Beam loss distribution \textcolor{\MODCOLOR}{and residual radiation level}
  along the beamline at the present running conditions with 480~kW.
Horizontal axis is the distance from the extraction point.}
\label{fig:loss_dist} 
\end{figure}

The beam loss is measured by fifty of the beam loss monitors 
installed along the primary beamline. The monitor is 
a wire proportional counter filled with an Ar-CO$_{2}$ mixture.  
Fig.~\ref{fig:loss_dist} shows the beam loss distribution
along the primary beamline. 
The residual radiation dose is also regularly measured. 
It has been found that the residual dose is within 
the manageable level although 
the residual radiation does at the most upstream part of 
the preparation section is large even 
at 485~kW operation.
One of these causes could be 
the small aperture of the steering magnet placed at the most
upstream of the preparation section (PV1) and the beam halo
from the MR causes the beam loss.

The assumed maximum beam loss at the primary beamline is
750~W point loss at any location of the preparation section,
1~W/m line loss at the arc section and 250~W point loss at
any location of the final-focus section 
to determine the radiation shield thickness. 
It is not meant to operate the primary beamline with the above beam loss 
continuously since the equipment is designed based on 
the hands-on maintenance scenario with semi-remote or quick action devices.

\subsubsection{Normal-conducting magnets}
The preparation section and the final-focus section consist of
11 and 10 normal-conducting magnets, respectively.
The parameters are summarized in Tab.~\ref{tab:ncmag}.

\begin{table}[h] 
\centering 
\caption{Summary of normal conducting magnet parameters.}
\label{tab:ncmag}
\includegraphics[width=\textwidth] {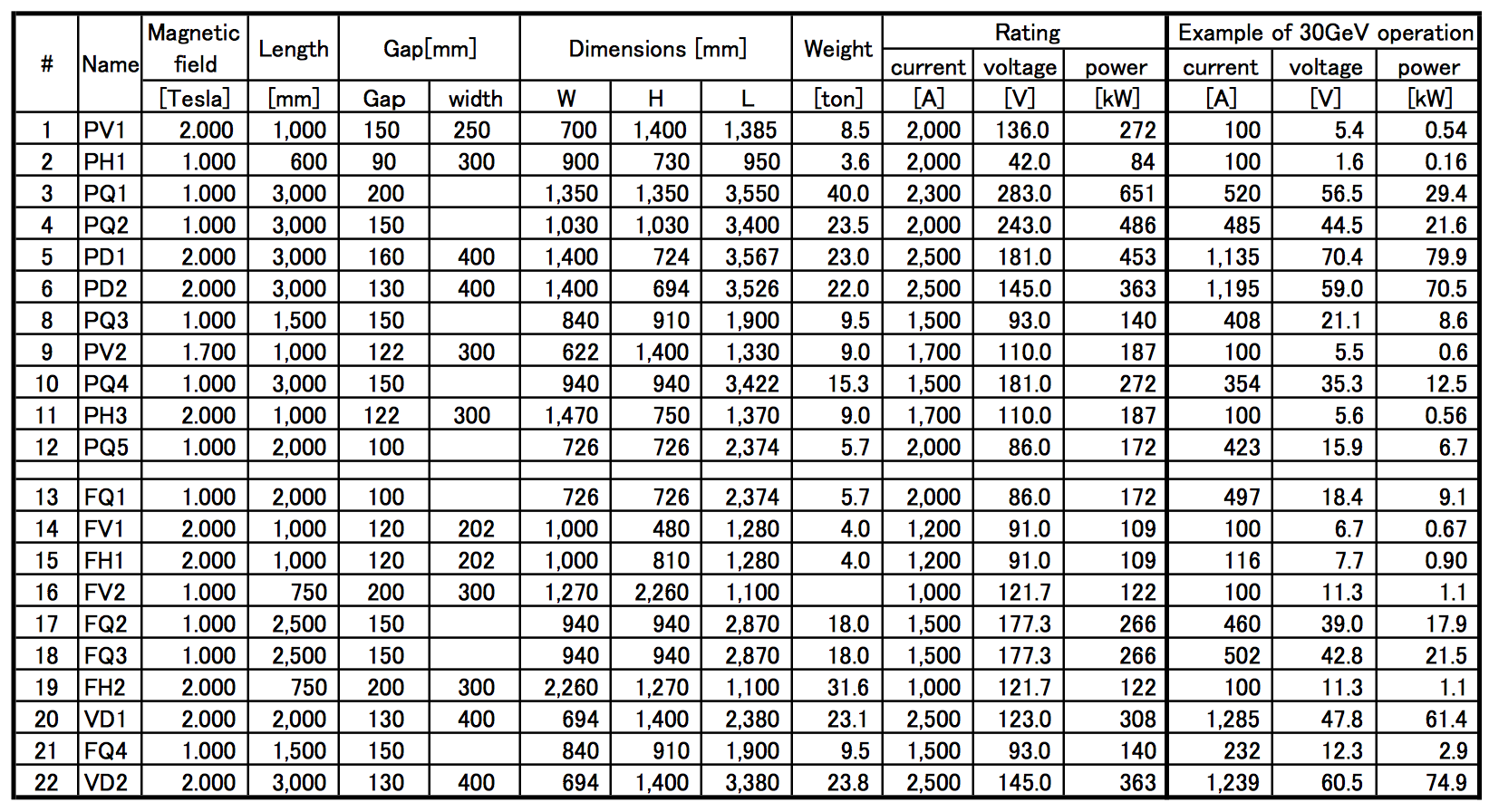}
\end{table}

The length, and hence the $BL$, of the preparation section magnets are designed for 50~GeV
proton beam while those of the final-focus section magnets are designed 
for 40~GeV beam in term of the capability of cooling water system and
electricity.
The original power supplies,
reused from the KEK 12~GeV PS, were capable of 50~GeV operation. 
This system was replaced in 2014, since we faced difficulty in
obtaining maintenance parts, by new power supplies designed for 30~GeV 
operation.
This change in the design proton energy originates from the change of
the MR energy goal.

At the design stage, large beam loss of about 1~MGy/year was forecasted
at the upstream part, close to the fast-extraction point.
Therefore upstream magnets are made with mineral-insulated coils or
polyimide-insulated coils. These have radiation resistance up to
10$^{11}$~Gy and $4\times10^8$~Gy, respectively~\cite{Hirose:2012nla}. 
On the other hand, downstream magnets are made of epoxy-insulated coils,
which has radiation resistance up to 10$^7$~Gy.
Beam loss at 480 kW operation at present is estimated to be order of
0.1~W/magnet,
which is two-orders of magnitude smaller than the assumption 
at the design stage.
Therefore even epoxy-insulated coils are expected to survive
the higher beam power operation.

\subsubsection{Super-conducting magnets}
The beam is bent by 80.7$^\circ$ using super-conducting 
combined function
magnets~\cite{Nakamoto:2010tvk,Sasaki:2010wpz,Okamura:2010zz,Makida:2010zz,Ogitsu:2010zz}
at the arc section.
These magnets are the world's first 
combined-function superconducting magnets,
consisting of 14 doublets (focus/defocus) and 3 pairs (normal/skew)
of steering magnets.
With an inner diameter of the coil of 173~mm,
they are capable of generating a 2.6~T dipole field with
a 19~T/m quadrupole field to bend 50~GeV protons towards 
the Kamioka direction with a radius of 104~m, 
together with 8000~A power supply.
The cooling power of the cryogenic system is 1.2~kW.

The heat load due to beam loss is assumed to be 1~W/m,
or 150~W in total for the 150~m long section,
which is small enough compared to the cooling power.
The beam loss of 1~W/m corresponds to 1~MGy irradiation on the
coil for 30-years operation.
Endurance of the material against this irradiation was established by
gamma-ray irradiation test. 

\subsubsection{Vacuum system}
In order not to have beam windows at the boundaries of the MR and
the super-conducting section, the beamline vacuum of the preparation
section and of the final-focus section is designed to achieve a level of
$10^{-6}$~Pa at the boundaries.
This is achieved by four ion pumps (IP) at each section.
At the beginning, the final-focus section was equipped with two
turbo-molecular pumps (TMPs) not to shorten lifetimes of the ion pumps
due to out-gassing from the most downstream monitor stack.
In 2013, the TMPs were replaced with IPs in order to avoid
to exhaust any fragments if the beam window at the end of the primary beamline
is broken. 

The beam ducts are designed to have an 81$\pi$ mm mrad aperture in both
the emittance of horizontal and vertical plane for dipoles and 
an 81$\pi$ mm mrad ellipse for quadrupoles except for a few magnets
(PV1 and FQ3) due to last-minute location swap and optics re-calculation,
respectively. The apertures of the beam ducts are summarized in
Tab.~\ref{tab:ducts}.

\begin{table}[h] 
\centering 
\caption{Summary of apertures of the beam ducts. The unit of inner aperture is millimeter (mm).}
\label{tab:ducts}
\includegraphics[width=0.4\textwidth] {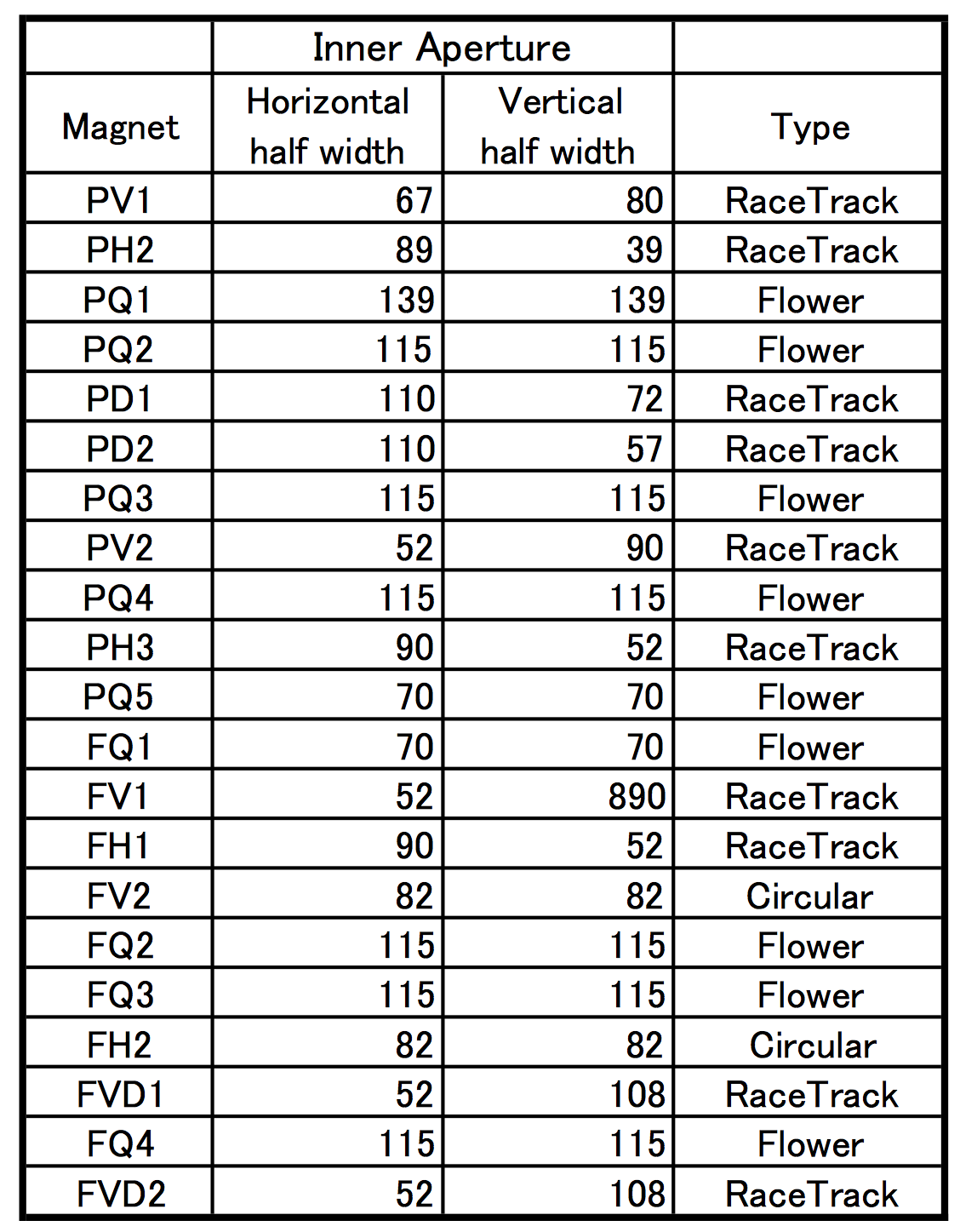}
\end{table} 

In order to withstand the thermal shock stress due to a direct
hit of a mis-steered beam, the beam duct material should be either
titanium or aluminum, 
\color{\MODCOLOR}
based on the simulation results summarized in Tab.~\ref{tab:ductsim}. 
\color{black}
There are still a few beam ducts made of
3 mm thick stainless steel in some steering magnets (PH3, FH1, FV1 and FV2)
due to financial reasons.  
\color{\MODCOLOR}
These will be replaced with the titanium beam duct 
before going beyond $2 \times 10^{14}$ ppp. 

\begin{table}[h] 
\caption{The simulation results of the thermal shock on the beam duct 
with direct one-shot hit of 50GeV 750kW beam with $3.3 \times 10^{14}$ 
ppp\cite{InterimJNU:2005}.}
\label{tab:ductsim}
\vspace{1ex}
\centering
\begin{tabular}{rrrrrr}
\hline\hline
Duct material & Thickness & Energy deposit & $\Delta T$ & Stress & Maximum Strength \\\hline
SUS &  5~mm & 150~kGy/pulse & 290~K & 1.0~GPa & $\sim 520$~MPa (SUS 304) \\
        &  3~mm & 120~kGy/pulse & 230~K & 840~MPa & \\\hline
Al-alloy &  5~mm & 60~kGy/pulse & 70~K & 100~MPa & $\sim 310$~MPa (Al-alloy 5056) \\
       & 3~mm & 60~kGy/pulse & 70~K & 100~MPa & \\\hline
pure-Ti   & 5~mm & 100~kGy/pulse & 190~K & 170~MPa & $\sim 550$~MPa (JIS-3) \\
       & 3~mm & 80~kGy/pulse & 150~K & 130~MPa & $\sim 390$~MPa (JIS-2) \\
\hline\hline
\end{tabular}
\end{table}
\color{black}

\subsubsection{Collimator}
The initial purpose of the collimators is to scrape off the beam halo. 
Four sections were reserved for the collimators at the preparation
section for later installation after obtaining the actual beam halo
characteristics.

Observations during beam operation indicate that beam loss is not large.
On the other hand, fast-extraction magnet failures occasionally happens.
This trip results in off-orbit beam coming into the primary beamline and
may hit the target off-center, or may hit the beam ducts,
or may hit the super-conducting magnets in the worst case.
Presently two collimators are installed, PC1 and PC4. 
The primary purpose of these collimators is not to scrape off the
beam halo but to block the off-center beam orbit in the case of
a magnet trip.
The aperture of the collimators, shown in
Tab.~\ref{tab:collimator}, are determined to block such off-center beam to 
protect important beamline components.
The effect was actually verified by scanning the beam position with very
low intensity beam.

\begin{table}[h] 
\caption{Summary of apertures of the collimators. The second column
is the distance from a gate-valve which placed the boundary between
MR and neutrino beamline.}
\label{tab:collimator}
\vspace{1ex}
\centering
\begin{tabular}{lccccc}
\hline\hline
 & & \multicolumn{2}{c}{Entrance half aperture} &
 \multicolumn{2}{c}{Exit half aperture} \\\cline{3-6}
\raisebox{1em}{Name} & \raisebox{1em}{Distance [mm]} & x [mm] & y [mm] & x [mm] & y [mm] \\\hline
PC1 & 28492 & 63 & 33 & 63 & 33 \\
PC2 & 31895 & \multicolumn{4}{c}{No installation plan at present}\\
PC3 & 36842 & \multicolumn{4}{c}{No installation plan at present}\\
PC4 & 46162 & 31 & 33 & 33 & 30 \\\hline
\hline\hline
\end{tabular}
\end{table}

Since the beam loss at the collimators is not large,
they are cooled by conduction to the shielding 
iron wall at present.

\subsubsection{Beam Plug}
A pair of beam plugs are installed at the upstream part of the
preparation section \color{\MODCOLOR} for the human safety purpose, namely \color{black}
to prevent the beam accidentally being injected to
the primary beamline when the downstream neutrino facility is being accessed. 

\begin{figure}[htb]
\begin{center}
\begin{minipage}{7.5cm}
 \includegraphics[width=7.5cm]{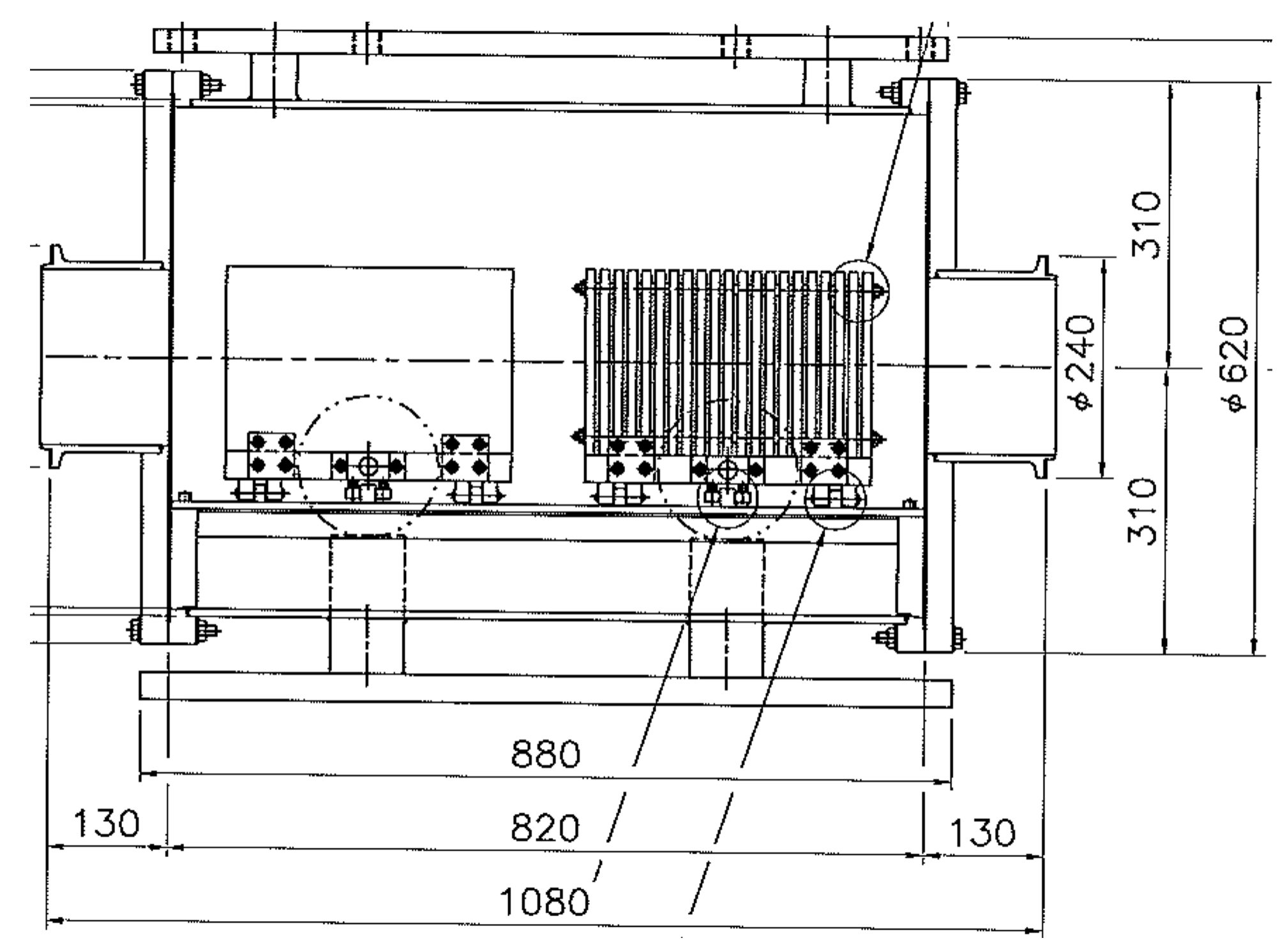}
 \caption{Design of the beam plug.}
  \label{fig:plug}%
\end{minipage}
\hfill
\begin{minipage}{7cm}
  \includegraphics*[width=7.5cm]{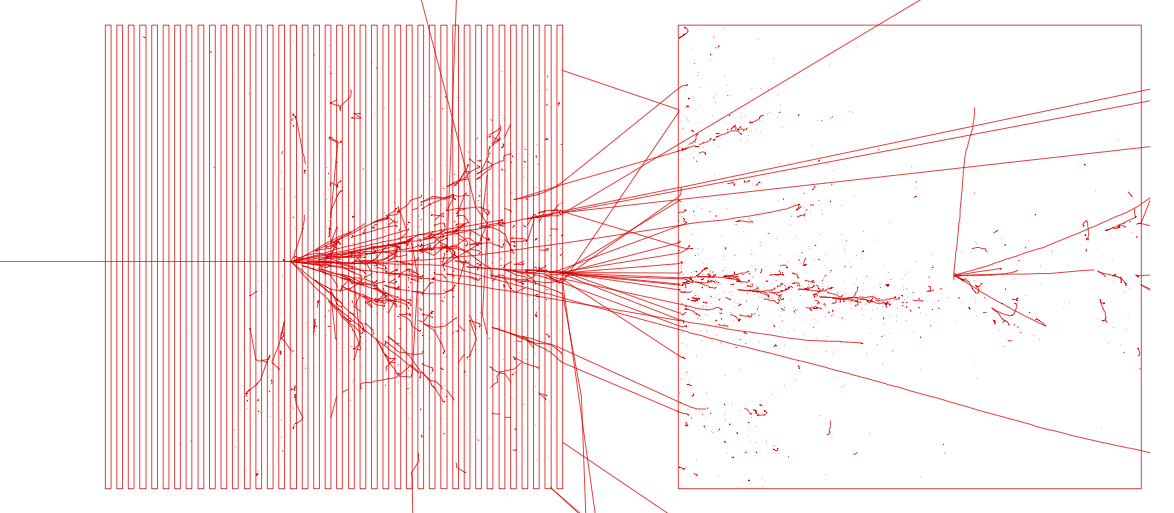}
   \caption{The shower development in the beam plug simulated by GEANT for 40~GeV proton hit.}
   \label{fig:plugsim}
\end{minipage}
\end{center}
\end{figure}

\begin{figure}[htb]
\begin{center}
\begin{minipage}{7.5cm}
 \includegraphics[width=7.5cm]{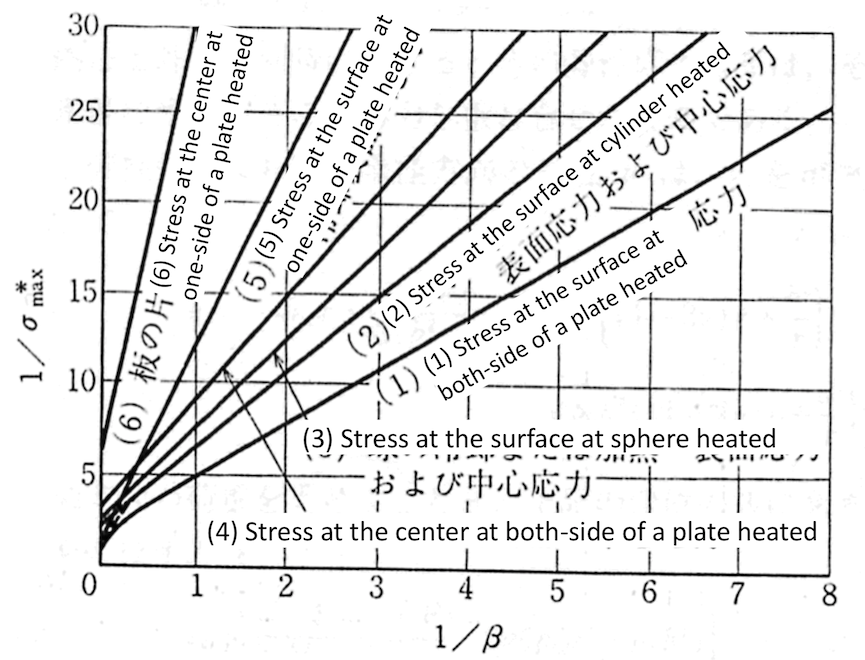}
\end{minipage}
\hfill
\begin{minipage}{7cm}
 \caption{Relationship between Biot's coefficient $\beta$ and maximum thermal shock stress $\sigma_{\textrm{max}}$.}
 \label{fig:geomf}%
\end{minipage}
\end{center}
\end{figure}

\color{\MODCOLOR}
The structure of the beam plug is shown in Fig.~\ref{fig:plug}. 
The upstream block is made of 20 layers of 10~mm-thick stainless steel plates with 5~mm gap, 
and the downstream block is made of 15 layers of 20~mm-thick stainless steel plates.
With this 50~cm-thick stainless steel, MARS simulation gives the radiation level less than 
0.1~$\mu$Sv at the end of the primary beamline 
when one shot of 50~GeV 750~kW beam hits the beam plug. 
This ensures the safety of the access to the target station when the MR is in operation, 
provided that both the blocks of the beam plug are inserted into the beam line and 
the safety magnets (two bending magnets PD1 and PD2) are kept off.
The blocks are made of thin plates to relax the thermal shock stress when the beam hits the beam plug. 
The upstream block has gaps between the plates to further relax the stress 
by defusing the beam distribution. 
The thermal shock stress was simulated using GEANT. 
Shower maximum, and thus the highest energy density, locates in the first block as shown 
in Fig.~\ref{fig:plugsim}. 
Using the expected beam spot of 2~cm$^2$ at the beam plug, 
if the blocks are hit by 30~GeV proton beam with $2\times10^{14}$ ppp, 
the expected protons per pulse at 750 kW operation, 
the maximum thermal shock stress is calculated to be 440~MPa, 
slightly below the tensile strength of stainless steel, 520~MPa. 
Here we used thin-plate approximation formula with adiabatic condition and 
sharp-edged circular beam spot;
\begin{equation}
\sigma = E\cdot\alpha\cdot\Delta T\cdot\frac{f(\beta)}{(1-\nu)}
\end{equation}
where $\sigma$ is the maximum thermal shock stress at the beam spot edge, 
$E$ is the Young's modulus, $\alpha$ is the thermal expansion coefficient, $\Delta T$ 
is the temperature rise, $\nu$ is Poisson's ratio, and $f(\beta)$ is a geometry factor to be 0.4 at 
the Biot's coefficient $\beta = \infty$, 
\color{\MODCOLORB}
with considering a stress at the surface at cylinder heated from the beam hit 
and then $1/\sigma^*_{\textrm{max}} = 2.5$ from Fig.~\ref{fig:geomf} which corresponding to $1/f(\beta)$. 
\color{black}
\color{black}

\graphicspath{{figures/main_primary}}

\subsection{Upgrade plan of the primary beamline} \label{sec:primaryup}

The primary beamline is basically
capable of accepting 1.3~MW beam power provided that
beam loss is kept low as the present level. 

\color{\MODCOLOR}
\subsubsection{Beam optics and larger aperture magnets}
\begin{figure}[h] 
\centering 
\begin{minipage}{7.5cm}
\includegraphics[width=7.5cm] {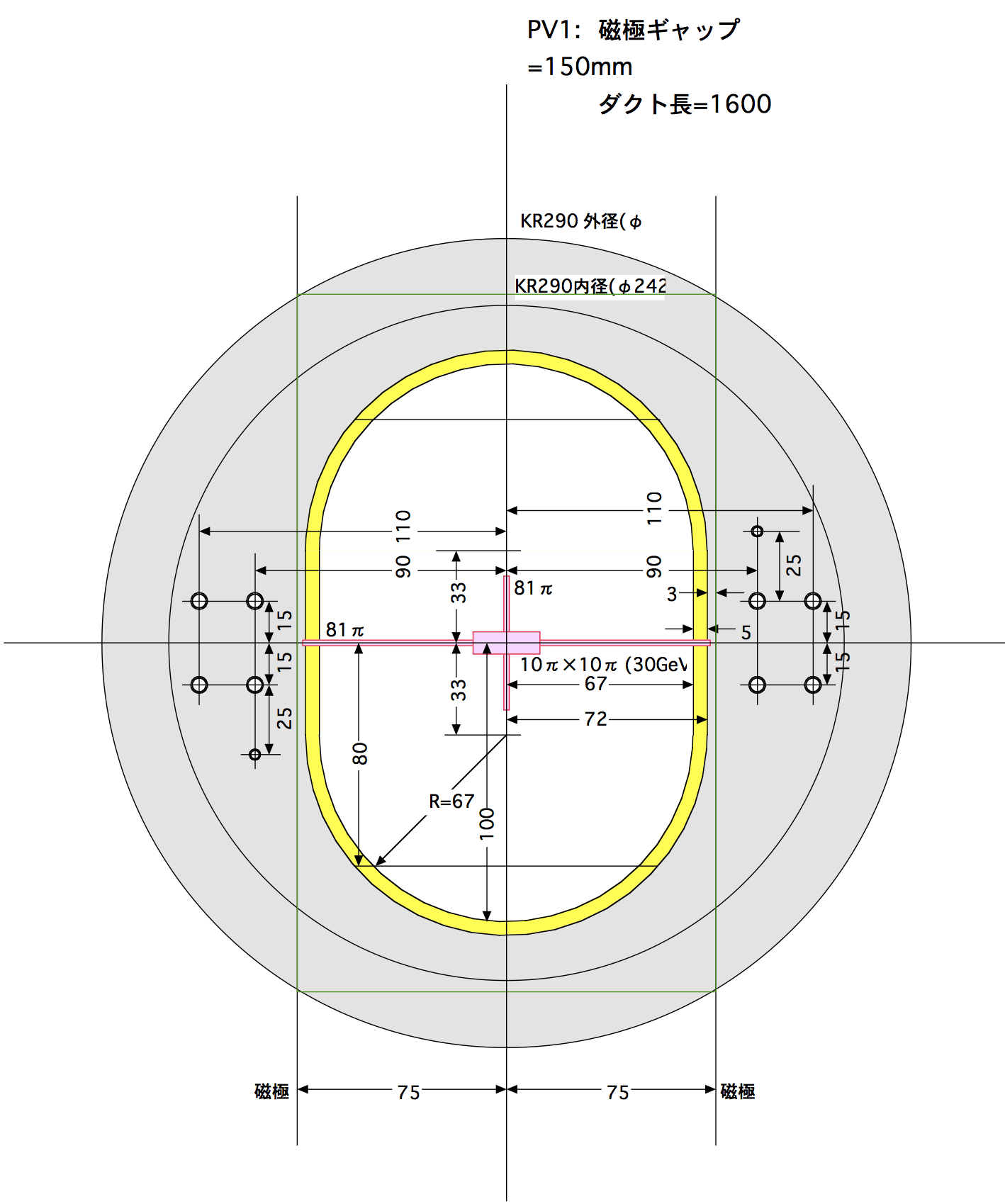}
\caption{Beam size and aperture of PV1.}
\label{fig:pv1}
\end{minipage}
\hfill
\begin{minipage}{7.5cm}
\includegraphics[width=7.5cm] {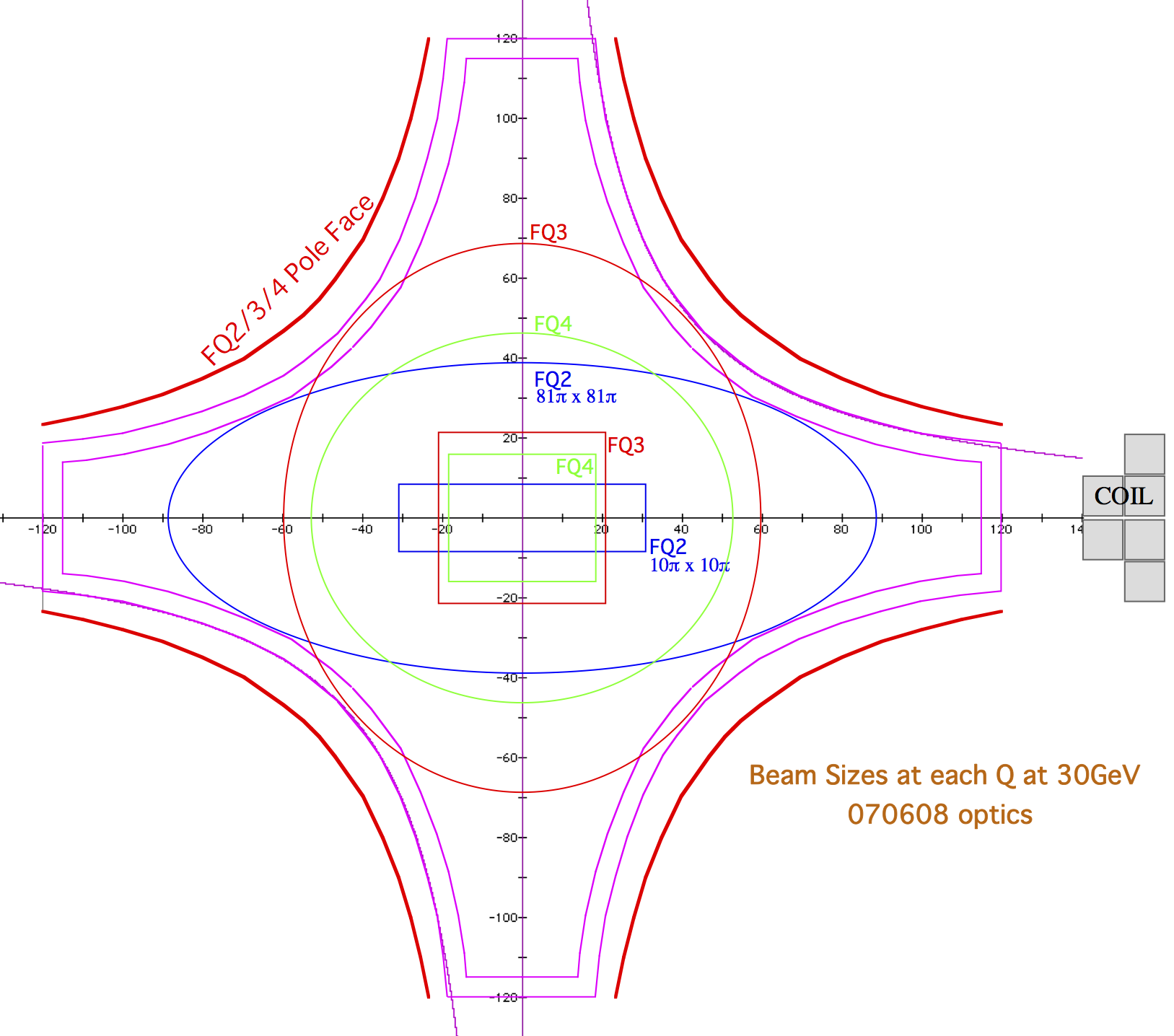}
\caption{81$\pi$ mm mrad beam size at FQ2, 3 and 4.}
\label{fig:fq2}
\end{minipage}
\end{figure}

\begin{figure}[h] 
\centering 
\includegraphics[width=0.8\textwidth] {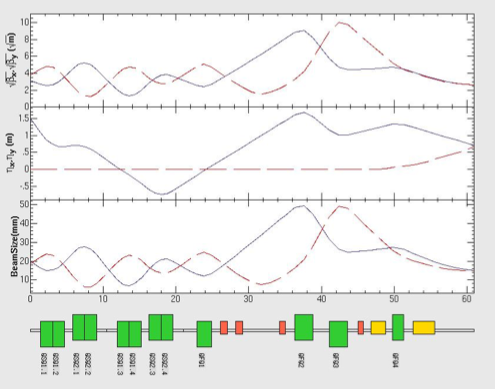}
\caption{Beam evenlope at FF section for 24$\pi$ mm mrad.}
\label{fig:ff24pi}
\end{figure}

The beam loss of the most upstream part of the preparation
section, and hence the residual radio-activation, is large even 
at 485~kW operation. One of these causes could be 
that PV1 magnet does not have 81$\pi$ mm mrad in the horizontal
direction as shown in Fig.~\ref{fig:pv1} and the beam halo from the MR causes the beam loss.
The final-focus quadrupole doublets, FQ2 and FQ3, may also need to
enlarge their aperture (present size is 150~mm) for the large emittance beam at the higher ppp. 
Figure~\ref{fig:fq2} shows the 81$\pi$ mm mrad beam size at 
FQ2, 3, 4 for the 10$\pi$ mm mrad emittance beam, 
and Fig.~\ref{fig:ff24pi} shows the beam envelope for the 24$\pi$ mm mrad emittance beam.
The aperture of the FQ3 is very tight even at 10$\pi$ mm mrad beam,
and the larger emittance beam may result in intolerable beam loss.
The quadruple magnets with the aperture of 200~mm such as PQ1 are one candidate. 

The beam halo could depend on the MR beam power as well as 
the MR larger aperture magnet installation plan. 
The MR plans to increase the beam power by installing the MR main power supply 
by 2021, and upgrading the MR RF by 2026. 
The beam halo should be evaluated with higher protons per bunch (ppb) even before the 
main power supply and the RF installation. 
The beam study before these installation is planned 
to evaluate the beam loss at the neutrino 
primary beamline caused by the beam halo. The ppb is gradually increased up to 
$4 \times 10^{13}$ ppb, which is the ppb at 1.3~MW, and the number of bunch is kept 
to be 2~bunch because the 8~bunch operation requires the RF upgrade. 

The design of the larger aperture magnets, PV1, FQ2 and FQ3, will be  
ready by 2020 based on the MR larger aperture magnet plan and the results of 
the beam study of the higher ppb beam. 
The fabrication of these magnets is planed to start on 2021 and therefore 
the summer of 2022 is the earliest installation. 
This strategy can accommodate the 1~MW beam, which is expected before 2022, 
because the present aperture magnet was able to accept the 510~kW (corresponding 
to $3.3 \times 10^{13}$ ppb) without any significant beam loss at the neutrino 
primary beamline.

The aperture of the super-conducting magnets are large enough to accept the beam
without significant beam loss even at 1.3MW. 

\color{black}

\subsubsection{Collimator}
The collimators have a much smaller aperture compared to
the magnets in order to block the off-orbit beam. 
Therefore they are very sensitive to the beam tail. 
At 485~kW operation, the beam halo is not significant. 
The collimators are not scraping off the halo and the heat load is
negligible. The 750~kW operation will be achieved by simply
increasing the repetition rate, and no significant increase
of the heat load is expected.
The 1.3~MW operation, however, needs an increase of the protons per pulse,
and the beam halo condition may drastically change.
\color{\MODCOLOR}
Our operation principle is, however, even at beyond 750kW, 
to keep the beam loss at the present level so as not to increase the residual radiation 
of the primary beam-line equipment for sustainable maintenance scenario. 
\color{black}

\subsubsection{Beam plug}
The present beam plug can withstand the beam injection up to 750~kW. 
For the larger ppp beyond 750~kW, we need to replace the first block with an Invar plate. 
An aluminum or a titanium plate also gives much smaller stress, 
but densities are too low to accommodate in the present vessel.

\color{\MODCOLORB}
Thermal stress for Invar was also simulated by GEANT. 
For bulk Invar, the maximum sigma is 308~MPa.
For 20 layers of 1cm thick Invar plates with 1cm gap, maximum sigma is 121~MPa
The yield strength of Invar is 400~MPa.
Stress at the actual configuration be between above two,
probably closer to 121~MPa.

\color{black}

\graphicspath{{figures/main_primary/beamMonitors}}

\subsection{Proton beam monitors} 
\label{sec:protonmonitor}

\begin{figure}[h] 
\centering 
\includegraphics[width=9cm] {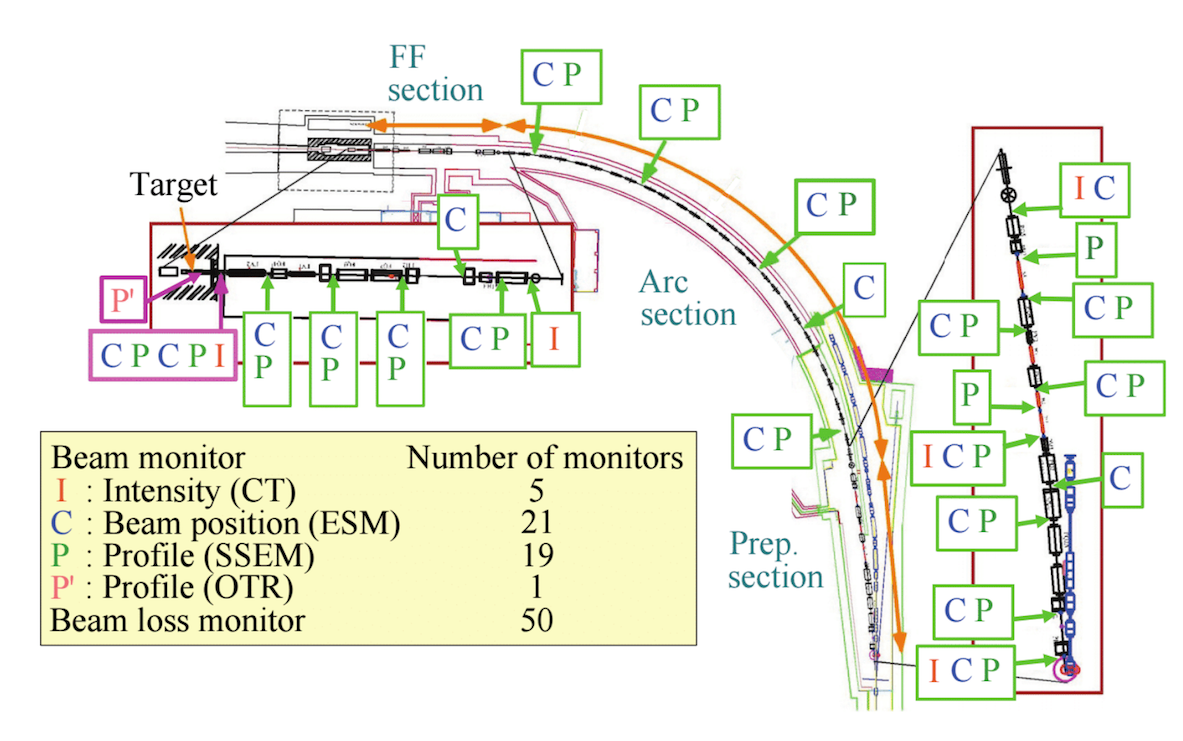}
\caption{Location of the beam monitors in the J-PARC neutrino beamline.}
\label{fig:beammonpos} 
\end{figure}

The proton beam conditions are continuously monitored by a suite of proton beam
monitors along the neutrino primary proton beamline, as shown in Fig.\ \ref{fig:beammonpos}
and described in Ref.\ \cite{Abe:2011ks}.  

Five Current Transformers (CTs) are used to continuously monitor the proton
beam intensity.  Fifty Beam Loss Monitors (BLMs) continuously measure the 
spill-by-spill beam loss and are used to fire an abort interlock signal in the case of a high loss
beam spill.  Twenty-one Electro-Static Monitors (ESMs) are used as Beam
Position Monitors to continuously monitor the beam position and angle.

So far, these monitors have been running well with the design precision and stability. 
These monitors were all designed to work continuously at high intensity 
(\(3.3\times10^{14}\) protons per pulse), 
and we intend to continue to use them stably with minimal hardware upgrades for the
foreseeable future.  Regular calibration, improvements in calibration methods,
and analysis improvements may be necessary for maintaining or improving the monitor 
stability or precision, and these will be carried out as needed.

The proton beam profile (beam position and width)
 is monitored bunch-by-bunch during beam tuning by a suite of 19 Segmented Secondary Emission Monitors
 (SSEMs) \cite{Abe:2011ks} distributed along the primary beamline, where only
 the most downstream SSEM (SSEM19) is used continuously. An Optical
Transition Radiation Monitor (OTR) \cite{Hartz:2010zz}, placed directly 
upstream of the production target, also continuously monitors the beam profile spill-by-spill.  
Other beam parameters, such as the beam angle, twiss parameters, emittance, etc, 
at the position of the baffle and target, 
are extrapolated by a fit to several downstream SSEMs and the OTR during beam
tuning.  The beam position and angle at the target is continuously monitored 
spill-by-spill by a fit to several ESMs, SSEM19 and the OTR during standard running, 
while the other proton beam parameters are extrapolated by scaling the data from SSEM-IN runs with
spill-by-spill SSEM19 and OTR information when the other SSEMs are OUT.

Potential issues with the SSEMs are described in Sec.\ \ref{sec:ssem} and a plan
to upgrade SSEM18 is described in Sec.\ \ref{sec:WSEM}.  Recent OTR issues and plans for
OTR upgrades are discussed in Sec.\ \ref{sec:OTR}.  An ambitious profile monitor
upgrade project -- development of a continuous, non-destructive Beam Induced
Fluorescence (BIF) beam profile monitor -- is also described in Appendix \ref{sec:BIF}.

\subsubsection{Segmented Secondary Emission Monitor (SSEM)} \label{sec:ssem}

\begin{figure}[h] 
\centering 
\includegraphics[width=7.6cm] {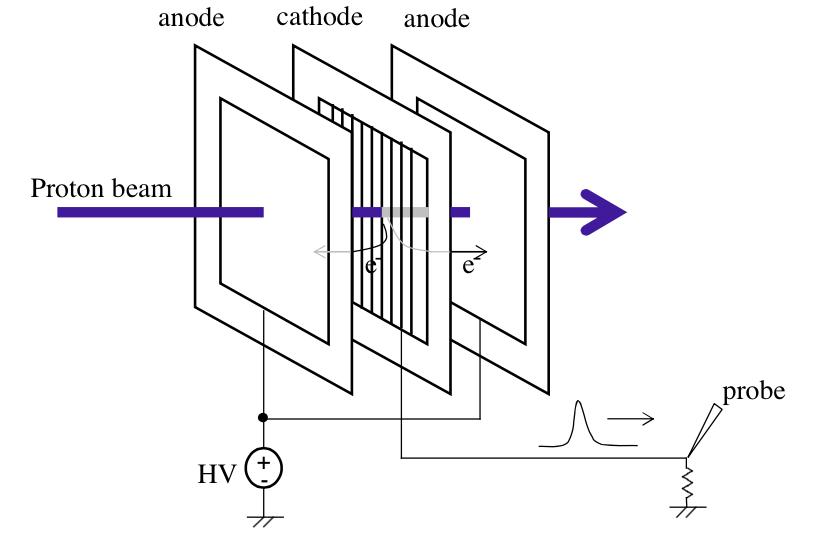}
~~~~
\includegraphics[width=5.0cm] {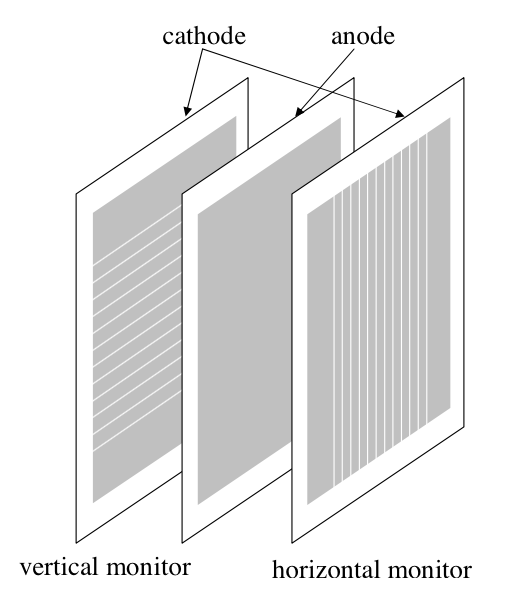}
\caption{Principle behind SSEM sensor design (left) and diagram of the SSEM sensor head
used in the J-PARC neutrino beamline (right).}
\label{fig:ssemdiagram} 
\end{figure}

Each SSEM sensor head consists of two thin (5~\(\mu\)m, \(10^{-5}\) interaction lengths) titanium foils stripped 
horizontally and vertically (to measure the vertical and horizontal beam profiles respectively), 
and an anode HV foil between them, as shown in Fig.~\ref{fig:ssemdiagram}.  The strips are hit by 
the proton beam and emit secondary electrons in proportion to the number of protons that 
go through the strip, and compensating charge in each strip is read out as a pulse with positive polarity. 
The proton beam profile is reconstructed from the resulting charge distribution from
all strips on a bunch-by-bunch basis.  The strip width
of each SSEM ranges from 2 to 5 mm, optimized according to the expected beam
size at the installed position, and the gap between strips is 1 mm. 

The precision of the SSEM position and width measurements are 0.07~mm and 0.2~mm respectively. 
The SSEM profile monitors are designed to work at high beam intensity (\(\sim\)\(3\times10^{14}\)~ppp), 
and so far there have been no major issues with the stability and precision of these
monitors.

Each SSEM causes 0.005\% proton beam loss while in the beam.  Therefore, the monitors can be remotely moved into
and out of the beamline to eliminate additional loss during standard running.  Eighteen of
the SSEMs are only used to check the beam profile during beam tuning or after
some expected parameter change, while the most downstream SSEM (SSEM19) and
the OTR, which are located inside the monitor stack and Target Station respectively, 
and therefore in a high-radiation environment already, are used continuously.  The beam width
continuously measured by SSEM19 is used to protect the beam window and target : if the
beam density (number of protons/beam spot size) at the target becomes 
$N_p/(\sigma_x\times\sigma_y) < 2\times10^{13}$ ppp/mm$^2$, a beam abort interlock is fired in
order to avoid potential damage.   The beam position, angle, width, etc.\ measured spill-by-spill
are also used as an input into the T2K neutrino flux prediction.

\begin{figure}[h] 
\centering 
\includegraphics[width=10cm] {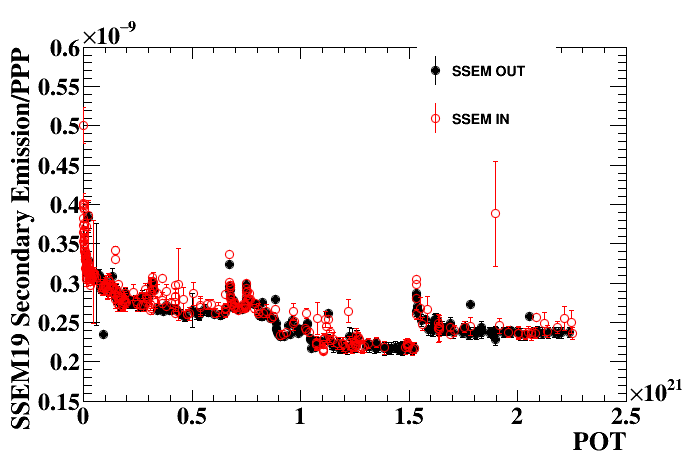}
\caption{Observed secondary emission of SSEM19 over the full T2K run so far.  Jumps in
normalized secondary emission appear to be correlated with changes in beam
  power.  Points are shown when the other SSEMs are OUT (black) and IN (red).} 
\label{fig:SSEM19deg} 
\end{figure} 

\begin{figure}[h] 
\centering 
\includegraphics[width=6cm]{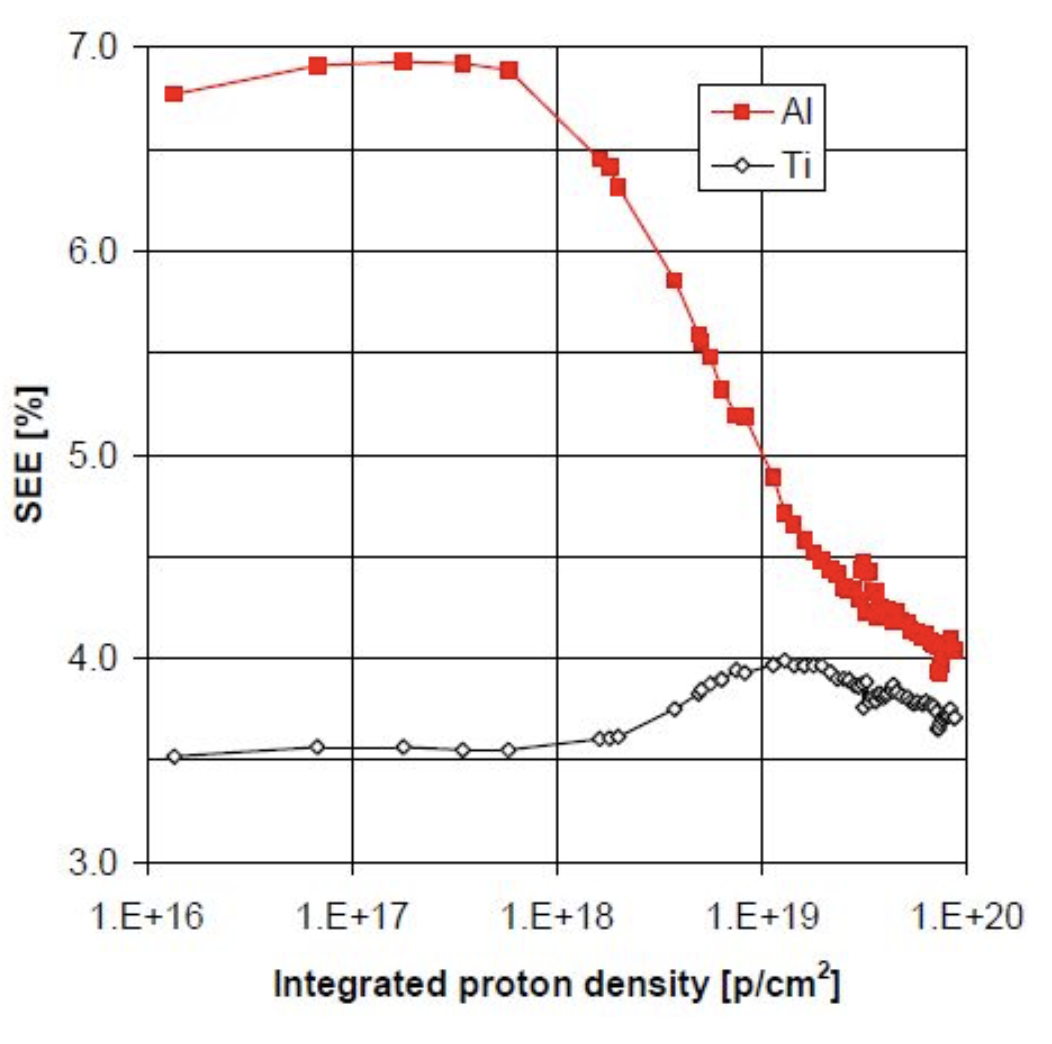}
\caption{Change in the secondary emission efficiency of aluminum and titanium
foils in the CERN SPS beam line \cite{Ferioli:1997ssem}.} 
\label{fig:cernssem} 
\end{figure} 

Since bunch-by-bunch (and spill-by-spill) information from SSEM19 is necessary for T2K,
potential degradation of the secondary emission
signal should be carefully monitored.  The recent light yield vs.\ POT for SSEM19
is shown in Fig.\ \ref{fig:SSEM19deg}.  A potential \(\sim\)20\% decrease in the secondary
emission of SSEM19 has been observed after an integrated \(2.3\times10^{21}\) POT (with an average 
beam spot size of 4~x~4~mm), although the secondary emission stability at stable beam power 
(after an initial burn-in period) appears to be very good.
The expected SSEM lifetime is not precisely known, however studies by both
J-PARC and CERN have indicated that the secondary emission efficiency of titanium is stable up to
\(10^{18}\)protons/cm\(^2\), as shown in Fig.\ \ref{fig:cernssem}.  
Although this integrated POT has already been exceeded at T2K, Fig.\ \ref{fig:SSEM19deg} 
shows that the SSEM19 secondary emission is basically stable after an integrated
\(1\times10^{21}\)~POT.  However,
SSEM19 should be periodically replaced if degradation begins to occur, and the
first SSEM19 replacement is currently planned for summer 2019 or 2020.

Since the SSEMs other than SSEM19 (SSEM1-18) have a relatively low total integrated
incident number of protons, no issue with degradation is expected for SSEM1-18.  
SSEM1-18 are also installed in the primary beamline, which is much easier
to access then the monitor stack (where SSEM19 is installed).  Therefore, the 
SSEM1-18 sensor heads can be relatively easily replaced by spares if any issues do
occur.

\subsubsection{Wire Secondary Emission Monitor (WSEM)} \label{sec:WSEM}

Beam loss due to interactions in material placed into the beam is proportional
to the volume of material in the beam.  On the other hand, secondary emission is
proportional to the surface area in the beam.  Therefore, the original SSEM beam
profile monitors used 5-\(\mu\)m-thick stripped foils to maximize interaction
rate and minimize beam loss.  Each monitor head consists of 3 foils.

\begin{figure}[h] 
\centering 
\includegraphics[width=6cm] {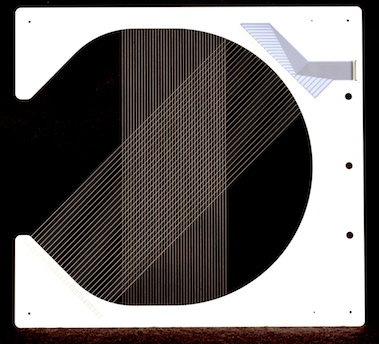}
  \caption{WSEM frame with mounted electron collection anode plane (vertical
  wires) and one set of cathode wires (diagonal wires).}
\label{fig:WSEM} 
\end{figure}

Beam loss from these profile monitors can be reduced by switching from foils to wires
intercepting the beam.  A new Wire Secondary Emission Monitor (WSEM), one plane
of which is shown in Fig.\ \ref{fig:WSEM}, has been
jointly developed with the monitor group FNAL as part of the US/Japan collaboration,
and prototype planes were built for the J-PARC primary neutrino beamline.  This monitor consists of
2 planes with 25\(\mu\)m diameter twinned pure Ti (Grade 1) wires with 3~mm pitch.
An anode plane between them, consisting of of 25\(\mu\)m single Ti wires with a 2- or 6-mm pitch, 
can be set to 100~V to sweep away electrons.  All wires were mounted to the ceramic frame 
under a tension of 20 g/wire.  The expected and measured signal size and beam loss
for the WSEM compared to the standard SSEM is given in Table~\ref{tab:wsempar}.

\begin{table}[h]
  \centering
  \caption{SSEM vs WSEM parameters calculated assuming 3mm diameter beam spot size and 
  measured during a beam test.  Note that test conditions differ slightly
  between the monitors, as discussed in the text.}
  \label{tab:wsempar}
  \begin{tabular}{l l llll } 
    \hline\hline
    Monitor & Strip Size & Area in & Measured & Volume in & Measured \\
    & & Beam (mm\(^2\)) & Signal (a.u.) & Beam (mm\(^3\)) & Loss (a.u.) \\
    \hline \hline
    SSEM & 2\(\sim\)5mm\(\times\)5\(\mu\)m & 7.07 & 60300 & 0.106 &  872 \\
    WSEM & 25\(\mu\)m\diameter x2 & 0.24 & 2300 & 0.007 & 112 \\ \hline
    Ratio & & & & & \\
    SSEM/WSEM & -- & 29.5 & 26 & 15.1 & 7.8 \\
    \hline\hline
  \end{tabular}
\end{table}

A prototype WSEM monitor has already been fabricated and installed for testing
in the neutrino primary beamline.  

The WSEM wires are mounted at 45\(\degree\)~with respect to the square ceramic frame in
order to allow for a c-shape cutout design, while the anode plane wires are
  mounted aligned to the square frame.  This c-shape allows for motion of
the monitor into and out of the beam during beam running (allowing for automated, periodic 
checks of the beam profile when used in non-continuous mode).  This means that
the monitor must be mounted at 45\(\degree\)~on the beampipe, requiring a mover also
mounted at 45\(\degree\).  For testing, the WSEM was mounted on a spare standard
J-PARC SSEM mover resting on a modified 45\(\degree\)~stage, however an improved
mover design, which moves the WSEM readout cables at the same time as the sensor
head and should allow for more reliable use of the WSEM, was also prepared in JFY2017.  

\begin{figure}[h] 
\centering 
\includegraphics[width=6cm] {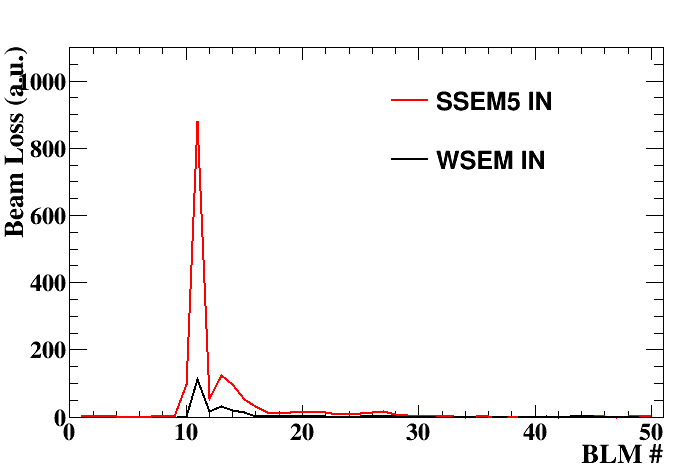}
\caption{Measured beam loss due to WSEM vs neighboring SSEM.}
\label{fig:WSEMloss} 
\end{figure}

\begin{figure}[h] 
\centering 
\includegraphics[width=6cm] {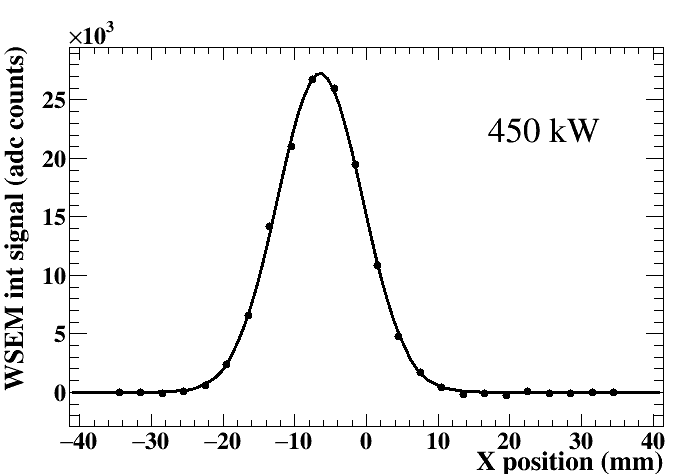}
\includegraphics[width=6cm] {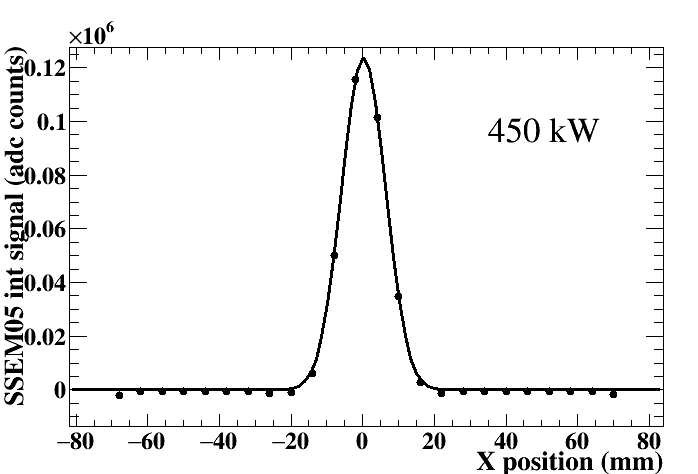}
\caption{Horizontal beam profile measured by the WSEM (left) and neighboring SSEM05 (right). Note that the 
horizontal axis ranges are different and the WSEM absolute alignment calibration
is not applied.}
\label{fig:WSEMprof} 
\end{figure}

\begin{figure}[h] 
\centering 
\includegraphics[width=6cm] {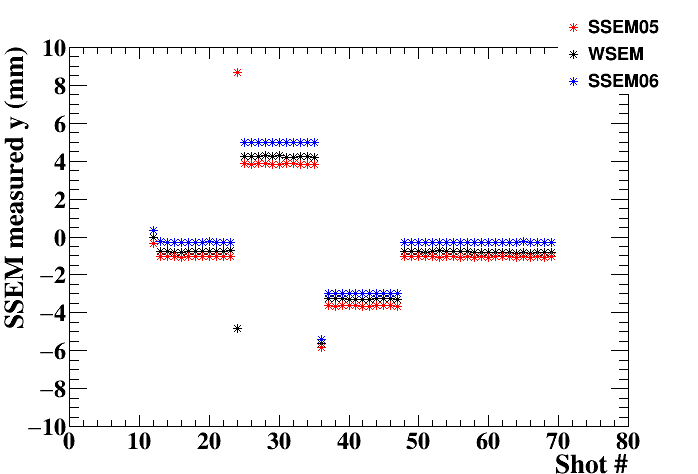}
\includegraphics[width=6cm] {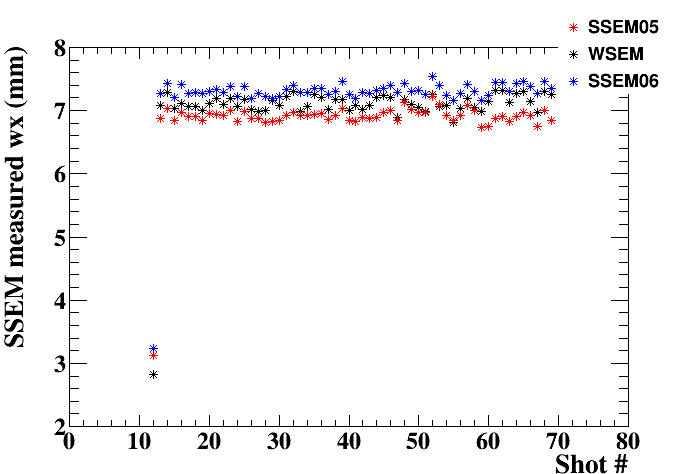}
\caption{Beam position (left) and width (right) measured by WSEM and neighboring SSEMs at low
beam intensity.}
\label{fig:WSEMvsspill} 
\end{figure}

In beam tests, the beam loss was measured to be \(\sim\)8x lower for the
WSEM than for the neighboring SSEM05, as shown in Fig.~\ref{fig:WSEMloss}.  However, the Beam Loss 
Monitor acceptance is different for the two monitors and is actually slightly
higher for the WSEM, such that the actual reduction in total loss is greater
  than a factor of 8.
The beam position and width measured by the WSEM
matches those measured by the neighboring monitors (SSEM05 and SSEM06), as shown
in Figs.~\ref{fig:WSEMprof} and \ref{fig:WSEMvsspill}, although a full analysis
procedure for the WSEM data is still under development.  

Beyond signal checks, extensive robustness
testing of the WSEM has also already been carried out -- the WSEM was left in the high intensity
(460\(\sim\)475~kW) proton beam for \(\sim\)160 hours without any observed issues.  
Heating of the wires in the J-PARC proton beam is expected to be very modest compared
to the temperature rise required for a substantial change in Ti yield strength, 
as confirmed by this beam test. 

Currently only SSEM19 can be used continuously, since it is in the Target Station monitor stack 
where radiation shielding and remote handling procedures are already in place. SSEM19 is essential 
for monitoring the beam profile spill-by-spill.
Since the beam loss due to the WSEM is \(\sim\)0.0005\%, this monitor
may be suitable to leave in the beam at all times even outside the Target
Station.  The residual radiation level due to using the WSEM continuously was 300\(\mu\)Sv/hr on 
contact directly downstream of the WSEM after 150 h in the beam 
(measured 6 hours after beam stop).  The residual radiation level on contact downstream of SSEM18 is
1.2mSv/hr after continuous running at 475~kW due to backscatter from the Target Station.

The present plan is to replace SSEM18 with a WSEM during summer 2018 
and use the WSEM as SSEM18 is currently used (ie periodically insert it during beam tuning).  
In the case that SSEM19 becomes unusable, the WSEM can be used
continuously.  If the performance of the WSEM proves to be better than an SSEM, we
may opt to replace other SSEMs with WSEMs in the future.
\textcolor{\MODCOLOR}{Further long-term testing of the WSEM performance will be
possible with the WSEM installed at the SSEM18 position.  As described in Sec.\ \ref{sec:primarymainte}, a 
new semi-remote maintenance scheme for the most downstream part of the FF
beamline will also be prepared within 2018.  With this new maintenance scheme, if there is any
trouble with the WSEM, replacement of the WSEM monitor head will be much easier at the 
SSEM18 position compared to at the SSEM19 position inside the monitor stack.}

\paragraph{WSEM Schedule} \label{sec:WSEMschedule}
The schedule for the WSEM testing and installation is as follows :
\begin{itemize}
  \item JFY2015 : WSEM prototype was designed and fabricated at FNAL
  \item JFY2016 : WSEM was mounted on J-PARC NU beamline and preliminary beam
    tests were performed
  \item JFY2017 : Further beam tests were carried out.  An improved mover for
    the WSEM was designed jointly with FNAL experts
  \item JFY2018 : New mover will be built.  WSEM will be installed in the SSEM18
    position during the summer shutdown, if schedule permits
  \item JFY2019 summer : WSEM will be installed in the SSEM18 position, if not
    done during 2018
\end{itemize}

\subsubsection{Optical Transition Radiation Monitor (OTR)} \label{sec:OTR}

The OTR uses optical transition radiation, 
light emitted from a thin metallic foil when a charged beam passes
through it, to form a 2D image of the proton beam directly upstream of the
neutrino production target.  

The OTR active area is a 50-\(\mu\)m-thick titanium-alloy foil, which is placed
at 45\degree~to the incident
proton beam. As the beam enters and exits the foil,
visible light (transition radiation) is produced in a narrow cone around the
beam. The light produced at the entrance transition is reflected at 90\degree~to
the beam and directed out of the Target Station (TS) He vessel by four aluminum 90\degree~off-axis
parabolic mirrors to an area with lower radiation levels. It is then collected by a charge
injection device (CID) camera to produce an image of the proton beam profile
spill-by-spill.

The precision of the OTR position and width measurements are better than 0.8~mm in X and 0.5~mm in Y.
According to mechanical strength simulations, the OTR is designed to work at high beam intensity 
and should be able to withstand \(3.3\times10^{14}\) ppp. \textcolor{\MODCOLOR}{So far there has been only one issue
with the OTR beam profile measurement.
The OTR position measurement in X drifted by 2~mm between 12 March and 1 April 2015, although
measurement of upstream proton beam monitors didn't drift during the same time period.
The drift is compatible with a movement of the CID camera or the OTR foil moving in
the direction of the proton beam.
The cause of the drift is still unknown.}

\begin{figure}[h]
  \centering
  \includegraphics[width=10cm] {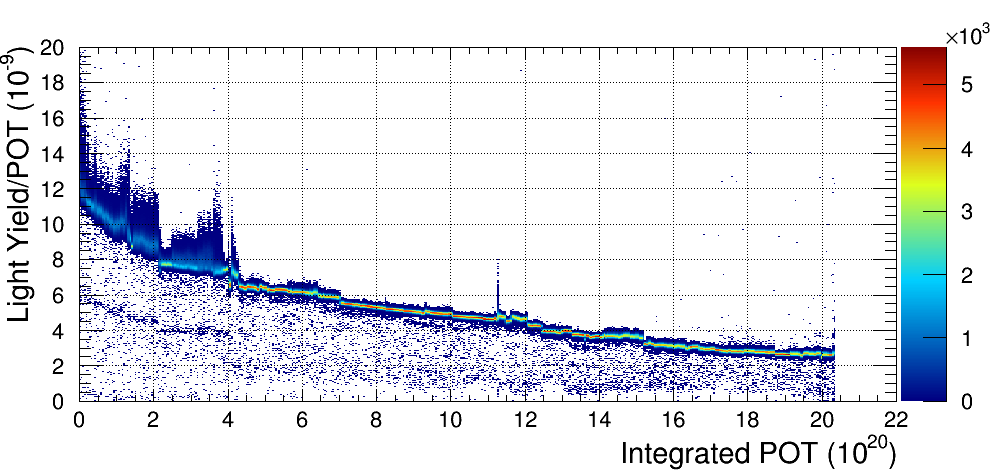}
  \caption{\textcolor{\MODCOLOR}{Observed OTR light yield degradation.}}
  \label{fig:OTRdeg}
\end{figure}

The recent light yield vs.\ POT for the OTR is shown in Fig.\ \ref{fig:OTRdeg} --
a decrease in the OTR light yield of \textcolor{\MODCOLOR}{\(\sim\)75\%} after an incident \textcolor{\MODCOLOR}{\(2.0\times10^{21}\)} protons has
been observed.
Potential degradation of OTR monitor foils can be mitigated by having multiple
available OTR foils, as described below.

\begin{figure}[h] 
\centering 
\includegraphics[width=5cm] {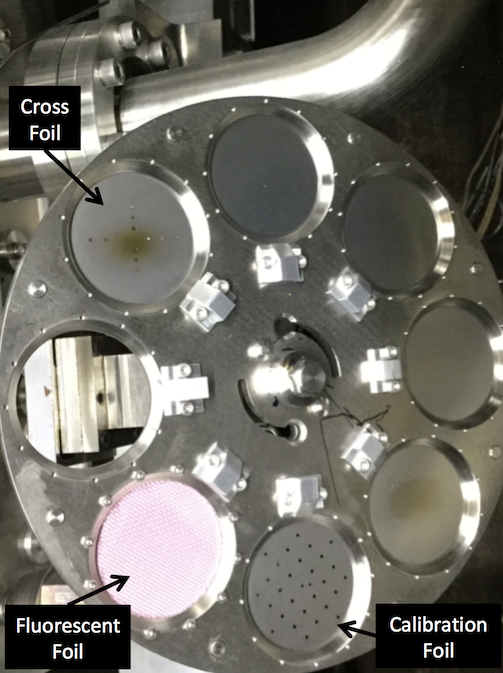}
~~~~
\includegraphics[width=5cm] {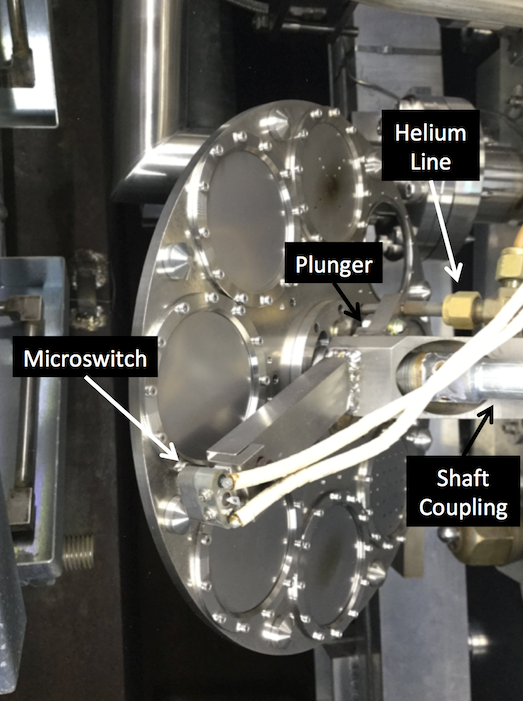}
\caption{Front (left) and back (right) of OTR disk.}
\label{fig:OTRdisk} 
\end{figure}

The OTR monitor consists of an arm holding a disk which is an 8-position carousel, shown in 
Fig.\ \ref{fig:OTRdisk}.  The OTR disk currently holds :
\begin{itemize}
\item Four 50-\(\mu\)m-thick Ti 15-3-3-3 (15\% V, 3\% Cr, 3\% Sn, 3\% Al) foils
\item One 50-\(\mu\)m-thick Ti 15-3-3-3 foil with 30 holes in a grid pattern for
calibration
\item One 50-\(\mu\)m-thick Ti 15-3-3-3 ``cross foil'' with twelve holes in a cross pattern
\item One ceramic (fluorescent) foil for low-intensity running
\item One empty position
\end{itemize}

The OTR disk is rotated by a long rigid steel shaft which couples to a flexible steel shaft that follows a 90
degree bend to the arm supporting the foil disk. A spline coupling is made at
the end of the arm to another flexible shaft which runs along the arm and
connects to the disk. The motor is connected to the shaft through a 100:1
gearbox, so that the disk rotates slowly.  The foil position
can be determined in principle by counting the number of motor steps using its
encoder, however, due to backlash in the flexible shaft system, alternate
methods of ensuring precise foil positioning are necessary. The primary method
to change foil positions is to run the motor until it is turned off by a
micro-switch (model 6302-16 from Haydon, designed for extreme conditions) 
engaged by a machined titanium button on the disk.  The switch position has been
adjusted so that it engages the button just before a foil is in the correct position. A
steel ball-bearing plunger mechanism, spring-loaded against the surface of the
disk, then falls into a matching machined depression in the disk \textcolor{\MODCOLOR}{flange}, which locks
it firmly into the correct position for each foil.  Most of these features can be seen in 
Fig.\ \ref{fig:OTRdisk}. When the
motor is turned on again to move to the next foil position, the torque of the
motor and shaft system is sufficient to start rotating the disk and bring the
ball bearing out of the depression against the spring force. The foil position
repeats with this method to 0.1~mm precision. A backup system uses a pressurized
helium gas line, which ends in a brass tube with an end face parallel to the
back of the disk about 0.1~mm from the surface (see Fig.\ \ref{fig:OTRdisk}). As the disk rotates
the He pressure is maintained until the correct foil position is reached, at which
point the tube end encounters a hole through the disk, reducing the pressure.
This method is precise only to \(\sim\)2~mm, but sufficient 
to double-check the position of a foil in the beam in the case of failure of the micro-switch. 
This same helium line can be used to remove any accumulated dust from the foils, since
the gas passing through the hole in the disk is guided by a custom nozzle 
to blow across the foil surface.

The OTR operation history is as follows :
\begin{itemize}
\item OTR-I
\begin{itemize}
\item Built and installed in 2008\(\sim\)2009
\item Stable operation 2009\(\sim\)2013  
\end{itemize}
\item OTR-II
\begin{itemize}
\item Built in 2009 and stored as spare of OTR-I
\item Assembled, calibrated, tested, and installed with the new Horn 1 in 2014
\item Operation between May 2014\(\sim\)present 
\item Minor problems from Mar. 2015 (see below)
\item Motor disconnected and OTR disk fixed at cross foil position since Jan. 2016 
\end{itemize}
\item OTR-III 
\begin{itemize}
\item Present spare system built in 2013
\item Assembled, calibrated and tested in 2014 together with OTR-II
\item Exchange rehearsal using the manipulator arm system in the remote maintenance area performed in Dec. 2015 
\item Disk flange and plunger are being re-designed (see below)
\item Ti foils to be modified 
\begin{itemize}
\item Additional holes will be drilled in the Ti
foils
\item However, the exact hole pattern is currently still under consideration
-- foil mechanical strength and usefulness for optical calibration must be balanced
\end{itemize}
\item Plan to remove OTR-II and install OTR-III in summer 2018
\end{itemize}
\item OTR-IV
\begin{itemize}
\item Built identical to OTR-III
\item Plan to use it as spare of OTR-III
\end{itemize}
\end{itemize}

Although the disk was originally designed for frequent rotations, there have been some recent 
issues with the rotation mechanism, as described below, and the 
disk is currently being kept in a single position.
The cross foil is now in use for beam running, since that foil may be used to
both measure the beam profile, and the holes allow for periodic check of the 
OTR alignment and calibration by back-lighting.  Future versions of the OTR will
have holes for periodic monitoring of the alignment on each of the Ti foils.

So far, the cross foil has seen \textcolor{\MODCOLOR}{\(\sim\)\(2.0\times10^{21}\)~POT}, as shown in Fig.\ \ref{fig:OTRdeg}.
\textcolor{\MODCOLOR}{Gradual degradation of the OTR light yield as a function of POT
has been observed. Most of this yield decrease is probably due to
radiation-induced darkening of a leaded glass fiber taper coupled to the CID camera,
and \(<\)20\% due to the visible darkening of the foil.}

\paragraph{OTR Disk Rotation Issue}

Although the OTR disk was originally designed for many rotations in order to
frequently switch between the different foils, issues with the disk
rotation mechanism have been observed. For OTR-II, it was found that 
after many disk rotations,
the plunger mechanism becomes stiff such that it becomes
difficult for the plunger to disengage from the hole, making rotations in one
direction by the motor impossible. \textcolor{\MODCOLOR}{However}, 
it was found that rotations where the disk is moved backwards worked fine.

\textcolor{\MODCOLOR}{One possible cause of this issue is that the plunger became rusted since the
ball and its housing was made of the common 304 stainless steel (SS).
Because of this, the ball may not be able to rotate properly and more torque
from the motor is needed to disengage it from the hole.
The non-rotating ball can also cause mechanical damage to its housing and the
disk flange made of titanium which is not as hard as 304 SS.}

Test bench tests have shown a similar issue in a non-radiation environment,
although those tests show that rotation in both directions is affected. 

A new custom plunger \textcolor{\MODCOLOR}{and disk flange have been designed and tested for OTR-III and OTR-IV. 
The new custom plunger ball and housing was made of 440C SS heat-treated to a hardness of 55-60 Rockwell C.
The ball now fits better to the plunger internal diameter and an external screw was included
to adjust the plunger's spring force.
The new disk flange was made of 440C SS to prevent mechanical damage during disk rotation.
The new design was tested and works stably for \(>\)100 rotations without any change in rotation torque.}

\paragraph{OTR Microswitch Issue}
The microswitch in OTR-II has not been working stably since Nov. 2014 : even if the
switch is engaged, sometimes the electric signal from the switch indicates that
it is not engaged.
\textcolor{\MODCOLOR}{The issue started in one foil position and it gradually expanded to all positions.}
The cause of this issue is currently unknown, but it may be due to a short
in the microswitch wiring near the OTR disk. This issue is under 
investigation and should be fixed in the OTR-III (and OTR-IV) design before installation.

\paragraph{OTR Plan}
With the introduction of calibration holes in each of the Ti foils, the OTR can
be used continuously at high beam power even without fixing the above disk rotation and
microswitch issues.  
Fixing
the above mechanical issues will make rotations during beam studies easier, as well as
allowing us to continue to use the OTR fluorescent foil at low beam power during 
tuning\textcolor{\MODCOLOR}{, though}.

\textcolor{\MODCOLOR}{Design upgrades to improve the observed light yield decease due to fiber-taper
darkening are currently being considered.  One
potential solution could be to use an easily replaceable (relatively cheap) fiber taper which would be
periodically replaced.  Another option could be to modify the optical focusing system
to eliminate the need for a fiber taper.  If degradation of the foil material or
light yield decrease due to the foil itself is also observed,}
the OTR disk could be frequently remotely exchanged during summer shutdowns.

Another OTR upgrade option could be to switch to a more robust foil material
which would maintain a high light yield beyond an integrated \(\sim\)\(6\sim12\times10^{20}\) POT, and
preliminary studies of various foil materials, such as carbon (graphite) or Ti grade 5 (Ti-6Al-4V), 
are currently underway.  However, materials studies by the RaDIATE collaboration 
indicate that the current foil material (Ti-15-3-3-3 alloy) exhibits good radiation 
damage resistance, suggesting that it may be better to continue to use this foil
type.

Finally, the OTR DAQ must also be upgraded to reduce the readout latency in order to run
with \(<\)2.48~s beam spill repetition rate, as described in Sec. \ref{sec:ctrldaq}.

\textcolor{\MODCOLOR}{There is a \(\sim\)30\% difference between the beam width measured by the
OTR and that measured by the SSEMs, as shown in Fig.\ \ref{fig:optics}.  It is believed that this
is an artificial broadening of the OTR light profile, possibly due to
background light caused by proton-beam-induced scintillation of the He inside the He vessel. 
OTR background studies are currently underway and will hopefully allow us to further understand this
issue.}

\subsubsection{Proton Beam Monitor Upgrade Summary} \label{sec:monupgradesummary}

The CTs, BLMs, and ESMs have all been used continuously and stably since the
beginning of T2K without
any major issues.  These monitors were all designed to work continuously at high intensity,
and should require minimal or no hardware upgrades.

The SSEMs have also been stably running throughout T2K, although periodic
replacement of the continuously used SSEM19 may be necessary at high beam power.  
SSEM sensor heads can be replaced as necessary without any design change,
although SSEM19 replacement does require access to the monitor stack.  A new
WSEM has been developed as an option for a profile monitor sensor head replacement which yields reduced beam
loss.  SSEM18 will be replaced with a WSEM during 2018 or 2019 in order to
allow SSEM18 to be used continuously if SSEM19 fails.

Various small improvements have been made to the OTR design, and the
currently-in-use OTR will probably be replaced during summer 2018.  Periodic
replacement of the OTR disk may also be necessary after 2021 or 2022.
The OTR DAQ will also be upgraded.

A new BIF non-destructive proton beam profile monitor is also being developed, as 
described in Appendix \ref{sec:BIF}.  Installation of a prototype BIF monitor
in the primary beamline will be carried out in 2018 and 2019.

\graphicspath{{figures/main_primary}}

\subsection{Maintenance senario of the primary beamline} 
\label{sec:primarymainte}

The original scenario of maintenance of the primary beamline components was
hands-on with quick action devices.
This includes electric and cooling water connection of
the normal-conducting magnets and
connection of the vacuum flanges.
For the removal and re-installation of the normal-conducting magnets,
remote hoisting tool and positioning keys were made.
Positioning keys were also discussed for the beam monitors,
but were not implemented at the beginning. 

\subsubsection{Normal-conducting magnets}
The design of the quick water connectors,
remote-hoisting tool, and key positioning scheme are described
in ref~\cite{InterimJNU:2005} in detail.
In addition, knife-switch type quick electric connectors were made
and installed.
These have been working well in the actual maintenance since 2009. 

\begin{figure}[htb]
\begin{center}
\begin{minipage}[t]{7.5cm}
 \includegraphics[width=7.5cm,clip]{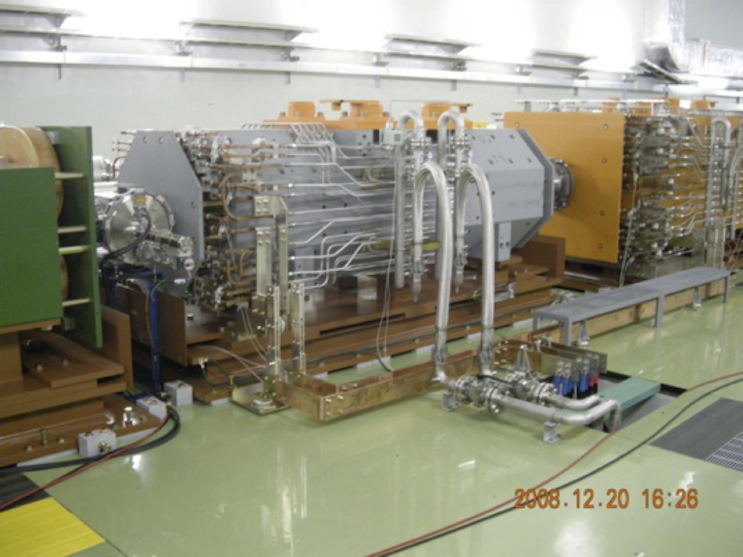}
  \caption{A whole view of PQ2 magnet, equipped with quick electric and cooling water couplers.}
  \label{fig:pq2}%
\end{minipage}
\hfill
\begin{minipage}[t]{7.5cm}
 \includegraphics[width=7.5cm,clip]{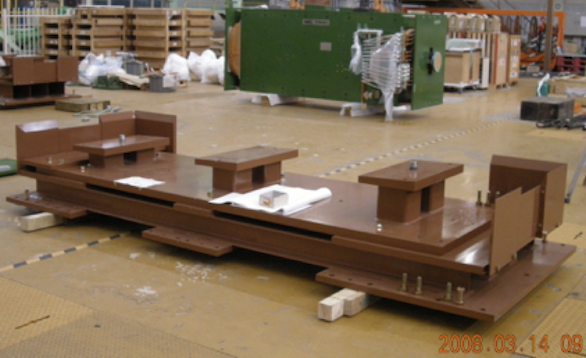}
  \caption{A magnet stage with key cones for precise magnet positioning.}
  \label{fig:magstage}%
\end{minipage}
\end{center}
\end{figure}

\begin{figure}[htb]
\begin{center}
\begin{minipage}[t]{7cm}
 \includegraphics[width=7cm,clip]{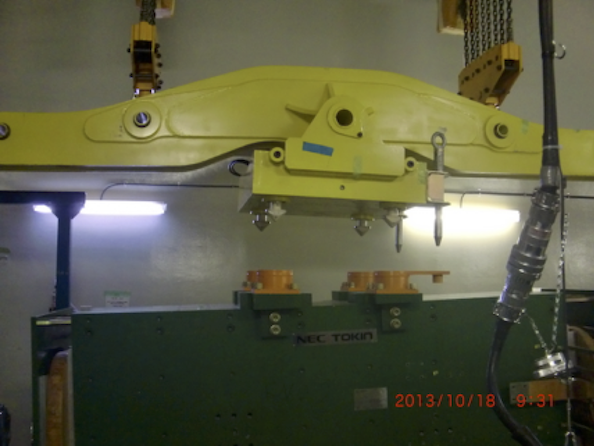}
 \caption{Twist-lock semi-remote hoisting attachment.}
 \label{fig:twistlock} 
\end{minipage}
\hfill
\begin{minipage}[t]{4cm}
 \includegraphics[width=4cm,clip]{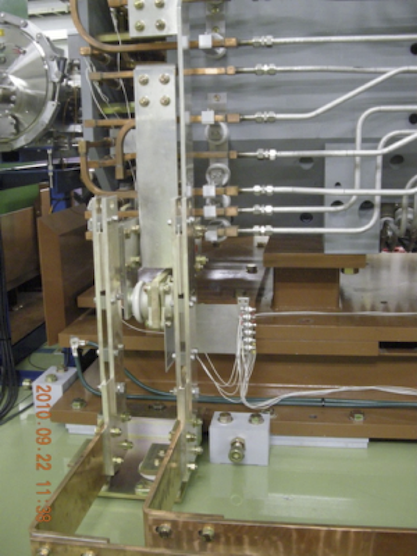}
 \caption{A knife-switch type electric connector for quick operation.}
 \label{fig:kinfeswitch} 
\end{minipage}
\hfill
\begin{minipage}[t]{4cm}
 \includegraphics[width=4cm,clip]{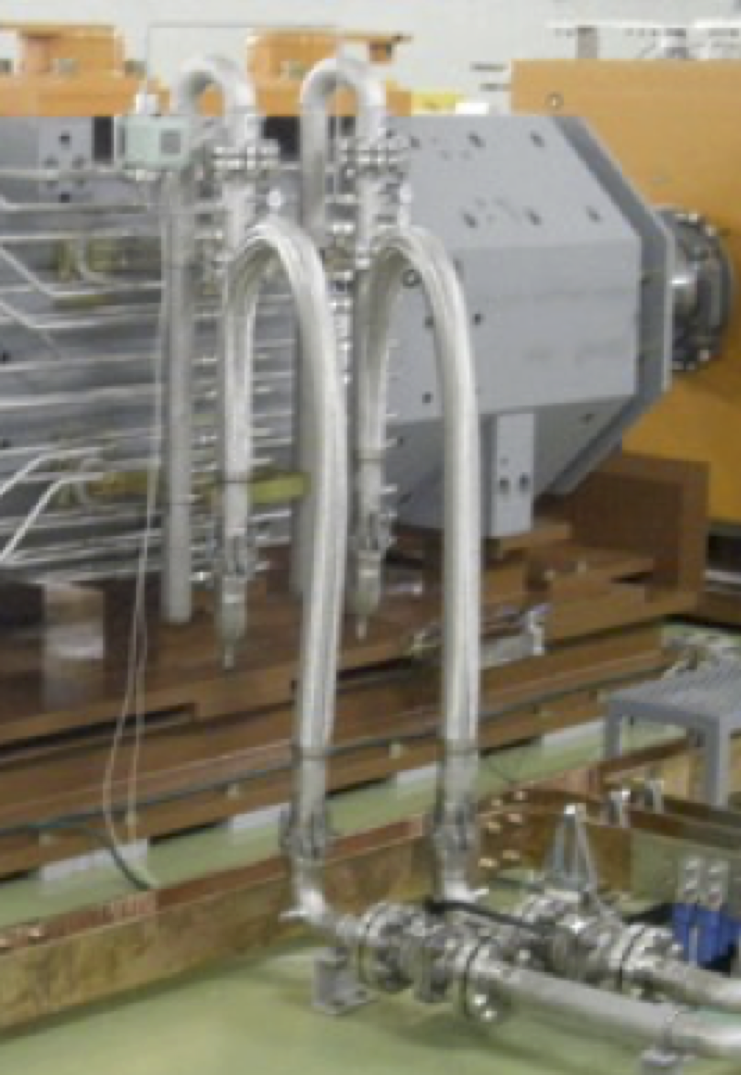}
 \caption{Quick-action water connectors.}
 \label{fig:waterconnector} 
\end{minipage}
\end{center}
\end{figure}

A whole view of one magnet (PQ2) is shown in Fig.~\ref{fig:pq2}.
The magnet stage has key cones as shown in Fig.~\ref{fig:magstage} 
while the magnets has holes to precisely mate with the keys to
reproduce the magnet position when it is removed off-beamline for some work.
Hoisting of the magnets are done with remotely-operational
twist-lock attachment as shown in Fig.~\ref{fig:twistlock}. 
The electric connectors are knife-switch type as shown in Fig.~\ref{fig:kinfeswitch}, 
and allow quick and easy connection/disconnection work,
assuring 2000~A current flow with spring contact.
Quick water couplers are cam-lock type with lever-action
as shown in Fig.~\ref{fig:waterconnector} with metal gasket sealing.

All these design are intended to finish work near
the beamline within several minutes.
Operation experience after 2009 has shown
that radio-activation is much less than we postulated at the design stage,
and hours of work is possible at present.
We think present scenario will work even after 1.3 MW operation has been realized. 

\subsubsection{Vacuum flange operation}
Vacuum flanges are one of the hottest of all the primary beamline components. 
In order to work on them, our initial design is not full remote system
such as pillow seals but quick flanges operational from some distance
(so-called semi-remote system).
Our nine-years operation has shown that this design works well mostly.

\begin{figure}[htb]
\begin{center}
\begin{minipage}[t]{7.5cm}
 \includegraphics[width=7.5cm,clip]{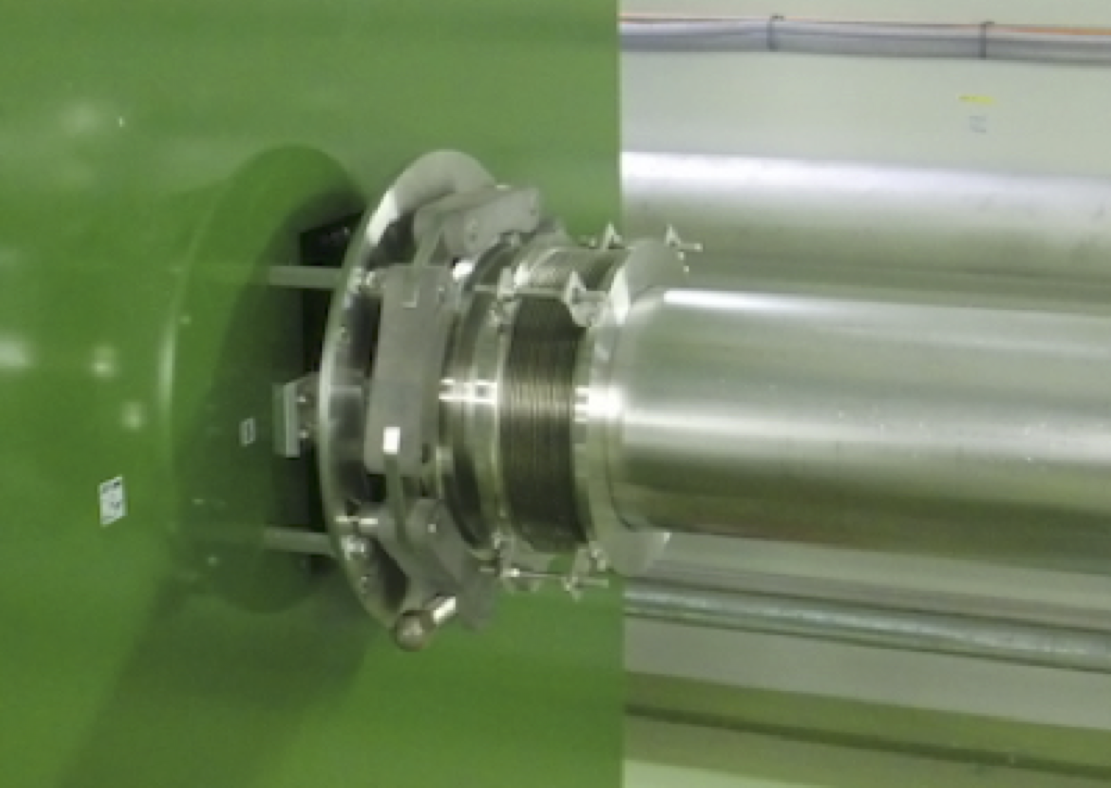}
  \caption{Semi-remote type chain clamp RH series to tighten the flanges from distance with a single screw.}
  \label{fig:chainclamp}%
\end{minipage}
\hfill
\begin{minipage}[t]{7.5cm}
 \includegraphics[width=7.5cm,clip]{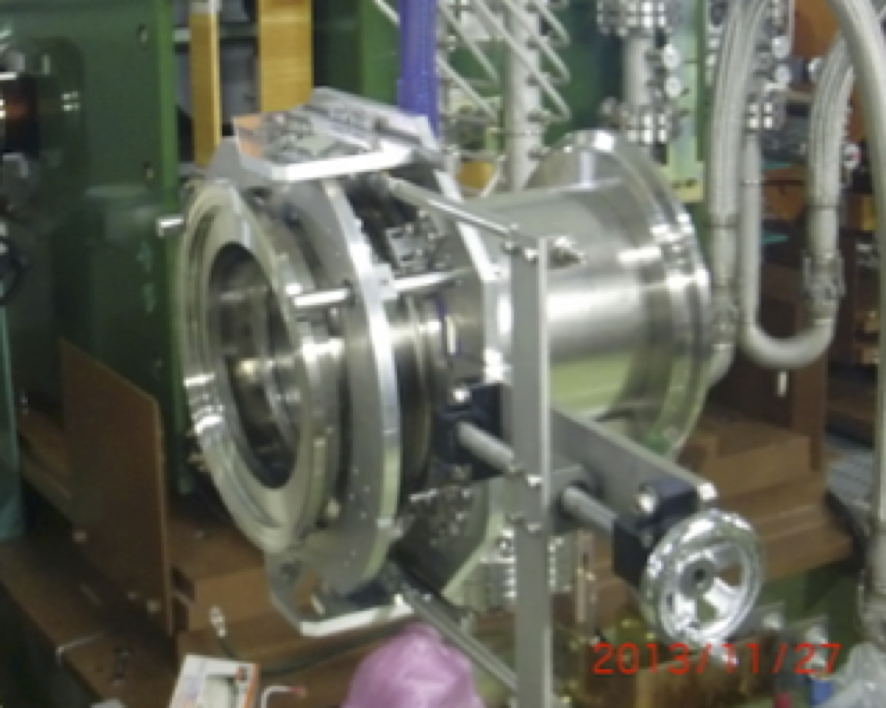}
  \caption{A single-screw mover to expand the bellows to meet the flanges quickly.}
  \label{fig:ssmover}%
\end{minipage}
\end{center}
\end{figure}

\begin{figure}[htb]
\begin{center}
\includegraphics[width=12cm,clip]{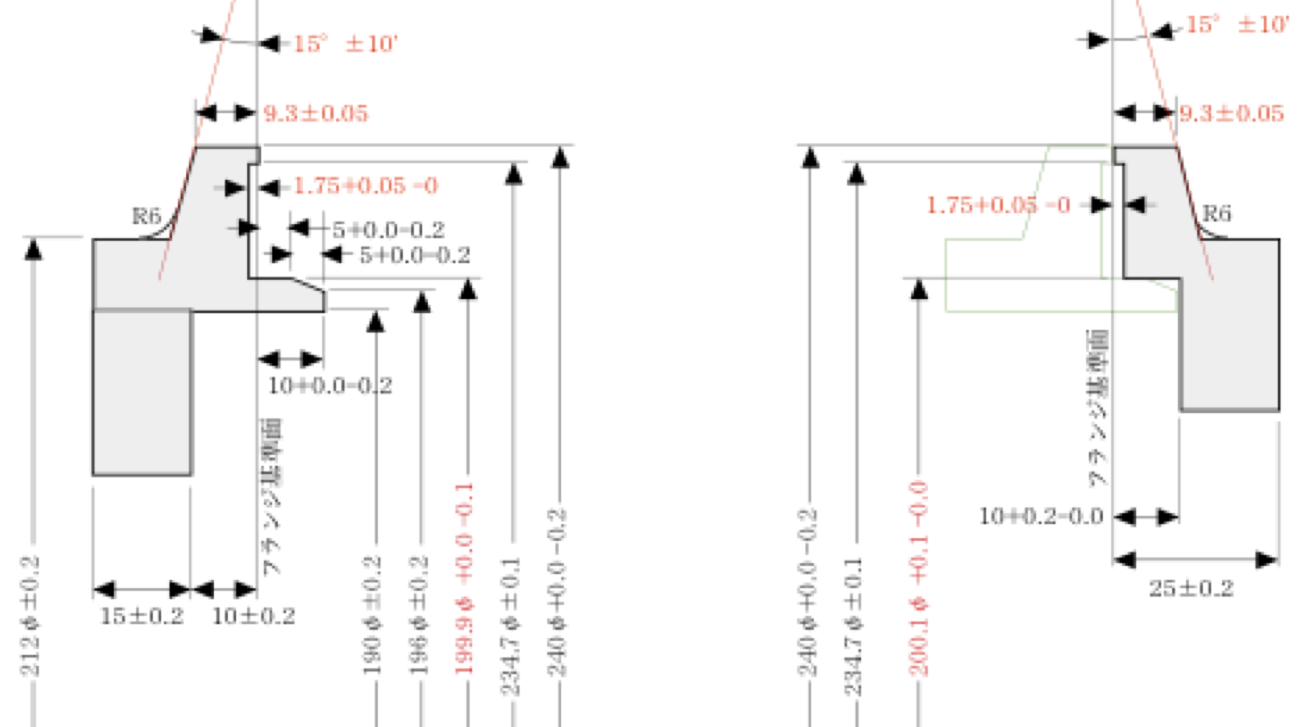}
\caption{A pair of male/female flanges to mate without fine adjustment by hands.}
\label{fig:flange}%
\end{center}
\end{figure}

\begin{figure}[htb]
\begin{center}
\begin{minipage}{5cm}
 \includegraphics[width=5cm,clip]{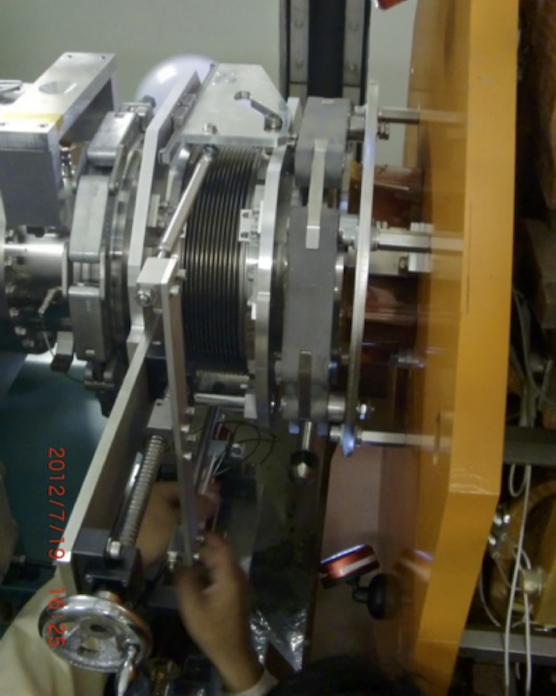}
\end{minipage}
\begin{minipage}{7.5cm}
  \caption{RH chain clamp and quick bellows movers are installed on the male/female flanges at the downstream of a Q magnet.}
  \label{fig:triplet}%
\end{minipage}
\end{center}
\end{figure}

The connection operation of the flanges consists of expansion of the bellows,
mating of the flanges, and fastening with a clamp.
A semi-remote type chain clamp,
RH series as shown in Fig.~\ref{fig:chainclamp},
are adopted for locations where large beam loss was expected, 
such as most-upstream part of the preparation section,
downstream part of the final-focus section, and the collimators.
This RH clamp enables fastening the flanges by rotating a single screw from distance.
Expansion of the bellows to meet the flanges are done with a mover as
shown in Fig.~\ref{fig:ssmover}.
By rotating a single screw from some distance,
this mover expands/shrinks the bellows quickly.
The mating flanges is achieved using male/female flanges as shown
in Fig.~\ref{fig:flange}.
Original idea was to install this triplet system for all the flange connection points.
Due to financial limitation, however, installation was limited to where
the highest activation was expected (Fig.~\ref{fig:triplet}), 
and addition to other places was to be done seeing the operation status.

For re-positioning of the monitors after removal for some work off-beamline,
positioning keys were to be installed as in the case of the normal-conducting magnets.
However due to lack of the resources, this was omitted at the initial installation,
and positioning has been done by hands so far.
Seeing the present radio-activation,
this will work even after 1.3~MW has been achieved except for one location,
the most-downstream part of the final-focus section.
Installation of positioning keys are definitely needed there.

\subsubsection{Maintenance scheme of the most-downstream part of the final focus section}
Due to the back-scattered radiation from the down-stream target station,
the most downstream part of the final focus section suffers build up of
the radio-activation.
At 470 kW operation, its radio-activity is $~$6 mSv/h on contact
at the highest device 6 hours after the beam stop, and $~$300~$\mu$Sv/h at
one foot distance one week after the beam stop.
Re-positioning of the monitors by hands shall not be possible at 1.3~MW operation.
Therefore installation of the positioning keys are indispensable. 

\begin{figure}[htb]
\begin{center}
\begin{minipage}[t]{7.5cm}
 \includegraphics[width=7.5cm,clip]{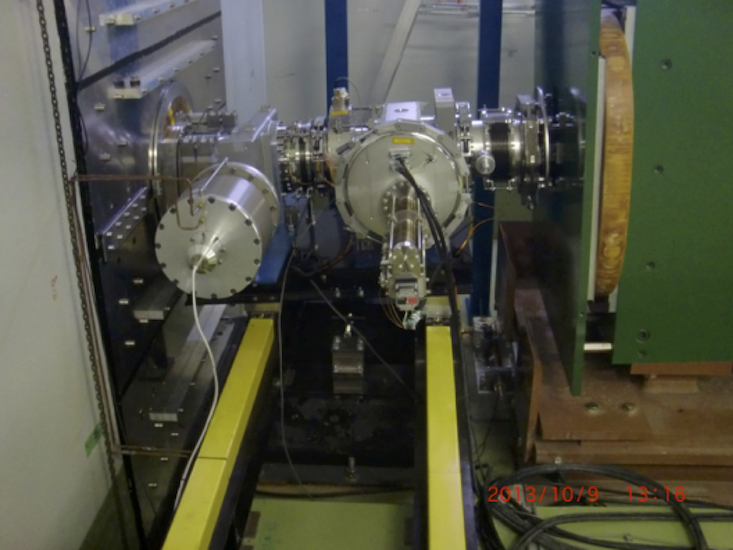}
  \caption{A pair of sliding rails installed at the most downstream part of the final focus section. %
  All the components on the moving stage are designed to be lined out on maintenance.}
  \label{fig:ff}%
\end{minipage}
\hfill
\begin{minipage}[t]{7.5cm}
 \includegraphics[width=7.5cm,clip]{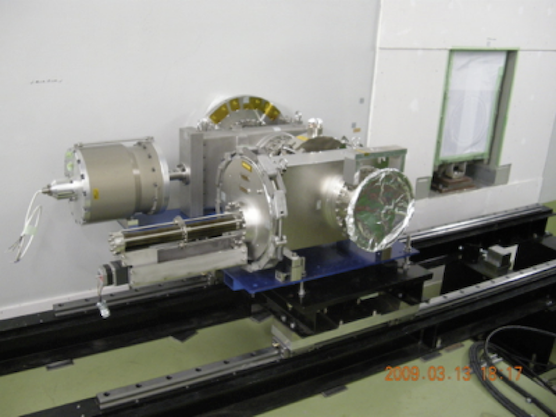}
  \caption{Monitor triplet (ESM, SSEM, CT) on the moving stage are lined out for maintenance.}
  \label{fig:lineout}%
\end{minipage}
\end{center}
\end{figure}

We expected high-radiation environment at the design stage,
and sliding rails are installed to line-out everything as a whole from
the beamline for maintenance as shown in Fig.~\ref{fig:ff}.
By disconnecting upstream RH flange and downstream pillow seal,
and shrinking upstream and downstream bellows,
all the components can be lined out off the high radiation area
for maintenance as shown in Fig.~\ref{fig:lineout}.
However, after operation of years,
it turned out that components on the stage themselves have gotten highly radioactive,
and moving out does not improve the radiation condition so much.
In addition, we found that moving out the pillow flange imposes
very difficult problem on the air isolation between the final-focus section
and the target station.
Therefore, we decided to discontinue this line-out scheme and
adopt quick operation with positioning keys on the beamline, using over-head chain hoists.

During the long shut-down period in 2018 summer-fall,
we plan to re-build stages of the currently-installed beam monitors and
a gate valve to have positioning key cones on the lower stage and
key holes on the upper stages,
and install quick bellows movers on the upstream bellows.
With already-installed RH chain clamp,
it is estimated that monitor triplet removal or
re-installation will not need more than one hour.
Even with these quick schemes, we need to cool down the section
at least one week for heavy maintenance. 

\subsubsection{Summary}
The residual dose of the primary beamline components is less than we expected at the design 
stage except for the most-downstream part of the final focus section.
For the components placed at the preparation section, the arc section and the final focus
section other than the most-downstream part, 
we think present scenario, which is the hands-on maintenance with the quick action devices
and the semi-remote devices, will work even after 1.3 MW operation has been realized.

For the components placed at the most-downstream part of the final focus section,
we plan to improve the maintenance scenario to adopt quick operation with positioning keys
on the beamline. We plan to perform these improvement works during the long shut-down
period in 2018 summer-fall. 
\color{\MODCOLOR}
We will keep the present strategy which is the quick handling scheme, while the robotic scheme 
could be one of future options. 
\color{black}

\graphicspath{{figures/main_primary}}

\subsection{Muon monitor (MUMON)} %
\label{sec:mumon}

\begin{figure}[htbp]
\centering
\includegraphics[width=8cm]{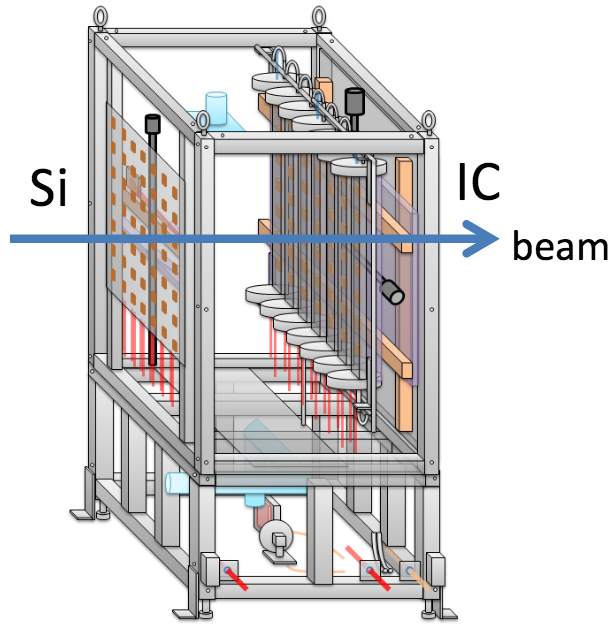}
\caption{Schematic view of the muon monitor.}
\label{fig:mumon}
\end{figure}

The secondary muon beam intensity and direction are monitored bunch-by-bunch by
a Muon Monitor (MUMON) \cite{Matsuoka:2010mumon}\cite{Suzuki:2014jyd} which is located directly behind the beam
dump.
The beam dump thickness is designed so that only muons whose momentum is higher than 5~GeV penetrate it. 
The muon monitor consists of two 7\(\times\)7 arrays of 25-cm interval
sensors : an array of Si PIN photodiodes and an array of ionization chambers, as shown in Fig.~\ref{fig:mumon}.
These muons are produced along with neutrinos, so the neutrino beam intensity and direction can be monitored by monitoring the muon beam.
It measures the muon profile center with a precision of 2.3 cm for horizontal and vertical directions, 
 which corresponds to a 0.28 mrad precision of the muon beam direction.
The resolution is better than 3.0 mm, which is equivalent to \(\sim\)0.025 mrad.
It also monitors the stability of the beam intensity with a resolution better than 3\%.
Two sensor-arrays are adopted for redundancy and so far the Si-sensor array has been used to select good quality spills for the T2K  analysis.

\begin{figure}[htbp]
\centering
\includegraphics[width=12cm]{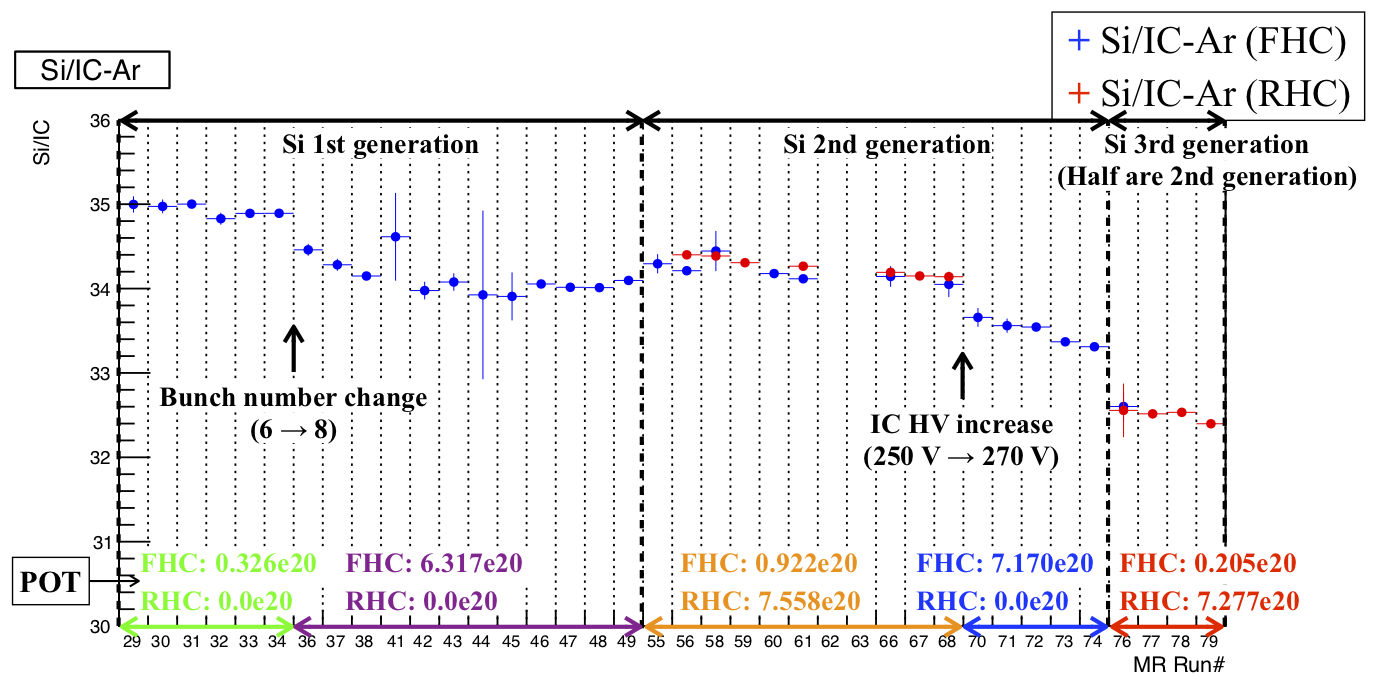}
\caption{Observed muon monitor Si sensor yield (normalized to IC yield) over T2K so far.}
\label{fig:mumondegradation}
\end{figure}

\begin{figure}[htbp]
\centering
\includegraphics[width=12cm]{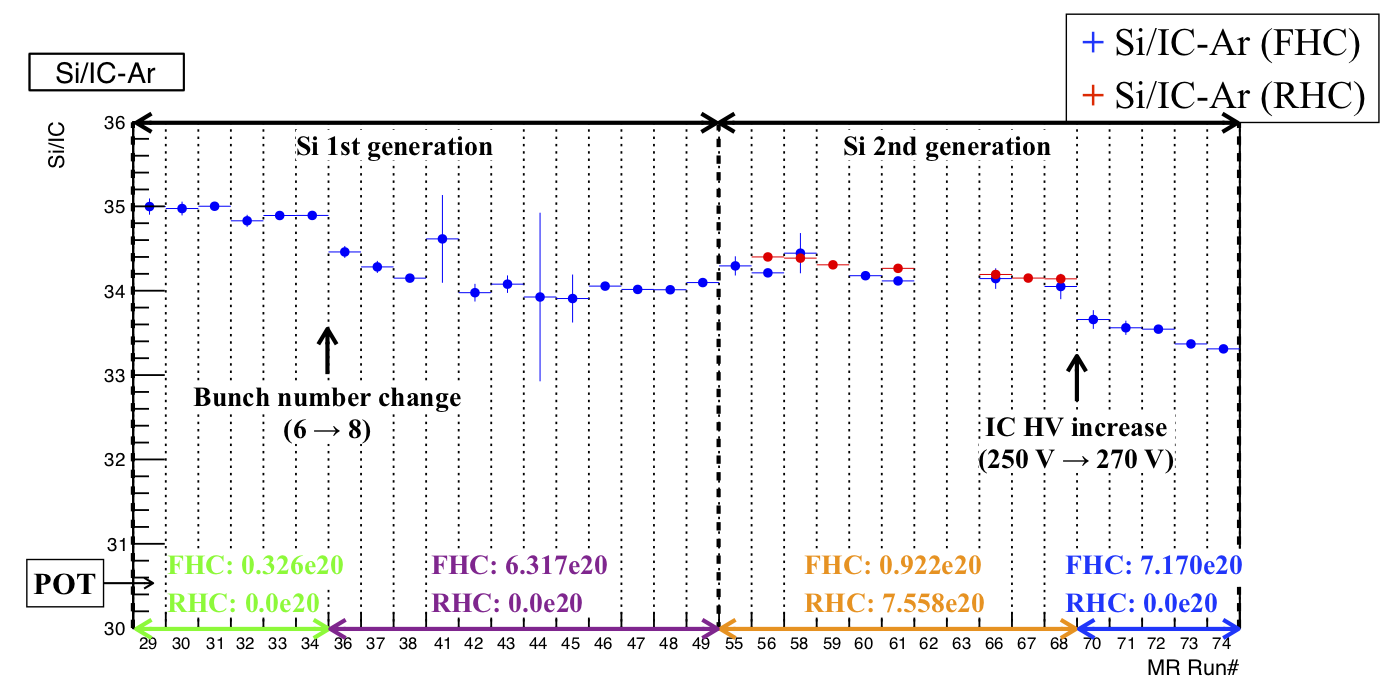}
\caption{Observed muon monitor Si sensor yield (normalized to IC yield) up to
  Run 74 (IC calibration was changed between Run 74 and Run 76).}
\label{fig:mumondegradationrun74}
\end{figure}

We have observed a $\sim 2\%$ signal reduction of the Si sensors so far.
This is demonstrated in Figs.~\ref{fig:mumondegradation} and \ref{fig:mumondegradationrun74} as a ratio of the signal yields of Si sensors and IC. 
From a test with a 100~MeV electron beam in 2009, it is known that the degradation of the Si PIN photodiode by radiation appears as non-linear response.
Non-linearity and reduction of the signal yield by 10\% at maximum were observed 
with an irradiation corresponding to a $8\times 10^{20}$~POT 250~kA neutrino-mode operation at the test.
To preserve the required precision, it is estimated from a MC study that the maximum allowable non-linearity is 5\% .
Therefore, the first set of the Si sensors were replaced in 2013 after the $7\times 10^{20}$~POT 250~kA neutrino-mode operation even though the signal reduction was much smaller than 5\%.
We conducted a linearity check with the second set of sensors after they were irradiated with
$8\times 10^{20}$~POT 250~kA neutrino-mode operation equivalent\footnote{Actual POT is $3\times 10^{20}$~POT in the neutrino-mode and
$8\times 10^{20}$~POT in the antineutrino-mode}.
No non-linearity was observed then.
This POT corresponds to 35 days operation in the  320~kA neutrino-mode at 750~kW.
Therefore, our baseline plan is to periodically check the radiation damage by the linearity measurement
and replace sensors when damage is found. The expected cycle of the check and replacement
is one or two months.
Since the cost of the Si sensors is relatively cheap ($\sim270,000$~Yen for one set), this plan is feasible
for the 750~kW operation with some extra cost for outsourcing for the replacement work.

\begin{figure}[htbp]
\centering
\includegraphics[width=9cm]{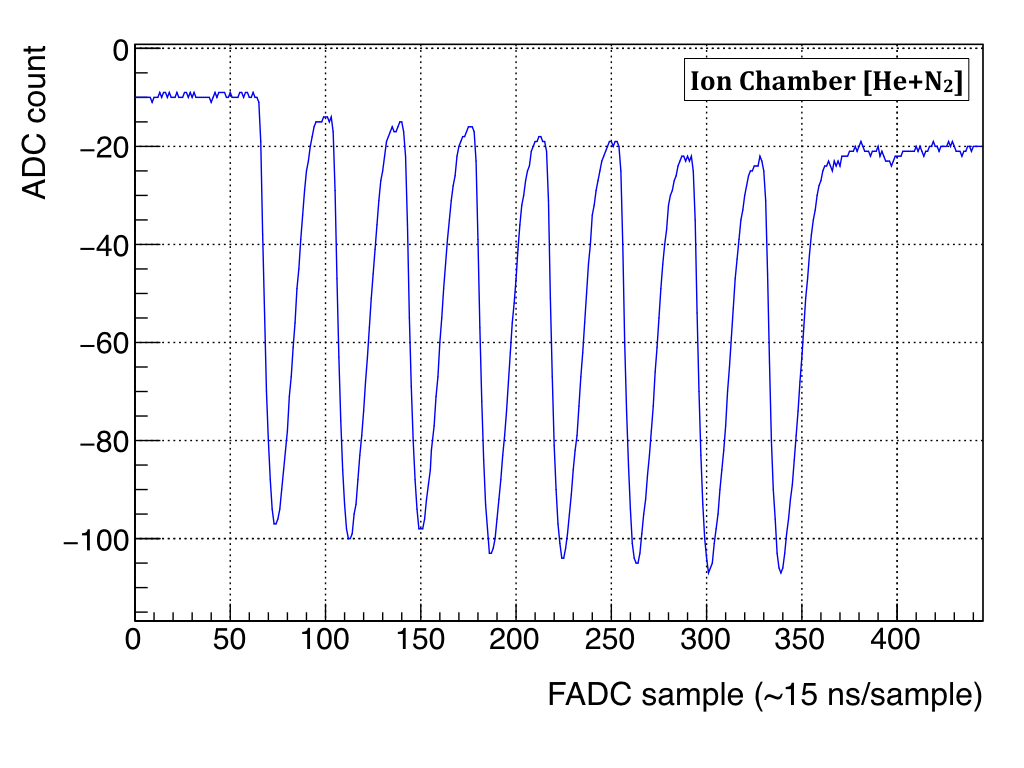}
\caption{Example waveform of the IC signal while the chamber is filled with He+1\% N\(_2\).
\label{fig:ICHE}}
\end{figure}

Two gas types have been used in the ionization chambers (IC): Ar with 2\% N\(_2\) is
used for low-intensity beam, while He with 1\% N\(_2\) is intended for use at
high intensity.
Experience has shown that Ar gas works well up to $2.5\times 10^{13}$~ppb in neutrino-mode and
$2.8\times 10^{13}$~ppb in anti-neutrino-mode, but at higher intensities non-linearities are seen
due to space-charge effects in the detector: at high intensity, too many
electron-ion pairs are created, and then many ions are accumulated because of
their slow drift speed, which distort the electric field inside the detector.
By increasing the high voltage between electrodes from 250~V to 270~V, the effect was slightly mitigated.
The He gas has been tested in-situ.  The space-charge effect is much smaller because He ions are less 
accumulated since their drift speed is relatively fast.
It is confirmed that the signal size is sufficient to provide the required resolution for the spill-by-spill monitoring if the beam intensity exceeds 300~kW (500~kW) for (anti)neutrino-mode.
However, it is found to have considerable after-bunch pileup induced by the movement of the He ions as shown in Fig.~\ref{fig:ICHE},
resulting in difficulty in the bunch-by-bunch measurement.

In summary, the 750~kW operation is feasible either with the Si sensors
or the He-gas based IC, though the former would require monthly exchange, and the latter would degrade bunch-by-bunch measurements. 

\subsubsection{Muon Monitor Upgrades}
Several sensor types have been tested in the J-PARC muon pit as possible upgrades.
Diamond detectors were tested \cite{HirakiThesis:2016}.
Even though detector-grade monolithic sensors have shown good performance, more yield degradation than Si-sensors was observed after irradiation.
SiC sensors (SG01XL-5 UV photodiode) were also tested, however the signal size for these commercially available sensors was very small.

\begin{figure}[h]
\centering
\includegraphics[width=14cm]{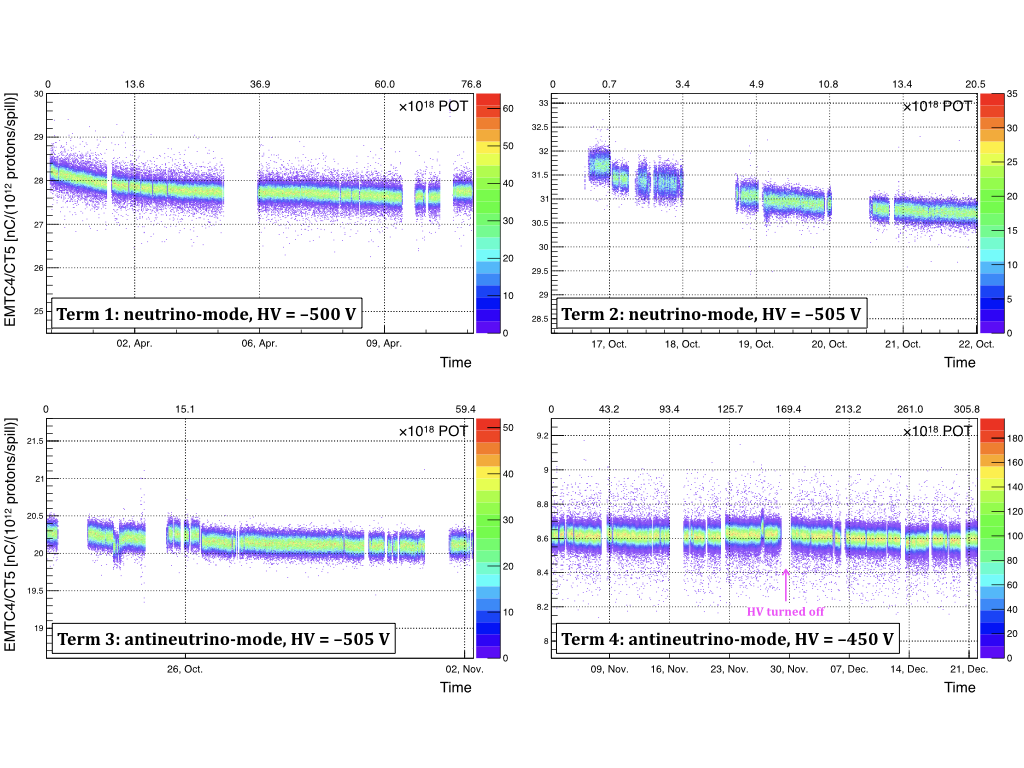}
\caption{Signal yield stability of the one of the EMT prototypes exposed to the muon beam in the muon pit.}
\label{fig:emt4stab}
\end{figure}

\begin{figure}[h]
\centering
\includegraphics[width=12cm]{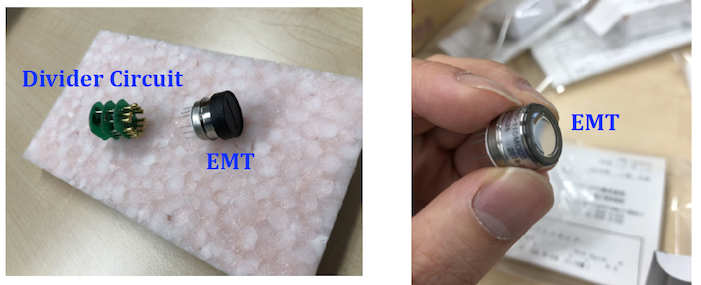}
\caption{One of the EMTs and a divider circuit to be tested in 2018.}
\label{fig:emt2018}
\end{figure}

Currently, Electron Multiplier Tubes (EMT), custom made by Hamamatsu, are under
investigation as a suitable upgrade sensor \cite{Ashida:2018mumonemt}. The EMT is equivalent
to a Photo Multiplier Tube (PMT) without a photocathode.
The principle is that muons hit a cathode or dynodes and cause secondary emission
of electrons, which then are multiplied by a set of dynodes.  Several of
these sensors have been installed in the muon pit, and have so far exhibited
reasonable properties, including a good signal-to-noise ratio, good time
response without a large tail pulse, and good intensity resolution.
Modest non-linearity was observed and considered to be due to the space-charge effect at the later-stage dynodes.
Figure~\ref{fig:emt4stab} shows the yield stability of one of the prototype sensors.
Some burn-in time has been observed like other similar secondary emission monitors.
For normal PMTs, a warm-up procedure is usually applied by a manufacturer in order to stabilize the 
alkali metals and Sb on the dynodes.
However, in our application, the photo-cathode is replaced by aluminum deposition and the warm-up procedure was not applied.
Several EMT sensors with different properties have been purchased from Hamamatsu (Fig.~\ref{fig:emt2018}) 
and \textcolor{\MODCOLOR}{have been under testing since} March 2018: PMTs with and without warm-up and EMTs with
and without Sb on the dynodes.
A so called ``tapered'' divider circuit is adopted for the HV supply to suppress the space-charge effect.
We expect that the non-linearity problem will be resolved by optimizing the divider circuit.
Long-term stability is key \textcolor{\MODCOLOR}{and will be checked by long-term
testing in-situ at the T2K muon pit.  Dedicated beam studies at another facility
may also be carried out.}

Another option, if a bunch-by-bunch measurement is required, is to re-configure the IC electrodes.
By reducing the 3mm gap between electrodes, the space-charge effect will be suppressed without 
compromising the signal size.  This can be achieved with a modest cost, but would require a few months work.

The plan for upgrades of the MUMON using these
sensors will be further refined based on tests of the various EMT sensors in 2018 \textcolor{\MODCOLOR}{and 2019}.
If the EMT option currently being investigated is proven to work, then the production and replacement can be done
in three months. The estimated cost is about 4,000,000 Yen.

\clearpage

\section{Secondary Beamline}
\label{sec:secondary}

\graphicspath{{figures/main_secondary}}

\subsection{Introduction}

\subsubsection{Overview}

The delivered proton beam strikes a neutrino production target to produce secondary particles,
mainly pions and kaons, which are focused along the beamline by magnetic horns and 
decay in flight into muon neutrinos and muons in a 94~m long decay volume. The purpose of
the secondary beamline is to produce high intense, narrow band neutrino beams 
with so-called off-axis method.
 
\begin{figure}
        \centering
        \includegraphics[width=0.8\linewidth]{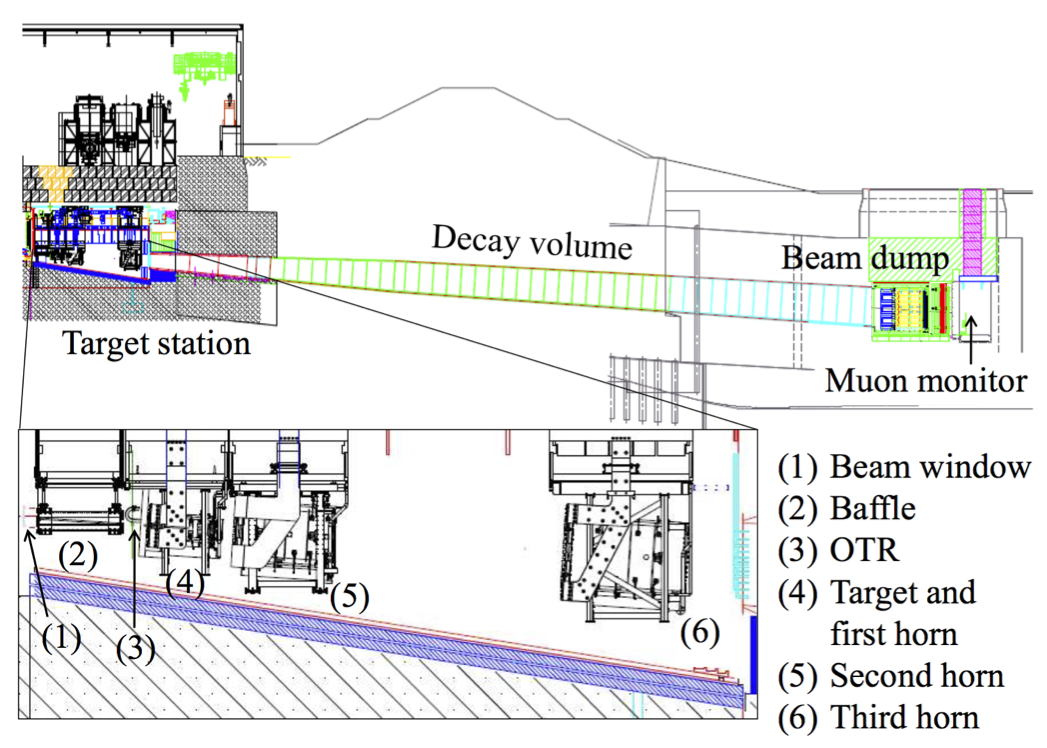}
        \caption{\small Side view of the overall secondary beamline. The proton beam enters from the
        left side of this figure.}
        \label{fig:secondary_beamline}
\end{figure}

\begin{figure}
        \centering
        \includegraphics[width=0.67\linewidth]{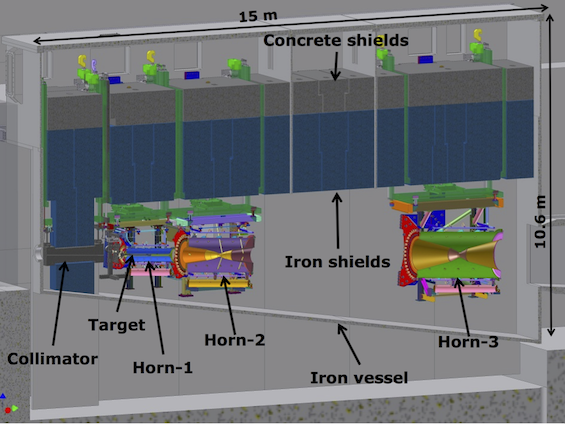}
        \includegraphics[width=0.32\linewidth]{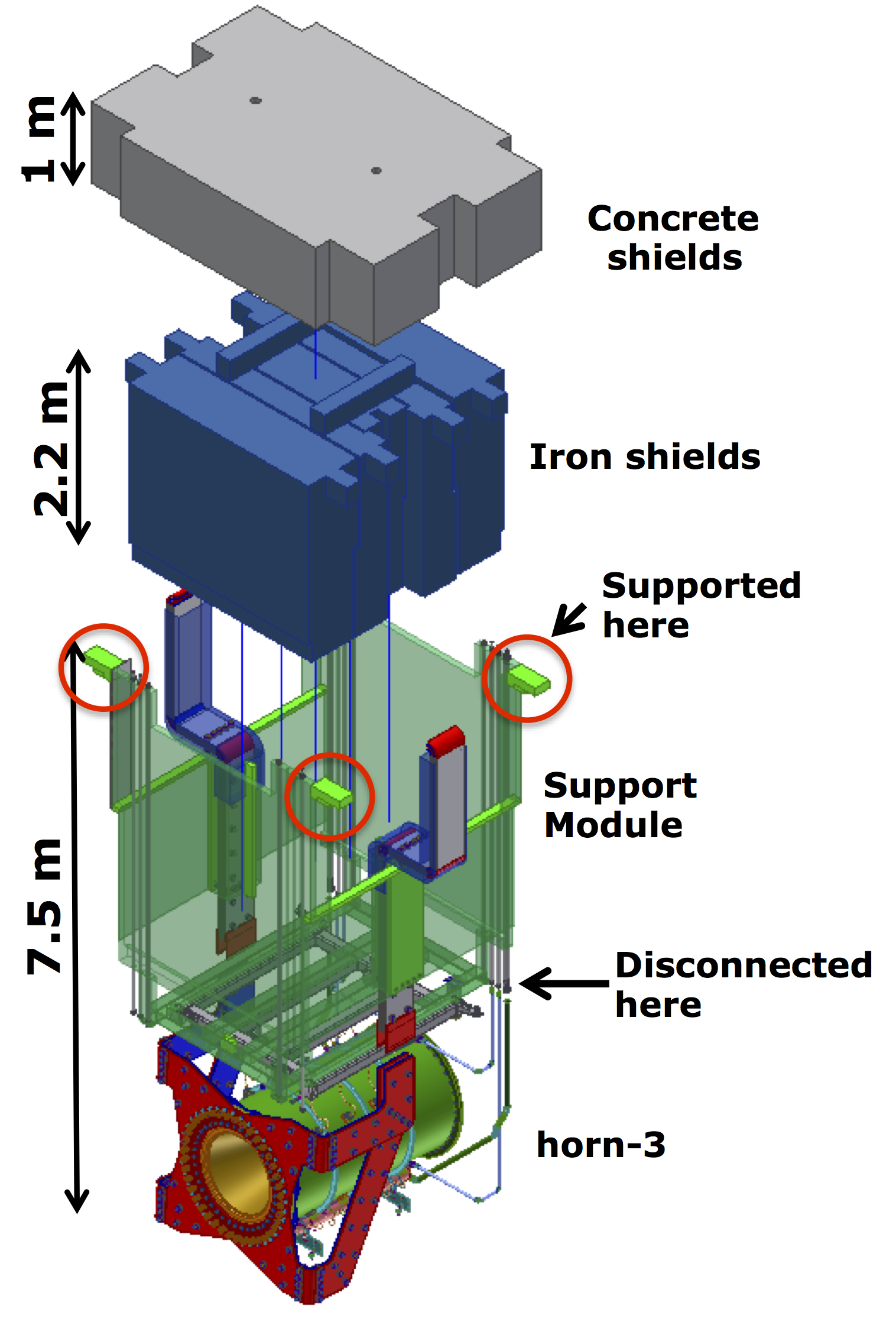}
        \caption{\small Schematic view of the secondary beamline equipment inside the
        helium vessel (left). The beam comes from left side. The horns are
        hanged by a box-shaped iron support structure called support module. Both iron and
        concrete shields are inserted in the inner space of the support module (right).}
        \label{fig:TS_helium_vessel}
\end{figure}

The secondary beamline consists of the Target Station, Decay Volume, Beam Dump, and Muon Monitors.
Fig.~\ref{fig:secondary_beamline} shows side view of the overall secondary beamline.
All the secondary beamline equipment other than the Muon Monitors are contained in a gigantic
helium vessel, whose volume is 1,500~m$^3$, where helium gas at atmospheric pressure is
filled in order to suppress pion absorption and to reduce production of tritium and nitrogen oxide (NOx).
The Target Station is a facility building to handle the target, magnetic horns, and 
related periphelalls, and the most upstream part of the large helium vessel is located
underground the Target Station. The helium vessel at the Target Station contains
a graphite collimator (called Baffle), the target, and the three magnetic horns
in order from upstream, as shown in Fig.~\ref{fig:secondary_beamline}. A detailed
schematic view of the secondary beamline equipment at the Target Station helium vessel
is shown in Fig.~\ref{fig:TS_helium_vessel}.
Shielding blocks, made of 2~m-thick iron and 1~m-thick concrete, are placed
on top of the beamline equipment.
Since all of these equipment become highly radioactive at $O(10\sim10^2)$~Sv/h on their surface
due to the high intensity beam operation, they are replaceable by using an automated
overhead crane and remote-controlled hoisting attachments.

The production target is made of a graphite rod with dimensions of 2.6~cm in diameter and 91~cm in length.
The graphite core is enclosed by a titanium container where high purity helium gas is filled.
The target is designed to survive thermal shocks by
beam exposure of 3.3$\times 10^{14}$ protons/pulse. High speed helium gas flow of $\sim$250~m/s provides
sufficient cooling for 750~kW beam operation. The target is placed inside the inner conductor of the most
upstream horn (horn1) to obtain a high focusing efficiency of secondary particles, and the upstream end
of the target is fixed by support frames of the horn1. If the target is broken, the horn1 including the broken target
is remotely transported to a maintenance area in the Target Station by the crane, and the broken target can be 
replaced with new one using a special exchange equipment developed by RAL that is operated by a remote 
manipulator equipped in the maintenance area.

The three magnetic horns (horn1, horn2, and horn3 in order from upstream) are placed downstream of the target.
Each magnetic horn consists of two coaxial (inner and outer) aluminum conductors which encompass a closed volume.
The horns are designed for 320~kA pulsed current to maximize focussing of the secondary particles with low momentum
and high emission angle. The maximum magnetic field of 2.1~T is generated at 320~kA operation.
At the rated current, neutrino flux at the far detector can be increased by a factor of 16, at
neutrino spectrum peak of 0.6~GeV, compared to that without horn focussing.
The aluminum conductors are cooled by sprayed water.

\begin{figure}
        \centering
        \includegraphics[width=0.8\linewidth]{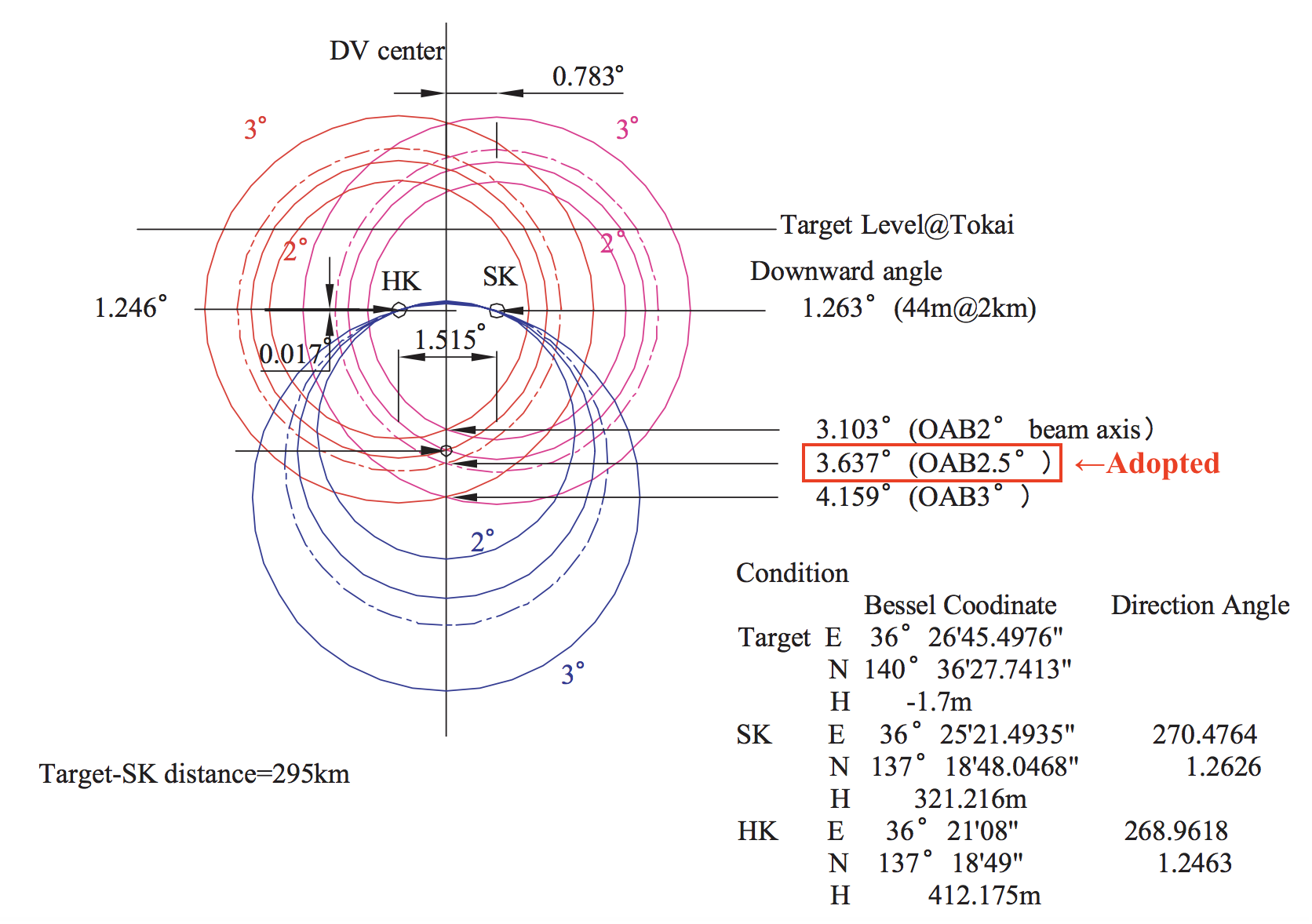}
        \caption{\small Schematic directional view of the Super-Kamiokande and Hyper-Kamiokande from the target~\cite{InterimJNU:2005}.
        The actual location of Hyper-Kamiokande can be different from this coordinate.}
        \label{fig:DV_geometry}
\end{figure}

\begin{figure}
        \centering
        \includegraphics[width=0.5\linewidth]{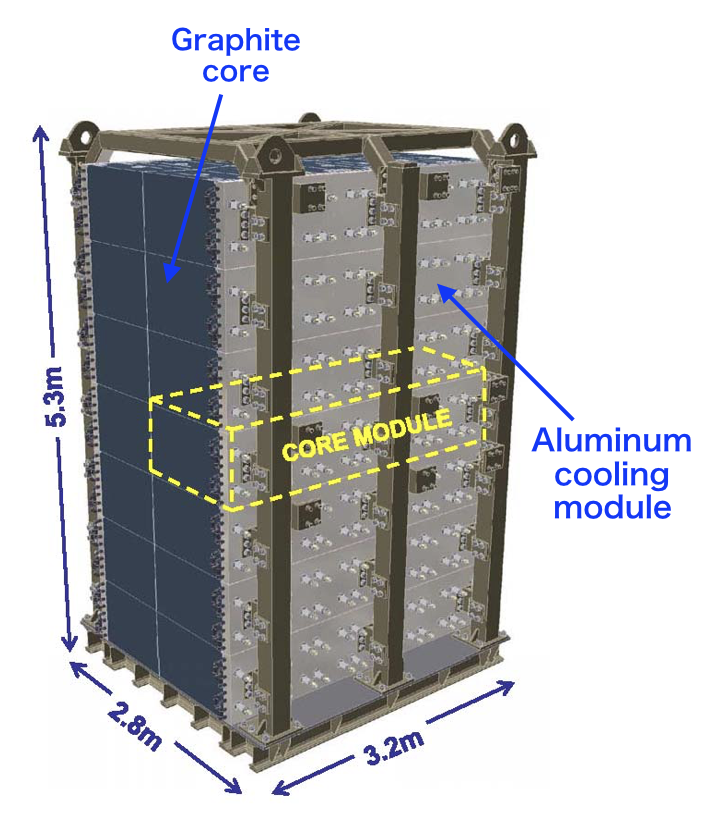}
        \includegraphics[width=0.3\linewidth]{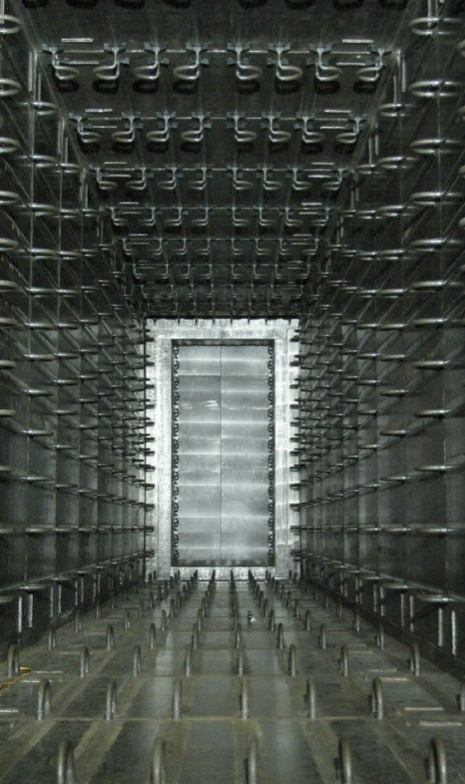}
        \caption{Schematic figure (left) and picture (right) of the Beam Dump.}
        \label{fig:BeamDump_pic}
\end{figure}

\begin{figure}
        \centering
        \includegraphics[width=0.4\linewidth]{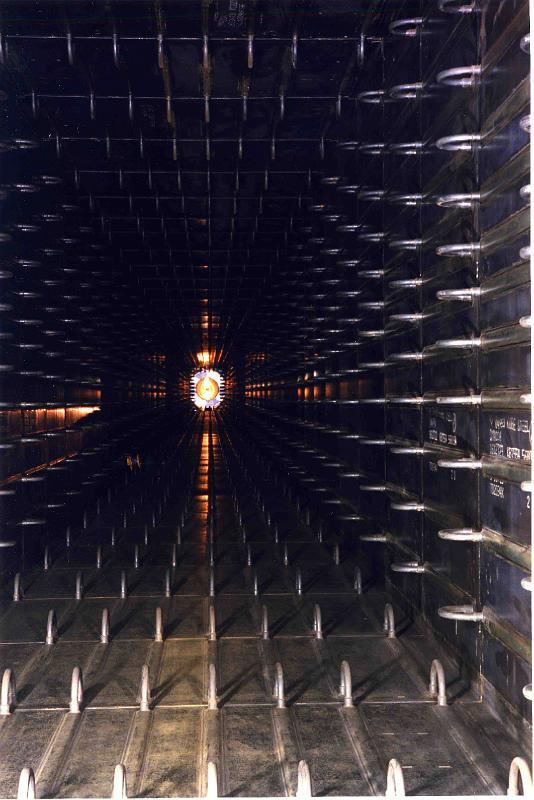}
        \caption{Picture of the Decay Volume.}
        \label{fig:DecayVolume_pic}
\end{figure}

The Decay Volume is a 94~m-long tunnel with a vertically elongated rectangular cross section allowing
variation of the off-axis angle to the far detector from 2.0$^{\circ}$ to 2.5$^{\circ}$.
The secondary beamline is directed 3.637$^{\circ}$ downward to have same off-axis angle of 2.5$^{\circ}$ 
to both Super-Kamiokande and Hyper-Kamiokande where it is proposed to be located at 8~km south of 
the Super-Kamiokande, as shown in Fig.~\ref{fig:DV_geometry}.
The Beam Dump, located at the end of the Decay Volume, is composed of graphite blocks with aluminum water
cooling modules attached, as shown in Fig.~\ref{fig:BeamDump_pic}.
The whole structure of the helium vessel is composed of 10~cm-thick iron plates where water cooling channels,
called plate coils, are welded on the wall to cool the wall and surrounding concrete shielding below 100~$^{\circ}$C, as shown in Fig.~\ref{fig:DecayVolume_pic}.
The helium vessel and the Beam Dump, which are inaccessible due to the high radioactivity after beam exposure,
are designed to survive thermal stress from 3$\sim$4~MW beam. For the Target Station helium vessel,
temperature increase of the iron wall should be limited to 30~$^{\circ}$C in order to suppress thermal expansion of
the wall below 1~mm, which can cause an uncertainty on alignment of the target and the horns.

In addition to these beamline components, there are three facility buildings for the neutrino beamline.
Equipment installed in each facility building are summarized in Tab.~\ref{tab:NU123}.
\begin{table}
        \centering
        \small
        \caption{\small Three facility buildings for the neutrino beamline.}
        \begin{tabular}{ll}
        \hline\hline
        Building & Installed equipment   \\\hline
        NU1      & Magnet power supplies for normal- and super-conducting magnets \\
                 & Cryogenic system for super-conducting magnets \\
                 & Water circulation system for magnets \\
                 & Control room for upstream beamline equipment and beam monitors \\\hline
        NU2      & Power supplies for normal-conducting magnets and magnetic horns \\
                 & Water circulation system for magnets \\
                 & Water disposal system for equipment in the Target Station \\
                 & Air ventilation system for downstream beamline tunnel \\
                 & Control room for downstream beamline equipment \\\hline
        NU3      & Air ventilation system for underground area (Beam Dump and Muon Monitors) \\
                 & Water circulation system for Beam Dump and downstream half of Decay Volume \\
                 & Water disposal system for Beam Dump and downstream half of Decay Volume \\
        \hline\hline
        \end{tabular}
        \label{tab:NU123}
\end{table}

Some additional information on the baffle and Beam Dump are described in the following part.

\begin{figure}
        \centering
        \includegraphics[width=0.5\linewidth]{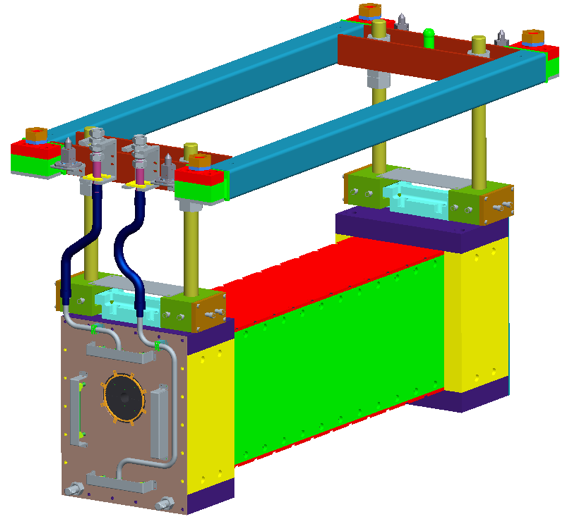}
        \caption{T2K Baffle.}
        \label{fig:baffle-1}
\end{figure}

\paragraph{Baffle}
The baffle is situated just downstream of the primary proton beam window and upstream of the target.
The main functions of the baffle are:
\begin{enumerate}
\item To protect the target, magnetic horns and the Beam Dump from a mis-steered beam
\item To reduce activation of components upstream of the target
\end{enumerate}
The core of the baffle consists of a large block (290$\times$400$\times$1700~mm) of Carbon Lorraine
2191, isotropic graphite. The block has a 30~mm bore for the proton beam drilled at the off-axis
angle of 2.5$^{\circ}$ (3.637$^{\circ}$ to gravity). To protect the graphite and reduce future
radiological contamination the graphite is clad in zinc coated steel plates. The baffle incorporates
thick steel shielding blocks which fit into the Target Station shielding to create a labyrinth seal
and reduce radiation backscattering from the target to the final focusing section. A model of
the baffle is shown in Fig.~\ref{fig:baffle-1}.

\begin{figure}
        \centering
        \includegraphics[width=0.5\linewidth]{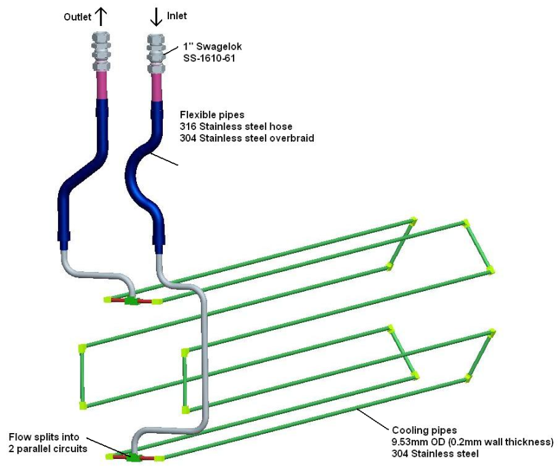}
        \caption{Baffle cooling circuit comprising of two parallel flows.}
        \label{fig:baffle-2}
\end{figure}

During normal operation at 750~kW it is estimated that the baffle will absorb around 2~kW of heat.
This heat is removed using water flowing through thin walled (0.2~mm) 304 stainless steel cooling
tubes embedded in the graphite. A good thermal contact is achieved using Hi-Therm HT-1205 thermal
interface material and clamping the pipes using the steel plates. The volume of water in the cooling
circuit has been minimized to reduce the activation and tritium production in the cooling water circuit.
The water splits into two parallel channels as shown in Fig.~\ref{fig:baffle-2}.
The baffle has an array of thermocouples around the bore at each end to monitor the baffle temperature
and to give an indication of the proton beam position. At 485~kW the maximum baffle temperature is
62~$^{\circ}$C at the downstream end and the ambient temperature around the baffle is 47~$^{\circ}$C.

In the event of a full beam strike on the baffle, around 700~kJ of heat is deposited.
Heat deposition calculation for the target, which has a slightly higher density graphite predicts
a maximum heat load of 200~J/g/spill. This will locally raise the temperature of the baffle by
around 200~$^{\circ}$C. Due to the low density, low thermal expansion and low elastic modulus properties
of graphite, it is one of the best materials to deal with this thermal shock as demonstrated
by the target.

At 1.3~MW the protons per pulse will remain the same as at 0.75~MW so the thermal shock and temperature
jump from a direct beam strike will remain the same. The steady-state temperature of the baffle will
rise as the integrated heat load and ambient temperature at the center of the baffle since the graphite
used for the baffle has a maximum service temperature of 427~$^{\circ}$C in air.

\begin{figure}
        \centering
        \includegraphics[width=0.5\linewidth]{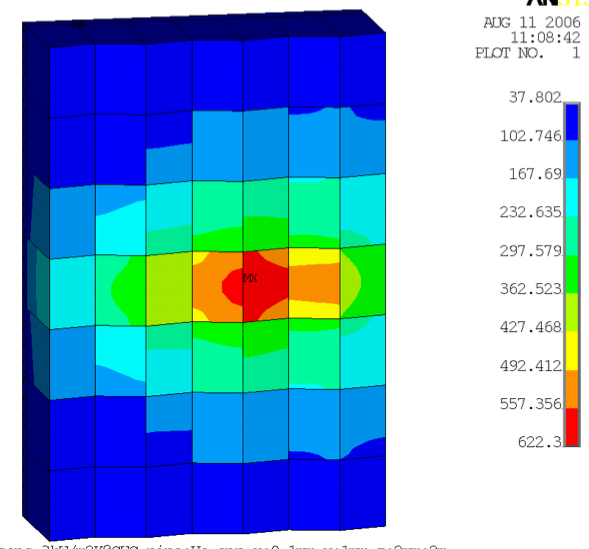}
        \includegraphics[width=0.5\linewidth]{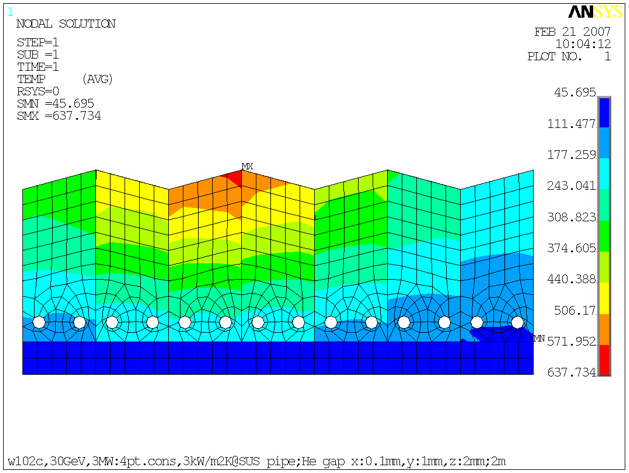}
        \caption{Plot of temperature distribution along center division of Beam Dump graphite
        core (top) and plan view of one half-layer (above) at 3~MW operation. Simulated using ANSYS
        Finite Element Analysis and MARS codes for a 30~GeV, 4.2~mm RMS proton beam disrupted
        by a two interaction length graphite target.}
        \label{fig:BD-1}
\end{figure}

\paragraph{Beam Dump}
The remnant proton beam is deposited in the Beam Dump at the downstream end of the Decay Volume.
The Beam Dump is required to dissipate approximately one third of the total beam power and the
core comprises a 3.2~m length of graphite followed by a total length of 2.4~m steel.
The Beam Dump core is contained within the helium vessel in order to minimize oxidation of
the graphite at the operating temperature, to minimize activation of the surrounding air and
to avoid the technical risk of a large beam window at the downstream end of the Decay Volume.
Due to its high radioactivity after operation and inaccessibility it is not possible to repair
or replace the Beam Dump. Consequently it must accommodate the full anticipated beam power and
survive the lifetime of the facility. The graphite core is 2~m wide to accommodate the majority
of the disrupted hadron shower and is 4.7~m high in order to permit the facility to accommodate
tuning of the off-axis angle between 2.0$^{\circ}$ and 2.5$^{\circ}$. As for the target, graphite
is the material of choice for the Beam Dump as it is robust to thermal shock and able to operate
at temperatures up to the level where oxidation from the trace oxygen in the helium atmosphere
may become an issue. Graphite has an excellent track record in this area e.g. for the CNGS
facility and many other beam intercepting devices. The T2K Beam Dump has 14 separate modules
arranged in seven layers to make up the full height, with the 2~m width divided in a zig-zag
along the central axis to reduce thermal stresses while blocking any clear beam path through
the center. Each half-layer comprises seven extruded graphite blocks bolted at the outer sides
to a cast aluminum cooling module which runs the full length of each layer. The cooling modules,
14 in total, incorporate steel cooling pipes into the aluminum castings. This external water
cooling arrangement minimizes activation of the water but dies mean that the core of the Beam
Dump will operate at an elevated temperature at high beam powers. Efficient heat transfer
between the graphite blocks and the cooling plate is realized by a flatness tolerance of 0.1~mm
for each of the cooling/loading surfaces, and a fastening and support system that permits
thermal expansion while maintaining the necessary surface-to-surface clamped contact.
Fig.~\ref{fig:BD-1} shows the results from a ANSYS Finite Element Analysis of the temperature
distribution of a half model of the Beam Dump using input from the MARS code. 
The simulation
modeled the effect of thermal conduction gaps resulting from thermal expansion between
the module layers, and due to the relative expansion between the aluminum cooling module and
the graphite blocks. An iterative process resulted in the selection of a realistic allowance
of a 0.1~mm helium gap between the graphite blocks and cooling module, a 1~mm horizontal helium
gap between the layers and a 2~mm gap between the separate blocks along the absorber length.
This resulted in a maximum temperature of around 620-640~$^{\circ}$C at the center at 3~MW
beam power. The radiation damage rate for the disrupted beam is sufficiently low that
a negligible degradation in thermal conductivity is anticipated for the lifetime of
the experiment.

\subsubsection{Operation status}

\begin{figure}
        \centering
        \includegraphics[width=0.9\linewidth]{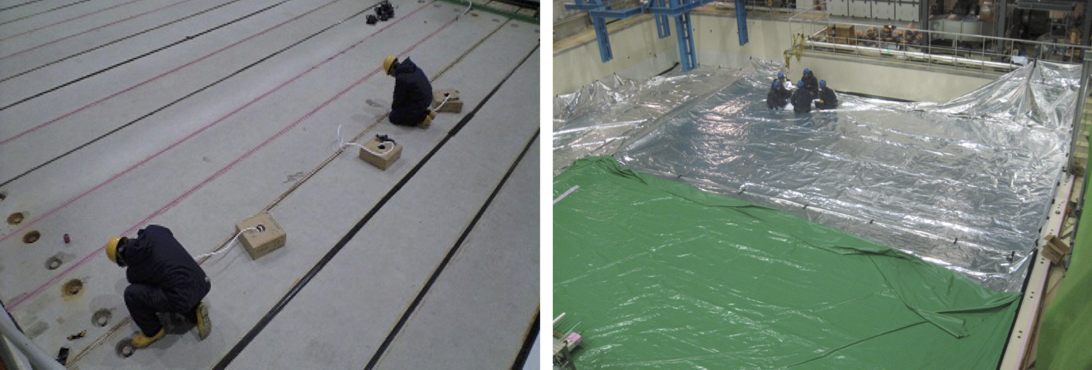}
        \caption{\small Pictures of the air-tight schemes in the Target Station ground floor.
        Left: Caulking between the concrete blocks. Right: Air-tight sheet on the concrete blocks.}
        \label{fig:TS_air_tight}
\end{figure}

Since the beginning of the beamline operation from 2010, the secondary beamline has been stably operated so far.
However, many problems, though not critical,
have occured and have been
solved. The most significant problem in the early stage around 2010-2013 was due to radioactive air exhaust.
Several short-lived radioisotopes such as $^{41}$Ar (its half-life is $\sim$2 hours) are produced by 
beam exposure around the outside of the helium vessel and gradually diffused to upward. 
In the Target Station the ground floor and underground area are separated by three layers of concrete 
shielding blocks. Since there are many gaps between the concrete blocks, 
the diffused radioisotope can reach the ground floor where the inner air is always ventilated with flow rate
of 13,000~m$^3$/hour to keep the inside pressure below the atmospheric pressure, and therefore the radioactive air
can be exhausted to the outside. In order to contain the radioactive air inside the underground area
all the gaps between concrete blocks are filled with caulking material and moreover a couple of layers of air-tight 
sheets are overlaid 
on the surface of the upper-most layer of concrete blocks as shown in Fig.~\ref{fig:TS_air_tight}. 
The air ventillation system was also modified to allow
a by-pass operation, by which the air flow to the ground floor is reduced to only 1/10 of all the flow
in order for the short-lived radioisotopes to stay underground and to decay. 
Thanks to these countermeasures the current radioactivity in the exhaust air is 0.5~mBq/cc level at 470~kW operation, 
while 9~mBq/cc is allowed based on the regulation. The radioactive air problem is not currently an issue to limit the beam power.     

Another challenging issue is erosion of water circulation loop made of iron.
The helium vessel is made of iron and its cooling water directly touches the iron surface that is gradually eroded.
There are several redundant filters in the circulation loop but they are often clogged by iron powder.
In the current operation scheme, filters are switched to the other ones once per a couple of weeks
and finaly whole set of filters must be replaced with new ones. The clogged filters must be treated as 
radioactive wastes. This is not the issue which limits the acceptable beam power, however,
frequent maintenance is needed in order to stably operate the beamline.

\subsubsection{Current acceptable beam power and upgrade plan}

The current acceptable beam power relevant to the secondary beamline components is summarized in 
Tab.~\ref{tab:current_acceptable_beam_power}.
\begin{table}
        \centering
        \small
        \caption{\small Summary of the current acceptable values for beam operation.}
        \begin{tabular}{llr}
        \hline\hline
        Component          & Limiting factor       & Current acceptable value           \\\hline\hline
        Target             & Thermal shock         & 3.3$\times 10^{14}$ protons/pulse  \\
                           & Cooling capacity      & 0.9~MW                             \\\hline
        Beam window        & Thermal shock         & 3.3$\times 10^{14}$ protons/pulse  \\
                           & Cooling capacity      & 0.75~MW                            \\\hline
        Horn               & Conductor cooling     & 2~MW                               \\
                           & Stripline cooling     & 0.75~MW                            \\
                           & Cooling capacity      & 0.98~MW                            \\
                           & Hydrogen removal      & 1~MW                               \\
                           & Operation current and cycle & 250~kA, 2.48~s               \\\hline
        Target Station helium vessel & Thermal stress & 4~MW                            \\
                           & Cooling capacity      & 1~MW                               \\\hline
        Decay Volume       & Thermal stress        & 4~MW                               \\
                           & Cooling capacity      & 1~MW                               \\\hline
        Beam Dump          & Oxidization of graphite blocks & 3~MW                      \\
                           & Cooling capacity      & 1~MW                               \\\hline
        Radiation          & Radiation shielding   & 0.75~MW                            \\ 
                           & Radioactive water disposal& 8.4$\times 10^{20}$~POT/year   \\\hline\hline
        \end{tabular}
        \label{tab:current_acceptable_beam_power}
\end{table}
The target and the beam window, where proton beams intersect, are originally designed for the intensity
of 3.3$\times 10^{14}$ protons/pulse, therefore the proposed 3.2$\times 10^{14}$ protons/pulse for 1.3~MW should
not be problem. The helium vessel iron structure is designed to survive the thermal stress from 4~MW beam.
The Beam Dump graphite core, whose limitation is due to oxidization of graphite blocks, 
is able to accept up to 3~MW beam with an assumption that oxygen contamination around the Beam Dump is kept below 
100~ppm. In most of the components, their cooling capacity is designed for 750~kW
and can be improved by increasing cooling water flow rate. 
Radioactive water disposal especially for tritium is the most limiting factor at this moment. 
Improvement on the disposal method should be realized even before 750~kW beam.

The equipment that cannot currently accept 1.3~MW beam will be upgraded according to the proposed
timeline for the accelerator upgrade. The upgrade items required for 1.3~MW are as following.
\begin{itemize}
\item Improvement on target helium cooling by increasing helium flow rate which requires higher pressure
tolerance of the titanium container up to $\sim$0.5~MPa.
\item Development of the new beam window which accommodate the higher beam power by optimizing window thickness and
material.
\item Upgrade of horn electrical system to increase the horn current to the rated current of 320~kA at 1~Hz operation.
\item Improvement of a hydrogen removal system for safe operation at higher beam power.
\item Improvement on horn stripline cooling.
\item Improvement on cooling capacity for all the secondary beamline components.
\item Improvement on radiation shielding by installing additional concrete blocks.
\item Upgrade of radioactive water disposal system.
\end{itemize}
The details of the upgrade for each component will be described in the later sections. 

\graphicspath{{figures/main_secondary}}

\subsection{Target
}

\subsubsection{Target Design Overview and Current Operation Status}
\label{sec:targetintro}

The main part of the production target is made of a graphite rod and
installed inside the first electro-magnetic horn so that the pion
produced at the target should be focused by the toroidal magnetic
field of created the electromagnetic horns. According to the
optimization to maximize the neutrino flux based on the Monte Carlo
simulation~\cite{Ichikawa:2012horn}, the length and diameter of the
graphite target are determined to be $\sim$900~mm ($\sim$2 interaction
length) and 26~mm, respectively, where the inner diameter of the
corresponding horn is 54~mm and the optimum beam size at the target is
$\sigma_{x}=\sigma_{y}=$4.2~mm. 

The material of the target is chosen to be robust against the
instantaneous heat generation due to the interaction with the
fast-extracted proton beam from J-PARC MR.  The polycrystalline
nuclear graphite used is a low Z, low modulus, low thermal expansion
coefficient and relatively high thermal conductivity refractory
material making it particularly resilient to the intensely pulsed
proton beam.  Several commercial isotropic graphite with enough
production size is surveyed, and the isotropic graphite, IG-430 by
Toyo. Tanso Co. ltd., is adopted because it has the higher thermal
shock resistance~\cite{TUedaThesis:2004}. The density and tensile
strength of IG-430 are 1.8~g/cm$^3$ and 37.2~MPa, respectively. The
heat load due to the power deposition in the neutrino target is about
20~kW for 750~kW proton beam power. The maximum temperature raise due
to 30~GeV $3.3\times10^{14}$~protons/pulse is estimated to be as about
200~K using MARS, and it corresponds to the equivalent stress of
7.2~MPa. IG-430 satisfies the requirements for the neutrino target
material, because the safety factor for this stress considering cyclic
fatigue is 3.5~\cite{TUedaThesis:2004}.

\begin{figure}
        \centering
        \includegraphics[width=0.75\linewidth]{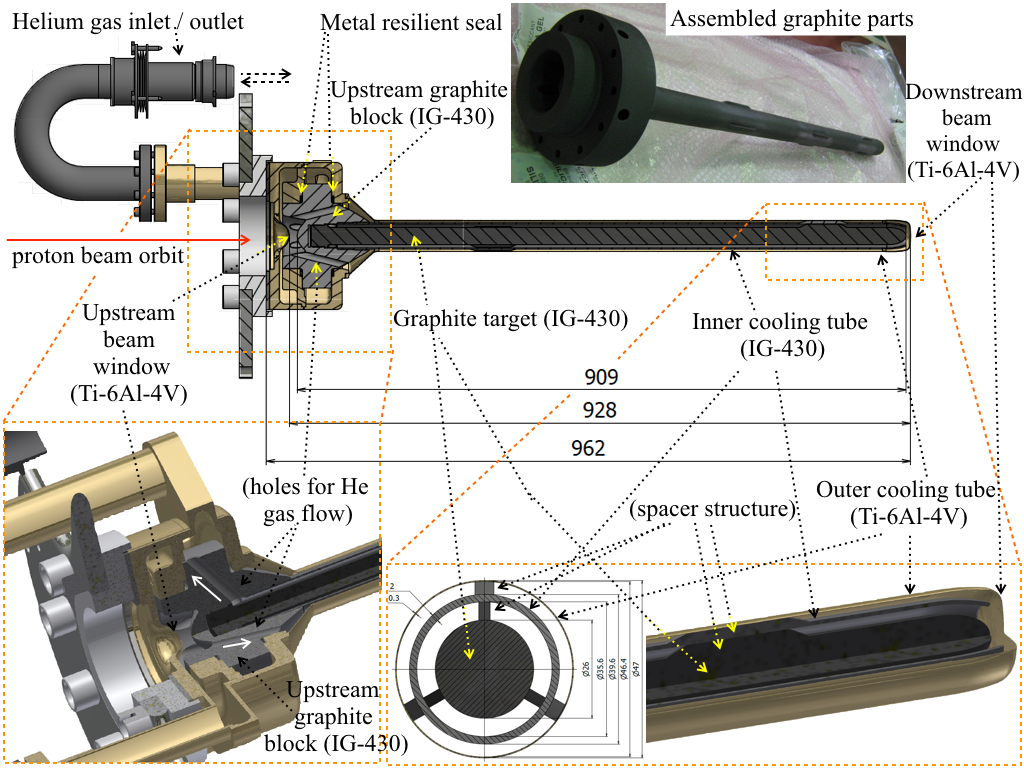}
        \caption{\small The picture of the graphite part of first version of the target, and the drawing of the structure of neutrino target shown in section.}
        \label{fig:targetver1structure}
\end{figure}

The expected radiation damage of the isotropic graphite used for 30~GeV
750~kW proton beam estimated using MARS is 0.25
DPA/year~\cite{Nakadaira:2008zz}. 
It \color{\MODCOLORB} could \color{black} cause that the target length
shrinks by $\sim$0.5\%(=5~mm) after 5 years beam
operation~\cite{JAERI:M91:153}. The property changes due to radiation
damage such as the degradation of thermal conductivity strongly
depends on the irradiation temperature, and it is relaxed when the
irradiation temperature is 400$\sim$800~$^\circ${}C. Less than 5\% of
the beam power is deposited as heat in the target, and this enables it
to be cooled by gaseous helium.  Helium cooling permits the graphite
to run at an elevated temperature thereby reducing the effects of
radiation damage, minimizes activation of the coolant and eliminates
any pulsed-beam induced shock waves that would be generated in an
incompressible liquid coolant such as water. Considering the dimension
change of the graphite and enabling that the target can be dismantle
of the horn relatively easier, the graphite part is sealed inside a
titanium alloy container and cantilevered within the bore of the first
magnetic horn. Fig.~\ref{fig:targetver1structure} shows the picture
of the graphite part and the drawings of the structure of the first
version (v1) of the production target.

In order to prevent oxidation of the graphite from any trace oxygen in
the target station, it is sealed within a thin titanium alloy
container which includes thin single skin entry and exit windows. The
target and its container walls are cooled by a single circuit of high
purity, high velocity helium. The alloy Ti-6Al-4V is used for the
container and windows since it has a relatively high strength and heat
capacity and low thermal expansion coefficient making it one of the
few structural materials able to withstand the shock wave stresses
generated within it by the pulsed proton beam. It is also known to
retain its mechanical properties albeit with a reduction of ductility
at proton fluence up to $10^{20}$~p/cm$^2$. However as with all
metals, titanium loses strength at elevated temperatures and it is
necessary for the helium to cool both the entry and exit windows
before cooling the target rod without generating an excessive pressure
drop over the complete circuit. 
The helium inlet enters an annular buffer volume in the target head
before flowing through a gap across the entry window. The window
profile and thickness minimizes the combination of stresses resulting
from the differential gas pressure across the window, the transient
thermal stress and the shock wave stresses resulting from the bunch
structure. A thick outer plate accommodates the pressure load and
tapers down to a thin ($<0.5$~mm thick) partial sphere at the centre
which is inverted to improve cooling and direct the helium flow. The
helium then flows through 6 angled holes in the graphite head to an
outer annular channel. The helium cools the titanium outer tube before
it performs a 180$^{\circ}$ turn as it cools the downstream window to
return along an inner annular coaxial channel separated from the outer
channel by a 2 mm thick graphite tube bonded into the graphite
head. The helium cools the target rod then flows out through 6 angled
holes in the graphite head which are interspaced between the 6 inlet
holes, to an outlet annular manifold. The graphite parts is fixed to
the Ti-alloy parts by bolts with the metal seals which require low
clamping force, Metal Resilient Seal by Mitsubishi cable industries,
Ltd.

The target assembly is supported as a cantilever from the upstream end
and is sufficiently rigid that the end deflection both due to gravity
and/or an off-centre beam pulse is less than 1 mm which is deemed
acceptable for uniformity of pion production. The target support is
accurately aligned and normal to the horn axis so that targets are
automatically installed on the horn centre line. The target supports
and helium pipe connections are isolated from the target itself. 
which is capacitively coupled to the horn. 
To avoid the target charging up,
the target is connected to the ground using the series of metal thin
film resistors ($\sim$4~M$\Omega$ in Total) that are embedded in the
ceramic parts. 
\color{\MODCOLORB} There is also a possibility that an electrical connection between the horn and the target 
by ionization of the intervening gas layer by beam although we didn't observe any discharge effect 
in the beam operation up to now.  \color{black}
The support system has been designed to permit failed
targets to be replaced as described in Sec.~\ref{sec:targetremoteexchange}.

The flow rate of helium gas for 750kW proton beam operation is
determined to 25~g/s (560~Nm$^{3}$/h) so that the maximum He
temperature at the outlet is less than 200~$^\circ$C
($\Delta{}T=170$~K). The steady temperature of the target and Ti-alloy
beam window is estimated to be 866~$^\circ$C and 114~$^\circ$C,
respectively, by computational fluid dynamics method using ANSYS CFX
considering the degradation of the thermal conductivity of the
graphite. The pressure drop at the target is estimated to be
0.055~MPa. The currently He circulation system has 30\% margin, and
maximum flow rate of He gas is 32~g/s that corresponds to the He velocity
at the target is about 230~m/s. As of 2014 June, the He flow rate
during the beam operation is set to about 14~g/s, because the J-PARC
MR beam power is $\sim$220~kW, and the measured temperature raise of He gas is
about 65~K. The compressor that is used for He circulation is based on
the Wing Compressor HB-2640TDBQ produced by Hori Engineering Co.,
LTD. The current circulation system is designed to supply helium with
the maximum pressure of 0.2~MPa at the compressor outlet.

The first target (v1a) was used from April 2009 until May 2013, and
exposed to $6.7\times10^{20}$ protons-on-target without significant
trouble. The first target is replaced with the second target (v1b)
that is made with same design as the first one. As of 2017 April, the
second target is exposed to \color{\MODCOLORB}$2.3\times10^{21}$\color{black} protons-on-target with
the maximum beam power of 500~kW. Because the produced neutrino flux,
and by-product muon flux has been stable so far, there is no
indication of the graphite core part so far. For the second target,
the trouble happened at the insulator part of the He tube. it will be
described in Sec.~\ref{sec:targethetubetrouble}.

\subsubsection{Target helium pipes and isolators}
\label{sec:targethetubetrouble}

\begin{figure}
        \centering
        \includegraphics[width=0.5\linewidth]{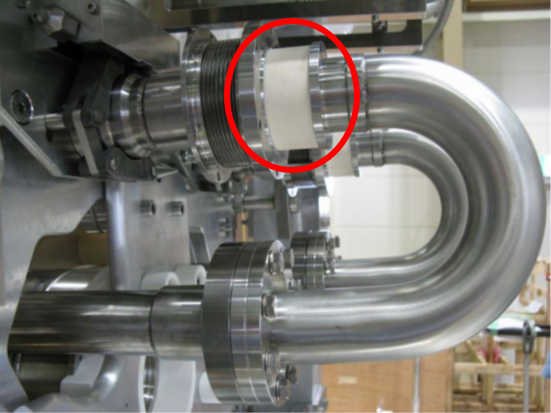}
        \caption{\small Target helium pipe (Location of helium leak highlighted)}
        \label{fig:tgt_utube}
\end{figure}

The first target operated very successfully and was exchanged along
with the first horn, and the second target has now experienced 2$\times 10^{21}$
protons. The only issue has been the failure of one ceramic isolator
in the hot helium return line from the target at around 450 kW beam
power. In June 2015 a helium leak was observed in the target cooling
system. The horn was moved to the Remote Maintenance Area (RMA) and
the source of the leak was confirmed as expected to be the ceramic
isolator in the outlet helium pipe as shown in
Fig.~\ref{fig:tgt_utube}. As the T2K target is electrically isolated
from the 1st horn it is necessary to electrically isolate the targets
helium inlet and outlet pipes. This is a challenging problem as the
isolators have to remain helium leak tight against the helium system
pressure in harsh environmental conditions (radiation and
radio-chemistry) and the hot helium return line in particular has to
withstand stresses from rapid thermal cycling resulting from beam
trips. The current method of joining ceramics is a diffusion bond
using a pure aluminum interlayer. The failed assembly was
successfully replaced with another diffusion bonded joint in an
operation not originally envisaged, using the target exchanger with a
suitable bespoke modification.

After considering the possible reasons for failure the following seemed the most likely candidates:
\begin{enumerate}
\item Failure of joint/ceramic from movement of stainless pipes (stress relieving)
\item Thermal fatigue failure of the diffusion bonded joint/ceramic.
\end{enumerate}
As the target pipes are fabricated using a 180$^\circ$ cold formed bend, it was considered that there may be high residual stresses in the stainless. The residual stress could relax over time and exposure to elevated temperatures and lead to movement. To test this some spare pipe from the same batch were measured and then exposed to 200~$^\circ$C for 64hrs. After heating the distance between pipe centers increased by 3.4~mm. As this is more than the lateral compliance of the bellows this was considered a possible reason for failure.

To mitigate this problem and more accurately fabricate pipes, welded mitre joints were used as an alternative to the pipe bends. Similar testing showed the welded mitres to be stable when heated. In addition to this, the components for the helium pipes were stress relieved after welding.

\begin{figure}
        \centering
        \includegraphics[width=0.5\linewidth]{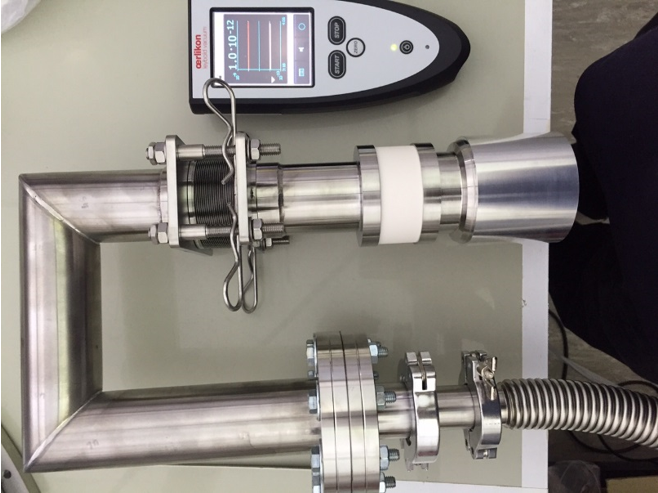}
        \caption{\small Replacement target pipe with mitre joints during helium leak testing}
        \label{fig:tgt_mitrejoint}
\end{figure}

To remove the possibility of stresses from welding affecting the diffusion bond, the diffusion bond cuff was welded to the bellows unit first. The resulting pipe used to repair the target in December 2015 is shown in Fig.~\ref{fig:tgt_mitrejoint}.

\begin{figure}
        \centering
        \includegraphics[width=0.5\linewidth]{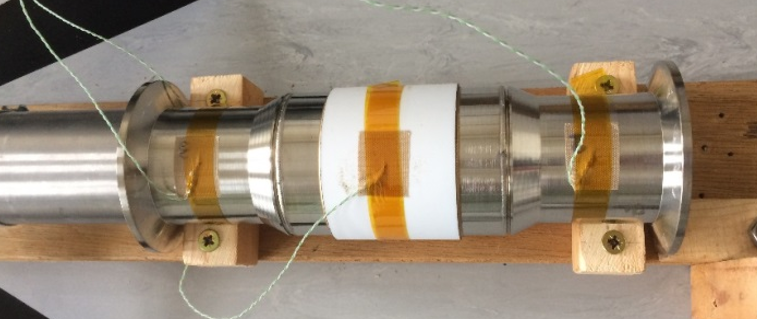}
        \caption{\small Brazed ceramic isolator during rapid cooling test}
        \label{fig:tgt_brazedceramic}
\end{figure}

Thermal fatigue of the ceramic joints is a more difficult and time consuming test to perform. When operating at 400~kW beam power, immediately after a beam trip the temperature of the helium gas falls at a rate of 50~$^\circ$C/min. This fast cooling rate can cause significant stresses in the joints and will become worse when operating at higher beam power. It is estimated in a typical year, the isolators may see around 5000 beam trips. An alternative brazed joint is being investigated and has performed well after several cycles of cooling from 180~$^\circ$C at the rate of 80~$^\circ$C/min (Fig.~\ref{fig:tgt_brazedceramic}). RAL is currently designing and building a thermal fatigue test rig to assess the long term performance of different joining techniques.

\color{\MODCOLOR}
For the operation 1.3~MW, the design criteria on the He temperature at the target outlet is $<200$~$^\circ$C, and it is not changed from the original 750~kW design. The thermal cycle stress is expected to be within the allowance of the current bonding technique. 
\color{black}

\subsubsection{Target Design changes}
\label{sec:tgtdesignchange}

The first and second targets (v1a and v1b) were manufactured by Toshiba and incorporated a bolted joint and a metal seal between the graphite head and titanium housing. RAL manufactured a version 2, currently a spare, which used a diffusion bonded aluminum wire. The latest target design iteration (v3 being manufactured by RAL) returns to the bolted joint and metal seal design in order to de-risk the target manufacture, simplify and take more control of the manufacturing process and streamline the schedule. The spacers between the target rod and inner graphite tube are streamlined for helium flow and are machined on the target rod to reduce the risk of failure and allow for a more precise fit in the tube.

The latest target has a 0.5~mm thick tube and beam windows. This small increase in thickness will allow the target to be fully evacuated safely and assist in the purification of the helium gas. During manufacture the 0.5~mm tube will be vacuum tested to ensure buckling does not occur. Latest beam window simulations have shown that 0.5mm titanium thickness has three times lower equivalent stress than 0.3~mm and is also valid for the target.

Finally the bore in the graphite head has been increased in size to allow the target rod to be inserted from the upstream side. This makes the design more resilient to an adhesive failure since the pressure differential will hold the target in position.

\begin{figure}
        \centering
        \includegraphics[width=0.5\linewidth]{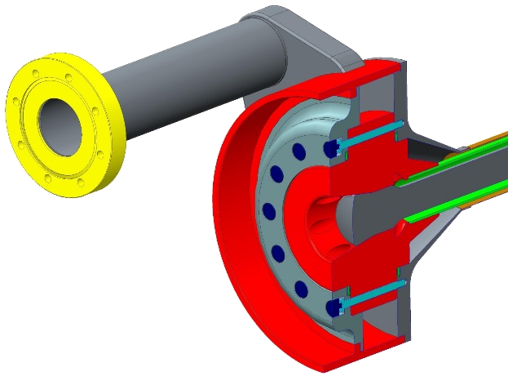}
        \caption{\small Version 3 target design schematics}
        \label{fig:tgt_v3design}
\end{figure}

\subsubsection{Upgrade to 1.3 MW}

Incremental developments to the target design are planned to enable it to operate at higher beam power, which will require higher helium pressure and result in higher temperature gradients. Tab.~\ref{tab:tgt_upgrade} shows the target pressure drop and operating temperature under a range of conditions. The top line of the table is the current design for 750~kW. It is shown that with a mass flow rate of 60~g/s the target core temperature is nominally the same 
\color{\MODCOLORB}($\Delta T \sim$50~$^\circ$C)\color{black} as the current design. Therefore oxidation of the graphite should be about the same as the current target if O$_2$ levels in the helium are similar. The table also shows how increasing the system pressure reduces the pressure drop and maximum velocity in the target. Keeping the pressure drop down is an important consideration for the helium compressors physical size and power requirements.

\begin{table}
  \centering
  \caption{Upgrade scenarios for 1.3~MW target}
  \label{tab:tgt_upgrade}
  \includegraphics[width=\textwidth] {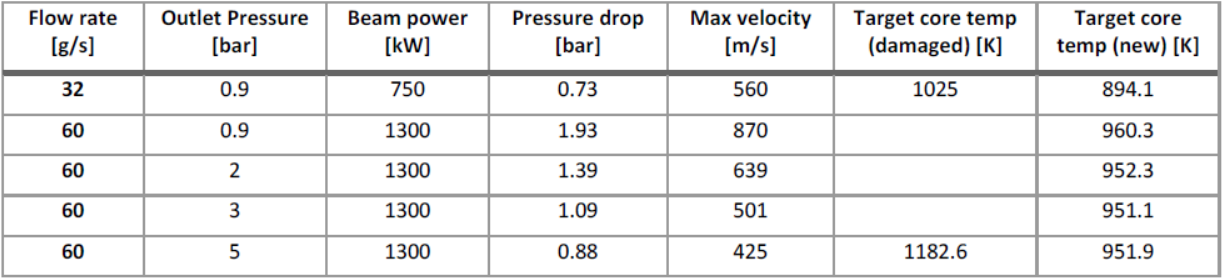}
\end{table}

\begin{figure}
        \centering
        \includegraphics[width=1.0\linewidth]{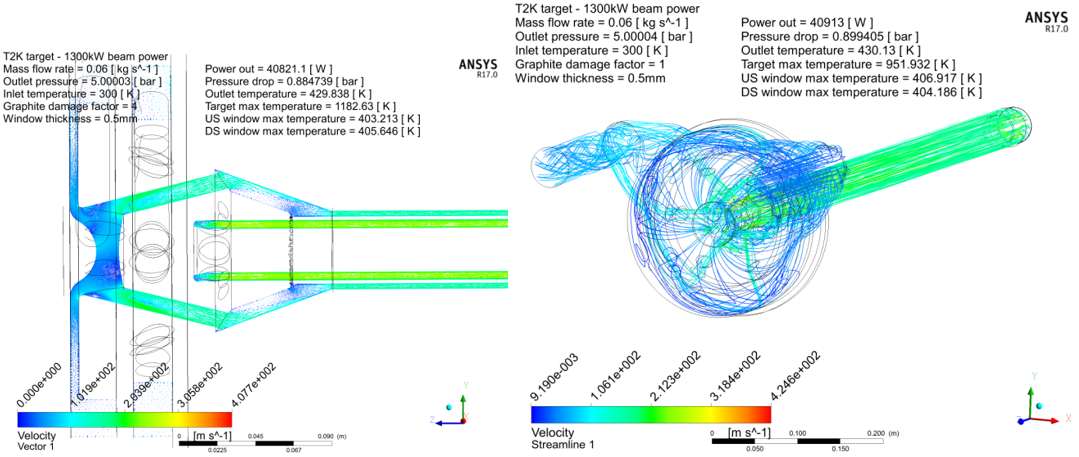}
        \caption{\small Velocity flow lines in current T2K target geometry operating at 5 bar outlet pressure and 1.3 MW beam power.}
        \label{fig:tgt_cfd}
\end{figure}

If the target outlet pressure is increased from 0.9 bar to 5 bar then it is possible to double the helium coolant mass flow rate without a significant increase in overall pressure drop. This means that in principle the existing target design may be able to dissipate the heat load deposited by a 1.3 MW beam. A velocity vector plot of a CFD simulation of this case is shown in Fig.~\ref{fig:tgt_cfd}. The objective will be to maximise the material safety factors in order to maximise target lifetime. New components and manufacturing methods will be prototyped and tested before the detailed design of the upgrade is completed and QA procedures developed. 

\begin{figure}
        \centering
        \includegraphics[width=1.0\linewidth]{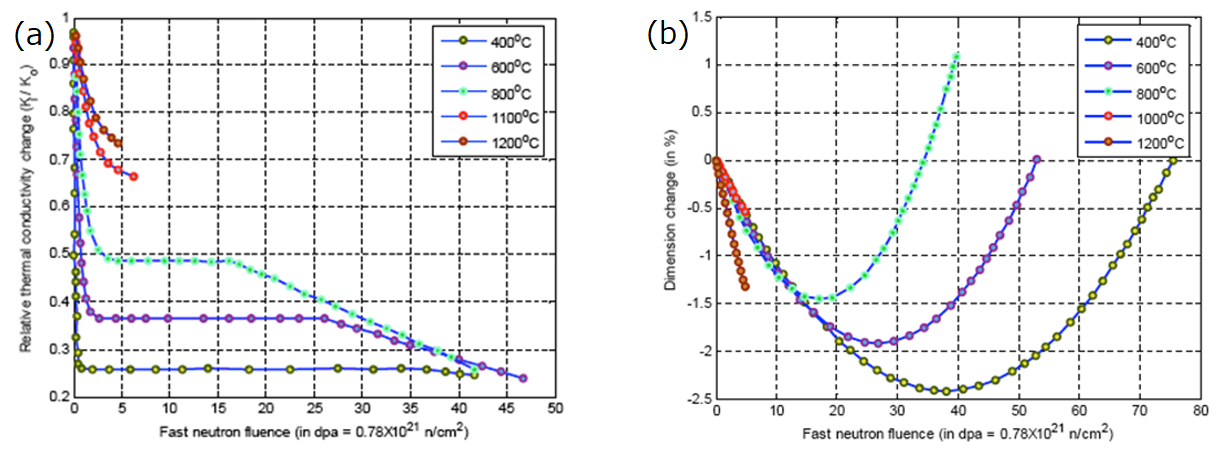}
        \caption{\small Fast neutron irradiated IG-110. (a) Reduction in thermal conductivity, (b) Dimensional change}
        \label{fig:tgt_graphitedamage}
\end{figure}

The elevated heat load also generates an increase in thermal gradients in the graphite, as shown in Fig.~\ref{fig:tgt_graphitedamage} for the casey where the thermal conductivity of the Toyo Tonso IG-43 graphite is reduced by a factor of 4 in line with radiation damage data for fast neutrons (for IG-110, similar to IG-430).

\begin{figure}
        \centering
        \includegraphics[width=0.8\linewidth]{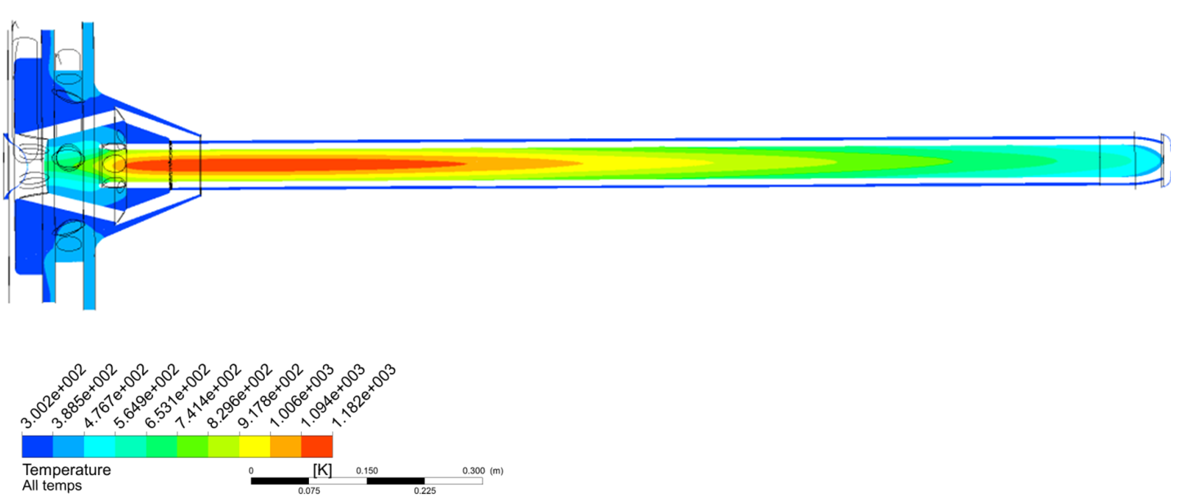}
        \caption{\small Temperature distribution operating at 5 bar helium pressure}
        \label{fig:tgt_5barsimtemp}
\end{figure}

\begin{figure}
        \centering
        \includegraphics[width=1.0\linewidth]{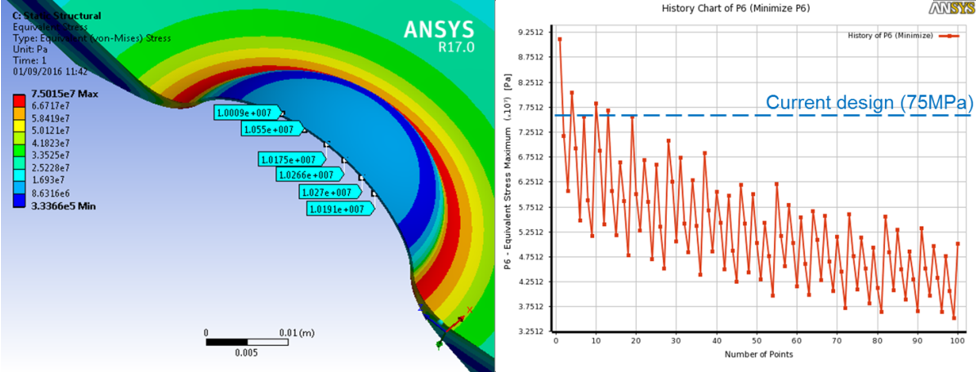}
        \caption{\small Von-Mises equivalent stresses in the existing upstream target window operating at 5 bar helium pressure (L) and the history chart (R) of the parameterised re-optimisation of the design}
        \label{fig:tgt_5barsimvms}
\end{figure}

This elevated pressure also generates an increase in the mechanical stresses in the titanium alloy enclosure, particularly the upstream and downstream windows. The upstream window has been studied and re-optimised using a parameterised model and a genetic algorithm. Figs.~\ref{fig:tgt_5barsimtemp} and \ref{fig:tgt_5barsimvms} show some results of these simulations. This shows that the stress can be halved with a relatively modest increase in the outer plate thickness, with no increase in the central dome thickness. The extra material is outside the beam footprint therefore will have no impact on pion production performance.

\begin{figure}
        \centering
        \includegraphics[width=0.5\linewidth]{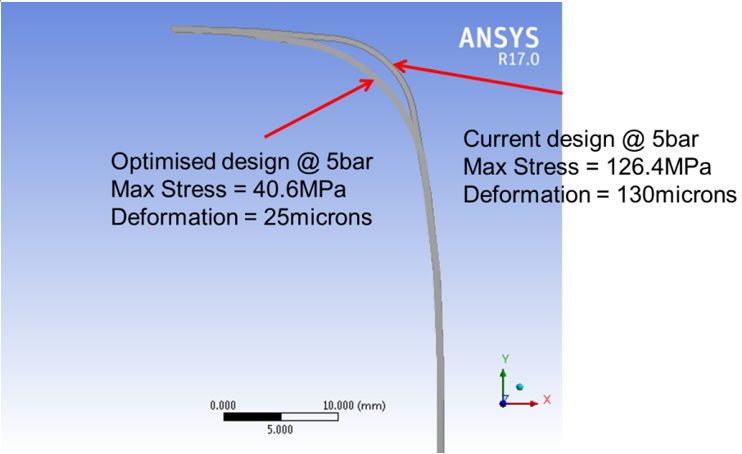}
        \caption{\small Design upgrade for profile of the target downstream window}
        \label{fig:tgt_dswindow}
\end{figure}

The downstream beam window is required to turn the high velocity helium flow by 180$^{\circ}$ without generating too high a pressure drop, and requires sufficient cooling at the centre despite a nominal zero helium velocity in that location. The current profile was found to have too high a stress generated by 5 bar helium pressure, and a re-optimisation has been carried out resulting in a new profile shown in Fig.~\ref{fig:tgt_dswindow}. This new profile reduces the pressure-induced stresses by a factor of three without any increase in the 0.5 mm material thickness.

As described in Sec.~\ref{sec:targetintro}, the current He circulation
system is designed to supply He gas up to 32~g/s with 0.2~MPa pressure
using the Wing Compressor HB-2640TDBQ produced by Hori Engineering
Co., LTD. It is necessary to upgrade the He circulation system so that
the flow rate of 60~g/s with 0.5~MPa pressure can be achieved for
1.3~MW beam. Because the compressor company produces the same type of
the compressor up to the flow rate of $\sim$70~g/s with 0.7~MPa
pressure as a catalogue product, no technical difficulty is expected.

\graphicspath{{figures/main_secondary}}

\subsection{Beam window}

The Target Station (TS) is a single helium environment maintained at atmospheric pressure, which contains the target, the baffle collimator, and the three magnetic horns etc. A beam window separates the TS chamber (1 atm He gas) from the vacuum of the primary beam line. In addition to withstanding this differential pressure, the window must survive intense heating and resulting thermal stresses from interaction with the pulsed proton beam.

\subsubsection{Window Design Overview}

\begin{figure}
        \centering
        \includegraphics[width=0.8\linewidth]{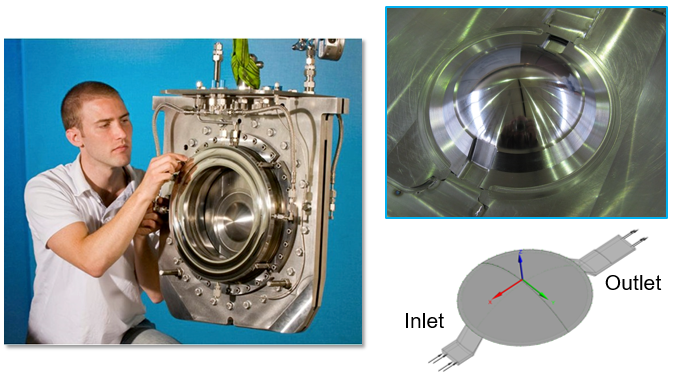}
        \caption{\small The beam window assembly. The 0.3~mm-thick doubled Titanium alloy domes 
                     are cooled by Helium gas flowing 2mm gap between them.}
        \label{fig:window_overview}
\end{figure}

Fig.~\ref{fig:window_overview} shows the beam window assembly. The window consists of two 0.3-mm-thick concentric partial hemispheres of a titanium alloy (Ti-6Al-4V), with forced convection helium flow through the 2 mm space between. A partial hemispherical shape has been chosen due to this shapes capacity to withstand pressure and its relative ease of manufacture. Sealing to the TS and the upstream beam monitor stack chamber is achieved using two inflatable bellows seals -“pillow seal”- for remote exchange capability.

Ti-6Al-4V was chosen as one of few viable candidate materials. When mitigating for thermal stress effects, a high ultimate tensile strength (UTS) and a low coefficient of thermal expansion (CTE) are desirable qualities. The UTS of Ti-6Al-4V at room temperature (RT) is extremely high (approximately 1,000 MPa) whilst CTE is relatively low (8.8$\times 10^{-6}$/K). Ti-6Al-4V also has a relatively low density (4.54), meaning that beam heating will be reduced compared with, say, inconel. The low thermal conductivity of titanium (7.1~W/mK) makes it unsuitable for high frequency beams, where the time-averaged power is high, but the low frequency of the J-PARC beam means that this is not a major concern. Beryllium was the other main candidate material, but titanium alloy was eventually chosen due to a slightly higher resistance to beam-induced thermal stress, and the toxicity of Beryllium may cause serious contamination problem in case of failure.

A thin window as possible was wanted initially so as not to disrupt the beam and to allow efficient cooling, but any thinner than 0.3 mm and pressure stresses would become significant. Fabricating the domes would also become much more challenging. 

\subsubsection{Stress and Cooling Analyses for 1.3 MW Operation}

The beam window will have three types of stress to cope with, each of which operates on a different timescale:
\begin{itemize}
 \item The static stress due to pressure, though this becomes insignificant by choosing a suitable shape and thickness.
 \item The quasi-static transient stress caused by the initial compression of the window material due to beam heating, 
          which is partially relieved between pulses as it is cooled.
 \item An elastic stress wave component that arises due to the short pulsed nature of the beam 
          and is most significant on the timescale of microseconds.
\end{itemize}
The elastic stress waves are an extra factor on top of the static and quasi-static stress and the magnitude of the stress waves will depend on the duration of the beam pulse in relation to the beam spot size. Note that, since the beam power is to be increased to 1.3 MW by reducing the beam cycle time and not by increasing the pulse intensity (the window was designed for original 3.3$\times 10^{14}$ protons-per-pulse), current design principles would basically work with rather minor upgrades suitable for high power operation. 

Due to the different timescales involved, which makes it impractical to study all phenomena together using finite element analysis because of time-step requirements, the quasi-static thermal stress and elastic stress waves have been examined in separate simulations.

\paragraph{Elastic Stress Waves due to a Single Pulse}

\begin{figure}
        \centering
        \includegraphics[width=0.8\linewidth]{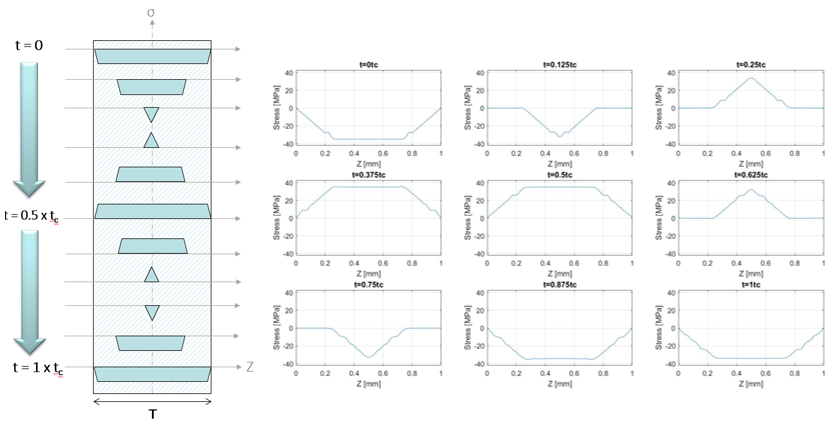}
        \caption{\small  (left) The schematic diagram illustrates the propagation of a stress wave through a window 
                    caused by a bunch at time t = 0. 
                   (right) Stress wave propagation along centre axis following one single 58 ns bunch at 1.3 MW beam power for 1 mm 
                   thick beam window ($t_c$ = 322 ns)}
        \label{fig:window_strw_schematic}
\end{figure}

With intense pulsed beams that create stress waves in targets and windows, certain combinations of material geometry and bunch structure can result in a type of “stress resonance” occurring. This is when stress waves produced by individual bunches constructively interfere with one another. In beam windows this can give rise to larger through-thickness longitudinal stress waves (those that reflect between inner and outer surfaces) than would otherwise occur. Consider a flat window of thickness $T$, as schematically illustrated in Fig.~\ref{fig:window_strw_schematic}. Upon interaction with a short proton pulse, compressive waves will initiate along the beam trajectory through window thickness, dissipate at the both surfaces, and be reflected as tensile wave. Characteristic time $t_c$ is the time taken for a stress wave to propagate through the window thickness and back:
\begin{equation}
t_c = \frac{2T}{c},  c = \sqrt{E/\rho}
\end{equation}
where $c$ is the speed of sound in the material, $E$ is the elastic modulus and $\rho$ is the density. The stress resonance will be maximum if $t_c$ is equal to the bunch spacing (and be minimum if $t_c$ is equal to half of bunch spacing). For 0.3 mm window, $t_c$ is ~100 ns, which corresponds to 10 MHz stress cycle.

\begin{figure}
        \centering
        \includegraphics[width=0.8\linewidth]{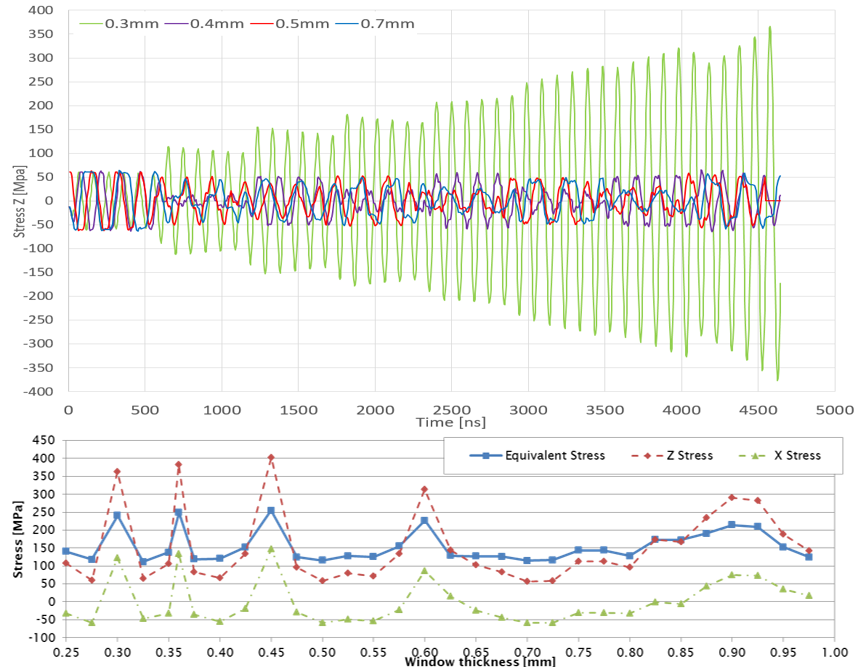}
        \caption{\small (upper) Stress in beam direction as function of time (ns) at window centre for 1.3 MW beam operation, 
                    with window thickness = 0.3, 0.4, 0.5, and 0.7~mm. 
                   (lower) Stress in the beam window as a function of thickness, after the final (8$^{\rm th}$) bunch of a full beam spill 
                    at 1.3 MW operation. Z Stress represents through thickness stress and X Stress shows radial stress, respectively.}
        \label{fig:window_strw_simulation}

\end{figure}

These wave profiles are evident in the ANSYS simulations using the latest beam bunch structure at 1.3 MW operation. To simulate the dynamic effect of a pulsed proton beam on the window, an ANSYS Multiphysics
model with coupled field elements has been created. To reduce the computational load, an axisymmetric model has
been approximated for simulations. Upper side of Fig.~\ref{fig:window_strw_simulation} shows stress in beam 
direction as function of time at window center for 1.3 MW beam operation with three different window thicknesses. Constructive interference of bunch structure (8 bunches) is visible for existing 0.3 mm thickness, where the maximum stress reaches to $S_Z = \sim270$ MPa. When the thickness of 0.3 mm was considered as optimal, original simulations had too few elements and too large time steps leading to sub-optimal thickness. Lower side of Fig.~\ref{fig:window_strw_simulation} gives the maximum accumulated stress as function of thickness. 
The plots exhibit nice agreement to each other. In order to increase the tolerance and safety margin for the bunched-beam induced stress wave resonances, the window thickness can be chosen to be at one of these troughs in the plot. 

In the real accelerator operation, there will be spill-by-spill fluctuation on beam bunch timing. Meanwhile, the analysis looks at the worst case scenario where the time between bunches is kept perfectly constant so that, at certain thicknesses, the stress waves can superimpose with each bunch. This is the constructive interference we see. If there was any fluctuation in bunch timing from bunch to bunch this would only improve things by disrupting the interference of consecutive bunches.

\paragraph{Transient Stress and Temperature over Multiple Pulses}

\begin{figure}
        \centering
        \includegraphics[width=0.8\linewidth]{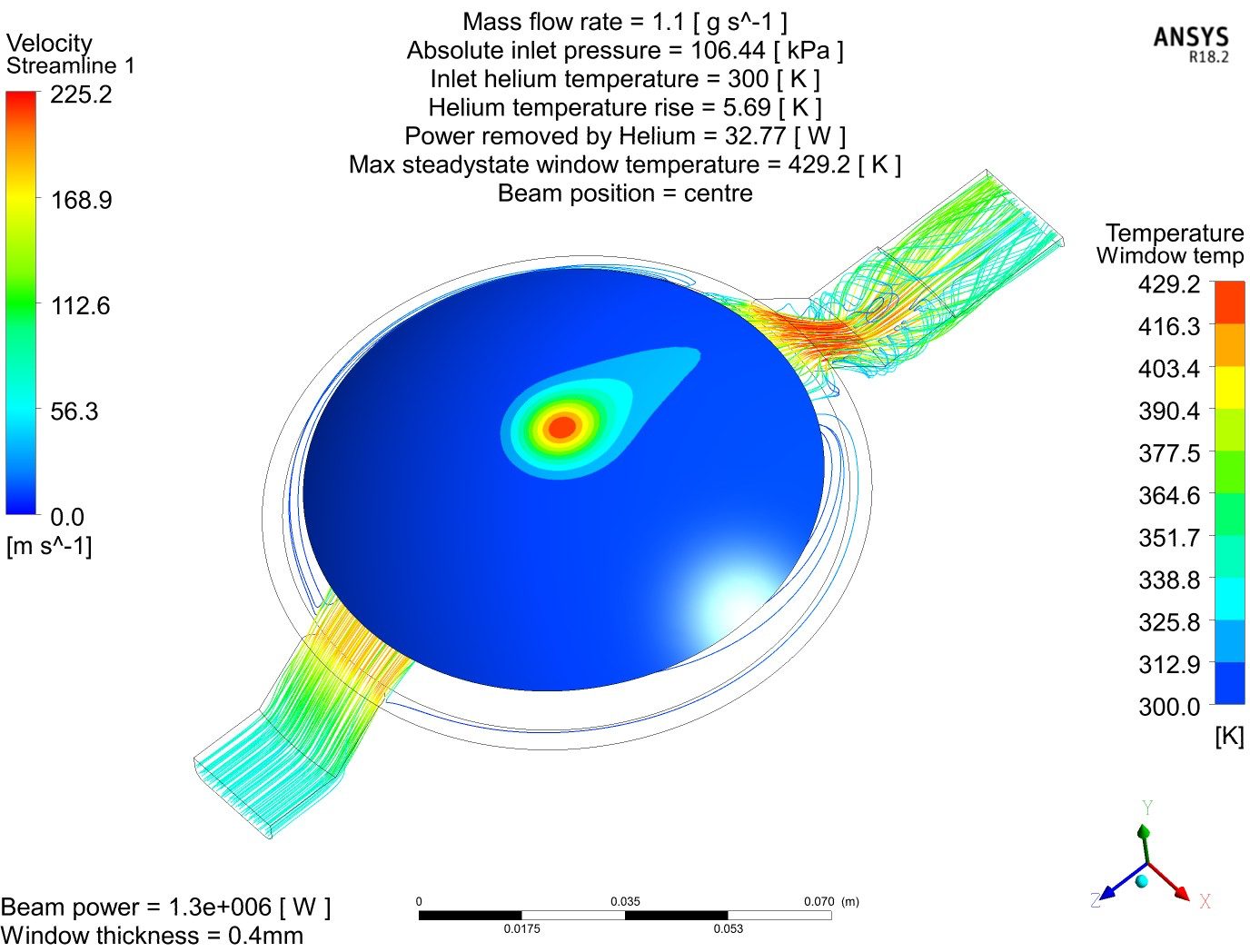}
	\caption{\small Helium flow streamlines for 0.4~mm-thick beam window by ANSYS CFD Analysis
	for 1.3~MW beam. Current operating mass flow rate of 1.1~g/s is used in this analysis.}
        \label{fig:window_stream_temp}
\end{figure}

Due to the relatively poor thermal conductivity of the titanium alloy, the window domes need to be cooled directly across the beam spot by helium gas flowing through the gap between two domes. A Computational Fluid Dynamics (CFD) FEA simulation of the window is shown in Fig.~\ref{fig:window_stream_temp}.

\begin{figure}
        \centering
        \includegraphics[width=0.8\linewidth]{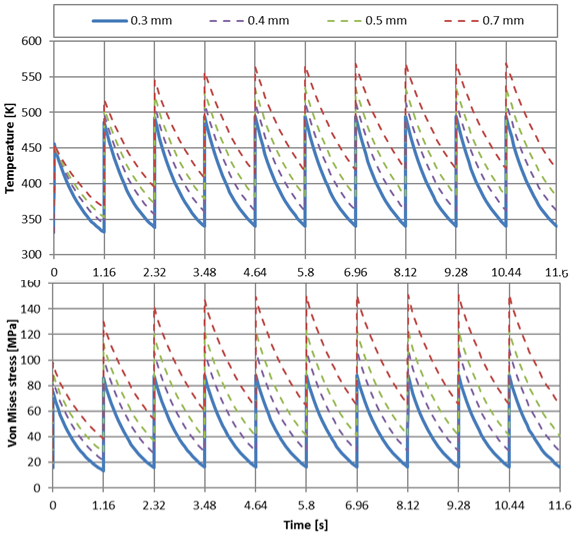}
        \caption{\small Temperature (upper) and equivalent stress (lower) of beam window with three typical thicknesses 
          as function of time for 10 successive beam pulses at 1.3 MW operation (3.2$\times 10^{14}$ ppp with 1.16 s rep rate).}
        \label{fig:window_temp_stress}
        \includegraphics[width=0.8\linewidth]{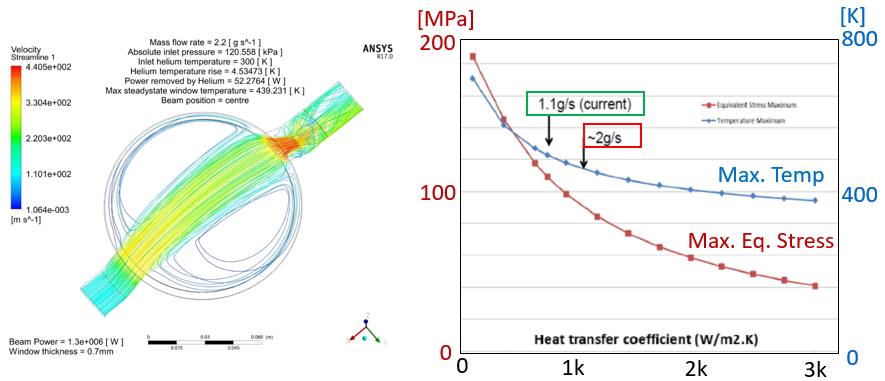}
        \caption{\small  (left) Helium flow streamlines with doubled mass flow rate 1.1$\rightarrow$2.2~g/s. Maximum velocity increases to 
         230$\rightarrow$440~m/s ($\times$ 2 higher), while pressure drop becomes 0.06$\rightarrow$0.2~bar ($\times$ 4 higher).
        (right) Maximum steady state temperature and equivalent stress maximum for 0.7-mm-thick window 
         as function of heat transfer coefficient at inner surface.}
        \label{fig:window_stream_flowrate}
\end{figure}

Because the pulse occurs over such a short time period, the resulting temperature jump will be independent of any cooling applied and of any material thickness. Both thermodynamic equations and ANSYS simulations have predicted a temperature rise of 156~$^\circ$C with each pulse. Fig.~\ref{fig:window_temp_stress} shows how the window heats up and begins to oscillate about an average temperature of about 150$\sim$200 $^\circ$C (dependent on window thickness) after only four to five pulses. This analysis assumes a heat transfer coefficient of 886 W/m$^2$K on the internal surface. This value has not been verified experimentally and using  ANSYS CFX 
as a realistic estimate for laminar helium flow between two flat plates of velocity at least 100 m/s. 

Fig.~\ref{fig:window_stream_flowrate} shows studies to evaluate how the mass flow rate affects on the cooling performance.  Left figure shows streamline with applying a double flow rate (2.2~g/s), where the pressure drop becomes four times higher (0.2~bar), and further increases would need to consider helium system pressurization. Right plots show maximum temperature and stress as function of heat transfer coefficient. As exhibited in the plot, even doubling the flow rate results in a tiny increase on the heat transfer coefficient, and thus improvements are small. 

\paragraph{Stress and Cooling Analysis Summary}

\begin{table}
  \centering
  \caption{Summary of the Stress and Cooling Analyses for 1.3 MW Operation.}
  \label{tab:window_strcool_analysis}
  \includegraphics[width=\textwidth] {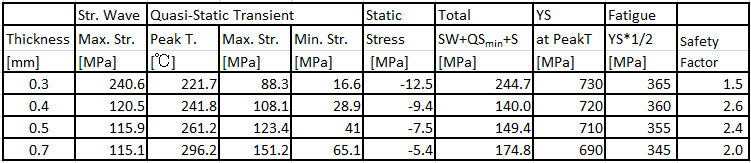}
\end{table}

Tab.~\ref{tab:window_strcool_analysis} summarizes results of the stress and cooling analyses for typical three window thicknesses, where estimated safety factors are also given.

\subsubsection{Operation / Spare Production Status}

The first beam window (ver.1) which was installed in October 2008 was replaced with an almost identical spare (ver.1a) during a scheduled shutdown in August 2017 as a precautionary measure after it had received 2.2$\times 10^{21}$ protons. Currently, ver.2 window is under fabrication at Rutherford Appleton Laboratory, with design upgrades to have better tolerance with the 1.3MW beam. 

\begin{figure}
        \centering
        \includegraphics[width=0.8\linewidth]{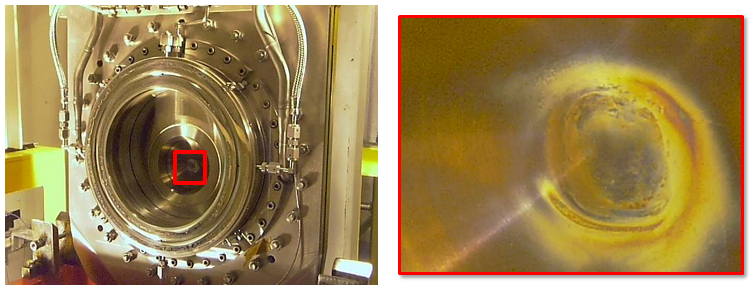}
        \caption{\small Replaced ver.1 beam window \color{\MODCOLORB}seen through the lead-glass window.\color{black} 
         Discolouration/visible damage at centre of the TS chamber side.}
        \label{fig:window_replaced}
\end{figure}

Although the first ver.1 window had operated perfectly without failure, the center of the window appeared to be discolored and/or damaged as shown in Fig.~\ref{fig:window_replaced}. The maximum radiation damage accumulated on the beam window Ti-6Al-4V domes is estimated to be approximately 2 Displacement Per Atom (DPA). 
Due to the “radiation hardening effect” as well known in the field of reactor / fusion material science, Ti-6Al-4V is expected to rapidly lose its ductility and to become highly brittle, so plastic deformation seems unlikely. It is to be noted that so far available neutron / proton irradiation data are only up to 0.3DPA, whereas our necessary study range is approximately 1 to a few DPA (2.4$\times 10^{21}$ protons is to be accumulated per one year for 1.3MW operation, which is similar to the one ver.1 window experienced). It is also possible that the discoloration arises from the radiochemical effects with the humidity in the helium. Naturally, it is priorities to realize Post Irradiation Examination (PIE) on the replaced beam window (this requires to solve regulation problems), and to understand the nature of radiation damage effect on the mechanical properties of Titanium alloys (especially high-cycle fatigue greater than 10 million cycles to be prototypic for the accelerator target/window application) in order to guide material development and to be able to predict a realistic lifetime /maintenance cycle for the beam window. These studies are now in progress under the aegis of RaDIATE international collaboration.

It is to be noted that radiation hardening is an increase in yield strength, loss of ductility, etc. These properties are important in the event of failure and would be important in the event that a material was subjected to a load which exceeds the yield strength. On the other hand, properties such as Young's modulus of elasticity, Poisson ratio, and coefficient of thermal expansion (CTE) are typically much less affected by radiation, and these are the properties required to calculate the linear elastic stress waves. This is the reason why radiation damage shouldn’t significantly affect the stress level. Also radiation damage isn’t expected to change the density of the material so the speed of sounds should be constant. This means that the characteristic thickness of the window to avoid resonance should also not change. So in short, if a material has become hardened and brittle it won't necessarily fail unless the stress level are high enough (or it suffer fatigue crack).

\subsubsection{Upgrade for 1.3~MW Operation}

\begin{itemize}
\item To increase the window thickness from existing 0.3 mm to 0.4 mm is a viable candidate, with sufficient tolerance 
to machining (enough plateau in Fig.~\ref{fig:window_strw_simulation}, while minimizing the resultant increase in  temperature, stress and activation of upstream beam-line area.
\item By considering variation of elastic modulus (speed of sound) due to the thermal cycle between 25 to 200~$^\circ$C (Fig.~\ref{fig:window_temp_stress}) the engineering tolerance on the thickness is recommended to be 0.39$\pm$0.01 mm (0.38$\sim$0.40 mm). This is a challenging machining tolerance, but achievable with regards to the past experience.
\item It is desirable to introduce better Titanium alloy with better radiation damage tolerance. This will occur in synergy with the RaDIATE collaboration program.
\item Current mass flow rate (1.1~g/s) already guarantees enough cooling capacity, as larger compressor was purchased than originally specified (0.08~g/s). To increase the flow rate is still desirable, but probably not a top priority.
\end{itemize}

\graphicspath{{figures/main_secondary/}}
       
\subsection{Horn}

\subsubsection{Overview}

\begin{figure}
        \centering
        \includegraphics[width=0.75\linewidth]{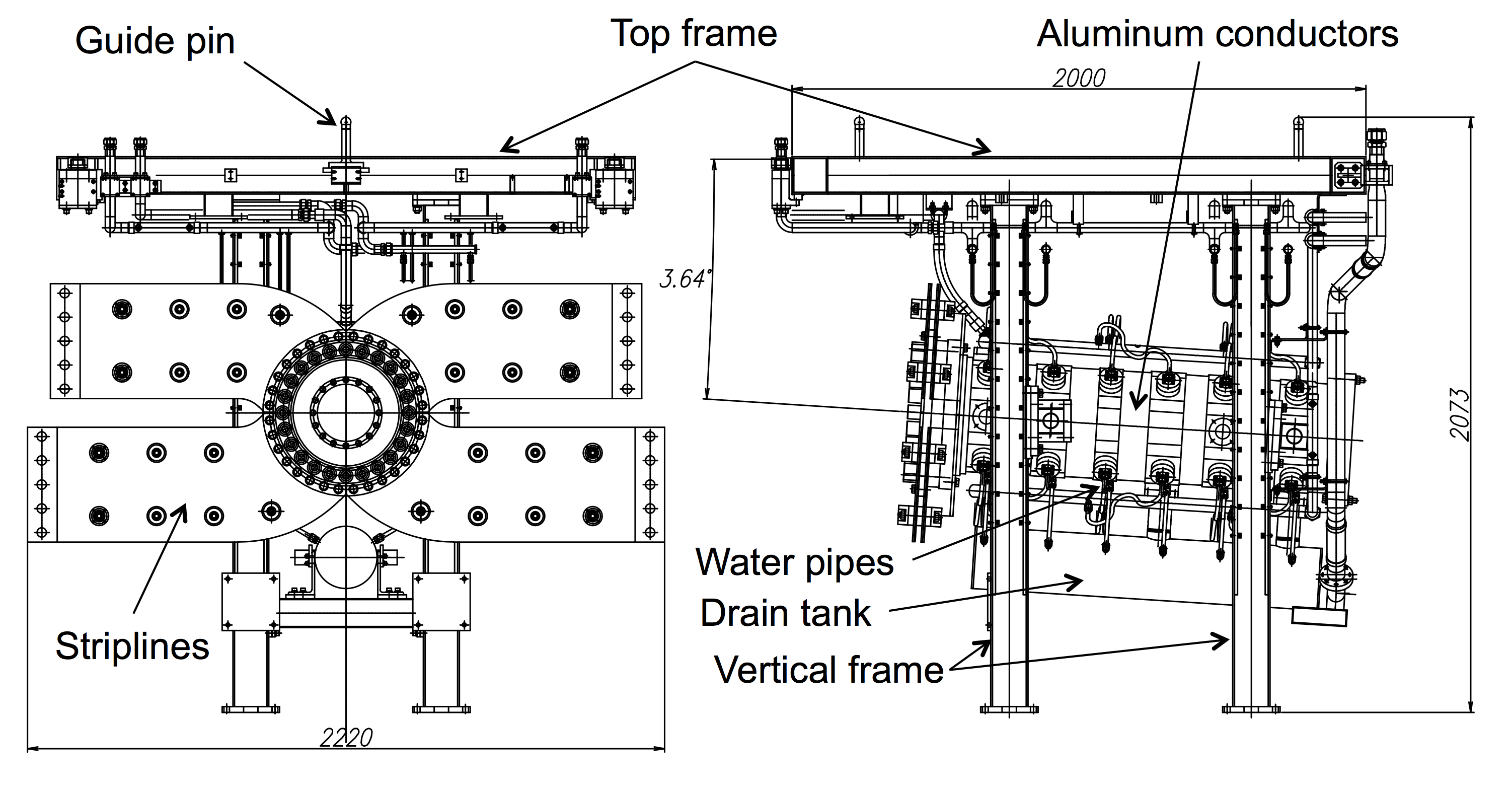}
        \includegraphics[width=0.75\linewidth]{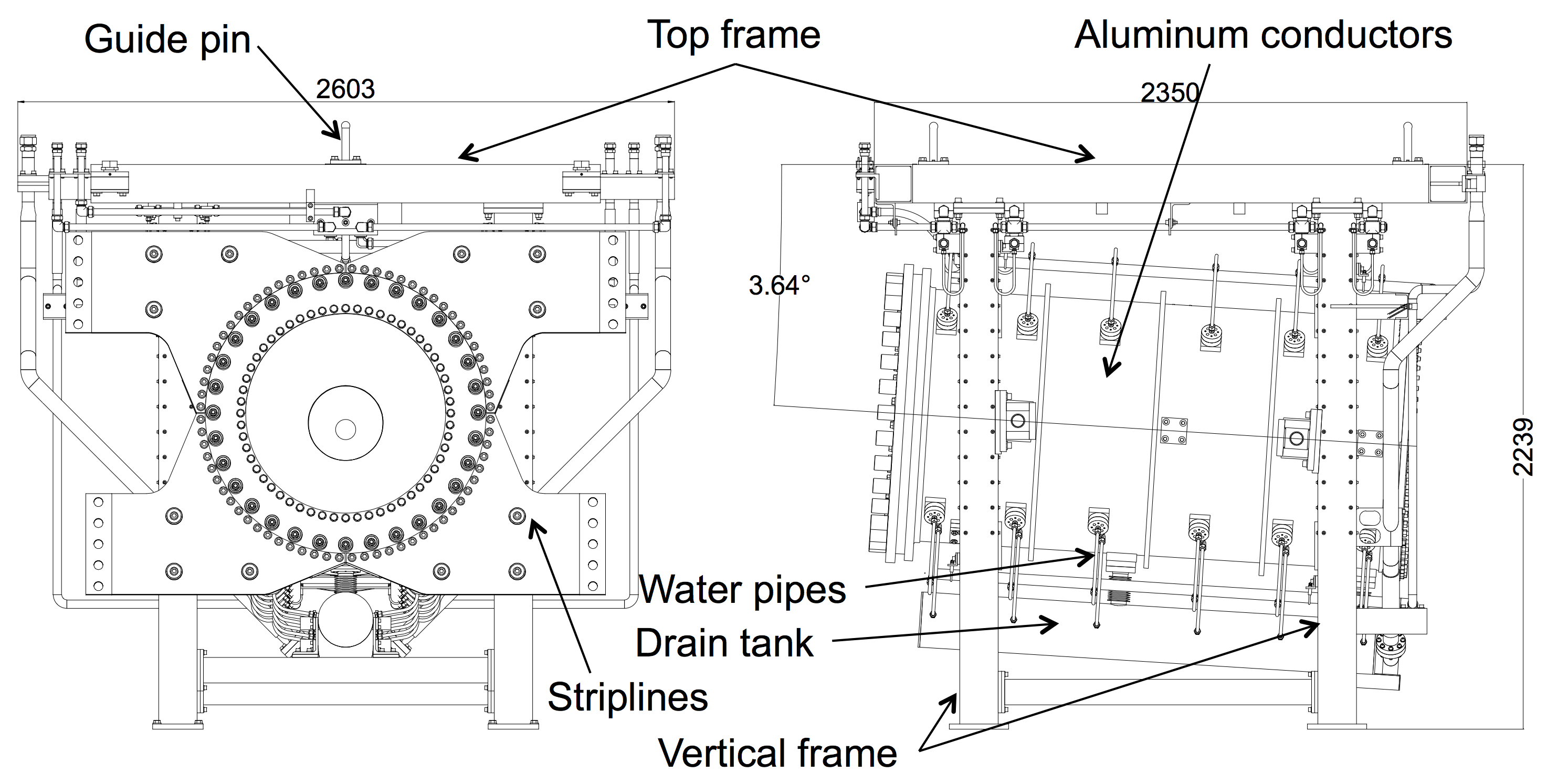}
        \includegraphics[width=0.75\linewidth]{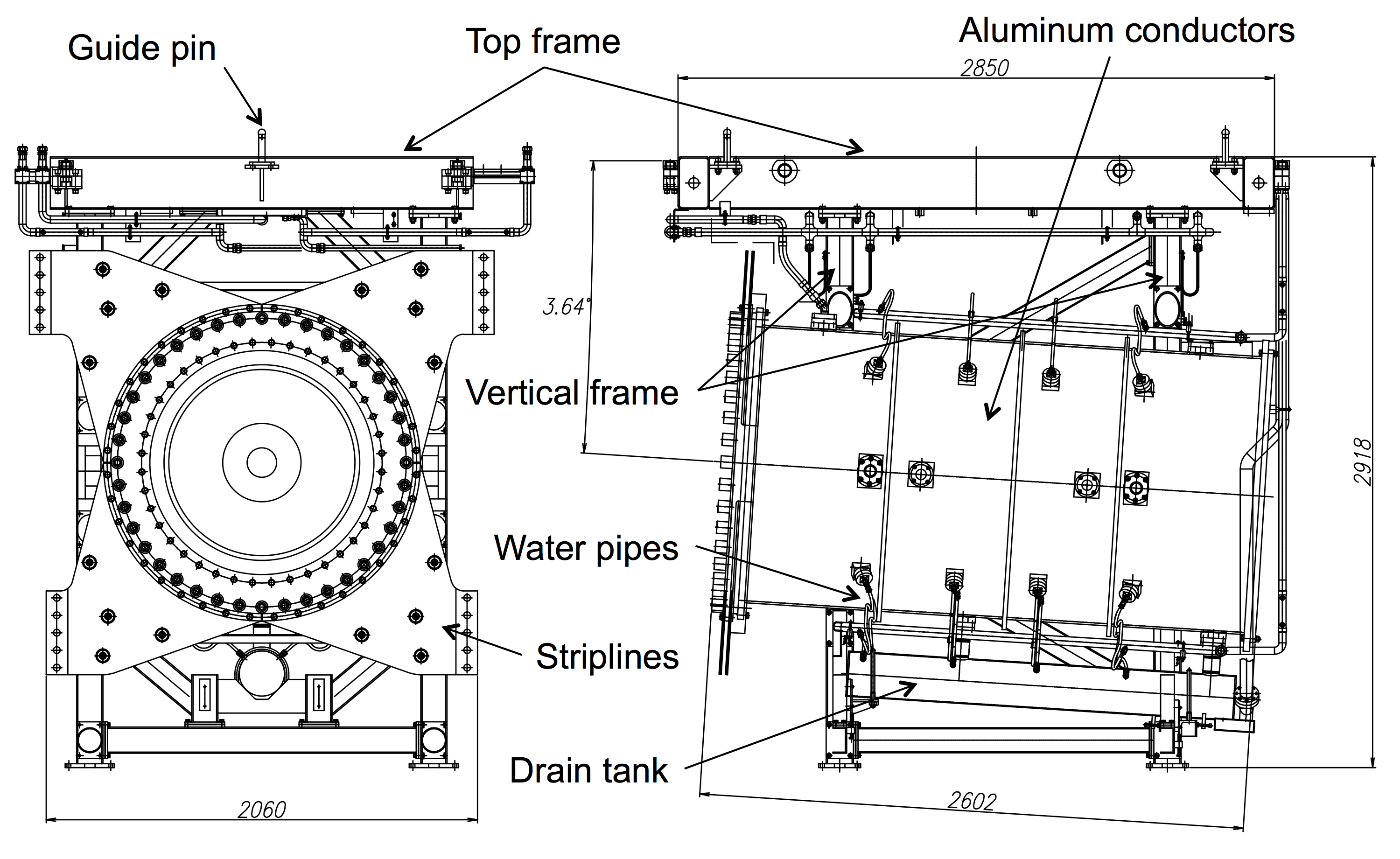}
        \caption{\small Drawings of T2K magnetic horns. Top: horn1, middle: horn2, bottom: horn3.}
        \label{fig:horn_schematic}
\end{figure}

\begin{figure}
	\centering
	\includegraphics[width=0.75\linewidth]{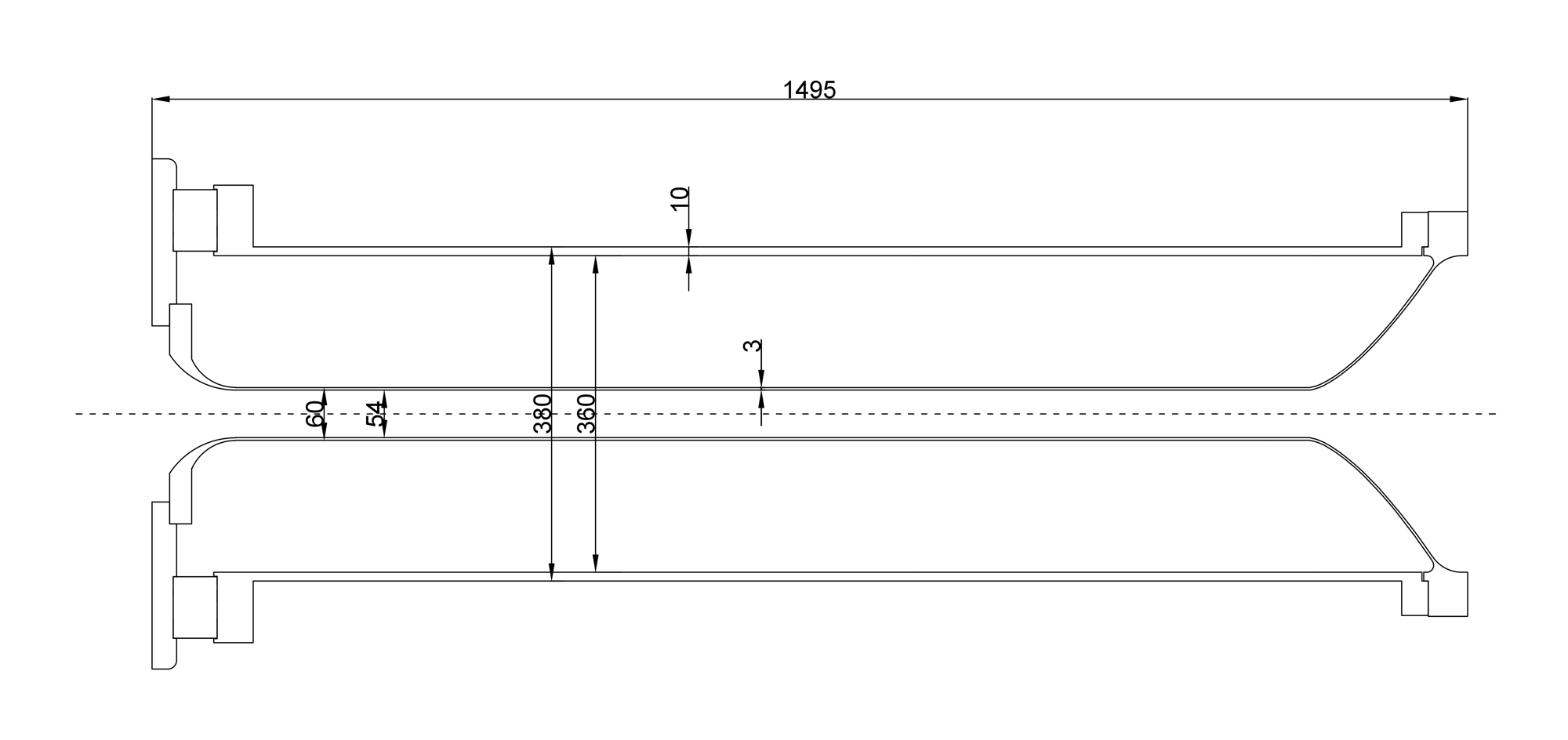}
        \includegraphics[width=0.75\linewidth]{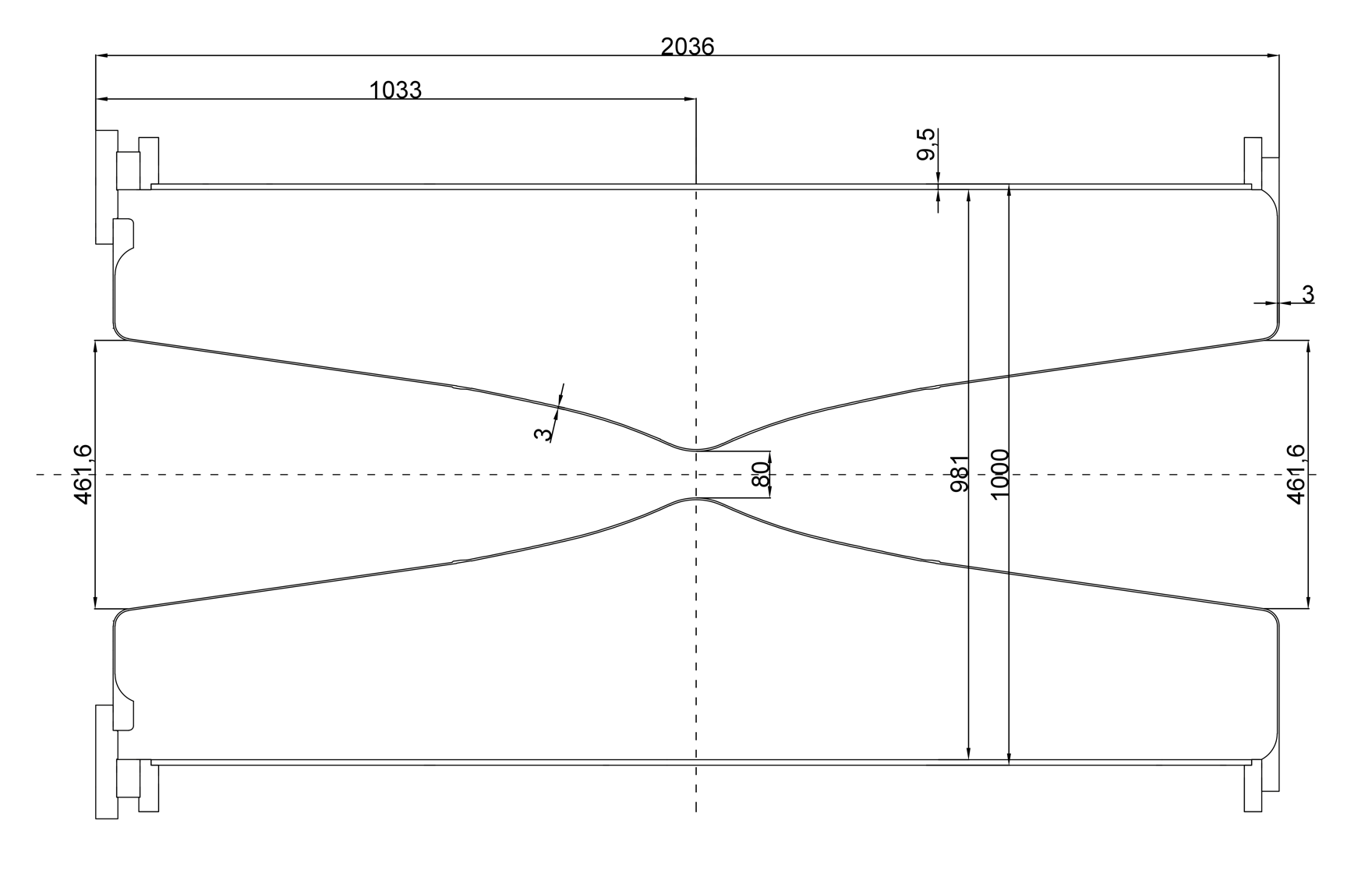}
        \includegraphics[width=0.75\linewidth]{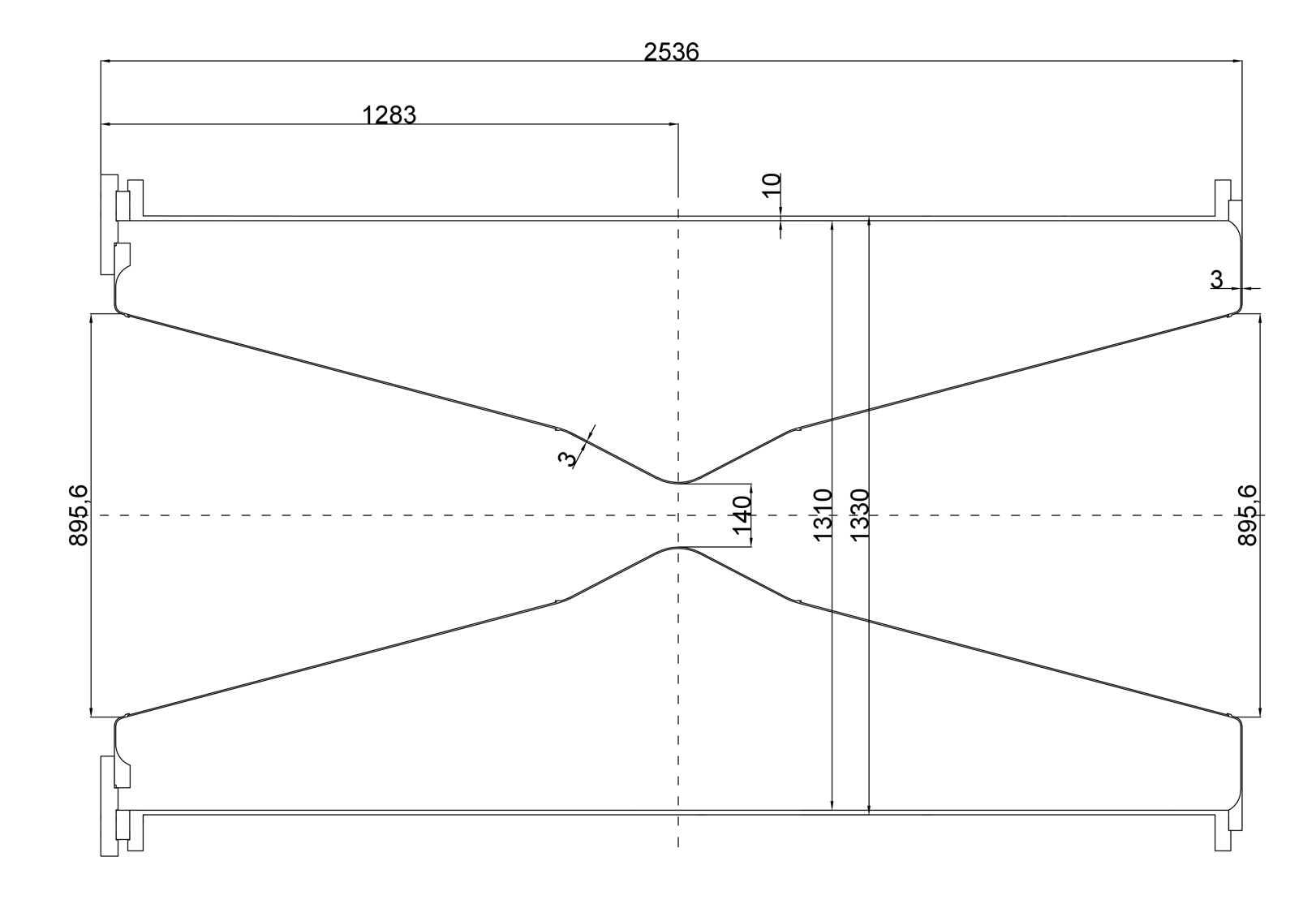}
	\caption{\small Cross section of T2K magnetic horn conductors. Top: horn1, middle: horn2, bottom: horn3.}
	\label{fig:horn_shape}
\end{figure}

T2K magnetic horns have a co-axial structure with inner and outer conductors,
made of an aluminum alloy A6061-T6 which is commonly used for a horn conductor material.
T2K uses three magnetic horns which are designed to be 
operated at 320~kA to efficiently focus secondary particles, produced at the target, with low momentum and 
high emission angle~\cite{Ichikawa:2012horn}. The drawings of the magnetic horns is shown in
Fig.~\ref{fig:horn_schematic}.
The magnetic horns are composed of aluminum conductors, their support frames, and plumbing.
The conductor shape is shown in Fig.~\ref{fig:horn_shape}
and typical dimmension is summarized in Tab.~\ref{tab:horn_dim}.
\begin{table}
\centering
\small
\caption{\small Typical dimmension of T2K magnetic horn conductors.}
\label{tab:horn_dim}
\begin{tabular}{lrrr}
\hline\hline
Parameters & horn1 & horn2 & horn3 \\ 
\hline\hline
Inner diameter (mm) & 54 & 80 & 140 \\
Outer diameter (mm) & 380 & 1,000 & 1,330 \\
Length (mm) & 1,495 & 2,036 & 2,536 \\ 
\hline\hline
\end{tabular}
\end{table}
The inner conductors are 3~mm thick to reduce interactions of secondary particles, while the outer
conductors are 10~mm thick.

\begin{figure}
        \centering
        \includegraphics[width=0.75\linewidth]{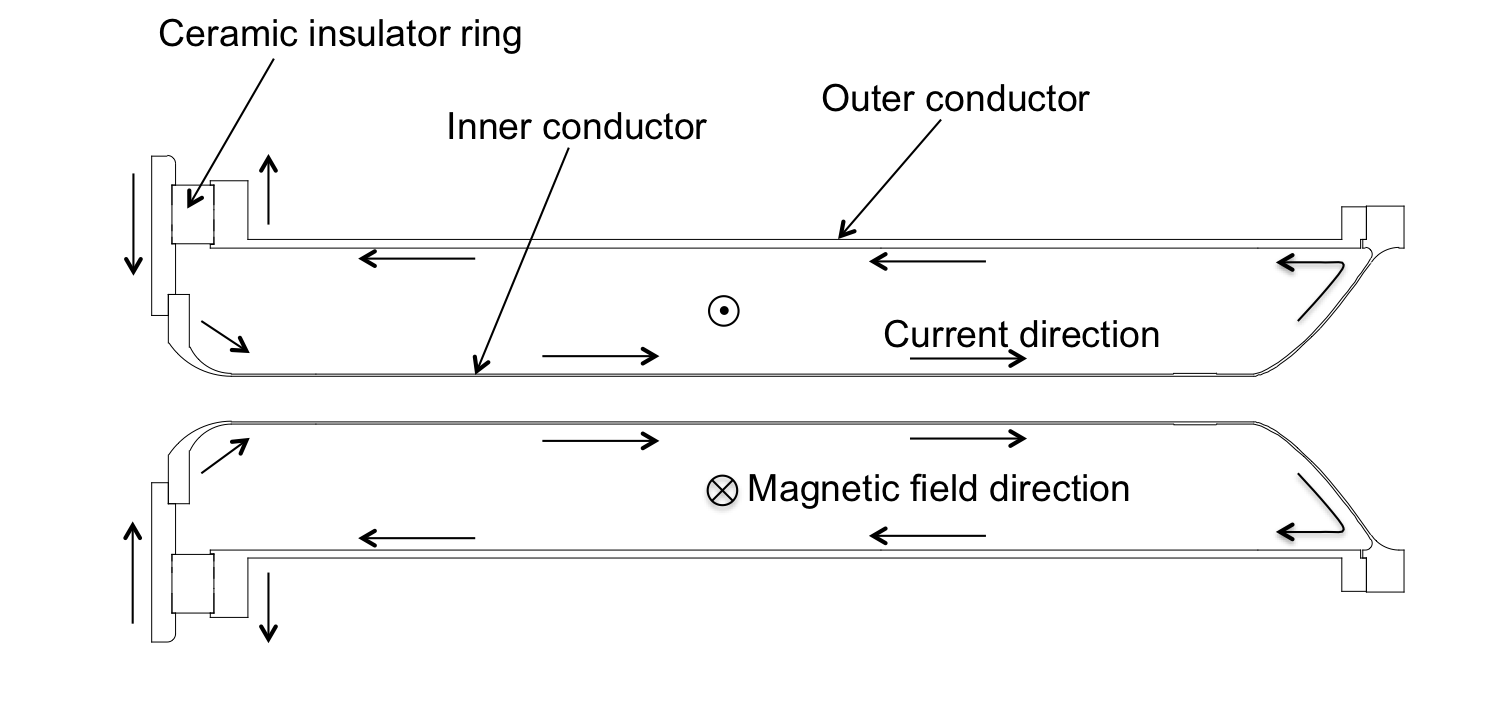}
        \caption{\small Cross section of horn1 conductors showing current and magnetic field directions.}
        \label{fig:horn_field}
\end{figure}

The 320~kA pulsed current is generated in the following way:
capacitor bank with 4~mF in the horn power supply is charged up to 6~kV within 1~s and then
discharged by turning on thyristor switches at the timing when trigger signal is recieved. 
32~kA pulsed current, whose pulse width is 2~ms, is output from the power supply and transferred to a pulse transformer
through power cables. The transformer amplifies the current by a factor of ten and the output 320~kA pulsed
current is transferred to the horn conductors through aluminum striplines.
In the region between the inner and outer conductors magnetic field of 2.1~T at maximum
is generated in the azimuthal direction as shown in Fig.~\ref{fig:horn_field}.
When the current flows from the inner conductor and returns through the outer conductor,
positive pions and kaons, which decay in flight to muon neutrinos, 
can be focused. With current flow in the opposite direction, negative pions and kaons 
can be focused to produce muon anti-neutrino beams.

\begin{figure}
        \centering
        \includegraphics[width=0.7\linewidth]{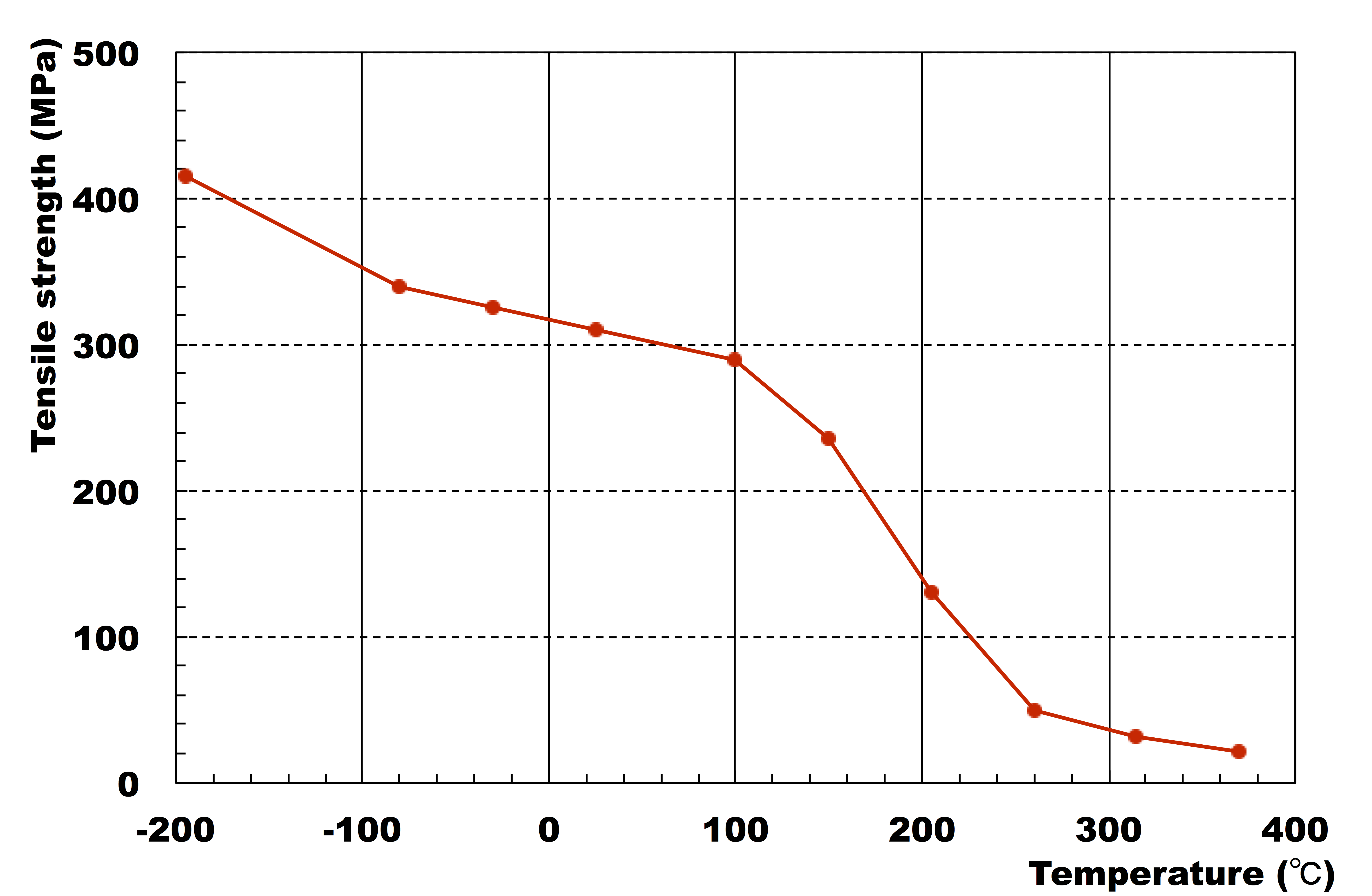}
        \caption{\small Temperature dependence of aluminum alloy A6061-T6 tensile strength.
        Data points are taken from \cite{AluminumHandbook}.}
        \label{fig:aluminum_strength_temp}
\end{figure}

The aluminum alloy A6061-T6 has the tensile strength of 310~MPa (at 25~$^{\circ}$C).
The strength is degraded by repetitive forces and
its fatigue strength is estimated to be 68.9~MPa for 97.5\% confidence that the material 
will not fail in 2$\times 10^8$ cycle. The material strength
is also degraded due to a corrosion compared to an air environment. 
An empirical factor of 0.43 is taken from the experience of MiniBOONE horn operation. 
Allowable stress on the horn conductors is 29.6~MPa by taking these reduction factors into account.
Another consideration is temperature effect on the material strength.
The strength changes depending on temperature as shown in Fig.~\ref{fig:aluminum_strength_temp}.
Since the strength degradation is large above 100~$^{\circ}$C, the allowable temperature
is set to be 80~$^{\circ}$C for the aluminum conductors so that the temperature 
effect can be small.

\begin{figure}
        \centering
        \includegraphics[clip,width=0.49\linewidth,bb=10 10 720 470]{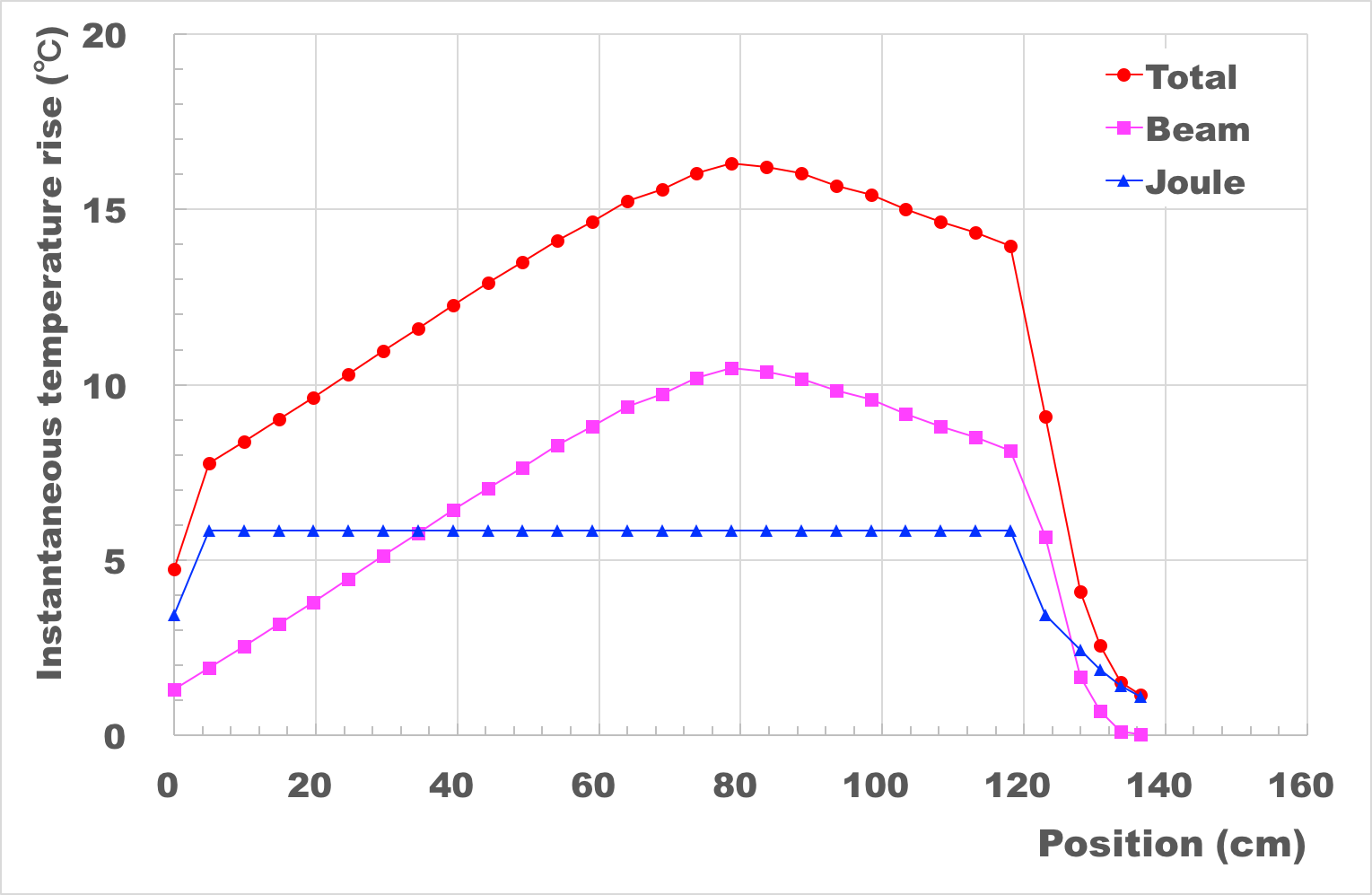}
        \includegraphics[clip,width=0.49\linewidth,bb=10 10 720 470]{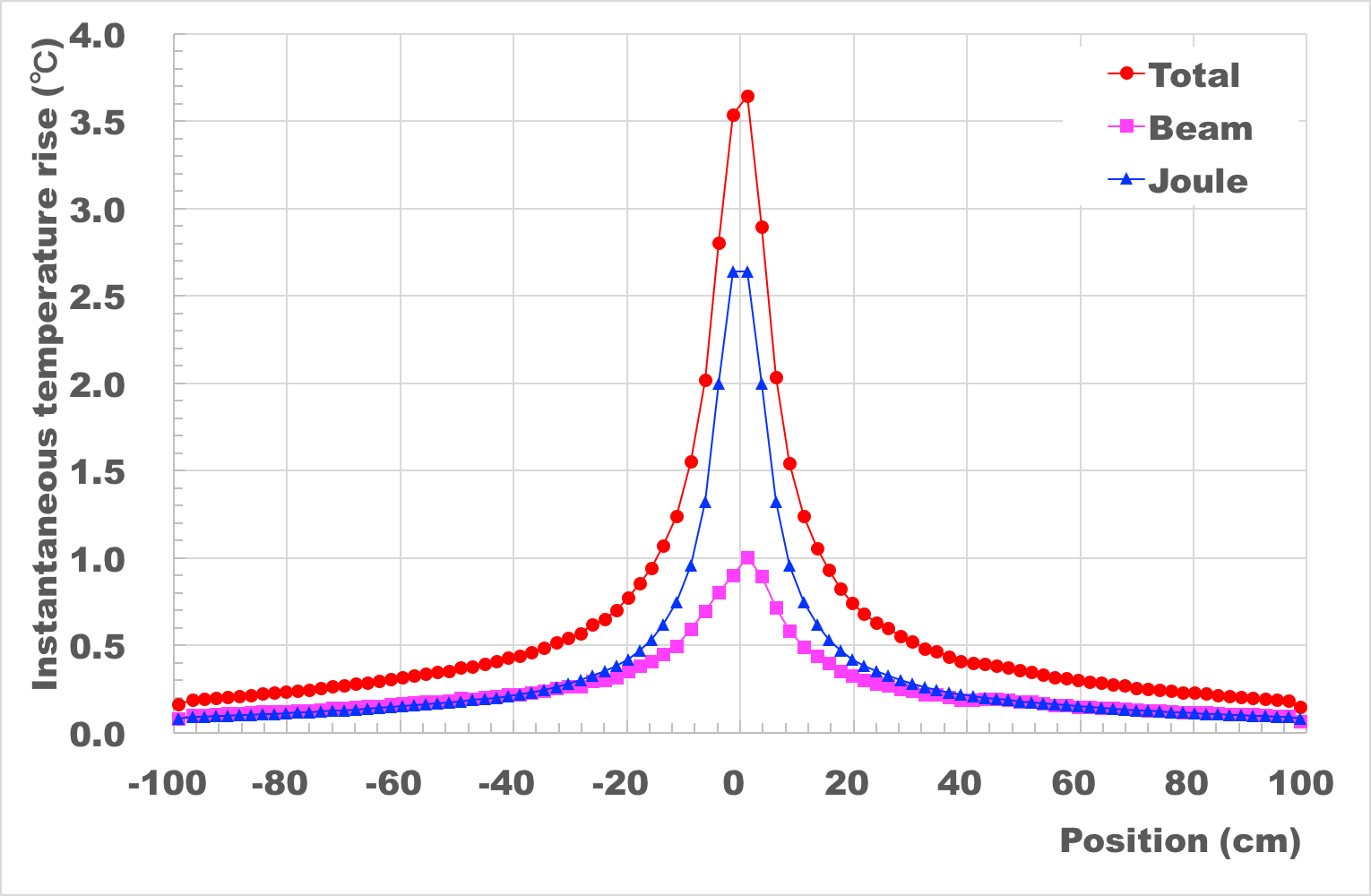}
        \includegraphics[clip,width=0.49\linewidth,bb=10 10 720 470]{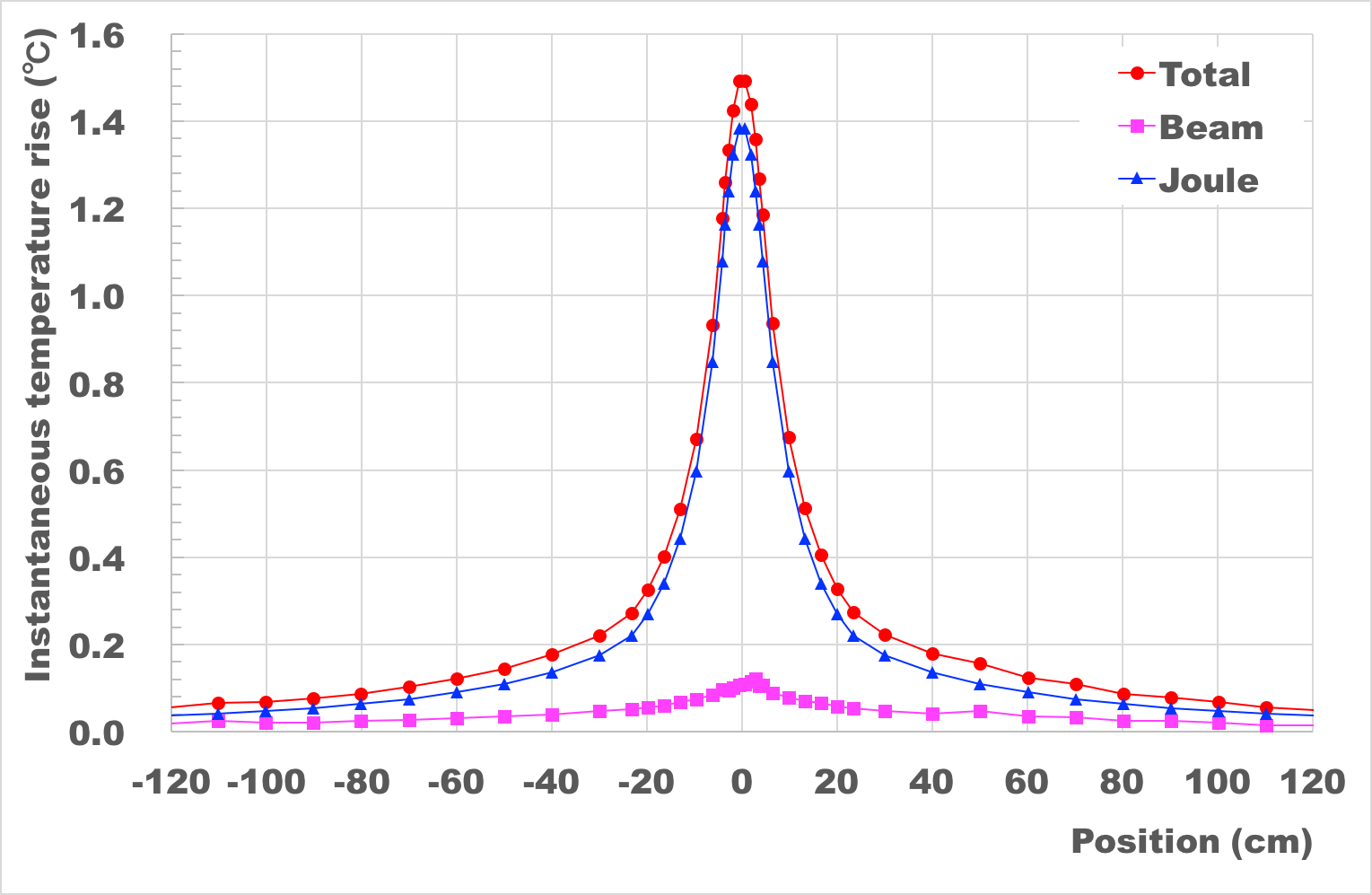}
        \caption{\small Instantaneous temperature rise at inner conductors for 1.3~MW case
        as a function of longitudinal position for horn1 (top left), horn2 (top right), and horn3 (bottom).
        The instantaneous temperature rise due to beam exposure and Joule heating are
        represented by rectangles and circles, respectively. The total instantaneous
        temperature rise is also indicated by dots.}
        \label{fig:heat_load}
\end{figure}

\begin{figure}
        \centering
        \includegraphics[width=0.49\linewidth]{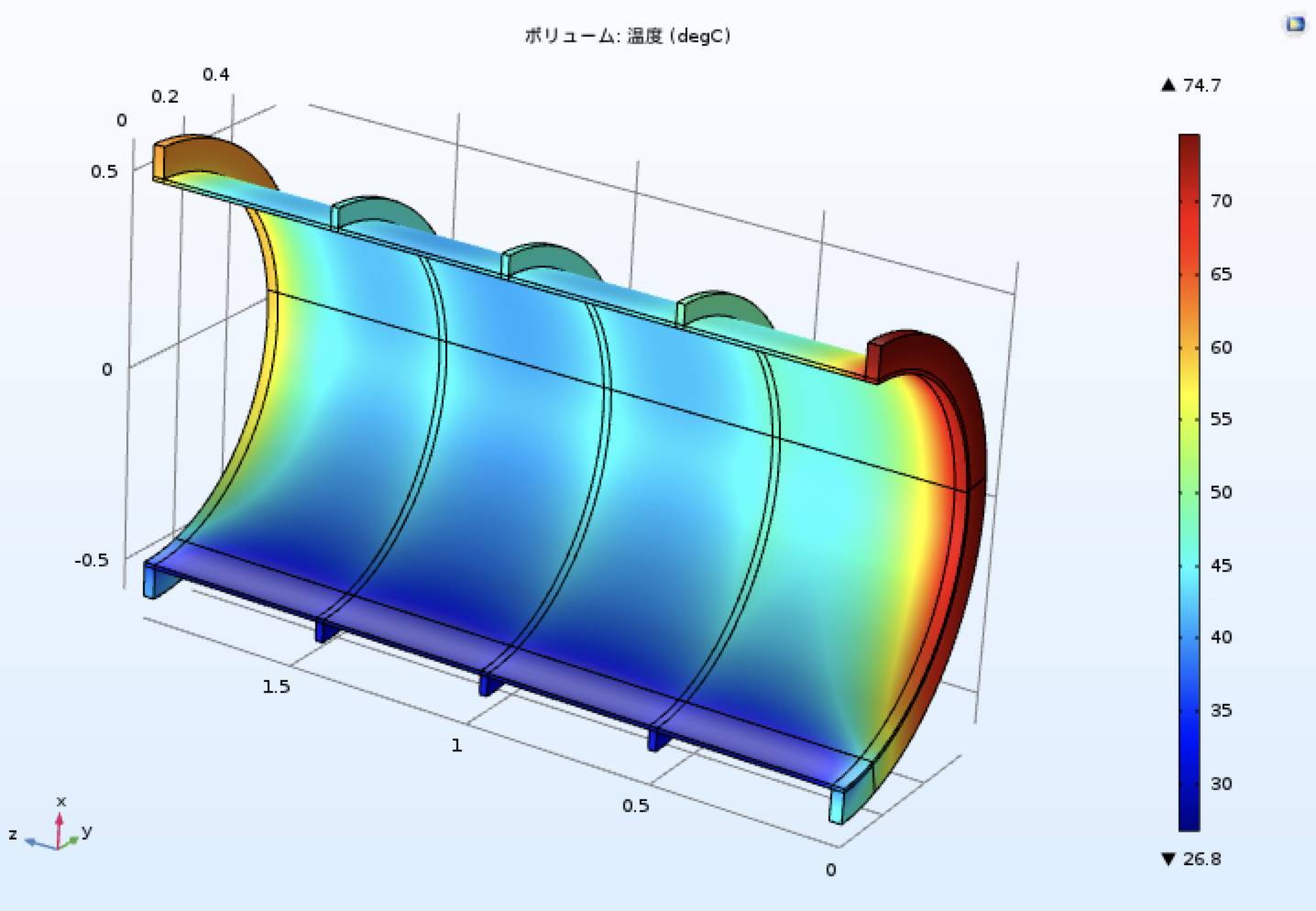}
        \includegraphics[width=0.49\linewidth]{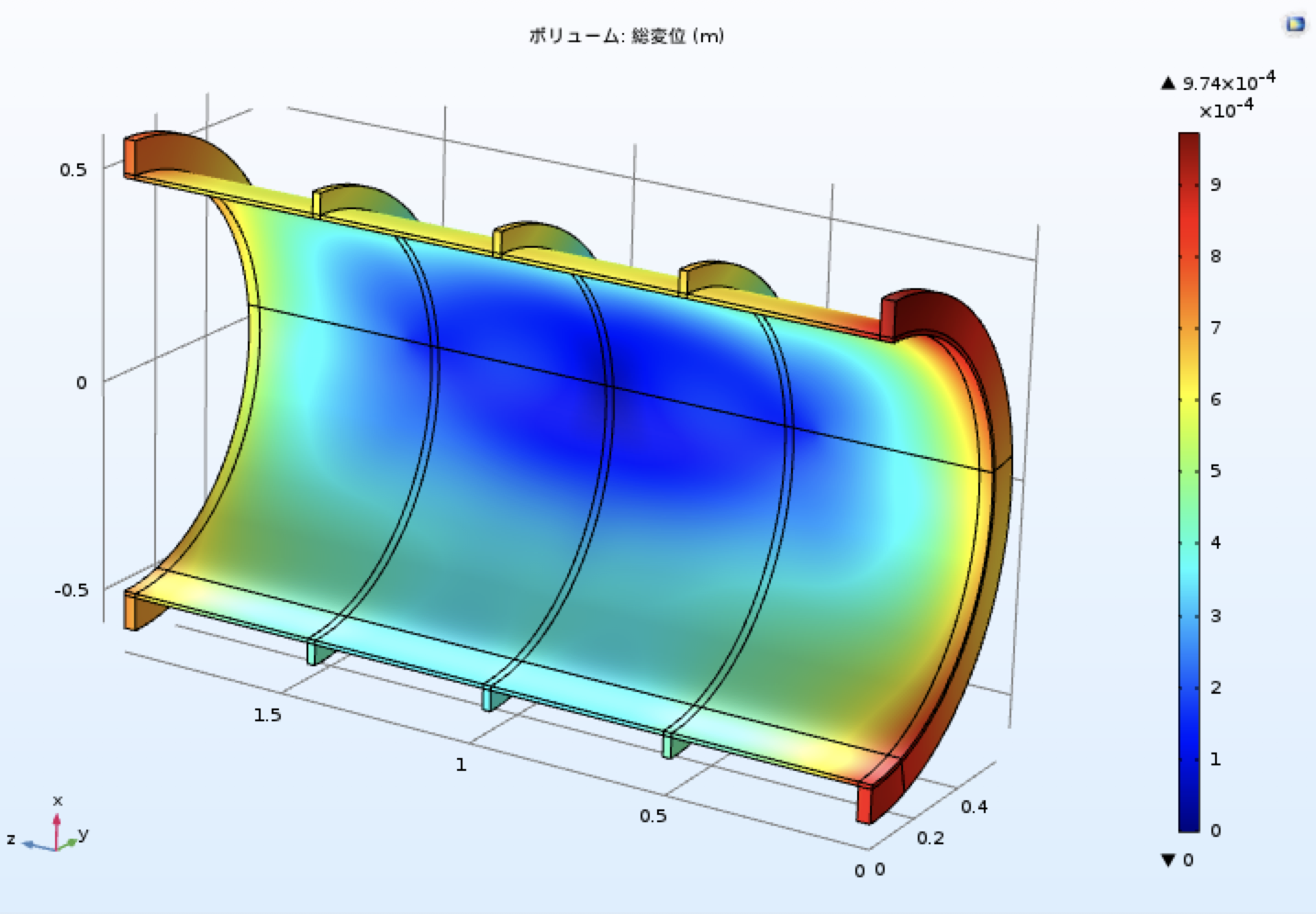}
        \caption{\small FEM simulation of cooling effect on horn2 outer conductor. Left : temperature at the outer conductor
        for 1.3 MW beam (in unit of $^{\circ}$C). Right : displacement of the outer conductor for 1.3 MW beam (in unit of m).}
        \label{fig:H2OCtemp}
\end{figure}

\begin{figure}
        \centering
        \includegraphics[width=0.4\linewidth]{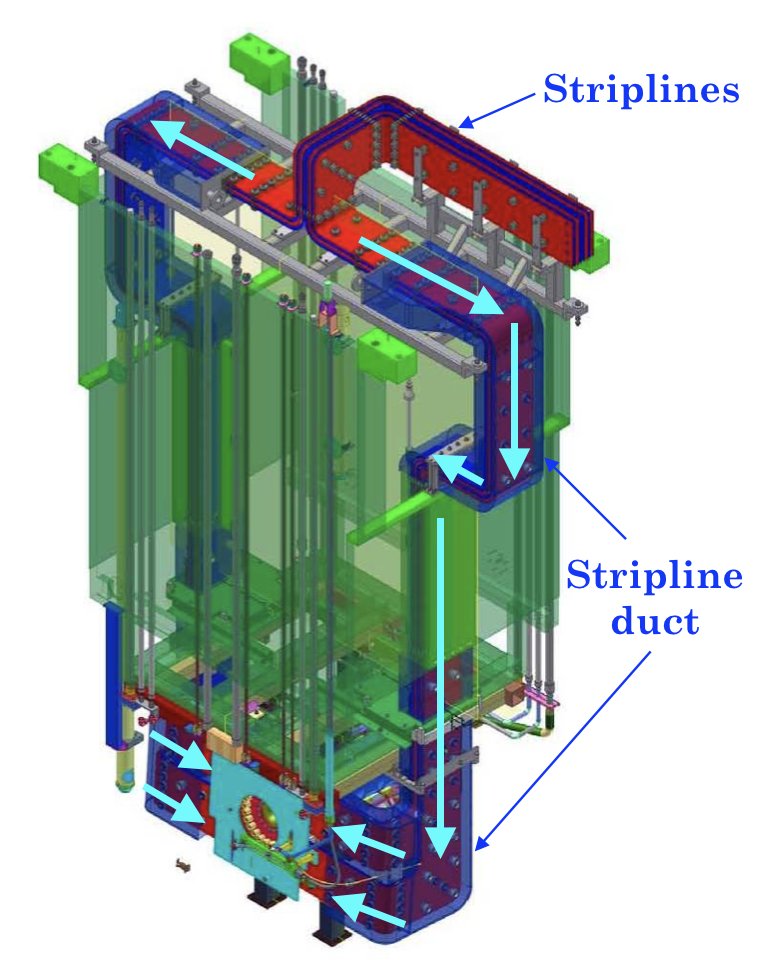}
        \includegraphics[width=0.59\linewidth]{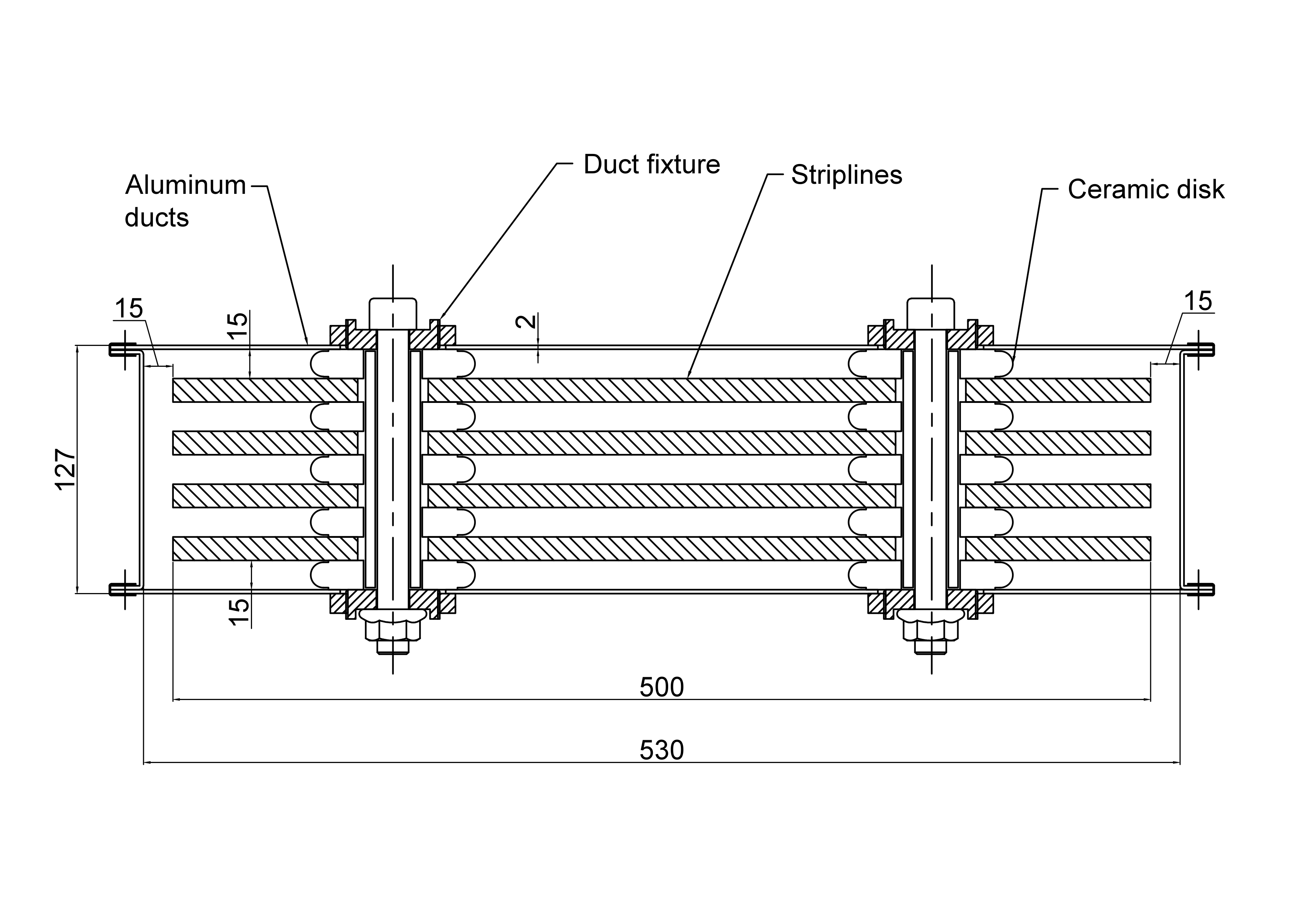}
        \caption{\small Schematic figures of the stripline helium cooling. Left: helium flow
        path for horn1. Right: cross section of the striplines and surrounding ducts.}
        \label{fig:stripline_duct}
\end{figure}

\begin{figure}
        \centering
        \includegraphics[width=\linewidth]{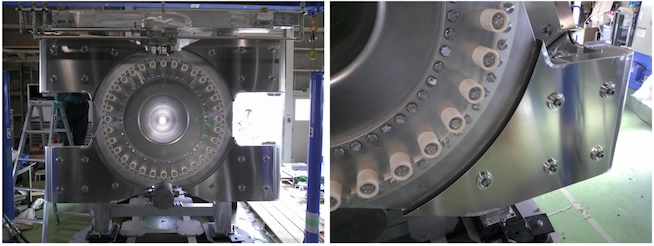}
        \caption{\small Pictures of the horn2 stripline ducts. Left: Front view of horn2. Right: Stripline
        duct near the horn conductor.}
        \label{fig:horn2_stripline_duct}
\end{figure}

The aluminum conductors suffer from heat deposition by beam exposure and Joule loss.
The heat deposition at the horn conductors for 1.3~MW operation is summarized in 
Tab.~\ref{tab:horn_heat_depo}.
\begin{table}
\centering
\small
\caption{\small Summary of heat deposit at each horn. Heat deposit from beam exposure is
based on beam intensity of $3.2\times 10^{14}$ protons/pulse for 1.3~MW. Joule heating
of each horn is estimated for pulse widths of 2.0~ms. 
The calculation of the total heat deposit in units of kW is based on 1.16~s cycle.}
\label{tab:horn_heat_depo}
\begin{tabular}{lrrrrrr}
\hline
Heat deposit & \multicolumn{2}{c}{horn1} & \multicolumn{2}{c}{horn2} & \multicolumn{2}{c}{horn3} \\ \cline{2-7}
 & Inner & Outer & Inner & Outer & Inner & Outer \\
\hline\hline
Beam  (kJ) & 12.1 & 10.5 & 2.8 & 9.7 & 1.0 & 1.9 \\
Joule (kJ) &  9.7 &  0.5 & 3.3 & 0.3 & 2.4 & 0.3 \\\hline
Total (kJ) & \multicolumn{2}{c}{32.8} & \multicolumn{2}{c}{16.1} & \multicolumn{2}{c}{5.6} \\
Total (kW) & \multicolumn{2}{c}{28.3} & \multicolumn{2}{c}{13.9} & \multicolumn{2}{c}{4.8} \\
\hline
\end{tabular}
\end{table}
Instantaneous temperature rise at the inner conductor is also shown in Fig.~\ref{fig:heat_load}.
Heat load at the inner conductor of horn1 is largest and instantaneous temperature rise
is estimated to be 16.3~$^{\circ}$C.
Although the total heat load at the outer conductors are comparable to or larger than
those at the inner conductors because of their large volume, local heat deposit at
the outer conductor is quite small (instantaneous temperature rise for horn1, horn2, and horn3
are 0.4~$^{\circ}$C, 0.08~$^{\circ}$C, and 0.01~$^{\circ}$C, respectively).
The inner conductors are cooled by water sprayed from nozzles attached at the outer conductors.
The sprayed water is collected into a drain tank located below the outer conductor and
then pumped upward by a suction pump located 8~m above.
From the past measurements at several bench tests, typical heat transfer coefficients
for the horn inner conductor cooling were measured to be 7.9~kW/m$^2\cdot$K (horn1),
1.0~kW/m$^2\cdot$K (horn2), and 1.3~kW/m$^2\cdot$K (horn3).
Then the maximum temperatures at the horn conductors for 1.3~MW operation are
estimated as summarized in Tab.~\ref{tab:max_temp_horn}.
\begin{table}
        \centering
        \small
        \caption{\small Estimation of maximum temperatures at horn1, horn2, and horn3 for
        1.3~MW operation. Numbers shown are in unit of $^{\circ}$C.}
        \begin{tabular}{lrrr}
        \hline
        Items                          & horn1 & horn2 & horn3 \\\hline
        Instantaneous temperature rise & 16.3  & 3.6   & 1.0   \\
        (beam exposure)                & 10.5  & 1.0   & 0.1   \\
        (Joule loss)                   & 5.8   & 2.6   & 0.9   \\
        Steady state temperature rise  & 19.1  & 22.1  & 5.8   \\
        Coolant water temperature      & 25.0  & 25.0  & 25.0  \\\hline
        Maximum temperature            & 60.4  & 55.3  & 31.8  \\\hline
        \end{tabular}
        \label{tab:max_temp_horn}
\end{table}
The estimated maximum temperatures are well below the allowable temperature of 80~$^{\circ}$C.
The cooling performance of the horns is satisfactory with 1.3~MW operation.
However, cooling capacity of the horn water cooling system should be increased for 1.3~MW,
as described in Sec.~\ref{sec:upgrade}.
\color{\MODCOLORB} 
On the other hand, the outer conductors are not cooled but running water at their bottom can cool the outer conductors.
In addition, the horn inner volume is occupied by water vapor which also has a cooling effect.
Using a horn1 prototype, measurement of cooling effect on the outer conductor was performed.
Heat corresponding to 750~kW was applied by heater to the outer conductor and temperature at
several points of the outer conductor was measured. The measured temperature difference between
top and bottom of the conductor was 6.4~$^{\circ}$C. Then FEM simulation was performed to estimate
the cooling effect from the running water and water vapor. To reproduce the result of the cooling measurement,
heat transfer coefficients were estimated to be 2~kW/m$^2\cdot K$ and 450~W/m$^2\cdot K$ from the bottom 
running water and water vapor, respectively. Since horn1 is rather small compared to the other two horns,
cooling effect from the water vapor is not small. For horn2, a simulation for outer conductor temperature at 1.3 MW operation 
was performed using 1~kW/m$^2\cdot K$ (bottom) and 100~W/m$^2\cdot K$ (others). The maximum temperature
at cylindrical part was approximately 45 $^{\circ}$C, however temperature at upstream thick flange was as high as 
75 $^{\circ}$C (Fig.~\ref{fig:H2OCtemp}).
Water cooled striplines described later will be attached to the upstream flange, so indirect cooling may be obtained or
even a direct cooling is considered as a purpose of the stripline cooling. Distortion from the heating was also simulated
to be at most 1~mm at top and bottom of the upstream flange (Fig.~\ref{fig:H2OCtemp}). 
The temperature depends on the heat transfer coefficient by water vapor, the actual temperature may change.
Once spare horn2 is produced, cooling measurement with the actual outer conductor is necessary. Additional forced
cooling of outer conductor will be adopted if necessary. 
\color{black}
Striplines near the horn conductors also suffer from heat load by beam exposure and Joule loss. 
The heat load at the striplines are summarized in Tab.~\ref{tab:stripline_heat}.
\begin{table}
        \centering
        \small
        \caption{\small Summary of the heat flux at the striplines by beam exposure and Joule 
        loss. Beam exposure is based on 3.2$\times 10^{14}$ protons/pulse for 1.3~MW. Joule
        heating is estimated for pulse width of 2~ms. The current helium flow speed and the
        acceptable beam power relevant to the stripline cooling are also shown.}
        \begin{tabular}{lrrr}
        \hline
        Heat flux per stripline plate & Horn1 & Horn2 & Horn3 \\\hline
        Beam heating (J/m$^2$)        & 164   & 1042  & 123   \\
        Joule heating (J/m$^2$)       & 50    & 24    & 18    \\
        Total (J/m$^2$)               & 214   & 1066  & 141   \\\hline
        Helium flow speed (m/s)       & 2.7   & 2.7   & 2.2   \\
        Acceptable beam power (MW)    & 2.10  & 0.75  & 2.04    \\\hline
        \end{tabular}
        \label{tab:stripline_heat}
\end{table}
The largest heat load is expected at the horn2 striplines because defocused secondary particles
by horn1 pass through the horn2 striplines.
The stripline conductors are covered by aluminum ducts and the conductor plates are cooled 
by forced helium gas flow through the ducts as shown in 
Figs.~\ref{fig:stripline_duct} and \ref{fig:horn2_stripline_duct}.
The helium gas inside the helium vessel is circulated by helium compressor with flow rate of
approximately 2000 m$^3$/hour. The stripline ducts work as the inlet plumbing to the helium 
vessel. The helium flow speeds at the exit of the stripline ducts where the highest heat
laod is expected are estimated from the measured flow rate at the inlet for each horn 
as shown in Tab.~\ref{tab:stripline_heat}. 
The acceptable beam power for horn2, 750~kW, is lower compered to the other horns 
($\sim$2~MW) due to the larger heat load.
Improvement on the stripline cooling is necessary and the details will be described
in Sec.~\ref{sec:horn_upgrade_plan}.

\subsubsection{Operation status}
\label{sec:operation_status}

The first generation of the T2K horns was installed in 2009. They had been operated
over 12 million pulses until 2013. The currently used second-generation horns, with some significant
improvements, were installed during 2013-2014 long shutdown period.
They have been already operated over 17.2 million pulses since 2014 (as of May 2018).
The current operation parameters are summarized in Tab.~\ref{tab:horn_ope_summary}.
\begin{table}
\centering
\small
\caption{\small Summary of current operation parameters of T2K horns (May 2018).}
\label{tab:horn_ope_summary}
\begin{tabular}{|l|c|c|}
\hline
Parameters & horn-1 & horn-2 \& horn-3 \\
\hline
Operation current & 250~kA & 250~kA \\
Operation cycle & 2.48~s & 2.48~s \\
Operation voltage & 4.8~kV & 6.5~kV \\
PS configuration & single & series \\
\hline
\end{tabular}
\end{table}

There were many problems since the beginning of the T2K operation, many of which are
relevant to high intensity beams even at the early stage.
The detailed description of development and operation experience of the first-generation
T2K horns can be found in~\cite{Sekiguchi:2015ghw}. Current operation status is described
in the following part of this section.
\medskip

\noindent\textbf{Corrosion \textcolor{\MODCOLOR}{and water leak problems}}
\medskip

The horn inner volume is filled with helium gas, however,
a large air contamination into the inner volume occurred in spring 2012.
Then, acidification of horn cooling water (pH$\sim$3.8) happened due to
nitrogen oxides by beam exposure to the contaminated air, and
aluminum conductors of the horns were corroded significantly.
In order to solve this problem, ion-exchanger system was introduced to
neutralize the cooling water and has been used during beam operation.
A remote pH measurement system, which can sample the cooling water and measure
its pH remotely, was also developed to monitor pH during beam operation.
Because of the corrosion problem, the first-generation horn1 had a small water leak and 
needed to be replaced with the second-generation.
After the replacement with the second-generation horns,
although a small air contamination sometimes occurs, the cooling water has been
kept around pH=6 and therefore no \textcolor{\MODCOLOR}{acidification of the cooling water} has been observed so far.

\begin{figure}
        \centering
        \includegraphics[width=0.8\linewidth]{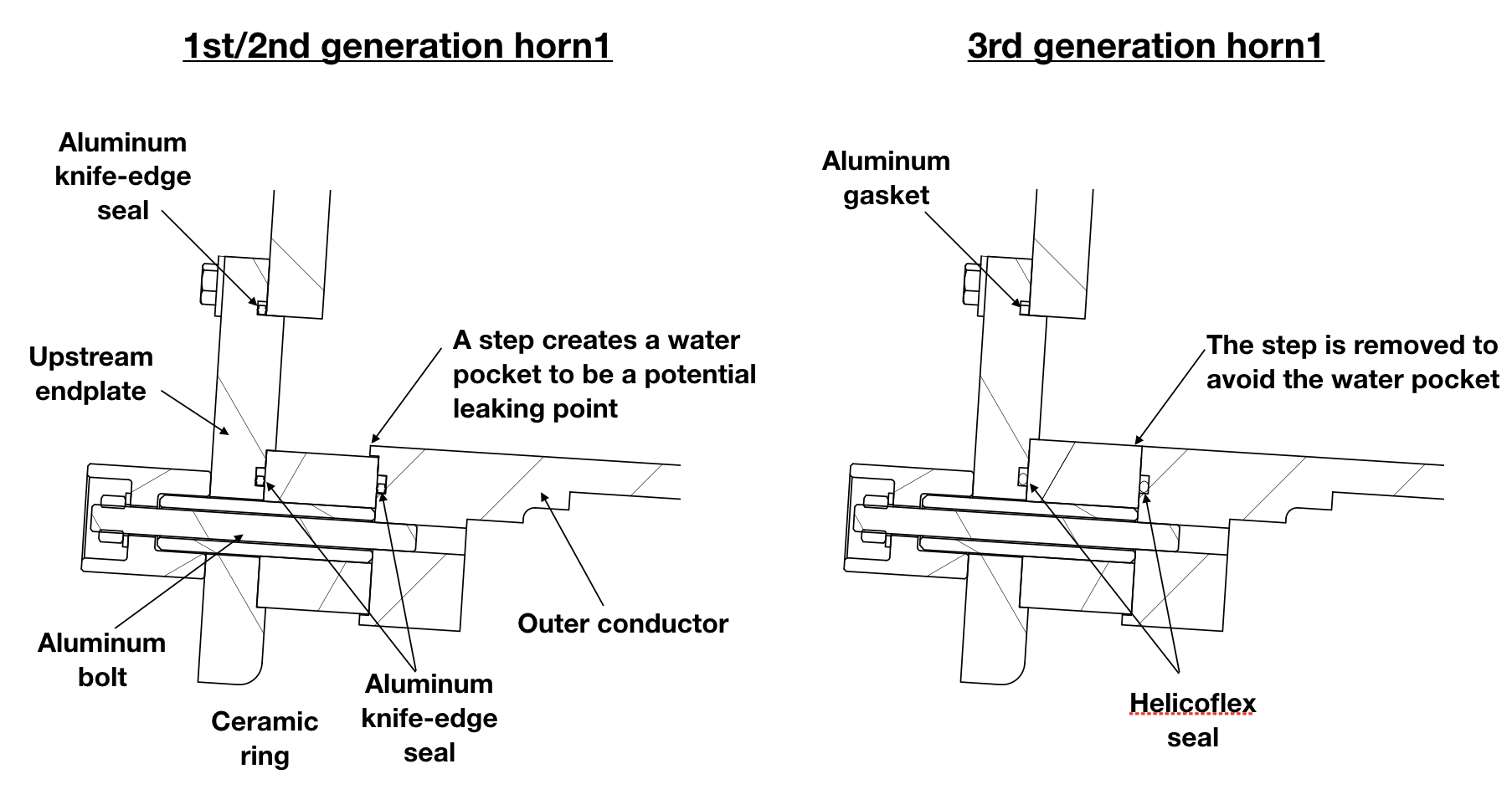}
        \caption{\small Drawings of upstream sealing structure of horn1 for 1st/2nd (left) and 3rd (right) generation horn1.}
        \label{fig:horn1seal}
\end{figure}

\color{\MODCOLOR}
The significant corrosion in the first generation horns resulted in a water leak from the horn1 after approximately 8 million pulses applied.
Aluminum knife-edge seals were used at the upstream end of the horn1 where the inner and outer
conductors were insulated with a large (510 mm in diameter) and thick (50 mm in thickness) ceramic ring and 
two aluminum knife-edge seals were placed on both sides of the ceramic ring, as shown in Fig.~\ref{fig:horn1seal}.
Since the inner diameter of the ceramic ring was a bit larger than that of the outer conductor, there was a step which created
a water pocket, resulting in an initial corrosion due to the acidic water, after no water circulation period in a long shutdown.
The water leak rate was approximately 5~L/day with only water circulation and doubled with beam operation of 220~kW.
There were 24 130 mm-long aluminum bolts at this upstream insulated joint, and
it is expected that a thermal expansion of the long aluminum bolts caused the increase of the leak rate
since the aluminum knife-edge seal did not have enough elasticity.
The leaked water dropped from the horn1 ran on the bottom of the helium vessel and collected at the most downstream
end where a drain pipe is connected and the leaked water can be drained at the machine room when a manual
valve is opened during beam off period. The significant amount of the leaked water existed during beam on in 2012-2013 run period
since the manual drainage could not be frequently performed due to the limited access during beam operation, and thus
the bottom surface of the helium vessel was significantly corroded, though no water leak from the helium vessel cooling water
was observed so far.

The second generation horn1 has the same upstream sealing structure as the first one.
It finally had a water leak of approximately 5~L/day with 485~kW beam, although no visible water leak
observed only with water circulation. The leak occurred in spring 2018 after 12.5 million pulses applied.
In this time only small amount of the leaked water was observed at the helium vessel
drain port since most of the leaked water could be evaporated due to helium circulation in the helium vessel and then
the water vapor in the helium circulation could be condensed after helium compressor. The condensed water was finally collected
in a drain tank connected to the helium circulation line. Thanks to the small leak rate and the water evaporation there was 
almost no extra damage of the helium vessel bottom surface.
After some water leak investigations, the leak point was identified to be the upstream insulation of horn1 as same before.
It is expected that the initial cause of the leak is due to a gap corrosion at the upstream end because of both the water pocket
and no water circulation period for two months during beam off time. Then both the thermal expansion and higher inner temperature 
(resulting in higher inner pressure) could accelerate the water leak rate during beam operation.
Based on the experience of the water leak, the third generation horn1 was already modified as shown in Fig.~\ref{fig:horn1seal}.
The sealing of the third generation horn1 was modified to use Helicoflex seal instead of the aluminum knife-edge seal.
The Helicoflex seal has a spring in its structure resulting in enough elasticity even for the case of thermal expansion of the bolts.
The step structure at the upstream ceramic ring was also removed to avoid the water pocket. This modification will greatly avoid
potential initial corrosion. The other consideration on the modification is the use of the aluminum bolts. Because of the large thermal
expansion coefficient and concern on a long term reliability of the aluminum bolts, an alternative is titanium bolt for this joint.
Its thermal expansion coefficient of 8.8$\times 10^{-6}$ is smaller by a factor of 2.7 and also similar to that of the alumina ceramic 
used for the upstream ceramic ring (7.3$\times 10^{-6}$). Therefore the effect by the thermal expansion should be much smaller 
than the aluminum bolts.
The aluminum bolts will be replaced with the titanium ones before installing the third generation horn1.
Although several countermeasures will be made, a water leak may be unavoidable for longer-term operation ($O(10)$ years).
In this sense the reinforcement of dehumidification inside the helium vessel should be considered in order to protect
the helium vessel and beam dump (especially to avoid oxidization of the graphite core blocks).
\medskip

\color{black}
\noindent\textbf{Hydrogen production problem}
\medskip

\begin{figure}
        \centering
        \includegraphics[width=0.9\linewidth]{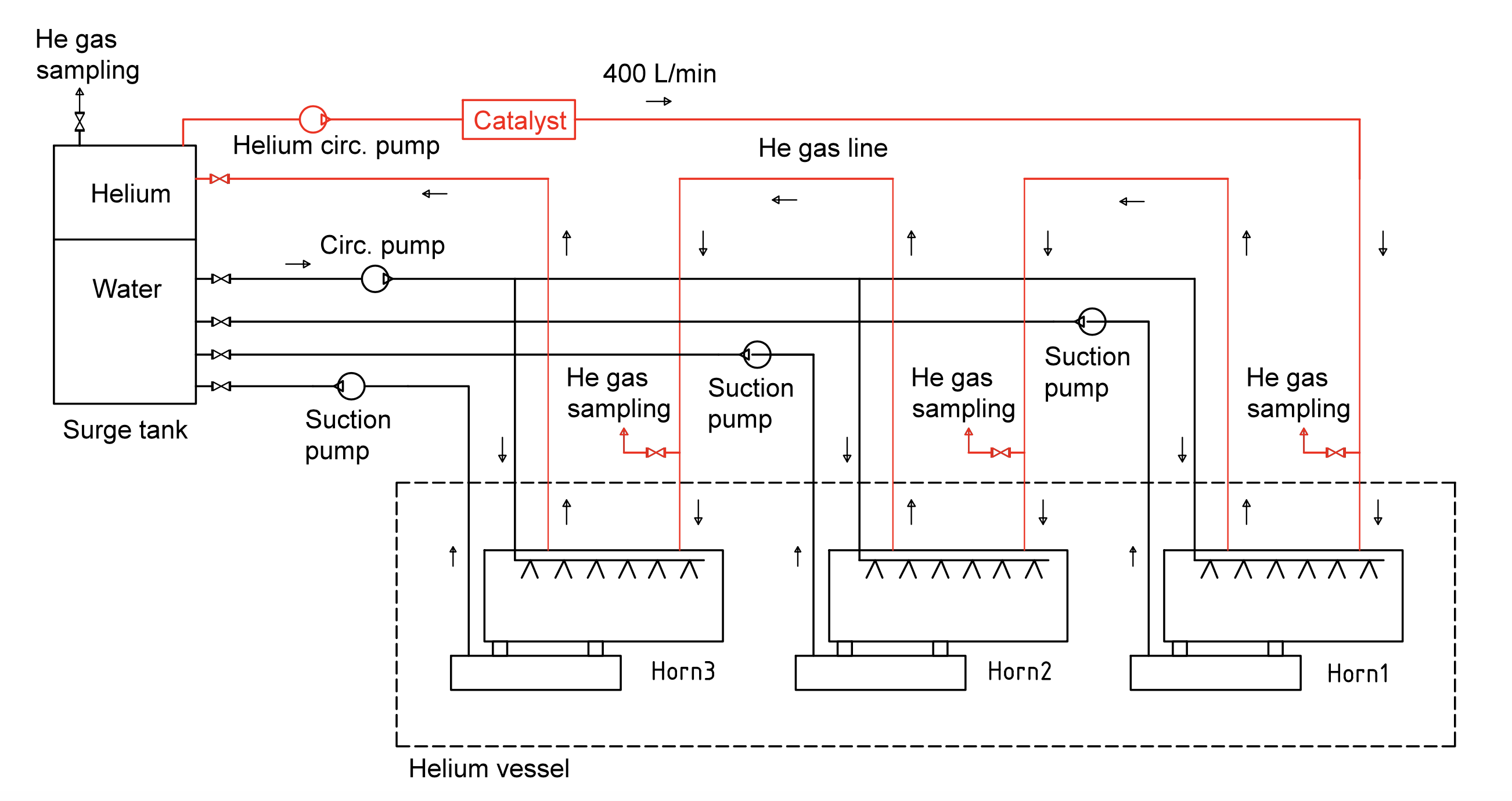}
        \caption{\small Schematic diagram of water circulation system for horns.
        The hydrogen recombination system is shown in red.}
        \label{fig:hydrogen_recombine}
\end{figure}

Hydrogen gas is produced by water radiolysis (i.e., decomposition of the cooling water 
by beam exposure). 
The hydrogen recombination system is adopted to the horn cooling water circulation system
as shown in Fig.~\ref{fig:hydrogen_recombine}.
The system has a catalyst (alumina pellet with 0.5\% paradium) which recombines hydrogen into water
(i.e., H$_2$ + 1/2O$_2$ $\to$ H$_2$O). 
The horn inner volume and the cover gas region of 
the surge tank are connected and the helium gas is circulated by helium circulation pump at flow rate of
\textcolor{\MODCOLOR}{400}~L/min. All three horns are connected in series in this helium circulation loop.
The total volume of helium gas is approximately 5.5~m$^3$. The helium gas is flown
through the catalyst that is kept at 60~$^{\circ}$C by heater in order to increase 
hydrogen recombination rate.
In-situ gas chromatography system can measure gas contamination in the helium atmosphere
by remotely sampling the helium gas from the circulation loop.
This recombination system can work well with presence of enough oxygen. However, it is found that the oxygen 
concentration is very low at $O$(10)~ppm level after starting the operation and therefore
the hydrogen concentration increases gradually. 
Typical hydrogen concentrations in several run periods are summarized in Tab.~\ref{tab:h2_history}.
\begin{table}
        \centering
        \small
        \caption{\small Summary of the hydrogen concentration in the past T2K runs. Run9 was divided into
        three periods depending on the configuration of ion exchanger operation.}
        \begin{tabular}{lcccccc}
        \hline
         Run period                            			& Run6     & Run7     & Run8   	& Run9-1  & Run9-2	& Run9-3 \\
         (Fiscal year)                         			& (2014)   & (2015)   & (2016) 	& (2017)   & (2017) 	& (2018)   \\\hline
         Ion exchanger configuration           		& old       	& new       & old    	& new	& new	& new	\\
         								& (single)	& (single)	& (single)	& (dual)	& (single)	& (single) 	\\
         Typical beam power (kW)              		& 330       & 390       	& 470    	& 450       & 475	& 485        \\
         \textcolor{\MODCOLOR}{Maximum H$_2$ concentration	 (\%)}	
      & \textcolor{\MODCOLOR}{0.4}  
      & \textcolor{\MODCOLOR}{2.5}		
      & \textcolor{\MODCOLOR}{1.5}		
      & \textcolor{\MODCOLOR}{4.0}		
      & \textcolor{\MODCOLOR}{1.0}		
      & \textcolor{\MODCOLOR}{2.4}		\\
         H$_2$ concentration (\%/10$^{19}$POT) & 0.094     & 0.482    & 0.117  	& 0.741    & 0.191 	&  0.106    \\\hline
         \end{tabular}
         \label{tab:h2_history}
\end{table}
In Run6, hydrogen concentration was kept quite small, although the beam power was 330~kW.
In Run7, after ion exchanger was replaced with new one, hydrogen concentration immediately increased,
however, from Run7 to Run8, the concentration was gradually decreased.
In late Run8 period, water conductivity was increased which indicated lifetime of the ion exchanger.
Then dual new ion exchanger configuration was adopted to increase the performance of ion exchange.
After that, the hydrogen concentration immediately increased to 4\% level. Finally,
the configuration was returned to the single operation and the concentration was reduced to half.
It is likely that the hydrogen concentration is strongly related to the ion exchanger operation.
There were many studies and investigations on water radiolysis from 1970's for nuclear reactors.
Main final products from water radiolysis are hydrogen, hydrogen peroxide, and oxygen, 
and production rate of oxygen is quite smaller than that of hydrogen or hydrogen 
peroxide~\cite{water_radiolysis1,water_radiolysis2}, which is consistent with the measured small oxygen concentration. 
\color{\MODCOLOR} 
Since the hydrogen peroxide is unstable and  naturally decomposed into water and oxygen (i.e., H$_2$O$_2 \to$ H$_2$O
+ O$_2$), the oxygen from the hydrogen peroxide can be a source of the hydrogen recombination.  
A bare hydrogen production rate at 485~kW beam power was measured to be 260~L (or 4.7\%) per 
10$^{19}$ POT without the operation of the hydrogen recombination system. 
The significant amount of oxygen of 115~L (or 2.1\%) per 10$^{19}$ POT was also measured, which 
indicated that the hydrogen peroxide was produced and then decomposed. As a result, the hydrogen concentration can be greatly 
reduced to 0.1\% per 10$^{19}$ POT by the recombination system.
However, the remaining hydrogen peroxide, although the measured concentration was approximately 10~mg/L at 485~kW operation,
can corrode the ion exchange resins due to oxidization of the resins and their lifetime became less than 1 year at this 
moment. This issue should be solved in near future, as explained in Sec.~\ref{sec:horn_upgrade_plan}.

\color{black}
Hydrogen explosion limit in air environment is 4\% and the limit can be higher in the helium
atmosphere. A risk of the hydrogen explosion may be very low with a lack of oxygen.
However, the hydrogen concentration is kept below 4\% in our operation to further reduce
the risk. To satisfy this requirement, a helium gas flushing is performed at weekly maintenance
days and the hydrogen concentration can be reduced to 0.1\% after the flushing.
Thanks to this every week flushing, the increase of hydrogen concentration was approximately 0.7\% at one week operation at 485~kW.
If we restrict the hydrogen concentration below 3\%, it corresponds to the acceptable beam power of 2~MW.
However, the hydrogen concentration strongly depends on ion-exchanger condition and so on,
and the existence of the hydrogen peroxide can reduced the lifetime of ion-exchangers.
Therefore further improvement of the hydrogen removal system including ion-exchanger system 
is really needed for safe and reliable operation even at 1.3~MW.
\medskip

\noindent\textbf{Power supply problem}
\medskip

The other big issue is due to the power supply.
\color{\MODCOLOR}
The history of the horn power supply configuration is summarized in Tab.~\ref{tab:PShistory}.
\begin{table}
        \centering
        \small
        \caption{\small History of the horn power supply configuration.}
        \begin{tabular}{lcccccccc}
        \hline
	Period & Horn           & \multicolumn{2}{c}{Configuration} & Current & \multicolumn{2}{c}{Voltage (kV)} & \multicolumn{2}{c}{Pulse width} \\
                    &                   & Horn1           & Horn2+3               & (kA)      & Horn1          & Horn2+3              & Horn1            & Horn2+3          \\ \hline
	Run1   & 1st-gen             & K2K-PS      & K2K-PS                   & 250       & 4.5              & 5.4                        & 2.4                 & 3.6		       \\
	Run2   & 1st-gen             & \multicolumn{2}{c}{T2K-PS1}           & 250       & \multicolumn{2}{c}{6.2}               & \multicolumn{2}{c}{4.6}               \\
	Run3   & 1st-gen             & \multicolumn{2}{c}{K2K-PS}                & 200       & \multicolumn{2}{c}{5.4}               & \multicolumn{2}{c}{4.3}               \\
	            & 1st-gen             & \multicolumn{2}{c}{K2K-PS}                & 250       & \multicolumn{2}{c}{5.4}               & \multicolumn{2}{c}{4.3}               \\
	Run4   & 1st-gen             & \multicolumn{2}{c}{K2K-PS}                & 250       & \multicolumn{2}{c}{6.7}               & \multicolumn{2}{c}{4.3}               \\
	            & 1st-gen             & T2K-PS1      & K2K-PS                          & 250       & 4.1             & 5.4                         & 2.2                 & 3.4		       \\
	Run5    & 2nd-gen           & T2K-PS2      & K2K-PS                          & 250       & 5.3              & 4.5                        & 2.2                 & 3.4		       \\
	Run6    & 2nd-gen           & T2K-PS2      & K2K-PS                          & 250       & 5.3              & 4.5                        & 2.2                 & 3.4		       \\
	Run7    & 2nd-gen           & T2K-PS2      & T2K-PS2                     & 250       & 5.3              & 6.3                        & 2.2                 & 2.8		       \\
	Run8    & 2nd-gen           & T2K-PS2      & T2K-PS2                     & 250       & 5.3              & 6.4                        & 2.2                 & 2.7		       \\
	Run9    & 2nd-gen           & T2K-PS2      & T2K-PS2                     & 250       & 4.8              & 6.5                        & 2.0                 & 2.7		       \\ \hline
        \end{tabular}
        \label{tab:PShistory}
\end{table}
\color{black}
From the beginning of the T2K operation, the horns have been operated at 250~kA
because the power supplies used for K2K magnetic horns \textcolor{\MODCOLOR}{(K2K-PS)}, whose rating current was 250~kA,
were refurbished for the T2K horns. The initial electrical configuration was that
one power supply operated horn1 and another
operated both horn2 and horn3 that were connected in series.
In fall 2010, one new power supply \textcolor{\MODCOLOR}{(T2K-PS1)} was developed to operate all three horns connected in series
at 320~kA~\cite{Koseki:2014hornPS}. 
Although the rating current of \textcolor{\MODCOLOR}{T2K-PS1} was 320~kA, the horns were
still operated at 250~kA for half a year. 
However, it had suffered from a serious damage due to a malfunction of IGBT switches
inside the power supply in the end of 2011, just after the recovery from the Great East Japan Earthquake
in 2011, and it took whole one year to recover it. During the recovery period,
one of the old \textcolor{\MODCOLOR}{K2K-PS} had been setup again and used for the three-horn series 
operation at 250~kA.
After the recovery of \textcolor{\MODCOLOR}{T2K-PS1}, two-power-supply configuration (\textcolor{\MODCOLOR}{T2K-PS1} for horn1
and \textcolor{\MODCOLOR}{K2K-PS} for horn2 and horn3 in series) was adopted again in order to reduce
their operation voltage since a high voltage operation such as 10 kV can often cause
some problems on semiconductor switches such as IGBTs and Thyristors and lower voltage operation
can significantly reduce a risk of their malfunction. The other new power supply \textcolor{\MODCOLOR}{(T2K-PS2)} and new
low-impedance striplines were then
developed for 320~kA and 1~Hz operation in 2014 and two \textcolor{\MODCOLOR}{of T2K-PS2} were produced
and installed in spring 2014 (one for horn1, and another for horn2 and horn3 in series). 
All striplines inside the helium vessel were also replaced 
with new one. They are currently under operation at 250~kA.
As described in Sec.~\ref{sec:horn_upgrade_plan}, three-power-supply configuration
is proposed for 320~kA operation at 1~Hz.

\begin{figure}
        \centering
        \includegraphics[width=0.7\linewidth]{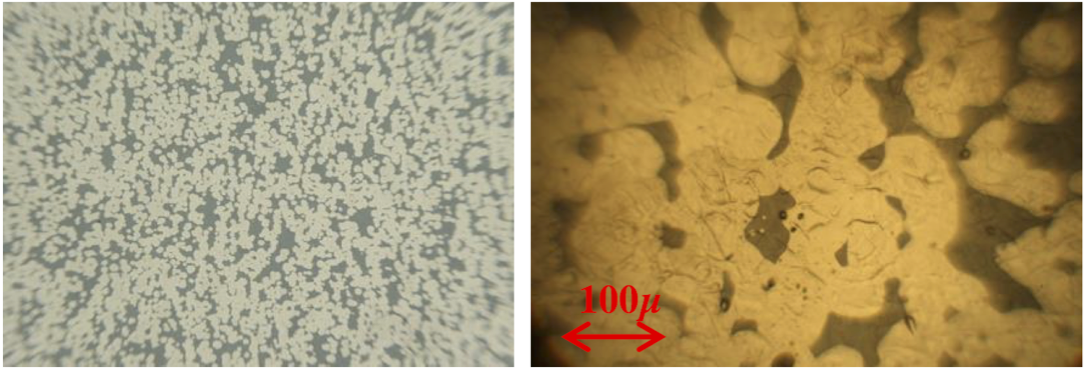}
        \caption{\small Pictures of the corroded surface on the aluminum metalized electrode
        observed in the visual inspection. In left picture,
        gray part shows the original electrode and white part shows the aluminum dioxide.
        Right picture is a zoom up of the electrode.}
        \label{fig:capacitor_inspection}
\end{figure}

During the run period from October 2016 to April 2017,
the horn current has been constantly decreased by 1.4\%.
The pulse width also gradually got narrowed by 1.2\% in this period.
After the physics run, the capacitance of each capacitor in the power supply
was measured. The measured capacitance was clearly decreased by 5$\sim$10\%
compared to the original value. This amount of the decrease was outside
of the specification, $\Delta$C/C $<$ 5\% at 2$\times 10^8$ shots,
although the number of operated shots was at most 1.25$\times 10^7$ at that time.
One of the capacitors that showed about 10\% decrease was investigated
by the manufacturing company. In the visual inspection, many spots of aluminum dioxide 
were observed on the aluminum-metalized electrode, as shown in Fig.~\ref{fig:capacitor_inspection}. 
It is thought that a very small quantity of 
the aluminum metalized electrode was lost during each operation cycle with
polarity inversion due to electro-corrosion of the electrode.
The solution to this phenomenon has been developed in recent three years.
It is proved that the addition of Zn in combination with aluminum for
the metallization is of great help to reduce the phenomenon~\cite{capacitor_test_CERN}. 
For the countermeasures to this problem, all the capacitors must be replaced with
the new ones with Zn and aluminum metallization. The details will be described in
Sec.~\ref{sec:horn_upgrade_plan}.

\subsubsection{Requirements and prospect for 1.3~MW operation}
\label{sec:upgrade}

The proposed operation condition for 1.3~MW beam is proton intensity of 3.2$\times 10^{14}$
protons/pulse and 1.16 s cycle. 
Since the horn operation must be synchronized with the accelerator cycle,
the high repetition rate operation of the horns is necessary.
The old power supplies (the K2K PS and the new T2K PS for three-horn configuration)
could not be operated at higher than 0.5~Hz. Therefore we have developed a 
new power supply that can be operated at 1~Hz. Another important factor is
to achieve the original 320~kA operation because the current horn operation at 250~kA gives
10\% less neutrino flux at the far detector and it also increases a contamination
of wrong-sign neutrino background (i.e., anti-neutrinos in neutrino beam operation or neutrinos
in anti-neutrino beam operation) by 5$\sim$10\% depending on neutrino energy.  
The new horn electrical system, including power supply, transformer, and striplines, 
has been developed to satisfy 320~kA operation at 1~Hz. The details will be described
in Sec.~\ref{sec:horn_upgrade_plan}.

Horn conductor cooling performance is sufficient for the 1.3~MW operation
as described in the previous section.  
The horn water circulation system was designed based on heat load of 56~kW for the 750~kW beam.
Rated heat capacity of the system is 83.8~kW which is based on water flow rate of 200~L/min at heat exchanger.
\color{\MODCOLORB}The measured heat load at 470 kW operation is 42.2 kW, which is actually larger than expectation.
Based on the measured values, the total heat load at 1.3 MW is estimated to be 127.2 kW.\color{black}
Since rated heat capacity of the current heat exchanger in this system is 122~kW,
\color{\MODCOLORB}the heat exchanger will be replaced with high spec one and water flow rate at the heat exchanger should also \color{black}
be increased up to 300~L/min. Therefore the secondary and tertiary water circulation system should also be upgraded
to increase their water flow rate. 
Heat load, heat capacity, and flow rate at the heat exchanger are summarized in Tab.~\ref{tab:heat_capacity}.
\begin{table}
        \centering
        \small
        \caption{\small Summary of heat load, heat capacity, and flow rate at the heat exchanger for
        current, 750~kW, and 1.3~MW conditions.}
        \begin{tabular}{lccc}
        \hline
        Beam power                          &  470~kW & 750~kW & 1.3~MW  \\
                                            &(current)&(design)&(upgrade)\\\hline
        \color{\MODCOLORB}Heat load at equipment (kW)         & 42.2    & 56.0   & 116.7\color{black}    \\
        Heat load at pumps (kW)             & 10.5    & 10.5   & 10.5    \\
        \color{\MODCOLORB}Total heat load (kW)                & 52.7    & 66.5   & 127.2\color{black}   \\
        Heat capacity (kW)                  &         & 83.8   & 122     \\
        Flow rate at heat exchanger (L/min) & 250     & 200    & 300     \\\hline
        \end{tabular}
        \label{tab:heat_capacity}
\end{table}
The details of the cooling capacity upgrade will be described in Sec.~\ref{sec:water_cooling_upgrade}.

\begin{figure}
        \centering
        \includegraphics[width=0.7\linewidth]{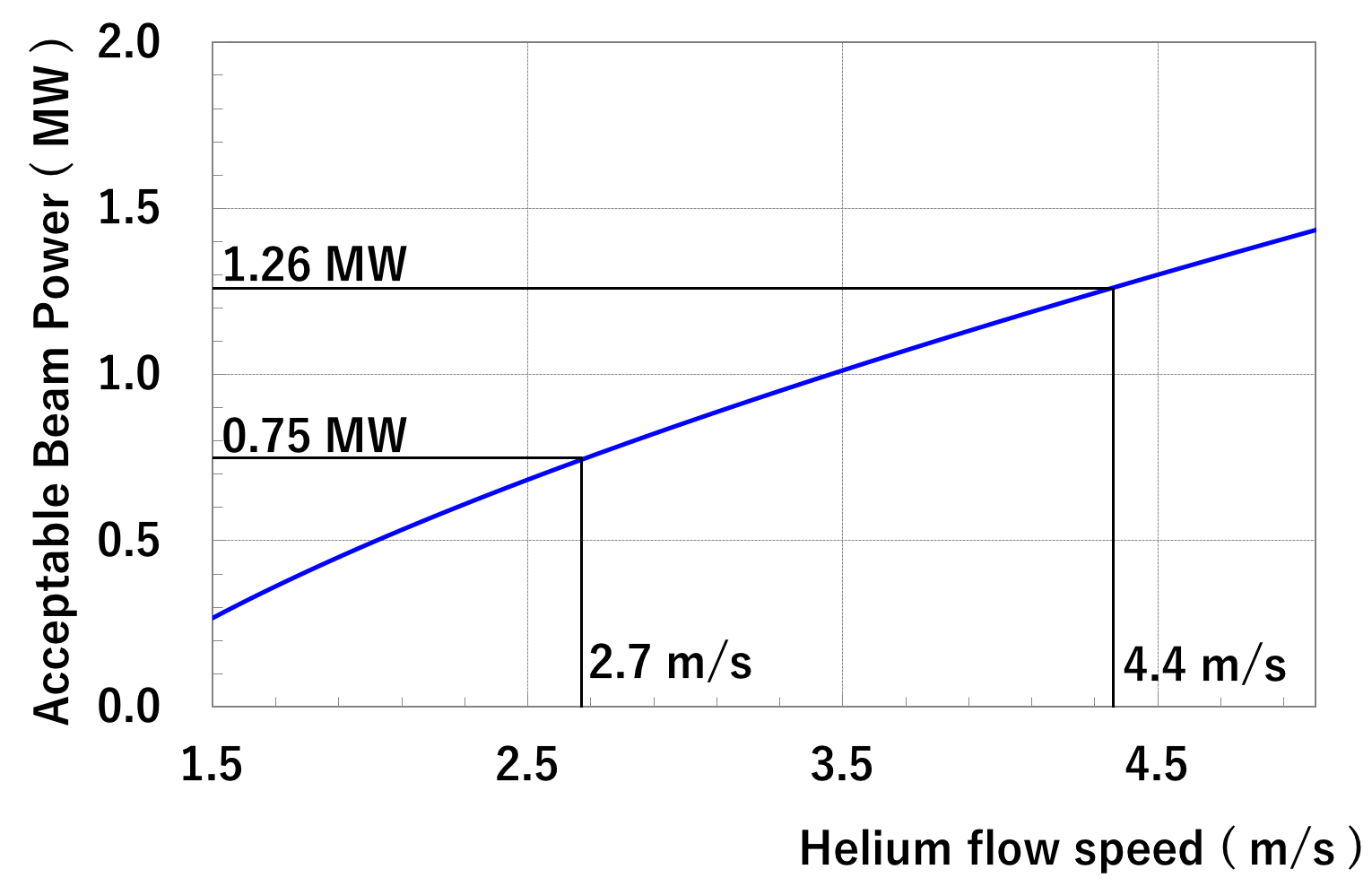}
        \caption{\small Acceptable beam power related to the horn2 stripline cooling 
        by forced helium gas flow as a function of helium flow speed.}
        \label{fig:stripline_cooling_vs_flow}
\end{figure}

The stripline cooling with the forced helium flow can accommodate only 750~kW beam with the current
helium flow rate. The flow rate can be improved by using dual helium compressors.
As a trial all the helium flow was applied only to the horn2 stripline ducts and
the helium flow rate was measured. Based on the measured flow rate 
the helium flow speed of 4.4~m/s could be achieved with the existing helium compressor.
In this case, the acceptable beam power can be improved to 1.26~MW as shown in
Fig.~\ref{fig:stripline_cooling_vs_flow}.
With the actual dual compressor configuration it may be feasible to adjust the helium flow rate 
for each horn and to increase the flow rate for horn2 a bit more to accept 1.3~MW beam.
It is however difficult to obtain much higher cooling performance by the forced helium flow scheme
for future higher beam power than 1.3~MW.
A totally different cooling scheme, a water cooling method, is under development.
The new water cooled striplines will be adopted to the horn2 for the beam power over 750~kW.
The details will be described in Sec.~\ref{sec:horn_upgrade_plan}.

The concentration of hydrogen produced by water radiolysis is as high as 0.7\% at 
1 week operation at 485~kW.
Thanks to the weekly helium gas flushing the hydrogen concentration is kept below
the requirement of 3\% at this moment. However, the hydrogen production rate depends
on condition of ion-exchanger operation and lifetime of the ion-exchanger is affected by
the existence of hydrogen peroxide. Therefore, hydrogen removal scheme should 
be modified to achieve safe and reliable operation for 1.3~MW or higher beam power.

\subsubsection{Upgrade plan}
\label{sec:horn_upgrade_plan}

In this section, details of the upgrades toward 1.3~MW are described.
The upgrade items are listed below.
\begin{itemize}
\item Upgrade of the horn electrical system for 320~kA and 1~Hz operation. 
\item Improvement of the hydrogen removal system.
\item Improvement of the stripline cooling method.
\end{itemize}
\medskip

\noindent\textbf{Electrical system upgrade}
\medskip

\begin{figure}
        \centering
        \includegraphics[width=0.7\linewidth]{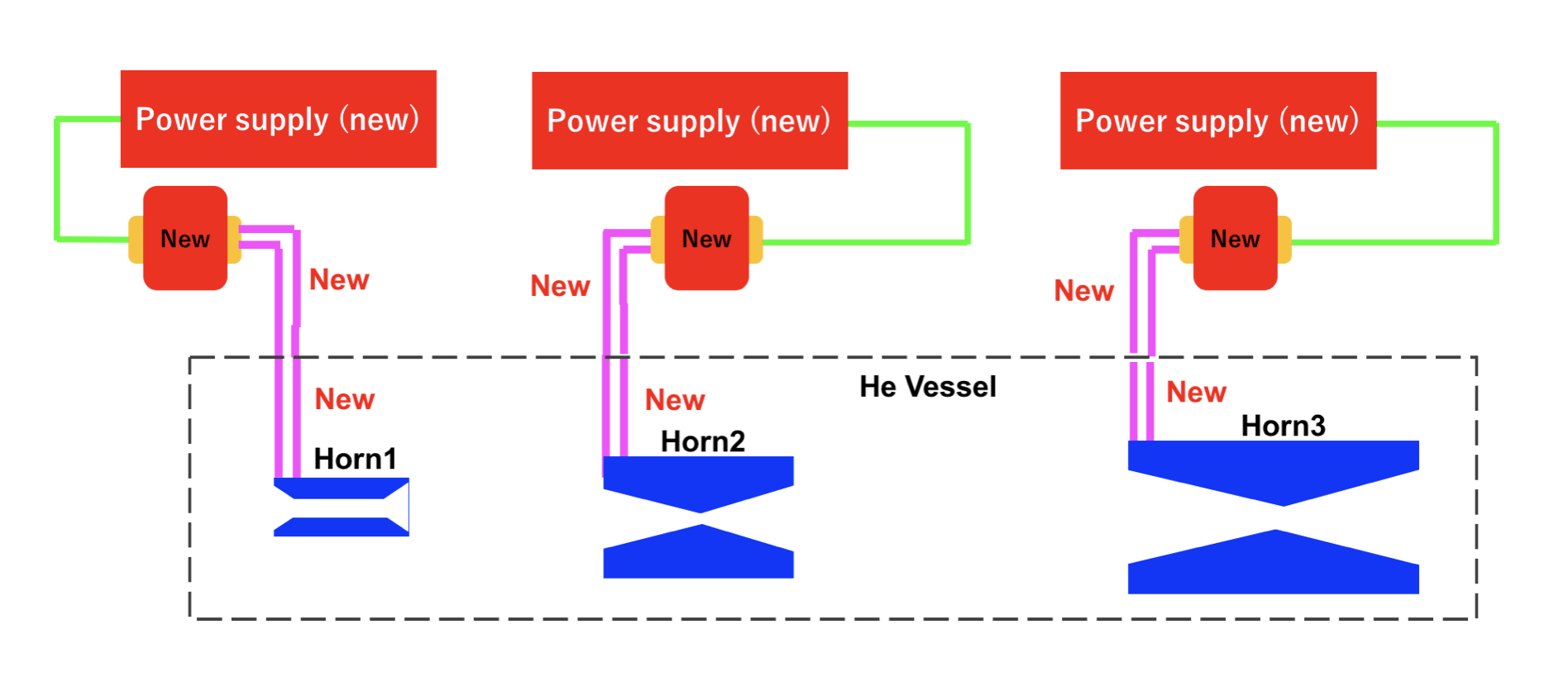}
        \caption{\small Schematic figure of the three-power-supply configuration for
        320~kA and 1~Hz operation.}
        \label{fig:3PSconfig}
\end{figure}

The 1~Hz operation requires a shorter charging time, 
while 320~kA operation requires higher operation voltage than that for 250~kA.
Lower voltage operation is desirable to reduce a risk of failure.
Therefore input load for one power supply should be as small as possible.
To meet these requirements, the following are adopted.
\begin{itemize}
\item Single horn operation by one power supply (i.e., three-power-supply configuration instead of
      the current two-power-supply configuration), which can greatly reduce the input load for
      power supply  
\item Energy recovery system (i.e., recycling electrical charge returned from the horns)
      to shorten charging time
\item New striplines that have lower inductance and resistance than the current ones
\item New transformer that has the rated current of 320~kA. Three transformers are needed.
      Because of the limited space, these transformers should be compact compared to the existing
      K2K transformers. 
\end{itemize}
A schematic figure of the three-power-supply configuration is shown in Fig.~\ref{fig:3PSconfig}.
The stripline structure was already modified to reduce both inductance and resistance.
The original and modified parameters for the striplines are summarized in Tab.~\ref{tab:stripline_parameter}.
\begin{table}
        \centering
        \small
        \caption{\small Summary of parameters for the original and modified striplines.}
        \begin{tabular}{lccl}
        \hline
        Parameter                 & Original      & Modified  &      					\\\hline
        Width (mm)                & 400            & 500         & (near horns) 			\\
	                                  & 400            & 400         & (at support module) 	\\
	                                  & 400            & 600         &(above support module) 	\\
        Thickness (mm)         & 10   	     & 12           &               				\\
        Gap (mm)                  & 20              & 15           &               				\\\hline
        \end{tabular}
        \label{tab:stripline_parameter}
\end{table}
Comparison of the input loads between the old and new electrical circuits is also shown in
Tab.~\ref{tab:comp_input_load}.
\begin{table}
        \centering
        \small
        \caption{\small Comparison of the input load between the old and new electrical
        circuits.}
        \begin{tabular}{lrrrrrr}
        \hline
        \multicolumn{7}{c}{Old configuration} \\\hline
        Components      & \multicolumn{2}{c}{horn1} & \multicolumn{4}{c}{horn2 + horn3} \\\cline{2-7}
                        & L ($\mu$H) & R (m$\Omega$) & \multicolumn{2}{c}{L ($\mu$H)} & \multicolumn{2}{c}{R (m$\Omega$)} \\\hline
        Horn            & 0.47       & 0.100         & \multicolumn{2}{c}{0.46+0.53} & \multicolumn{2}{c}{0.035+0.023} \\
        Striplines      & 0.28       & 0.100         & \multicolumn{2}{c}{0.60} & \multicolumn{2}{c}{0.210} \\
        Transformer     & 0.30       & 0.040         & \multicolumn{2}{c}{0.30} & \multicolumn{2}{c}{0.040} \\\hline
        Total           & 1.05       & 0.240         & \multicolumn{2}{c}{1.89} & \multicolumn{2}{c}{0.308} \\\hline\hline
        \multicolumn{7}{c}{New configuration} \\\hline
        Components      & \multicolumn{2}{c}{horn1} & \multicolumn{2}{c}{horn2} & \multicolumn{2}{c}{horn3} \\\cline{2-7}
                        & L ($\mu$H) & R (m$\Omega$) & L ($\mu$H) & R (m$\Omega$) & L ($\mu$H) & R (m$\Omega$) \\\hline
        Horn            & 0.47       & 0.100         & 0.46       & 0.035         & 0.53       & 0.023         \\
        Striplines      & 0.15       & 0.056         & 0.17       & 0.060         & 0.18       & 0.065         \\
        Transformer     & 0.25       & 0.025         & 0.25       & 0.025         & 0.25       & 0.025         \\\hline
        Total           & 0.87       & 0.181         & 0.88       & 0.120         & 0.96       & 0.113         \\\hline
        \end{tabular}
        \label{tab:comp_input_load}
\end{table}

\begin{figure}
        \centering
        \includegraphics[width=\linewidth]{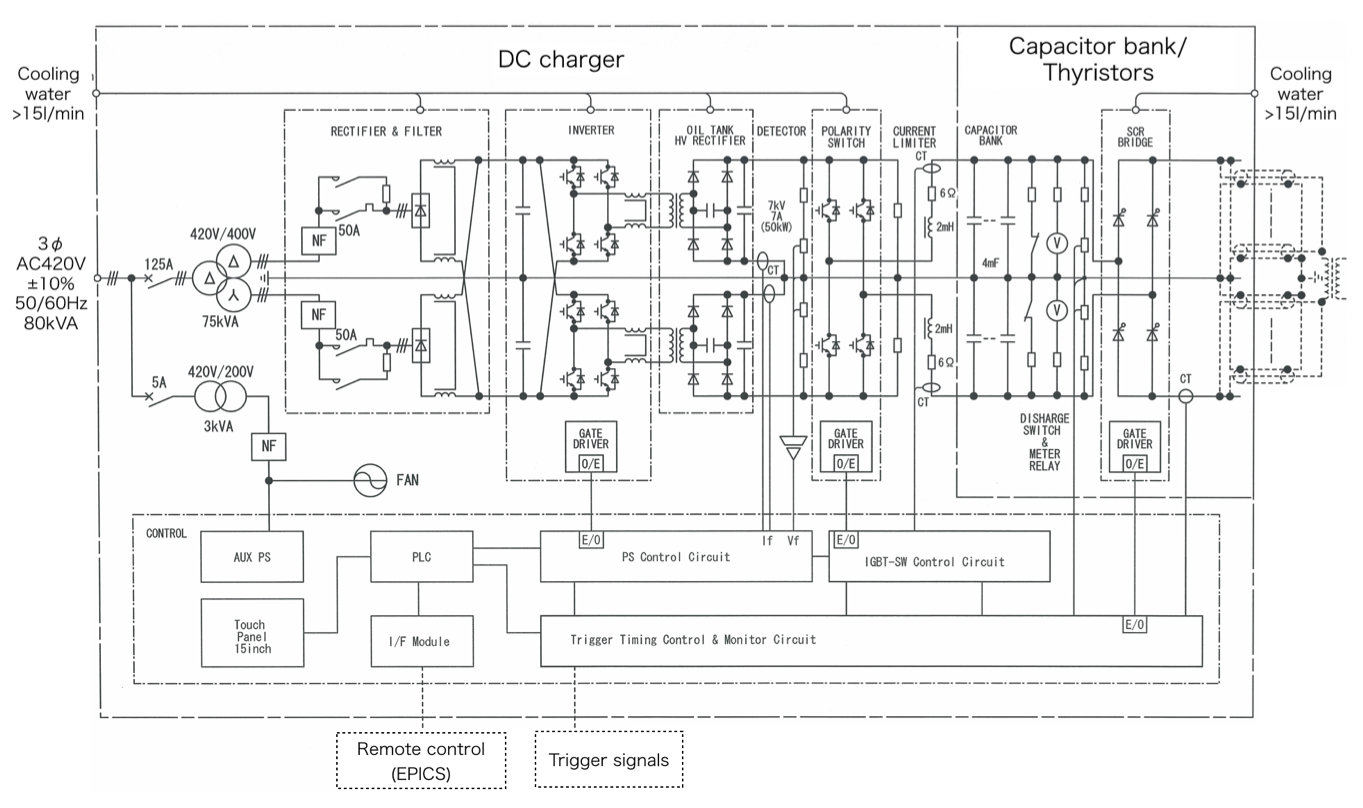}
        \caption{\small Schematic diagram of the power supply circuit.}
        \label{fig:PS_diagram}
\end{figure}

Specification of the new power supply is summarized in Tab.~\ref{tab:PS_spec}.
\begin{table}
        \centering
        \small
        \caption{\small Summary of specification of the new power supply.}
        \begin{tabular}{lc}
        \hline
        Item            & Value  \\\hline
        Rated operation voltage  & 7~kV \\
        Rated charging current   & 7~A  \\
        Charging unit            & 50~kW \\
        Rated operation cycle    & 1~Hz \\
        Total capacitance        & 4~mF \\
        Capacitor bank configuration & \\
        ~~~~~(original design)        & 2S16P (0.5~mF$\times$32) \\
        ~~~~~(modified)               & 2S24P (0.335~mF$\times$48) \\
        Pulse width              & 2~ms\\
        Rated output current     & 32~kA \\
        Stored energy            & 98~kJ  \\\hline
        \end{tabular}
        \label{tab:PS_spec}
\end{table}
Schematic diagram of the new power supply is shown in Fig.~\ref{fig:PS_diagram}.
The energy recovery system with full-bridge circuits was already adopted in \textcolor{\MODCOLOR}{T2K-PS1}~\cite{Koseki:2014hornPS}. 
The polarity of capacitor bank is alternatingly
changed pulse by pulse but the output current should have the same polarity at every discharge.
A full-bridge IGBT circuit, ``polarity switch'', is employed between charger and capacitor 
bank to control
the polarity of charging current. Also used is a full-bridge thyristor circuit to control
the polarity of output current to be same at every pulse. Many semi-conductor switches
are used in this system. As happened in the previous T2K power supply, any malfunction of
such high voltage semi-conductor switches can cause a critical damage on power supply, which
results in a long downtime of experiments. In the new power supply, a safety
system, ``current limiter'', is adopted to avoid a large current flow to the charging circuit by introducing
an inductive load of 2~mH between the polarity switch and the capacitor bank.
The current limiter can successfully reduce a reverse current flow to below 600~A, 
therefore, IGBTs are protected by this limiter.
Circuit simulations were performed based on these new parameters.
The obtained operation parameters are summarized in Tab.~\ref{tab:HornPS_simulation}.
\begin{table}
        \centering
        \small
        \caption{\small Simulated operation parameters for the new horn electrical system.}
        \begin{tabular}{lccc}
        \hline
        Parameter             & horn1    & horn2     & horn3       \\\hline
        Operation current     & 323~kA   & 323~kA    & 323~kA      \\
        Operation voltage     & 5.85~kV  & 5.72~kV   & 5.91~kV     \\
        Returned  voltage     & 4.60~kV  & 4.78~kV   & 5.00~kV     \\
        Voltage recovery rate & 78.6~\%  & 83.6~\%   & 84.6~\%     \\
        Pulse width           & 2.00~ms  & 2.01~ms   & 2.08~ms     \\
        Charging time         & 0.71~s   & 0.54~s    & 0.52~s      \\\hline
        \end{tabular}
        \label{tab:HornPS_simulation}
\end{table}
For all horns, the operation voltage is expected to be 5.7$\sim$5.9~kV for 320~kA operation.
The pulse width is also expected to be 2.0$\sim$2.1~ms.
Thanks to the low input load, the returned voltages are 79\%$\sim$85\% of the operation
voltages and thus the charging time can be reduced to less than 0.71~s. 
It is expected that the requirement of 320~kA and 1~Hz can be satisfied with the new configuration.

Two \textcolor{\MODCOLOR}{of T2K-PS2} and the new striplines inside the helium vessel
have been produced and installed in 2014. They have been operated stably so far.
One of the new transformer was also produced and installed in 2017. 
All the necessary upgrades were made for horn1 and 
then 320~kA operation of horn1 was tested. The measured operation parameters were
as following: the measured peak current was 321.4~kA with the charging voltage of 6.05~kV,
and the measured pulse width was 1.98~ms. The measured voltage (pulse width) was slightly 
higher (narrower) than the expectation, which is consistent with the observed $5\sim 10$\%
decrease of the capacitance. A short-term continuous operation for 4 hours was also
performed, although the operation cycle was 2.48~s, and steady-state temperatures at several 
places of the new transformer were measured. 
The maximum temperature was measured to be 35.3~$^{\circ}$C
at the secondary copper busbar, whereas the cooling water temperature was 26~$^{\circ}$C. 
For 1~Hz operation the maximum temperature is expected to be at most 49.1~$^{\circ}$C,
which is well below the temperature limit of 60~$^{\circ}$C at the transformer.
Therefore the feasibility of 320~kA and 1~Hz operation is confirmed.

\begin{figure}
        \centering
        \includegraphics[width=\linewidth]{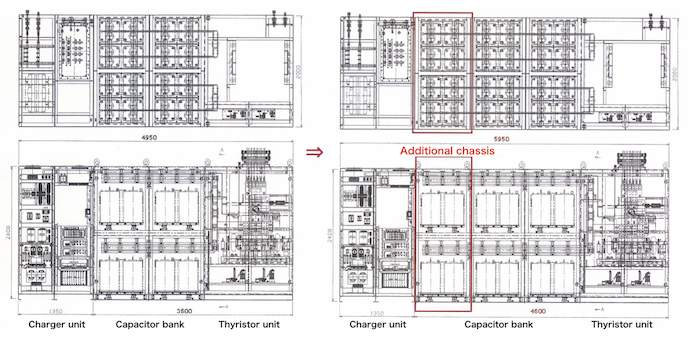}
        \caption{\small Drawings of the power supply in the current (left) and modified (right)
        configuration. Upper and lower figures show the top and side views, respectively.}
        \label{fig:NewPS_chassis}
\end{figure}

Remaining upgrade items are production of one \textcolor{\MODCOLOR}{T2K-PS2}, two new
transformers, and new striplines outside the helium vessel (for horn2 and horn3). 
All the capacitors with the countermeasures against the electro-corrosion must be produced and installed to the
existing power supplies. The new production method requires about 30\% thicker electrode,
which results in larger volume of the single capacitor by about 50\%. 
Since the new capacitors should be
installed to the existing power supplies, the size of the capacitor should be almost same as
the old ones. Therefore the capacitance of the single new capacitor is changed from
500~$\mu$F to 335~$\mu$F and the number of the capacitors is accordingly increased from 32
to 48 pieces per one power supply in order to have the same total capacitance of 4~mF. 
To contain the additional 16 capacitors, the additional chassis will be attached to 
the existing power supply as shown in Fig.~\ref{fig:NewPS_chassis}.

\begin{figure}
        \centering
        \includegraphics[width=0.7\linewidth]{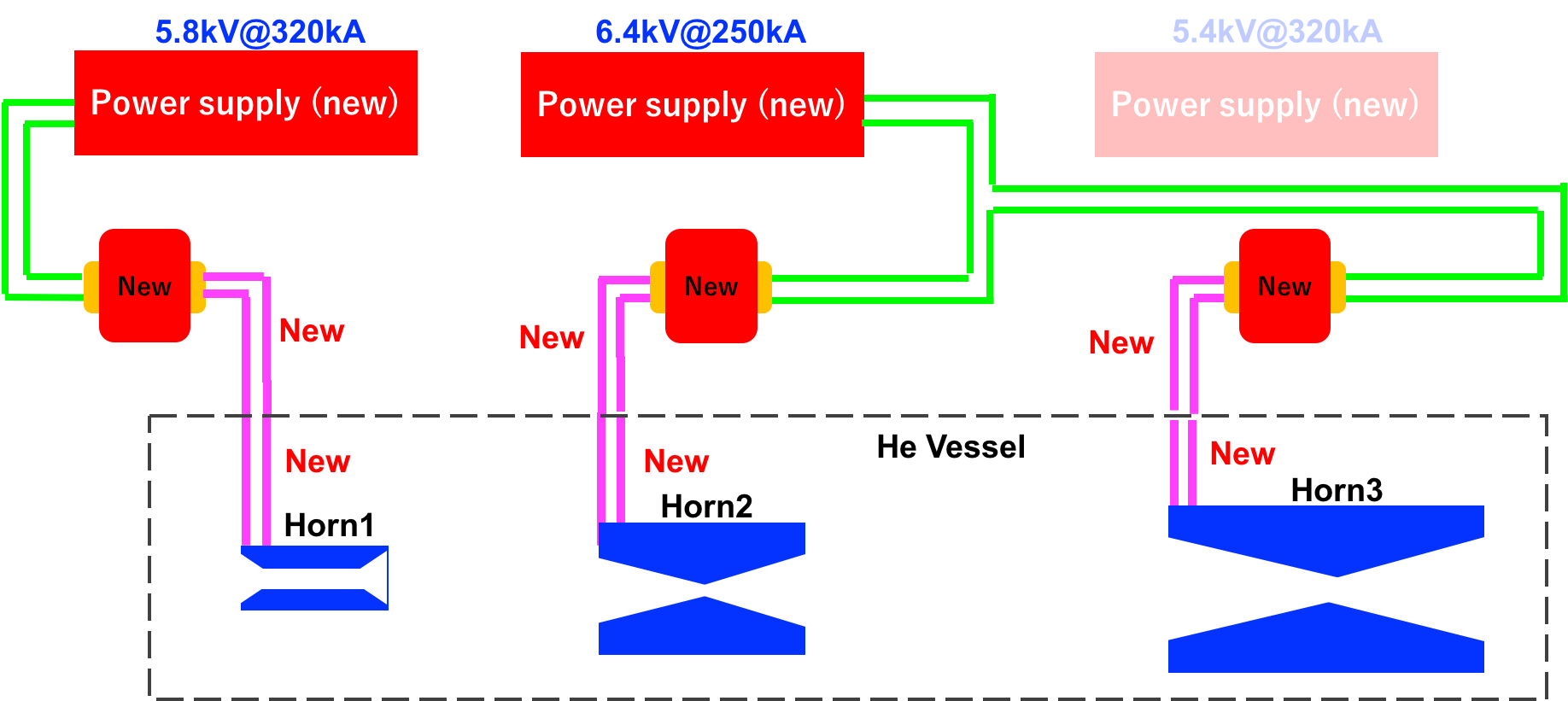}
        \caption{\small Schematic figure of two-transformer series configuration.}
        \label{fig:two-trans-series}
\end{figure}

\color{\MODCOLOR}
Those T2K-PS2 need to be operated for more than 10 years.
In case of some problems in the power supply, replacement of parts will be necessary.
It is very important to have enough spare parts, especially for some semiconductor switches,
which will be no longer in production sometime in future.
The other concern is that once three-power-supply configuration is realized, if one of them is broken due to some reasons,
any scheme or strategy to operate three horns by two power supplies must be considered.
It is very difficult to modify the striplines outside the helium vessel to the series configuration.
The alternative method is to connect two transformers in series in the primary circuit by changing cable connection.
The schematic figure of the configuration is shown in Fig.~\ref{fig:two-trans-series}.
A circuit simulation of this configuration was performed.
In order to keep the operation voltage below its rated voltage of 7kV (or a bit lower for safety),
250~kA operation with 2.8~ms pulse width can be feasible at 6.4~kV operation voltage, but 320~kA operation cannot be done.
This operation parameters are almost same as the current ones. Therefore reliable operation can be expected.
In this case, neutrino flux at the far detector decreases by 3$\sim$4\% compared to all horns at 320~kA.
The actual setup to realize the configuration is explained as following.
There is a cable exchanger in the power supply room in NU2 to change the configuration of
pairs of power supply and horn. For example, one power supply can selectively operate horn1 or horn2+3,
and another can operate vice versa. It can also flip the direction of current flow to change
the horn polarity. It is currently for two power supply configuration but will be modified for 
three power supply configuration later. Then series connection of two transformers can be performed at this cable exchanger.
Horn1 should be operated by one power supply anytime because heat load at horn1 is higher than other two and thus Joule heating
should be reduced, but horn2 and horn3 can be operated in series in this sense. If the power supply for horn1 is broken,
then one of two other power supplies is connected to horn1 by changing cable connection at the cable exchanger and another
is connected to horn2 and horn3 in series.
\color{black}
 \medskip
 
\noindent\textbf{Improvement of hydrogen removal system}
\medskip

As described in Sec.~\ref{sec:operation_status}, the acceptable beam power
related to the hydrogen production is enough for 1.3~MW operation, although
the weekly flushing is necessary. However, the current issue on the hydrogen removal
to be fixed to achieve a safe and reliable operation \textcolor{\MODCOLOR}{is removal of} 
hydrogen peroxide.
The hydrogen peroxide can damage ion
exchangers by oxidization in which process hydrogen may be produced as indicated from the measurements, 
thus it must also be removed in addition to hydrogen. The hydrogen peroxide can
be decomposed naturally into water and oxygen (i.e. 2H$_2$O$_2\to$2H$_2$O + O$_2$),
\color{\MODCOLOR}
however, the remaining hydrogen peroxide does affect the ion exchangers even though its concentration is not so high.
In order to avoid this degradation, a newly developed palladium-overlaid ion exchanger will be adopted
since the beam operation from Spring 2019 in addition to the usual ion exchangers.
The palladium can work as a catalyst for the decomposition of the hydrogen peroxide.
The performance of this new ion exchange resin was well tested with radioactive water of spent fuel pool in Tsuruga-2
nuclear power plant where hydrogen peroxide from water radiolysis is contained 
and its long-term stability and reliability were well established~\cite{Izumi}.
The utilization of the palladium-overlaid ion exchangers are expected to help reliable ion exchanger operation even at
1.3~MW.
\color{black}
Oxygen from the water radiolysis may be dissolved in the cooling water due to relatively high
solubility. With existence of such dissolved oxygen, rate of the water radiolysis is increased
because beam exposure to oxygen creates superoxide anions (O$^-_2$) which can work as an
accelerator of the water radiolysis.
In order to handle these problems, the following items will be considered.
\begin{itemize}
\item \textcolor{\MODCOLOR}{Palladium-overlaid ion exchangers which decompose the hydrogen peroxide will be adopted}. 
\item A degasifier system to remove the dissolved oxygen in the cooling water.
\end{itemize}  
With these improvements, the water cooling system with hydrogen removal scheme can be
operated with higher reliability and stability for 1.3~MW.
\medskip

\noindent\textbf{Improvement of stripline cooling}
\medskip

\begin{figure}
        \centering
        \includegraphics[width=0.8\linewidth]{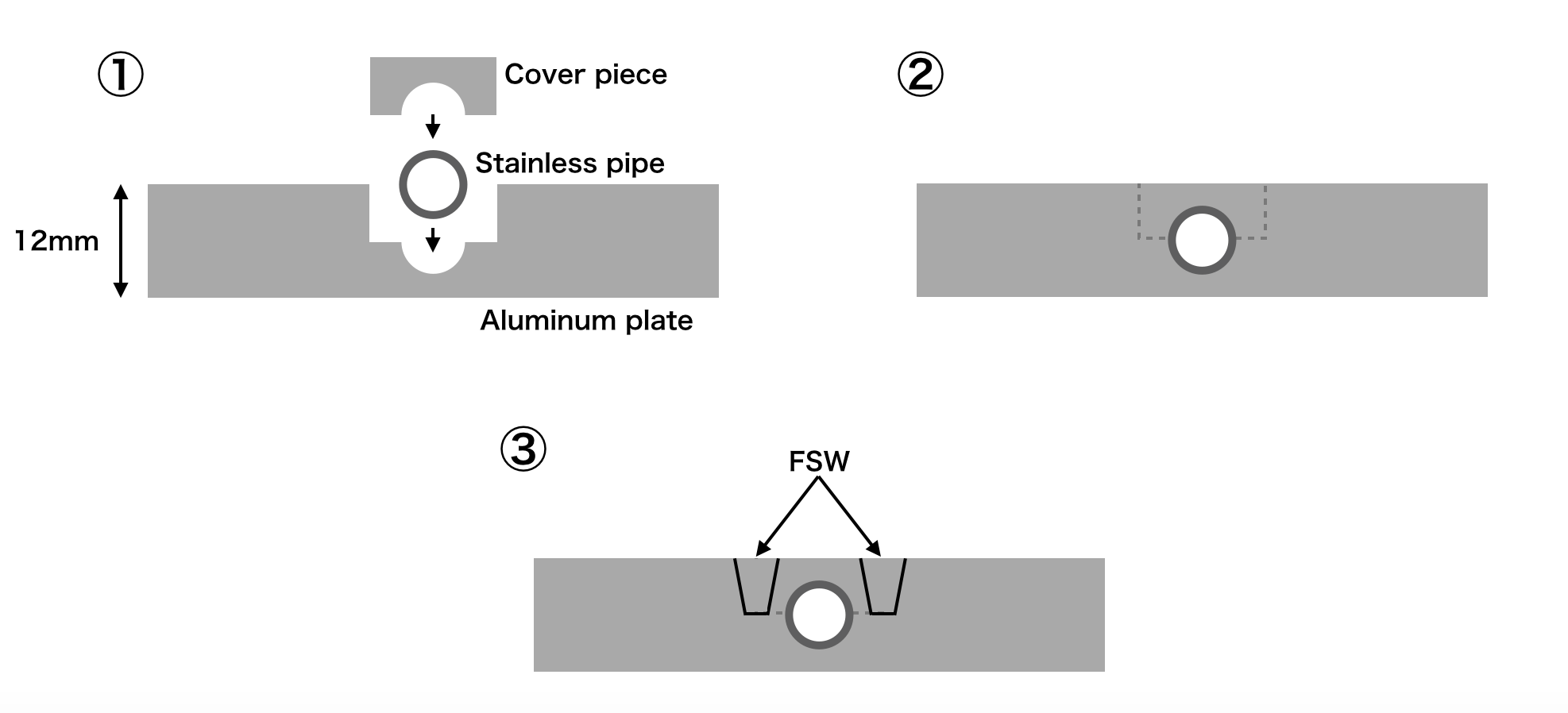}
        \caption{\small Schematic figure showing water cooling method by FSW technique.}
        \label{fig:FSW}
\end{figure}

\begin{figure}
        \centering
        \includegraphics[width=0.59\linewidth]{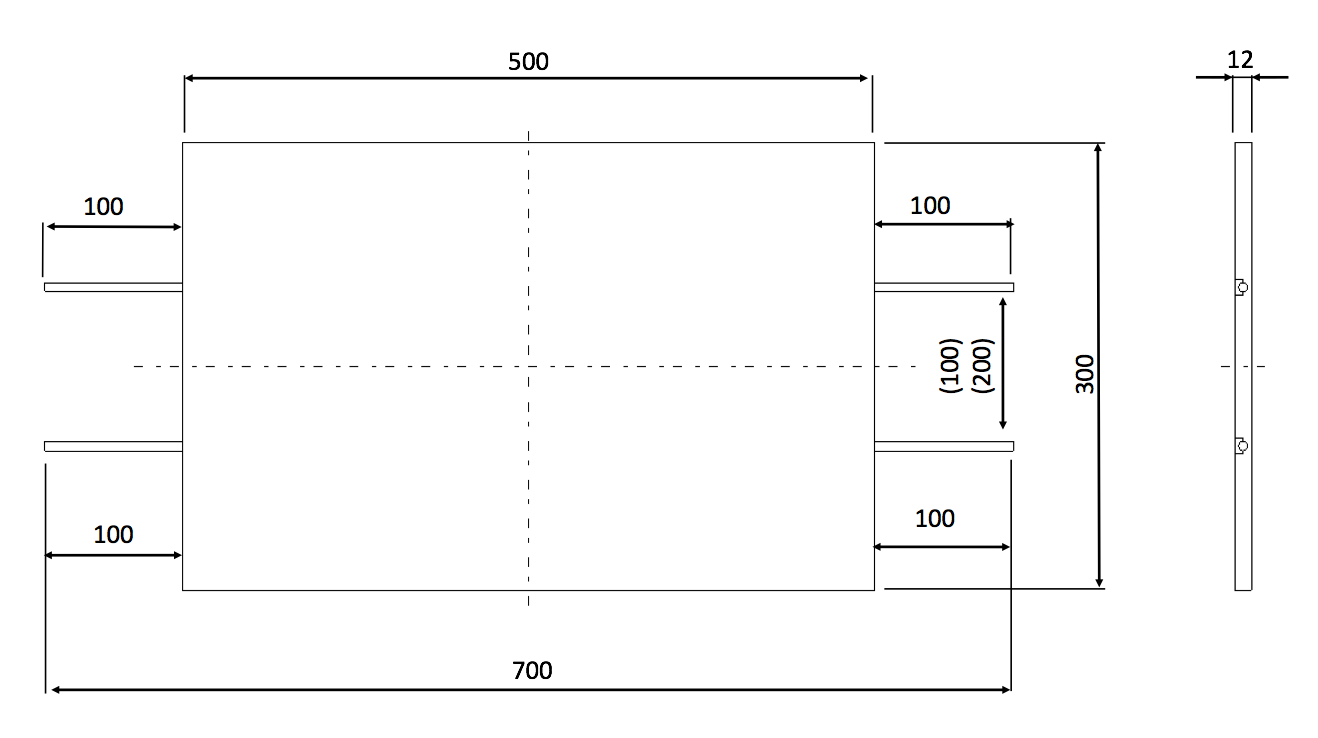}
        \includegraphics[width=0.4\linewidth]{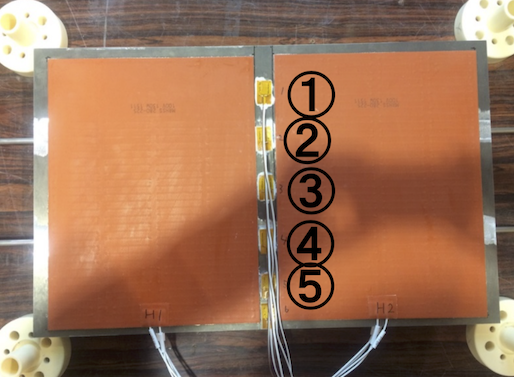}
        \caption{\small Drawing (left) and picture (right) of a test piece of water-cooled stripline. 
        Two 1/4 inch stainless pipes are embedded in 12~mm-thick aluminum plate by FSW.
        In the right picture, brown sheets are heaters attached on the aluminum surface. 
        Five RTDs are attached at the center of the test piece. 
        Two RTDs, No. 2 and 5, are located above the water pipes.}
        \label{fig:wcs_test_piece}
\end{figure}

\begin{figure}
        \centering
        \includegraphics[width=0.5\linewidth]{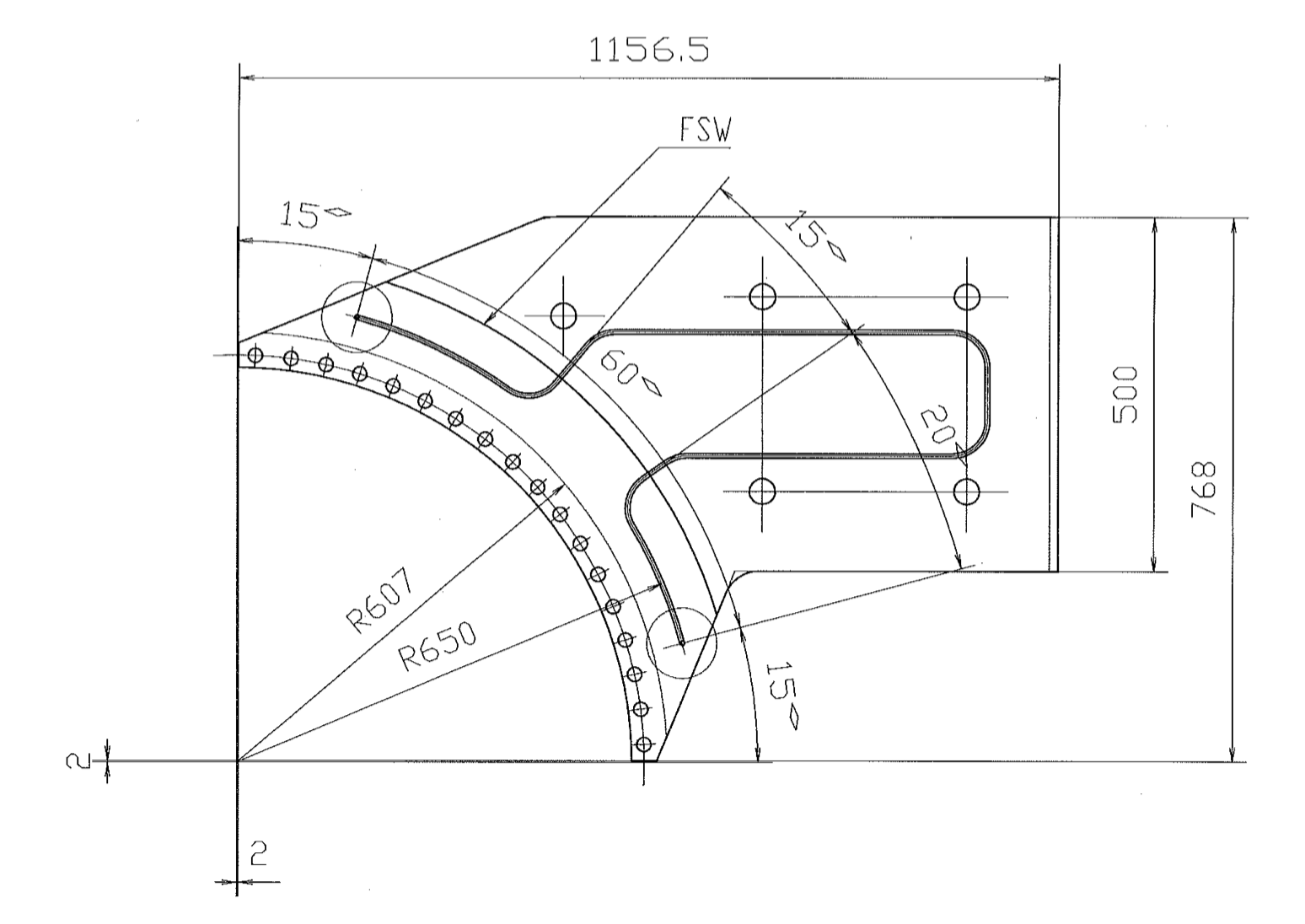}
        \includegraphics[width=0.45\linewidth]{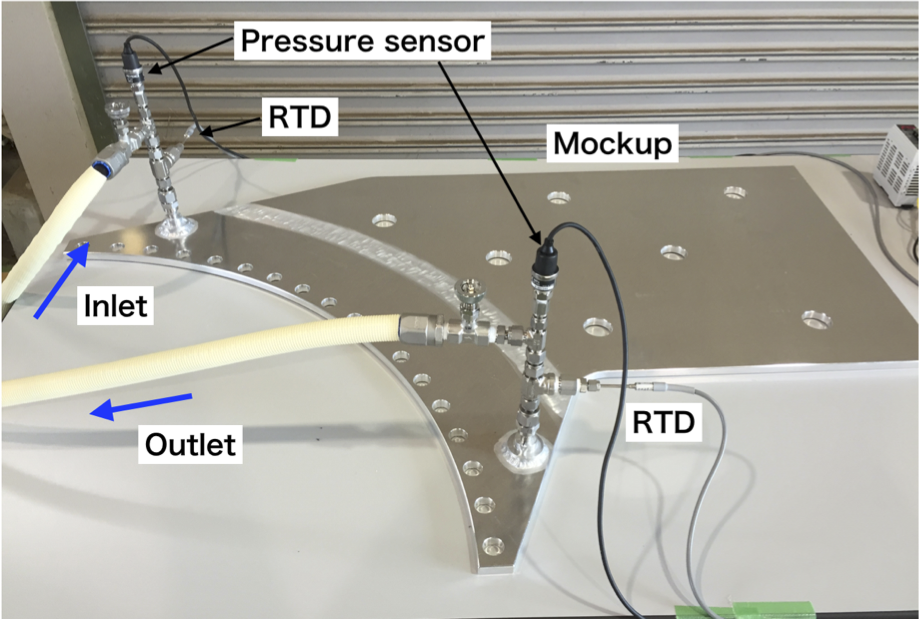}
        \caption{\small Drawing (left) and picture (right) of the mockup of the water 
        cooled striplines. }
        \label{fig:wcs_mockup}
\end{figure}

\begin{figure}
        \centering
        \includegraphics[width=0.6\linewidth]{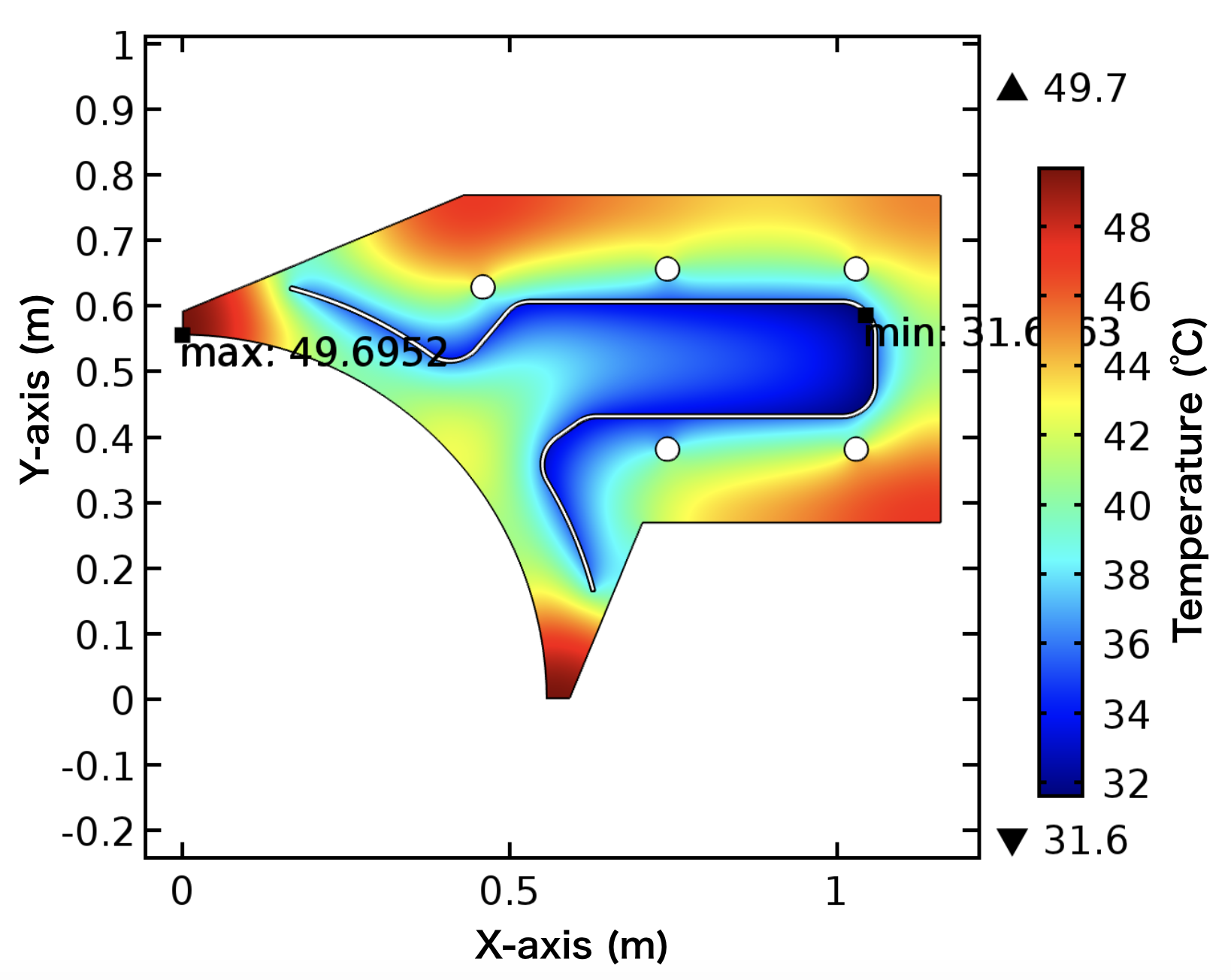}
        \caption{\small Estimated temperature distribution at the water cooled stripline for
        1.3~MW beam operation.}
        \label{fig:wcs_simulation}
\end{figure}

\begin{figure}
        \centering
        \includegraphics[width=0.6\linewidth]{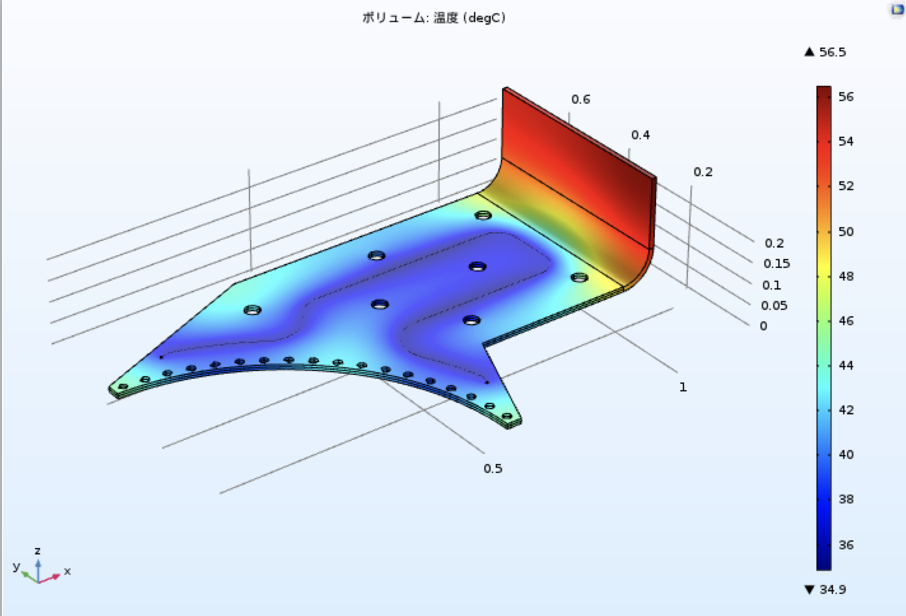}
        \caption{\small Estimated temperature distribution (in unit of $^{\circ}$C) at the water cooled stripline for
        1.3~MW beam operation from three-dimensional thermal analysis.}
        \label{fig:wcs_simulation_3d}
\end{figure}

The stripline ducts near the horn conductors, where the heat load is largest, 
are open-end and there is a difficulty in estimating the exact helium flow rate.
In addition, there is another difficulty to achieve much higher acceptable beam power 
by the forced helium flow scheme,
as shown in Fig.~\ref{fig:stripline_cooling_vs_flow}. 
Another cooling scheme is necessary to achieve much higher cooling performance.
A new water cooling scheme is developed by using a welding technique called
Friction Stir Welding (FSW) method, as shown in Fig.~\ref{fig:FSW}.
A 12~mm thick aluminum plate is machined to have a groove for a stainless pipe.
A 1/4 inch stainless pipe that is put in the groove is covered by an aluminum piece
and both ends of the cover piece are welded by FSW. The FSW technique has been used
for welding of the stripline plates and is well established technique.
A merit of this technique is that a cooling path can be flexibly placed in two-dimensional
plane. Cooling performance of this technique was investigated with a simple test piece
as shown in Fig.~\ref{fig:wcs_test_piece}.
Heaters were attached to the surface of the test piece and cooling water was flown
through the stainless pipes at flow rate of 1$\sim$4.3~L/min.
Effective heat transfer coefficients at some water flow rates were estimated by measuring
temperatures on the aluminum surface with platinum resistance temperature detectors (RTDs).
Figure~\ref{fig:wcs_test_piece} shows picture of the test setup and the measured
effective heat transfer coefficients are summarized in Tab.~\ref{tab:wcs_test_piece}.
\begin{table}
        \centering
        \small
        \caption{\small Summary of the measured effective heat transfer coefficients at
        several water flow rates.}
        \begin{tabular}{cc}
        \hline
        Water flow rate & Effective heat transfer coefficient \\
        (L/min) & (kW/m$^2\cdot$K) \\\hline
        1.0 & 2.38 \\
        3.0 & 3.48 \\
        4.3 & 4.05 \\\hline
        \end{tabular}
        \label{tab:wcs_test_piece}
\end{table}
The effective heat transfer coefficient depends on water flow rate. Since the diameter of
the stainless pipe that is embedded in 12~mm-thick aluminum plate is only 1/4 inch 
(inner diameter is 4.35~mm), the water flow rate achieved in this test was only 4.3~L/min.
A mockup of the water cooled stripline was produced to check the actual water flow rate
as shown in Fig.~\ref{fig:wcs_mockup}.
Water flow rate was measured by varying inlet water pressure. The measured parameters are
summarized in Tab.~\ref{tab:wcs_mockup_test}.
\begin{table}
        \centering
        \small
        \caption{\small Summary of the measured parameters in the mockup water flow test.}
        \begin{tabular}{ccc}
        \hline
        Inlet pressure       & Water flow rate         & Pressure drop       \\
        (MPa)                & (L/min)                 & (MPa)               \\\hline
        0.20                 & 3.2                     & 0.11                \\
        0.25                 & 3.9                     & 0.16                \\
        0.30                 & 4.5                     & 0.20                \\
        0.35                 & 5.0                     & 0.25                \\\hline
        \end{tabular}
        \label{tab:wcs_mockup_test}
\end{table} 
The maximum water flow rate of 5.0 L/min was achieved at 0.35~MPa pressure with the water
circulation pump used in this test, however, significant pressure drop was observed.
The values of the pressure drop are consistent with the calculation. Since a long path length
with the narrow pipe caused the large pressure drop, a use of
thinner (0.8~mm thick) stainless pipe is being considered.
In this case the pressure drop can be reduced from 0.20~MPa to 0.13~MPa for 0.3~MPa inlet pressure.
Further optimization of a water path will be performed.
Estimation of stripline temperature at 1.3~MW operation was performed using two-dimensional FEM simulation.
In case that heat transfer coefficient of 3~kW/m$^2\cdot$K is adopted to the water cooling path
and input heat load for 1.3~MW is applied to the stripline, temperature distribution at the stripline
was calculated as shown in Fig.~\ref{fig:wcs_simulation}.
The maximum temperature is estimated to be 49.7~$^{\circ}$C, which is well below the allowed
temperature of 80~$^{\circ}$C. 
\color{\MODCOLOR}
Then the FEM simulation was extended to three-dimensional analysis as shown in Fig.~\ref{fig:wcs_simulation_3d}.
In this simulation the bent section was also included to check the temperature around this section.
Although the cooling channel cannot be embedded in this section, it is covered by the stripline duct and
the forced helium cooling is also available. Heat transfer coefficients of 3~kW/m$^2\cdot$K and 10~W/m$^2\cdot$K 
were used for the water-embedded surface and stripline outer surface, respectively.
The value of 10~W/m$^2\cdot$K is expected from the current helium flow speed of 2.7~m/s, as shown in Tab.~\ref{tab:stripline_heat}.
The maximum temperature in this case was estimated to be 56.5~$^{\circ}$C at the bent section because of longer distance
from the water channel. The estimated temperature at the flat section was almost same as that from the two-dimensional analysis.
Based on this thermal analysis, with the assumed heat transfer coefficient of 3~kW/m$^2\cdot$K,
the acceptable beam power is estimated to be 2.45~MW.
\color{black}
Great improvement on the stripline cooling performance can be achieved with the proposed water-cooled striplines.

\begin{figure}
        \centering
        \includegraphics[width=0.6\linewidth]{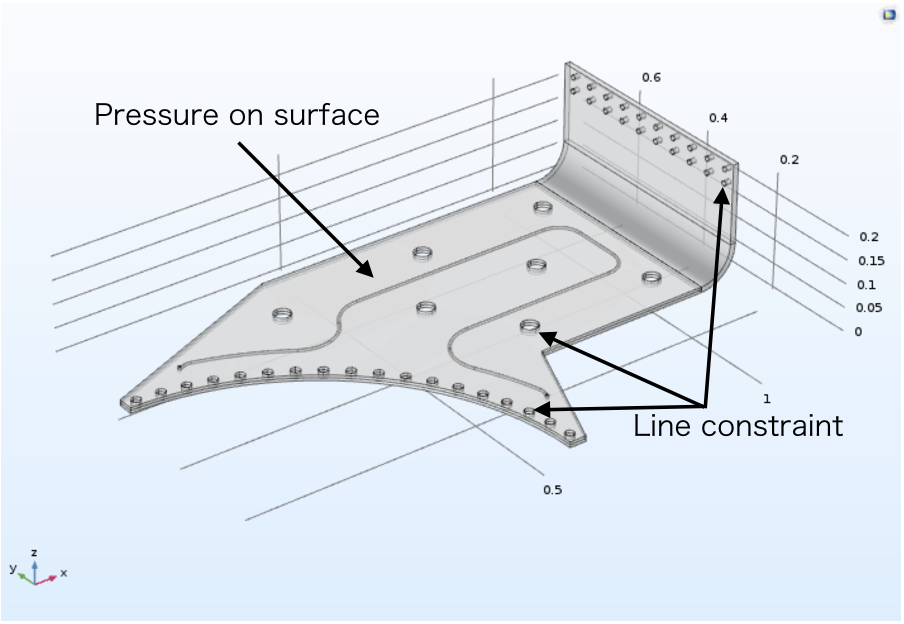}
        \caption{\small Input condition for the stress analysis of the water-cooled striplines.}
        \label{fig:wcs_simulation_model}
\end{figure}

\begin{figure}
        \centering
        \includegraphics[width=0.49\linewidth]{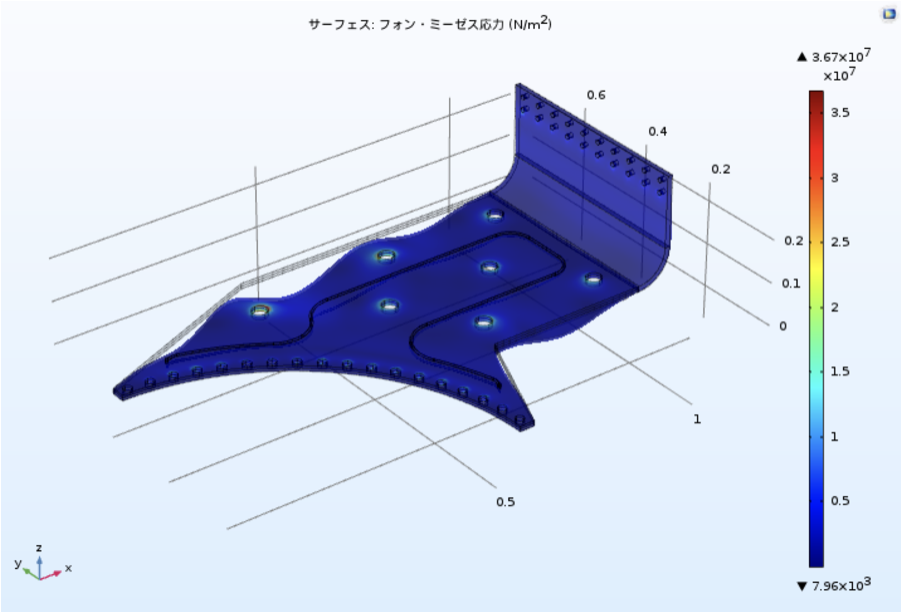}
        \includegraphics[width=0.49\linewidth]{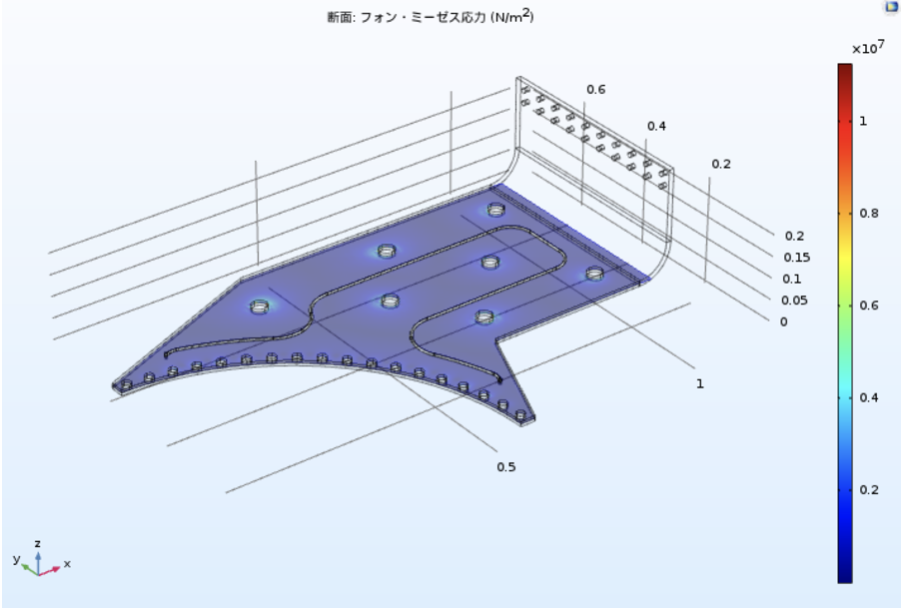}
        \caption{\small Simulation results of the stress analysis due to Lorentz force (in unit of N/m$^2$).
        Stress on the stripline outer surface (right) and at middle plane in half thickness (right).}
        \label{fig:wcs_simulation_stress}
\end{figure}

\begin{figure}
        \centering
        \includegraphics[width=0.6\linewidth]{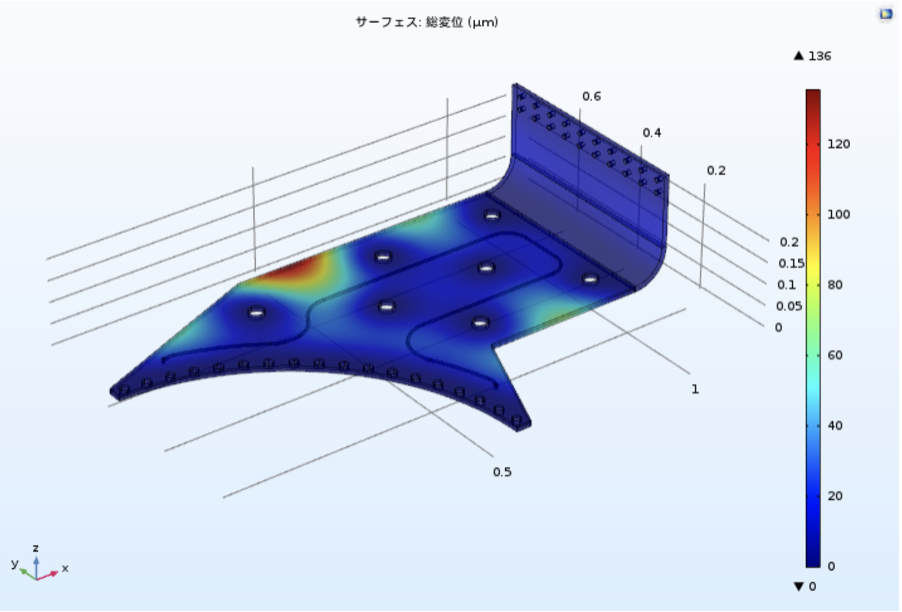}
        \caption{\small Estimated deformation at the water-cooled striplines (in unit of $\mu$m).}
        \label{fig:wcs_simulation_deformation}
\end{figure}

\color{\MODCOLOR}
Another study performed is to check the mechanical stress for the water-cooled striplines.
The embedded water pipes may be broken if they are suffered from large stress, therefore a FEM simulation was performed to check
stress and displacement on the striplines. Pressure of 32.2~kPa due to Lorentz force by 320~kA operation (80~kA per stripline conductor)
was applied on the stripline surface. Edge of each bolt hole was fix with a line constraint (which means no movement was allowed).
Figures~\ref{fig:wcs_simulation_model}  and~\ref{fig:wcs_simulation_stress} show the input condition and the results of the stress analysis,
respectively. 
The maximum stress was observed at the edge of large bolt holes where the stripline plates were fixed
with ceramic disks and bolted joints. In reality, the conductors were fixed not at line edge but in the area near the edge.
Thus the maximum stress estimated in this analysis would be overestimated. In any case the stress around the water pipe
was less than 1 MPa (or 1$\times 10^6$~N/m$^2$) and there is no concern on the stress at the water pipe.
Deformation of the water-cooled striplines was also calculated, as shown in Fig.~\ref{fig:wcs_simulation_deformation}.
The maximum displacement of 136~$\mu$m was observed at the edge of the stripline plate since this point was
far from the fixation points. A part of the water tube was deformed by approximately 60~$\mu$m.
This large deformation may affect the water pipe and fixation points should be optimized in later analyses.

\color{black}
Remaining study items for the water-cooled striplines are as following.
\begin{itemize}
\item Tolerance against vibration due to Lorentz force should be well considered. 
For example, a special care should be paid to the design of inlet and outlet connection structure. 
In order to investigate the vibration tolerance, operation test with mockup water-cooled striplines
will be performed.
\item Design of water plumbing connecting to the water-cooled striplines will be done.
Since several-hundred voltage is applied to the striplines, electrical insulation must be considered.
\end{itemize}
The water-cooled striplines will be adopted to only horn2 because of its lower acceptable beam power
than the other horns. There is no spare horn2 yet and it will be available around JFY2020.
Spare horn1 will be completed in JFY2018, thus a current testing of the water-cooled striplines
is planned with the spare horn1. Although the shape of the mockup water-cooled stripline for horn1
is different from the actual one for horn2, it is supposed that very useful outputs can be obtained
from the current testing with horn1, such as investigation of the vibration tolerance and plumbing
design around the water-cooled striplines (insulation structure) and so on. The current testing
with spare horn1 will be performed in \textcolor{\MODCOLOR}{JFY2018 and JFY2019}. 
After that, current testing with horn2 will be performed in JFY2020 and JFY2021.
\color{\MODCOLOR} 
In the current testing, temperature measurements will be performed to investigate the actual temperature
distribution as a function of water flow rate. Vibration measurements will also be performed to investigate
the deformation around the water pipes and frequency modes at several positions in order to check whether
any critical vibration cannot be observed or not. Any feedback will be applied to the later mockups and actual
striplines for horn2.
\medskip

\color{black}

\noindent\textbf{Horn production}
\medskip

For a high power beam operation, robustness of the beamline equipment is quite important.
Especially in case that the horns are broken due to some reasons,
it is of great importance to prepare spare horns in order to reduce downtime of the experiment.

The horns, under production or to be produced from now, are the third-generation T2K horns.
The new striplines and additional plumbing of He circulation line for hydrogen removal were already
adopted to the second-generation.
The horn1 will be minor upgraded with improved sealing structure against a thermal expansion.
The frames and fixture of the target attached to the horn1 need some minor modifications.
The horn2 needs some improvements, for example, 
modification of horn conductors to reduce displacement at the inner conductor and 
the water-cooled striplines. No modification is basically applied to the horn3. 

KEK has all the responsibilities relevant to the whole T2K horn system.
Production of the first- and second-generation horn1 and horn3 was managed by KEK and performed in Japan.
The first- and second-generation horn2 were produced by a T2K collaboration institute, 
University of Colorado Boulder (CU).
Since CU has a strong expertise and rich experience on horn productions, the third-generation horn1
is now under production at CU. With KEK's strong expertise and rich experience on horn operation,
many minor feedbacks were applied to the design and a production quality has been improved in 
the third-generation. CU is now in charge of the horn1 and horn2 production. The horn3 will be
produced in Japan, since it is too large to deliver it from the United States to Japan.
However, knowledge and experience on the horn production can be adopted to the spare horn3 production,
even though it is produced in Japan.

\subsubsection{Schedule}

\begin{figure}
        \centering
        \includegraphics[width=\linewidth]{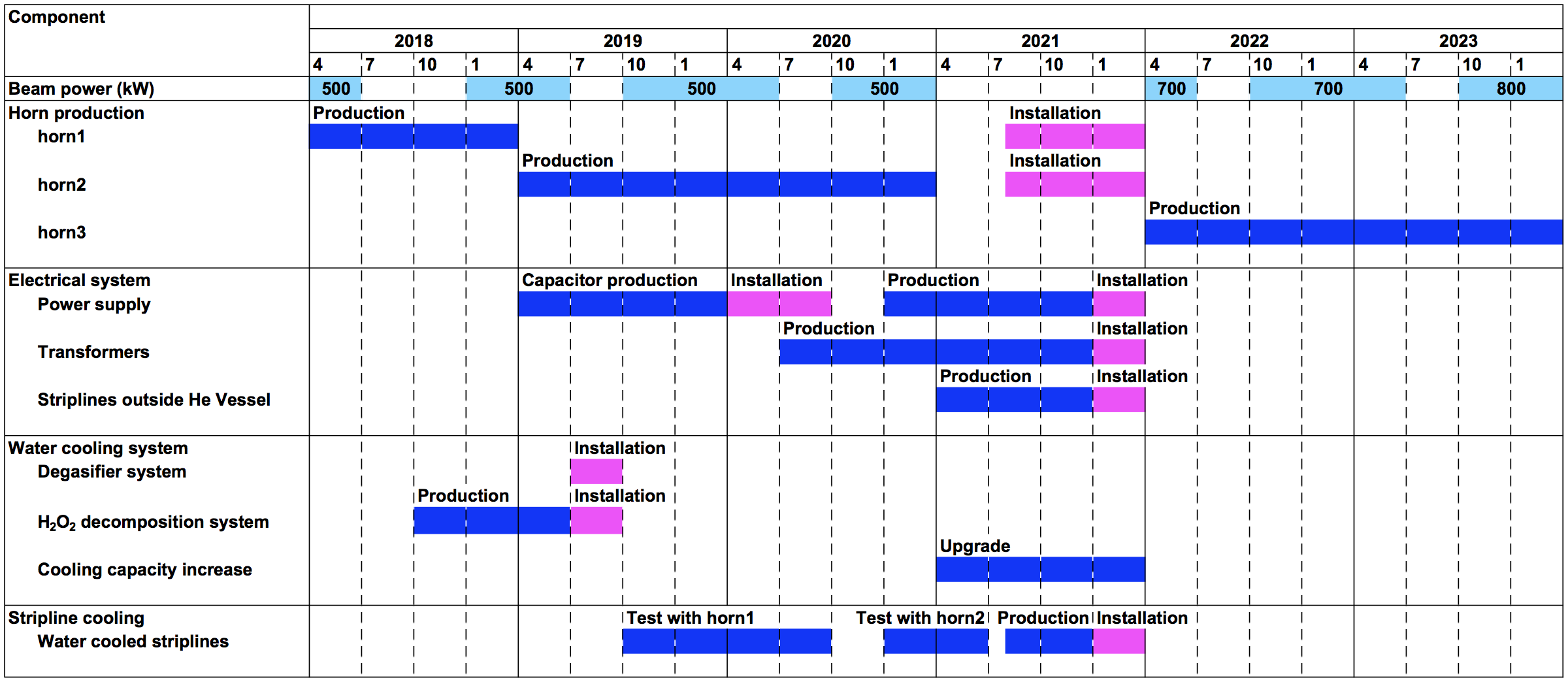}
        \caption{\small Summary of the upgrade schedule for the horn system.}
        \label{fig:horn_schedule}
\end{figure}

In this section, upgrade schedule relevant to the horn system is described.
The upgrade items, described in the previous sections, are summarized in Fig.~\ref{fig:horn_schedule}.
Spare horn1 is now being produced in Colorado and will be completed around \textcolor{\MODCOLOR}{fall} 2018.
Then it will be delivered to Japan. Design of target fixation frames and target He plumbing
will be modified to accommodate the new target. Their production and installation will be
done in Japan in FY2018. The upgraded target will be installed to the spare horn1. 
The horn1 with the new target will be installed during the long shutdown in JFY2021.
Horn2 will be produced by FY2020 in Colorado
and then will be delivered to Japan. The spare horn2 will be operated with the mockup water-cooled
striplines for their test. The horn2 with the actual water-cooled striplines will be finally installed in FY2021.
Spare horn3, without any modification planned so far, will be produced in Japan in FY2023 or later . 
After some current testing, it will be installed in FY2024 or later.

The upgrade of horn electrical system will be completed when MR magnet PS upgrade is completed in FY2021.
The new capacitors will be produced in FY2019 and will be installed to the existing power supply in FY2020.
One new power supply, two new transformers, and new striplines outside the helium vessel will be produced and installed 
in the shutdown period in FY2021.

In the horn water cooling system, the critical issue on water radiolysis should be solved
as soon as possible. Degasifier system and hydrogen peroxide decomposition system should be established
in FY2019. The improvement of cooling capacity of the horn water circulation system
will be achieved in FY2021.

Therefore, all the upgrades related to the horn system will be completed by FY2021.
However, it is technically feasible to complete these upgrades a year earlier if
a supplemental budget is allocated in a timely manner.  

\graphicspath{{figures/main_secondary/}}
       
\subsection{Remote maintenance of secondary beamline equipment}
\subsubsection{Overview}

\begin{figure}
        \centering
        \includegraphics[width=0.95\linewidth]{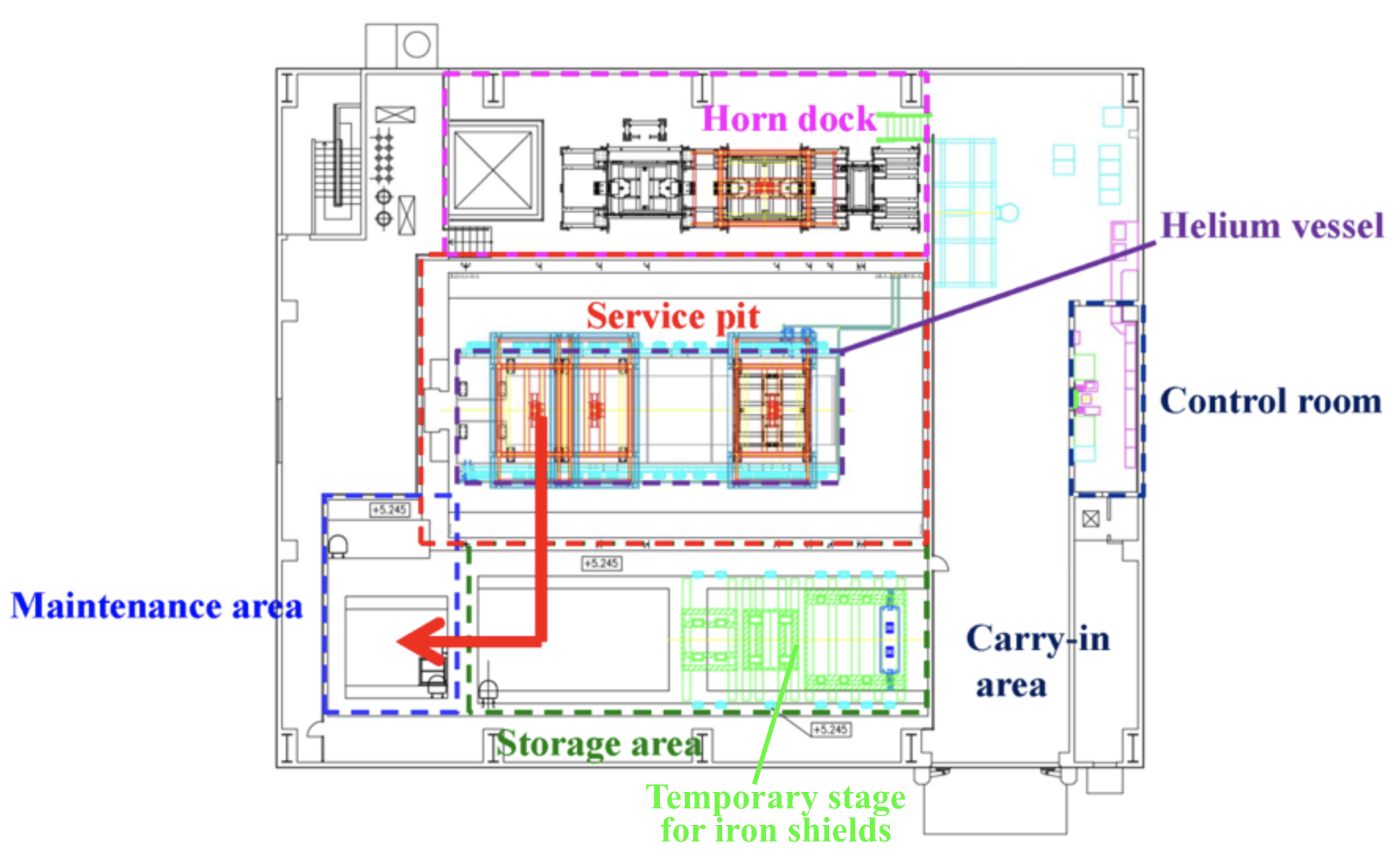}
        \caption{Top view of the Target Station.}
        \label{fig:TS_top_view}
\end{figure}

The secondary beamline devices become highly radio-active by beam operation 
and must be handled with great caution. In particular, it is not possible to handle 
the devices around the target by hand directly, and they have to be handled by remote control
for their maintenance or replacement. The Target Station, where several devices to be handled by
remote control are located, has designated systems for the remote maintenance.
Tab.~\ref{tab:remote_device} shows the devices maintained remotely in the Target Station.
\begin{table}
        \centering
        \small
        \caption{\small Devices maintained remotely in the Target Station.}
        \begin{tabular}{lcc}
        \hline\hline
        Device          & Readiness of remote handling system   & Experience of actual remote handling  \\\hline
        Beam monitor    & Under development                     & No                                    \\
        Beam window     & Yes                                   & Yes                                   \\
        Baffle          & No                                    & No                                    \\
        Target          & Yes                                   & Yes                                   \\
        Horns           & Yes                                   & Yes                                   \\
        \hline\hline
        \end{tabular}
        \label{tab:remote_device}
\end{table}
Fig.~\ref{fig:TS_top_view} shows the top view of the Target Station.
The Target Station consists of the following area.
\begin{itemize}
\item Service pit where the helium vessel with the target and the magnetic horns is installed
\item Machine room (underground floor below the horn dock area) where the cooling systems for the target, 
      the magnetic horns, the helium vessel, and the decay volume is installed
\item Horn dock area where the new magnetic horns are tuned
\item Maintenance area where the devices handled remotely are replaced
\item Storage area where the radioactive devices are stored
\item Carry-in area where the devices are carried into the Target Station from outdoor
\end{itemize}
Because there are the concrete shields above the service pit, the maintenance area, and the storage 
area at the beam operation, it is necessary to remove the concrete shields at first for maintenance 
of the devices on the beam line.
All devices can be accessed only from the top.

\begin{figure}
        \centering
        \includegraphics[width=0.8\linewidth]{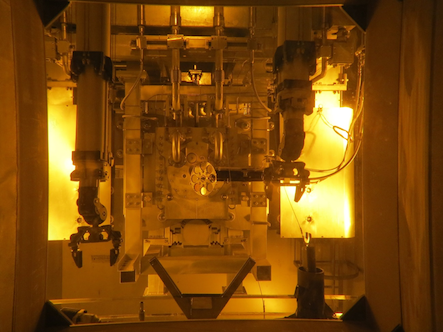}
        \caption{\small Photo of the maintenance area seen through the lead-glass window.
        In this picture, horn1 is set and can be seen from the front side.}
        \label{fig:manipulator_photo}
\end{figure}

There are two manipulators and a lift table for the replacement work in the maintenance area. 
Fig.~\ref{fig:manipulator_photo} shows the photo of the maintenance area through the lead glass.
Devices maintained remotely in the Target Station are carried from the beam line to the maintenance 
area and replaced there. The used devices are put into the casket and stored with the casket in the 
storage area. Thus every remote maintenance system consists of following three components.
\begin{itemize}
\item Handling machine which carries the device from the beam line to the maintenance area and installs 
      it into the beam line again
\item Exchanger which replaces the device with new one
\item Casket in which the used device is stored
\end{itemize}
The handling machine is attached to the crane and controlled from the crane control room in the carry-in 
area remotely.

The crane in the Target Station has 3-dimensional control system and the crane operator operates 
the crane with 3-dimensional coordinate. All motion systems (traveling, traversing, lifting, rotating) 
are duplicated. So if a motor breaks, the work can be continued by switching to another motor. 
Because all control boards of the crane are in the crane control room, they are not only 
shielded from radiation during remote maintenance work, but also they can be repaired in this room 
when they break.
The operator operates the crane and the handling machine while watching not only the coordinate and 
the values of sensors (tilt, load, and so on) but also the camera images. For example, we use 40 cameras 
when the magnetic horn is carried from the beam line to the maintenance area.

\color{\MODCOLOR}
We considered beforehand how to cope with a trouble for every procedure and wrote them on a manual.
There is never a case that a bolt galled at our remote maintenance operation,
but we will make a list of the parts where such a trouble seems to occur,
and consider a corrective action for the trouble.
\color{black}

\subsubsection{Target}
\label{sec:targetremoteexchange}

\begin{figure}
        \centering
        \includegraphics[width=0.8\linewidth]{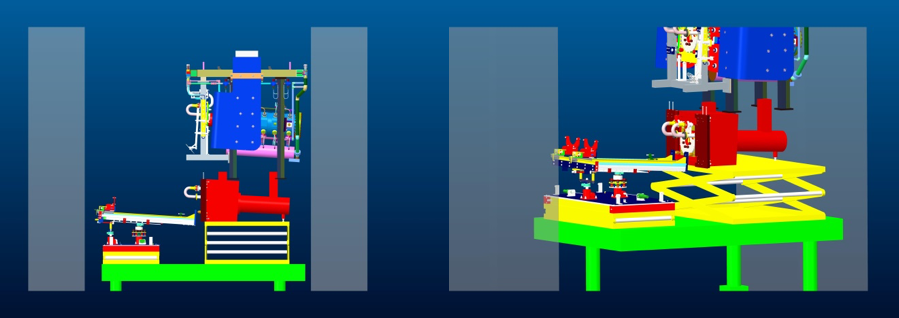}
        \caption{\small Horn, target exchanger and installation/disposal cask installed on lift tables in 
        Remote Maintenance Area of Target Station.}
        \label{fig:target_remote1}
\end{figure}

\begin{figure}
        \centering
        \includegraphics[width=0.8\linewidth]{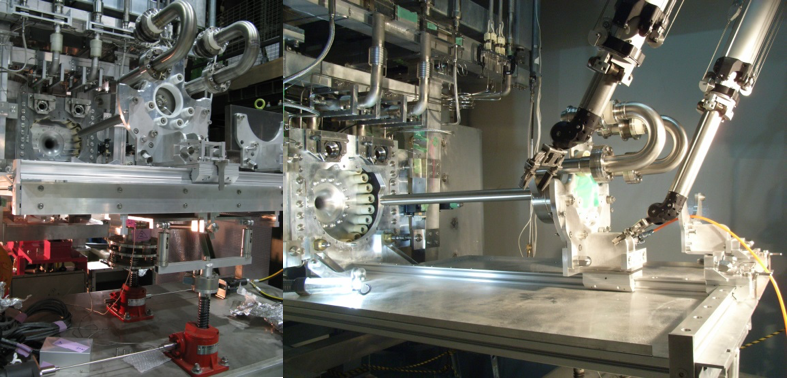}
        \caption{\small Photographs (Left) of target on exchanger docked to horn, showing red screw jacks and 
        suspension protection used to dock to horn at correct height and angle, and (Right) target 
        installation using  manipulators in remote maintenance area of Target Station.}
        \label{fig:target_remote2}
\end{figure}

\begin{figure}
        \centering
        \includegraphics[width=0.8\linewidth]{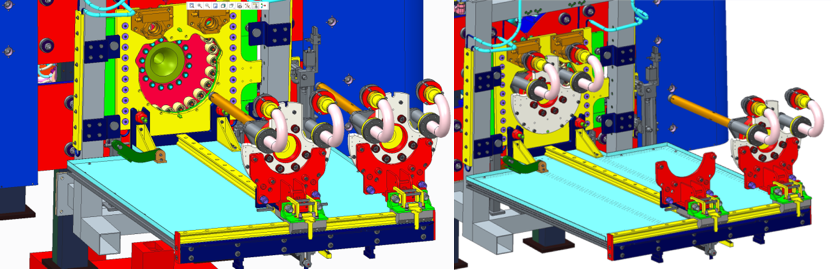}
        \caption{\small CAD models of target exchange procedure showing the longitudinal Z-rail used to 
        install/withdraw targets and a cross-rail (X-rail) to exchange.}
        \label{fig:target_remote3}
\end{figure}

\begin{figure}
        \centering
        \includegraphics[width=0.8\linewidth]{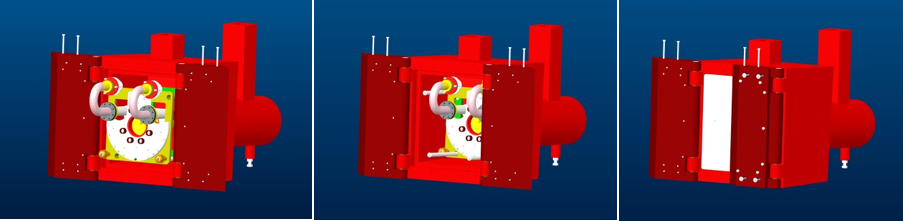}
        \caption{\small Target as installed in shielded Installation/Disposal cask (Left), pushed back on 
        internal rail system (Centre) and containment and shield door closure (Right).}
        \label{fig:target_remote4}
\end{figure}

\begin{figure}
        \centering
        \includegraphics[width=0.8\linewidth]{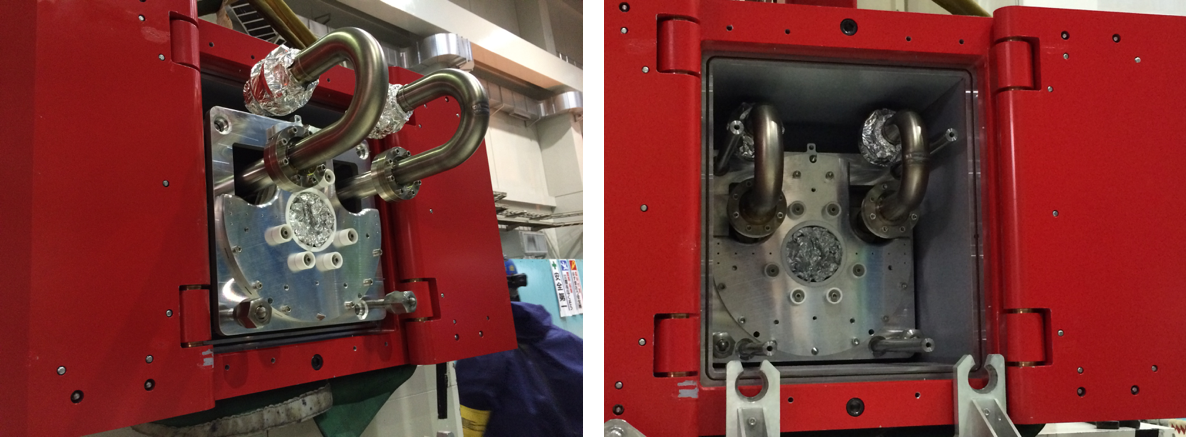}
        \caption{\small Pictures of the shielded Installation/Disposal cask. Left: target is set inside
        the cask. Right: target is fully contained.}
        \label{fig:target_remote5}
\end{figure}

As the beam power increases, target failures due to radiation damage are expected to occur more frequently 
than magnetic horn failures. A failed target can be replaced within the 1st magnetic horn in the Remote 
Maintenance Area (RMA) of the Target Station, permitting the horn to be re-used. In order to do this, 
a replacement target is first installed in the RMA and loaded onto a bespoke target exchange mechanism 
mounted on an independent lift table. The horn containing the failed target is then lowered into the RMA 
beneath its support module. Fig.~\ref{fig:target_remote1} shows a 3D CAD model of the horn, 
the target container and the target exchanger installed in the Remote Maintenance Area. 
The target exchanger is then raised on its lift table and carefully docked to the horn as shown in 
Fig.~\ref{fig:target_remote2} using the master-slave manipulators which are incorporated in the RMA. 
The failed target is removed from the horn and replaced with the new target using the target exchange 
mechanism as shown in Fig.~\ref{fig:target_remote3}.
The horn complete with the replaced target is then lifted from the RMA and re-installed in the beam line. 
Fig.~\ref{fig:target_remote4} shows the failed target installed in shielded cask which is then loaded 
into a larger shielded cask for storage in the morgue and eventual disposal.
Fig.~\ref{fig:target_remote5} shows pictures of the shielded Installation/Disposal cask. 

Access to the Remote Maintenance Area is only possible when the beam is shut down for a long enough period 
for the top layers of concrete shielding to be removed, which is a costly and time consuming process. 
Consequently opportunities to rehearse and develop the target exchange procedure have been relatively limited. 
It has never been possible to entirely exclude personnel from the 'active' side during the installation and set-up of 
non-activated components for fully realistic rehearsal purposes. Nevertheless every individual procedure for the 
remote target replacement as described above has been tested in the Target Station Remote Maintenance Area 
using the master-slave manipulators to perform manual operations. The only exception is the target disposal 
sequence, where a failed target is installed in its shielded cask, for which a complete rehearsal was not possible 
due to a lack of time available. Confidence in the use of the target exchange mechanism has been gained by 
using it to replace a leaking target helium pipe as described in Sec.~\ref{sec:targethetubetrouble}.

\subsubsection{Beam window}

\begin{figure}
        \centering
        \includegraphics[width=0.7\linewidth]{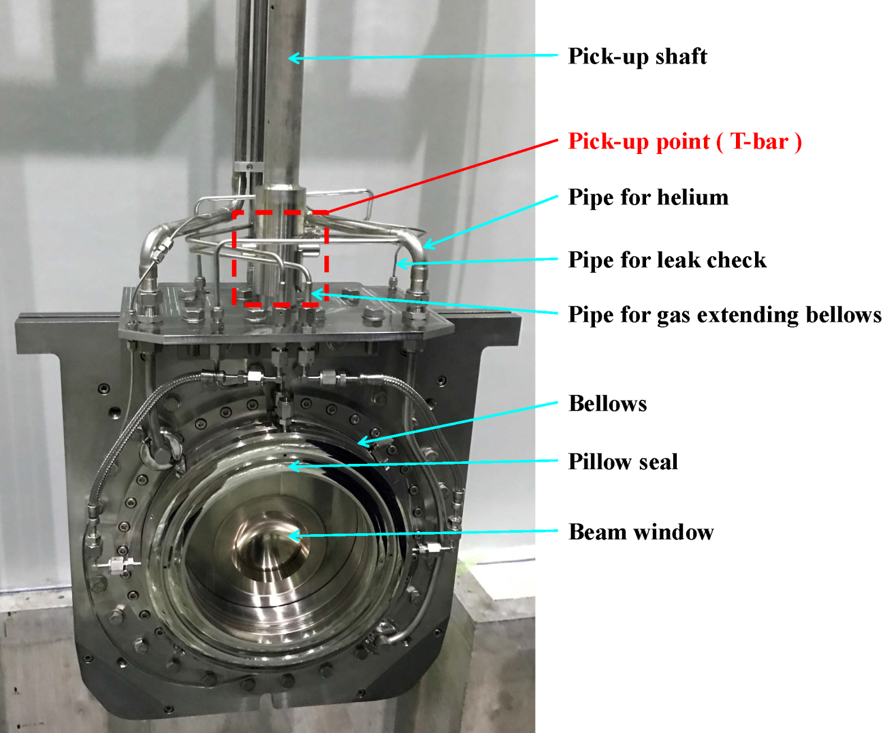}
        \caption{\small Picture of the beam window.}
        \label{fig:beam_window_1}
\end{figure}

The beam window is a partition which separates the primary beamline and the helium vessel, 
and is made of hemisphere-like double partitions made of titanium. It is cooled by helium 
flow between these double partitions. It is connected to the monitor stack with a vacuum seal (called pillow seal) 
in order to be possible to be exchanged, and separates vacuum in the monitor stack 
(upstream side) and the helium vessel (downstream side).
Fig.~\ref{fig:beam_window_1} shows a picture of the beam window.

The pillow seal is a kind of metal seals, and composed of a pillow and a mirror flange.
The pillow is attached to a bellows which can expand by applying pressure from outside in order to
obtain a contact pressure on the mirror flange. A nominal pressure during the beam operation is +0.3~MPa.
In case that the beam window is replaced, the pressure is released to shorten the bellows
in order to make a gap between the pillow and the mirror flange.
The mirror flanges are attached to the monitor stack and the helium vessel.
The pillows can be exchanged with the beam window replacement, but the mirror flanges still remain.
So it is important to keep the surface of the mirror flanges clean.
Because vacuum is kept by metal touch between the pillow and the mirror flange,
bruises and pollution on the surface of the pillow and the mirror flange influence seal performance. 
Some visual inspection devices and cleaners for the mirror flanges are developed for this purpose.

The beam window has a pick-up point (called T-bar) at the top and is handled with the pick-up shaft here.
Because the beam window is installed on the beamline which is 4m below the service pit floor where
workers access, height of the pipes (2 pipes for cooling helium, 4 pipes for gas extending bellows and 
2 pipes for leak check) is also 4 m. The beam window is carried together with these pipes into the maintenance 
area and the pipes are removed from the beam window there.

A shield (called upper-shield) is placed above the beam window during beam operation in order to shield 
radiation from the beamline. The upper shield is removed before the replacement work 
and installed again after the replacement.

\begin{figure}
        \centering
        \includegraphics[width=0.5\linewidth]{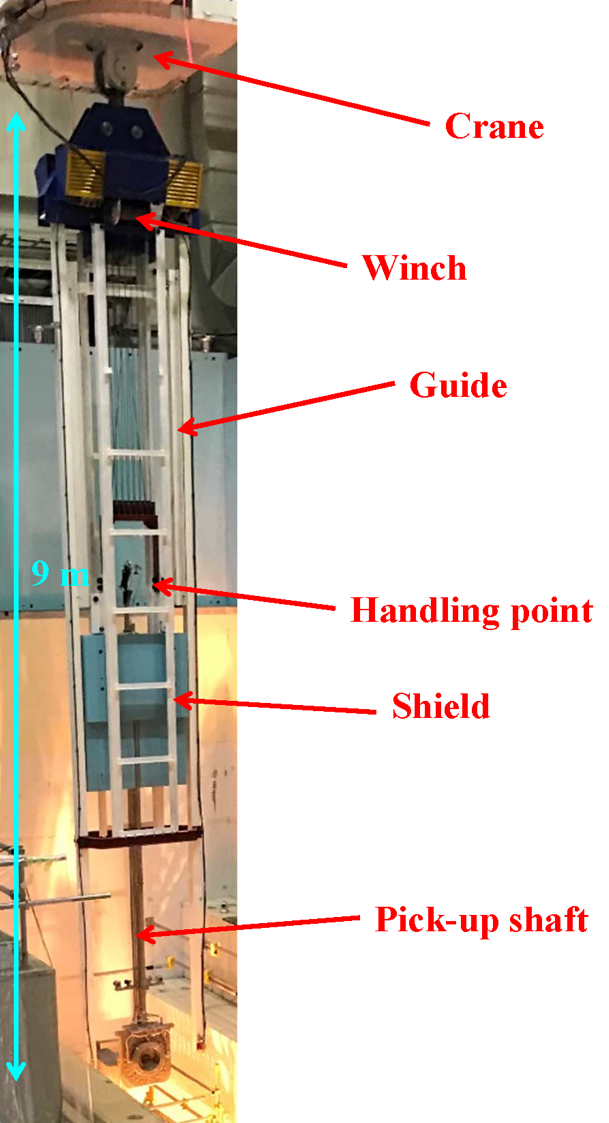}
        \caption{\small Picture of handling machine for the beam window.}
        \label{fig:beam_window_2}
\end{figure}

\begin{figure}
        \centering
        \includegraphics[width=0.6\linewidth]{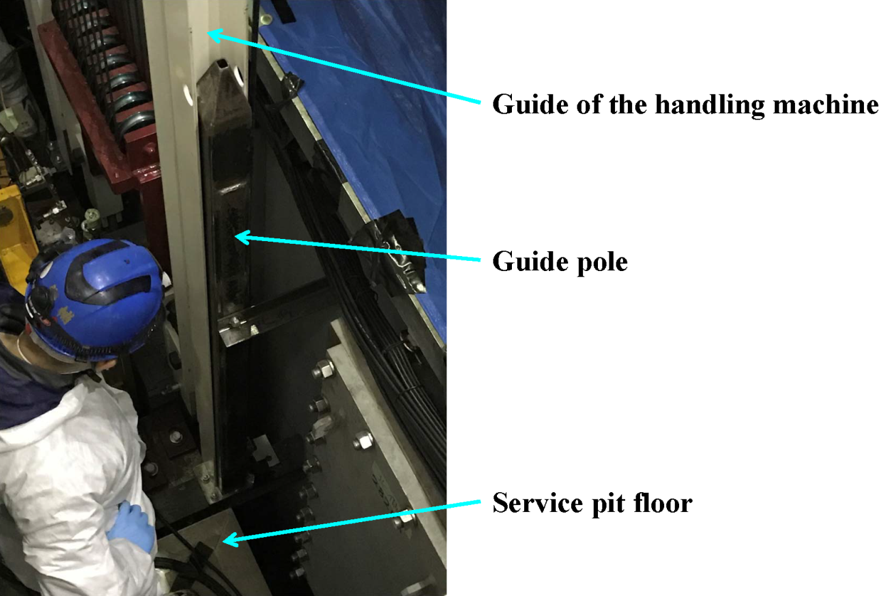}
        \caption{\small Guide of the handling machine touches down to the service pit floor.}
        \label{fig:beam_window_3}
\end{figure}

\begin{figure}
        \centering
        \includegraphics[width=0.6\linewidth]{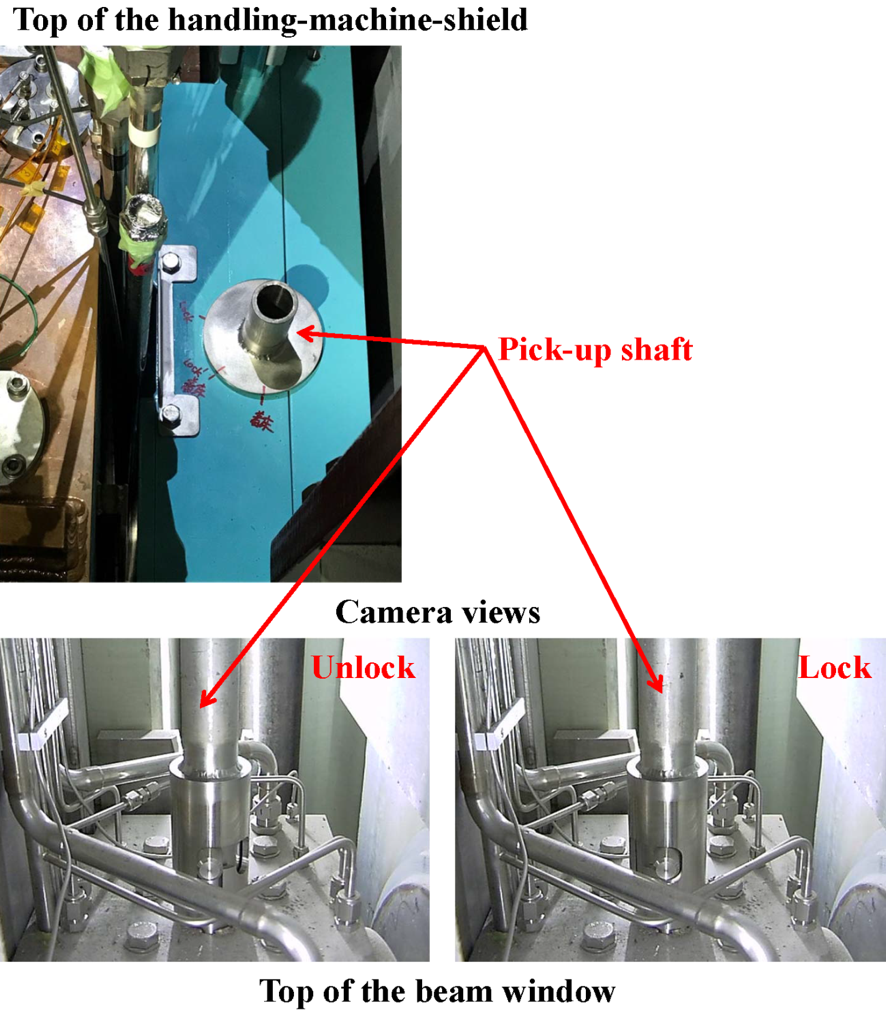}
        \caption{\small Beam window is picked up.}
        \label{fig:beam_window_4}
\end{figure}

Fig.~\ref{fig:beam_window_2} shows a handling machine for the beam window, which can be attached to the crane hook. 
The handling machine is very long and narrow,
because the beam window is 4 m below from the service pit floor. 
A set of guide poles, which are placed on the service pit floor, ensure the alignment of the handling
machine, as shown in Fig.~\ref{fig:beam_window_3}. 
Once the bottom of the handling machine touch down the guide poles and its alignment is fixed,
the pick-up shaft can vertically move along the guide frames by a winch. 
The handling machine can also handle the upper-shield. The sky-blue shield in Fig.~\ref{fig:beam_window_2} 
(called handling-machine-shield) shields radiation in the work after removing the upper shield and is 
removed from the handling machine when the upper-shield is handled.
The pick-up shaft for the beam window is hung on the top of the shield, and a worker rotates 
the shaft to lock or unlock the beam window over the shield, as shown in Fig.~\ref{fig:beam_window_4}
.
Once the beam window is picked up, the pick-up shaft is moved upward by the winch.
Then the handling machine with the beam window is transported to the maintenance area
and the irradiated beam window is inserted into a casket.
New beam window is thereafter installed to the beamline in the opposite procedure.

The first replacement work by remote control was done in August, 2017. There was no trouble. The leak 
rate of the new beam window after the replacement work is the same as that of the old beam window before 
the replacement work. The new beam window is used at the present beam operation.

\color{\MODCOLOR}
We plan to send the old beam window to the external facility after cooling.
We are considering the procedure of the making the sample (cutting the beam window, etc.)
and how to ship, and are preparing the devices (glove box, etc.).

We measured the radiation level of the old beam window through the casket wall (steel 100mm thickness).
The levels was 300$\mu$Sv/h after 132 days cooling and it was 70$\mu$Sv/h after 448 days cooling.
\color{black}

\subsubsection{Horn}

A remote exchange scheme for the magnetic horns is described in this section.
When one of the horns is broken, it is moved to the maintenance area. Then the horn itself
is disconnected from its support module and moved to the storage area and stored inside a cask for temperary storage.
Thereafter a new horn is set up in the maintenance area and connected to the support module.
The support module can be designed to be reused even if the horns are broken. After connection,
the new horn is moved to the helium vessel.

\begin{figure}
        \centering
        \includegraphics[width=0.8\linewidth]{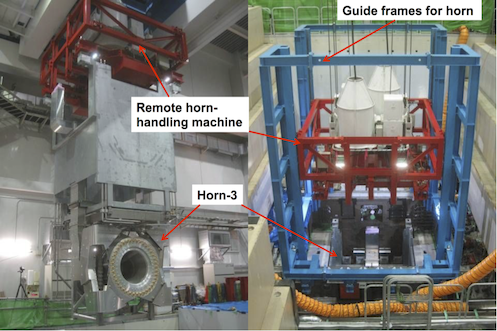}
        \caption{\small Pictures of the remote handling machine for the horns.
        The horn3 hanged by the special remote handling machine (left)
        and the remote handling machine lowering in the guide frame placed
        on top of the helium vessel (right).}
        \label{fig:horn_remote_pic1}
\end{figure}

\begin{figure}
        \centering
        \includegraphics[width=0.9\linewidth]{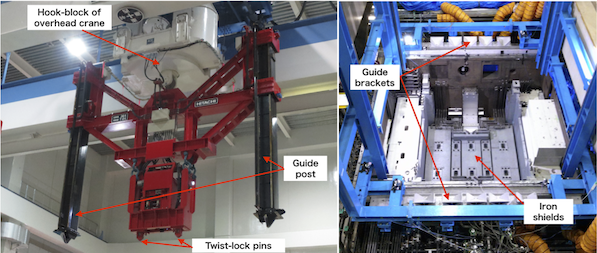}
        \caption{\small Pictures of the remote handling machine for iron and concrete shields (left) and its guide brackets located
        on the edge of the helium vessel (right).}
        \label{fig:shield_remote_handling}
\end{figure}

There are iron and concrete shields in the helium vessel, which are placed very tighly
with only 3~cm gaps between all the neighboring elements for radiation shielding.
During insertion into or extraction from the helium vessel, it is really important to
avoid for each component to hit the neighboring components even with such small gaps.
To ensure that the equipment is safely moved upward or downward, a guild system, composed
of guide frames and a special remote handling machine, was developed as shown in
Fig.~\ref{fig:horn_remote_pic1}.   
Two guide frames are placed both at the top of the helium vessel and at the maintenance area.
The remote handling machine has several guide rollers at each corner which ensure
a smooth movement along the guide frame. Before the horn transportation, the guide frames are aligned very precisely based on
the position of the horns and horizontal position within a few mm precision can be secured during
the upward and downward movements. Since the horns and the support module are 10~m-high at maximum
and the center of gravity is horizontally off by a few cm from the supporting point, it is important
to adjust the perpendicularity. Some counter weights are attached at the top of the support module and its perpendicularity
when hung by the remote handling machine is adjusted by using an angle meter which can be monitored remotely even during the transportation.
There is also a dedicated remote handling machine for the iron and concrete shields as shown in Fig.~\ref{fig:shield_remote_handling}.
This machine has two guide posts and their corresponding guide brackets are located on the both edges of the helium vessel 
to ensure relative alignment between the shields and the remote handling machine.

\begin{figure}
        \centering
        \includegraphics[width=0.6\linewidth]{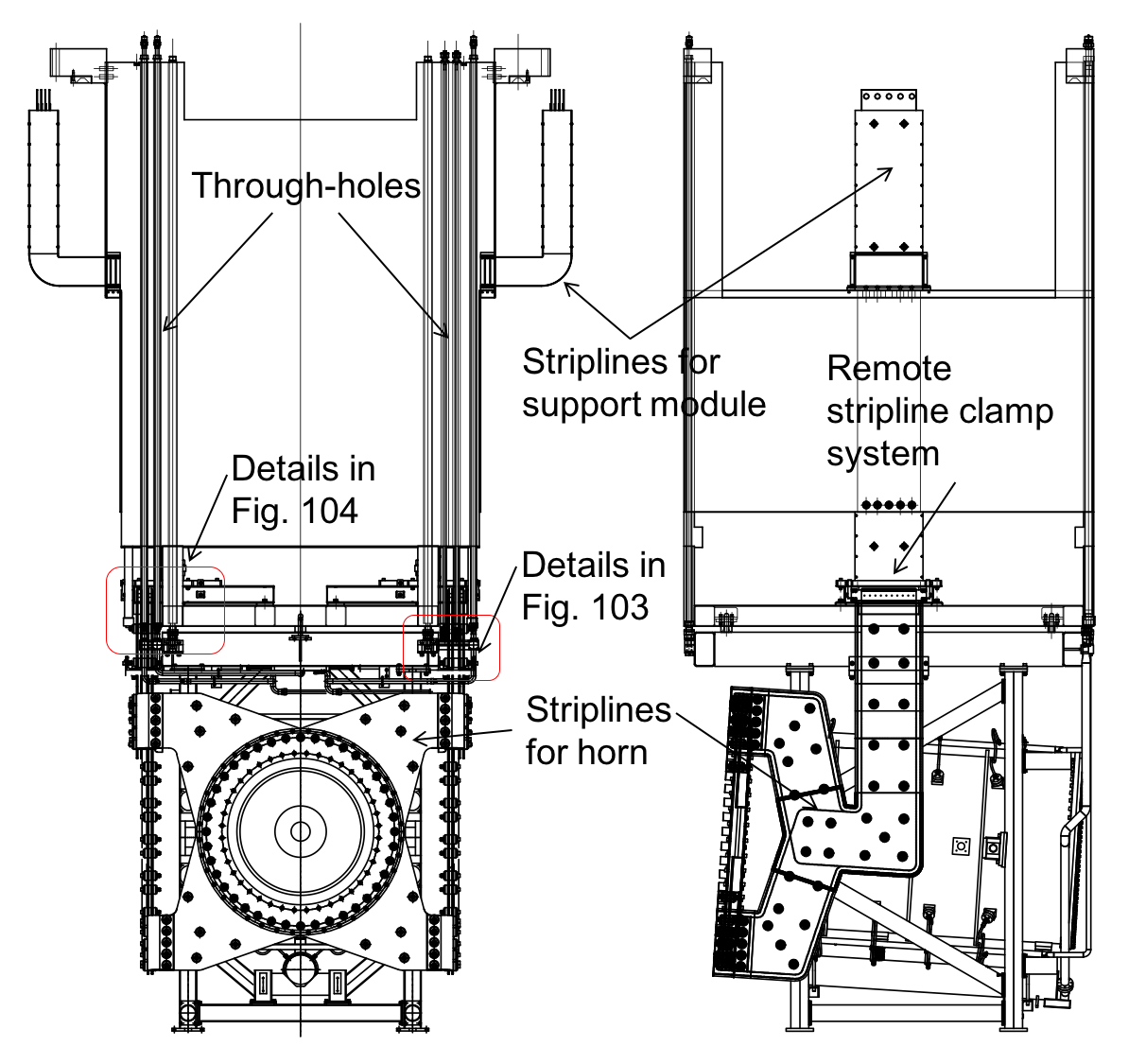}
        \caption{\small Drawing of the horn and its support module featuring their remote connection mechanism in front (left) and side (right) views.}
        \label{fig:remote_connection2}
\end{figure}

\begin{figure}
        \centering
        \includegraphics[width=0.67\linewidth]{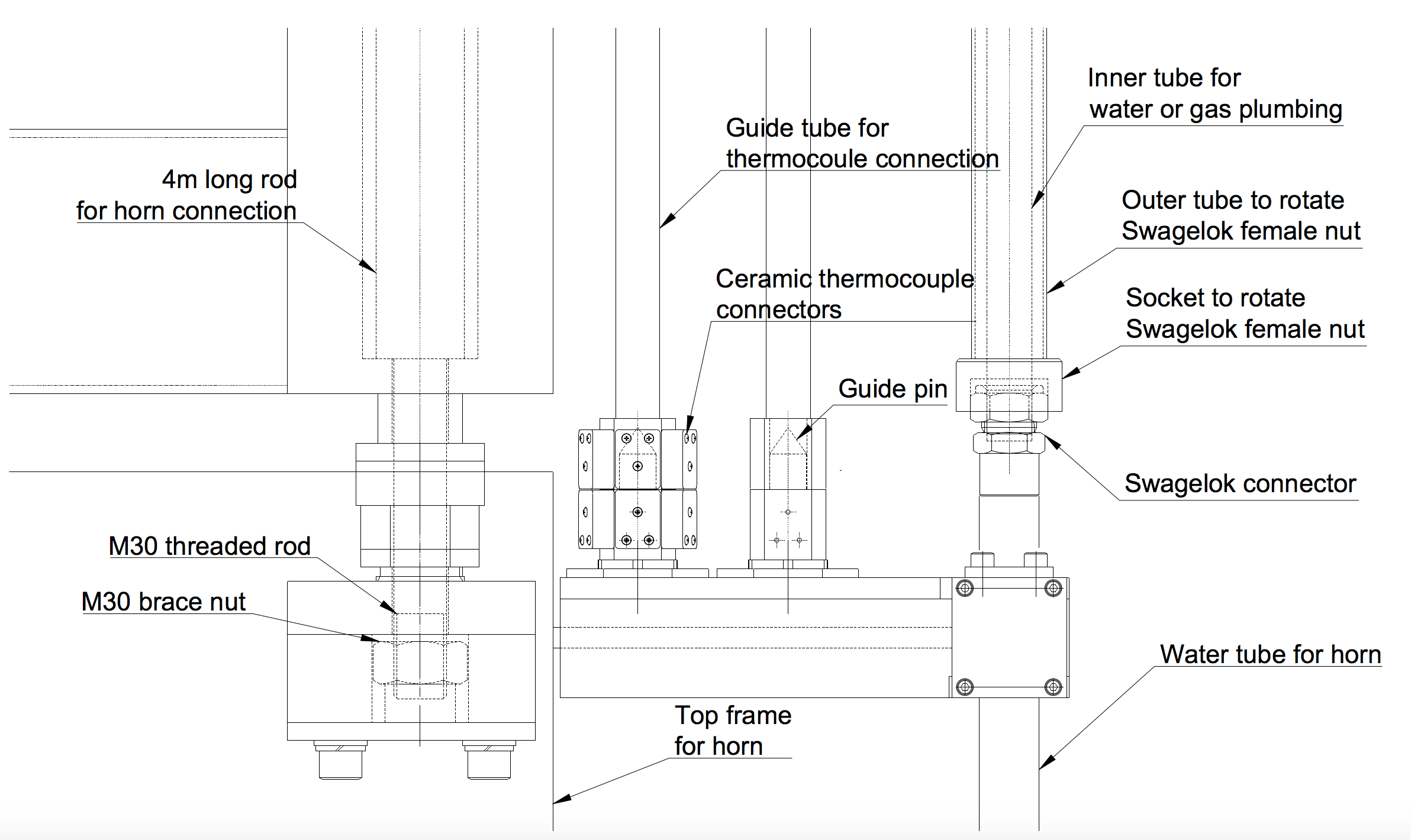}
        \includegraphics[width=0.25\linewidth]{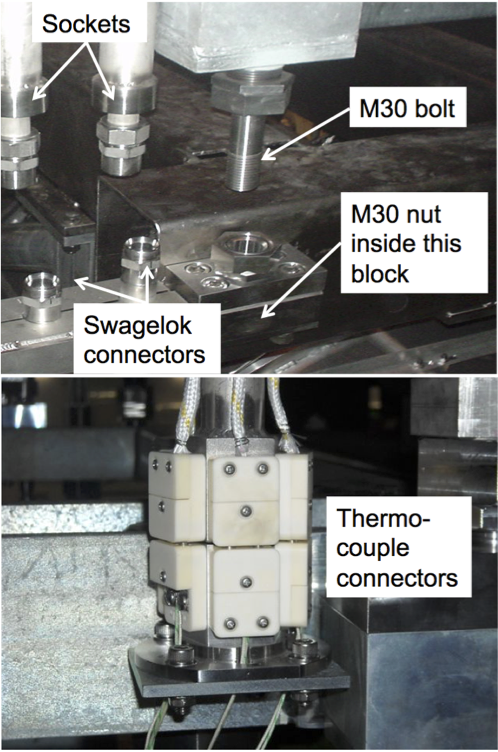}
        \caption{\small A drawing (left) and pictures (right) of the remote connection for the horn attachment, 
        the water/helium pipes, and the thermo-couples.}
        \label{fig:remote_connection1}
\end{figure}

\begin{figure}
        \centering
        \includegraphics[width=0.6\linewidth]{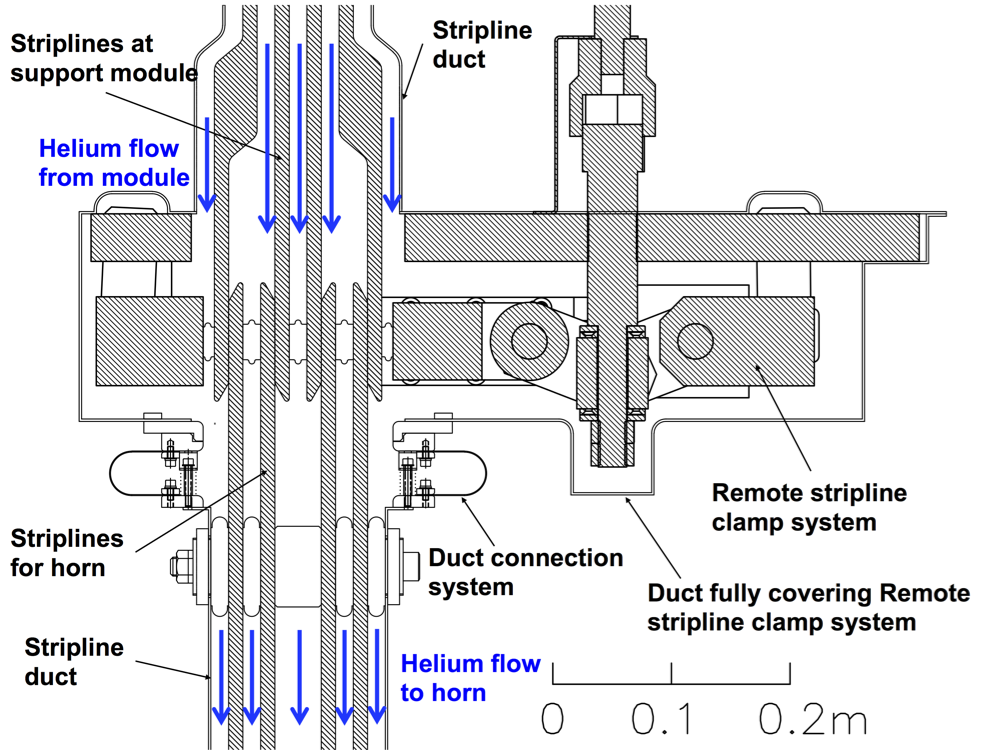}
        \caption{\small A drawing of the remote connection for the striplines.}
        \label{fig:remote_clamp}
\end{figure}

Once the irradiated horns are placed in the maintenance area, the exchange of the horns is performed.
The horns and their support module can be disconnected with a semi-remote manner. The remote connection/disconnection must be done
for the attachment of horn itself, the striplines, the water and helium pipes, and the thermo-couples. The detailes of those remote
connection mechanism are shown in Figures.~\ref{fig:remote_connection2}, \ref{fig:remote_connection1}, and \ref{fig:remote_clamp}.  
The basic concept for the horn remote connection is to perform the connection/disconnection of each item by using its dedicated
long shaft that is rotated by a worker who stands on the concrete shield inserted inside the support module. 
The support modules have many through-holes where the long shafts are inserted. 
For the horn attachment, 4 m-long stainless shafts that have M30 thread at the bottom end are adopted
in the support module and corresponding M30 brace nuts are located at the horn side. 
By rotating the long shaft from the top of the support
module, the connection of the horn to the support module is performed. 
For the connection of the water/helium pipes, commercial Swagelok connectors are used. The coaxial pipes penetrate the through-holes of
the support module. The inner pipe is used as water/helium plumbing and the outer used as a rotation tool for the Swagelok connectors.
The nut of the Swagelok connecter can be tightened/loosened by rotating the outer pipe from the top of the support module.
Ceramic connectors are used for the connection of thermo-couples. The connectors are aligned by using guide pins and guide holes.       
Striplines are also detachable at the bottom of the support module. Fig.~\ref{fig:remote_clamp} shows the remote
stripline clamp system where the rotation of the long shaft changes torque to a horizontal force.
It can clamp the stripline plates by 15 tons of force equivalent to 5~MPa contact pressure on the stripline surface.
The relative alignment between the horn and the support module is achieved within 0.3~mm accuracy 
by a set of guide pins and guide hole/slit.

\begin{figure}
        \centering
        \includegraphics[width=0.9\linewidth]{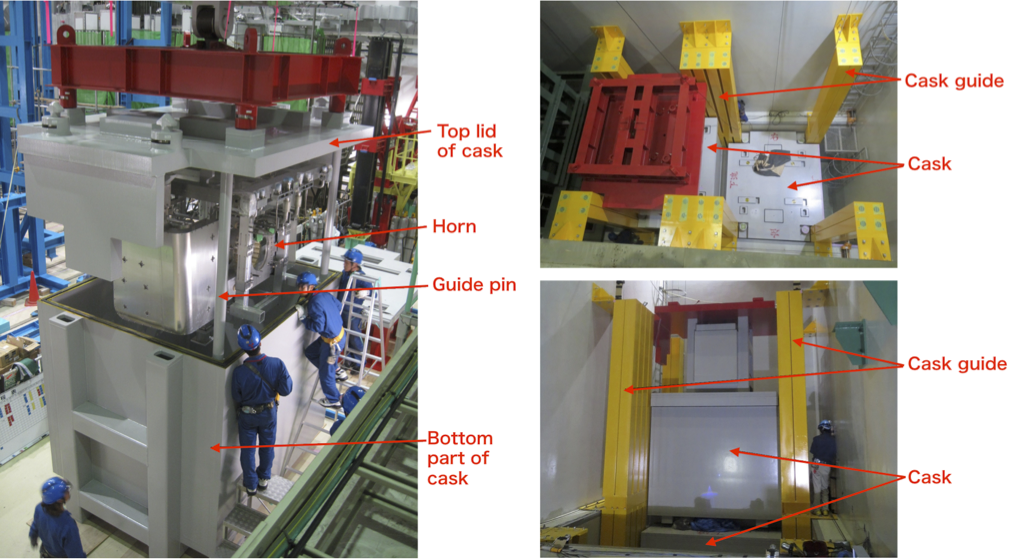}
        \caption{\small Pictures of the cask for the horns taken during an insertion test of non-irradiated horn (left) and
        cask and cask guide placed in the storage area (right).}
        \label{fig:horn_cask}
\end{figure}

The detached irradiated horn is then transported to the storage area and stored inside a cask made of thick iron plates.
A picture of the horn cask is shown in Fig.~\ref{fig:horn_cask}.
The top lid of the horn cask can hung the horn by using twist-lock system which manually operated on the lid.
The thickness of the lid is 30 cm to reduce radiation dose during the work.   
The horn hung by the top lid is moved to the storage area with remote operation of the crane.
For the insertion to the cask, the top lid is guided by guide columns to ensure the alignment as shown in Fig.~\ref{fig:horn_cask}.
Then a new horn is connected to the support module and transported to the helium vessel. 

In the past, the first remote transportation of the irradiated horn was performed in 2011 for inspection of the horns after the Great East Japan Earthquake.
There was a trouble occurred due to the perpendicularity and then the improvement described above was adopted. 
During the installation of the second-generation horns, remote exchange of all three horns
has been conducted without any problem. The remote maintenance of the horns was established.

\subsubsection{Beam monitors}

\begin{figure}
        \centering
        \includegraphics[width=0.8\linewidth]{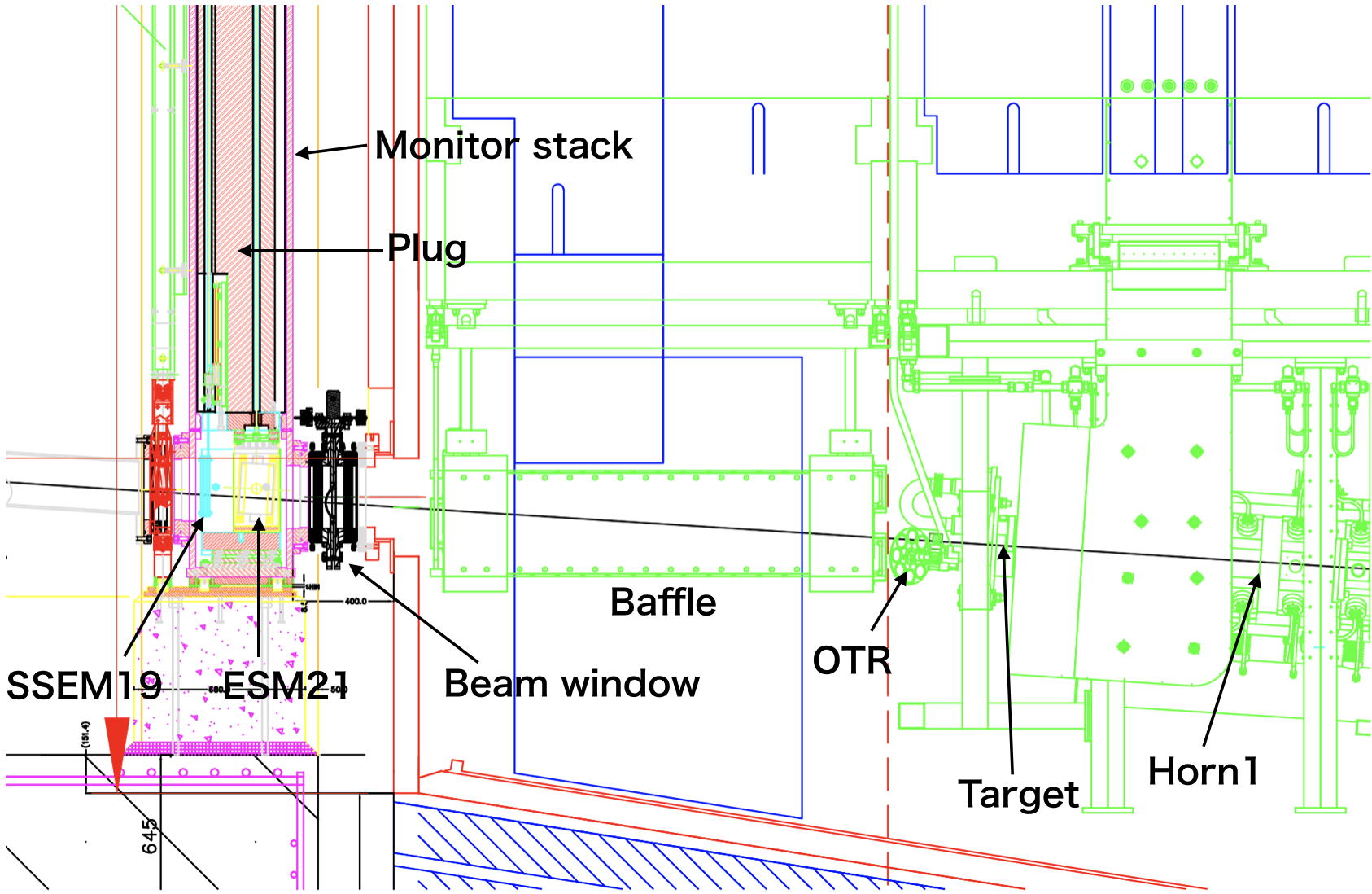}
        \caption{\small Side view of the three proton beam monitors in the Target Station.}
        \label{fig:TS_monitor}
\end{figure}

The three proton beam monitors, the beam profile monitor (SSEM19), 
the beam position monitor (ESM21), and OTR, are located in the Target Station, 
as shown in Fig.~\ref{fig:TS_monitor}.
SSEM19 and ESM21 are installed in the vacuum chamber 
(called monitor stack) set in the upstream side of the helium vessel. OTR is attached to the frame of 
the magnetic horn 1.

SSEM19 and ESM21 are attached to the bottom of the plug of the monitor stack and carried to the 
maintenance area with the plug. They are removed from the plug by manipulators there. The handling 
machine serves both for the plug and the beam window. The exchanger for SSEM19 and ESM21 is under 
development.

OTR is carried to the maintenance area with the horn 1 and exchanged by manipulators there. 
OTR-I was removed by manipulators from the horn 1 at the replacement work of the horn in 2014. 
But there is no experience in which we attached OTR to the horn 1 by manipulators.

\subsubsection{Disposal scenario of the irradiated equipment}

\begin{figure}
        \centering
        \includegraphics[width=0.8\linewidth]{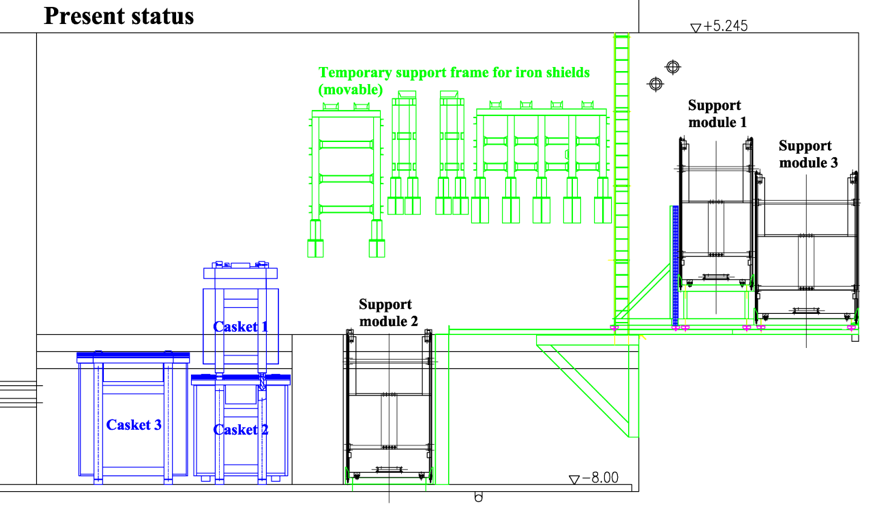}
        \caption{\small Current layout of the storage area in side view. Currently three caskets are
        stored in the storage area as shown in blue.}
        \label{fig:storage_area_current}
\end{figure}

\begin{figure}
        \centering
        \includegraphics[width=0.8\linewidth]{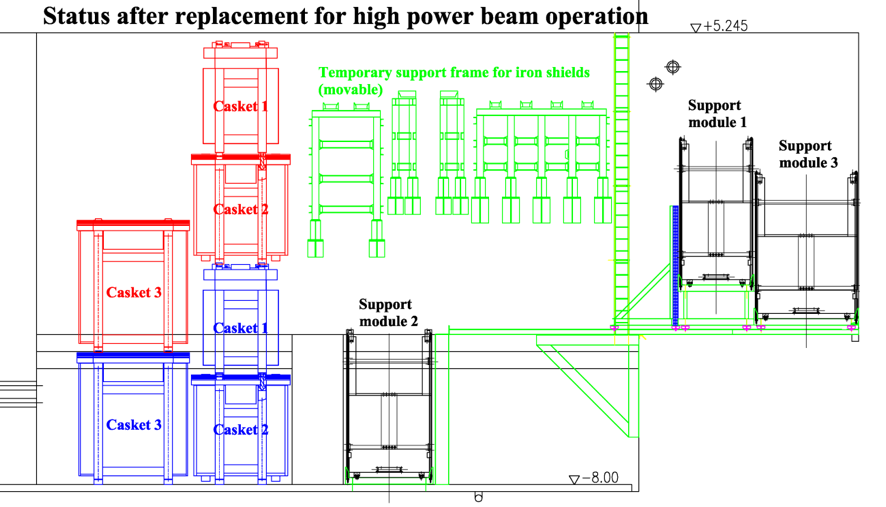}
        \caption{\small Future layout of the storage area in side view. After the third-generation horns
        are installed, the old three horns are stored in the storage area as shown in read.}
        \label{fig:storage_area_future}
\end{figure}

Used device replaced at the remote maintenance work is stored with its casket in the storage
area of the Target Station. The area is the size of 5~m in width, 18~m in length, and 13~m
in height. There are the magnetic horns with their caskets at the upstream side and the other
devices at the downstream side in the storage area. As shown in Fig.~\ref{fig:horn_cask},
the cask guide pillars to prevent the caskets from toppling over are installed in the storage
area, and the caskets are piled up on the correct location along them (maximum 4 tiers).
The storage area has the capacity containing 8 caskets for the horn1, 8 caskets for the horn2,
or 6 caskets for the horn3.
Fig.~\ref{fig:storage_area_current} shows the present status of the storage area (side view).
We replaced all three horns in JFY2013, and there are three caskets in this area.
Fig.~\ref{fig:storage_area_future} shows the expected situation after horn replacement
with third-generation is performed and the second-generation horns are stored in the storage area.
\color{\MODCOLOR}
We assume that the magnetic horn is exchanged every 5 years.
\color{black}
There may be some space for one or two more caskets. 
If we need further space for more caskets, we have to carry the oldest caskets to outside of the Target Station.
There is the building to store the radioactive devices from all facilities in J-PARC.
Used targets and used magnetic horns are carried into the building and stored after
several years storage in the Target Station.
\color{\MODCOLOR}
We will consider the details including consideration of the transport container
about transport from the Target Station to the storage building.
\color{black}

\graphicspath{{figures/main_secondary/}}

\subsection{Cooling capacity upgrade}
\label{sec:water_cooling_upgrade}

\begin{figure}
        \centering
        \includegraphics[width=0.8\linewidth]{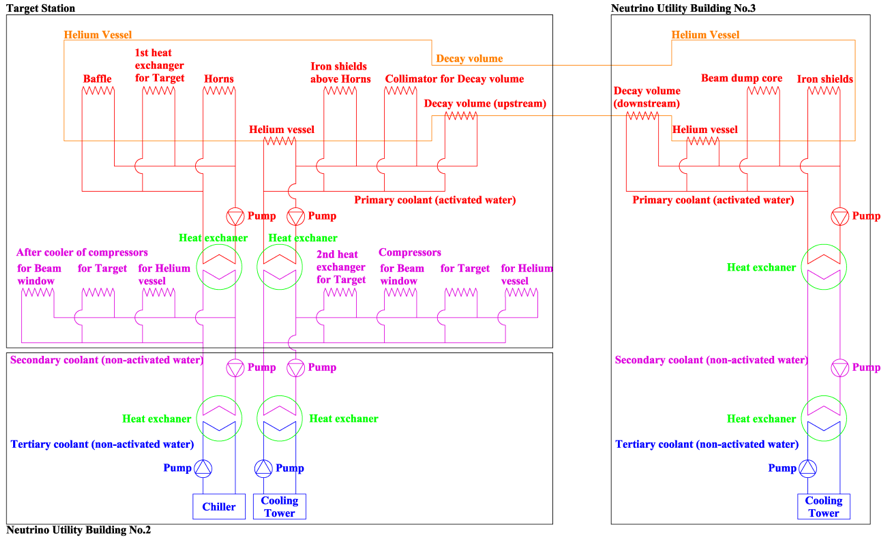}
        \caption{\small Schematic diagram of the water cooling system in the socondary beamline.}
        \label{fig:water_diag}
\end{figure}

Fig.~\ref{fig:water_diag} shows the schematic diagram of the water cooling system in the secondary
beamline.
There are two water cooling systems in the Target Station. 
One is for the magnetic horns, 
the target cooling helium gas, and the baffle (called “target-horn line”). 
And another is for the helium vessel, the iron shields above the horns, the upstream part of 
the decay volume, and the collimator for the decay volume (called “iron line”).
There is a water cooling system for the downstream part of the decay volume, the beam dump core, the iron shields in the 
beam dump, and the vessel for the beam dump in the neutrino utility building No.3 (called “NU3”).
Every system consists of three stages and the stages are coupled by the heat exchangers, 
so that the activated water does not mix into the tertiary coolant if one of the heat 
exchanger breaks.
The circulation pumps and the heat exchangers are in the machine room, which is next 
to the beam line area (called “service pit”) separated by a concrete shield wall. 
The target-horn line, the iron line in the target station, and the iron line in the NU3 have 
each one set of the pump and heat exchanger.

\begin{figure}
        \centering
        \includegraphics[width=0.8\linewidth]{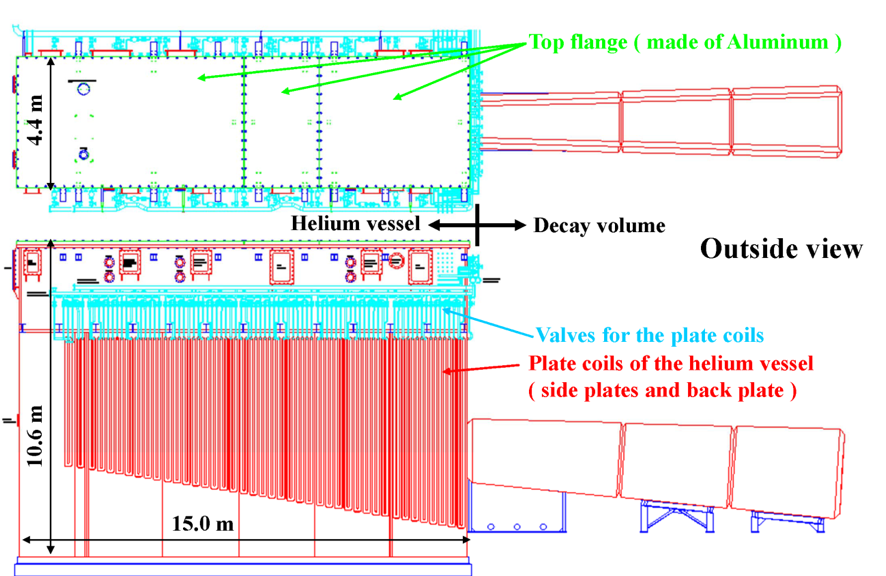}
        \includegraphics[width=0.8\linewidth]{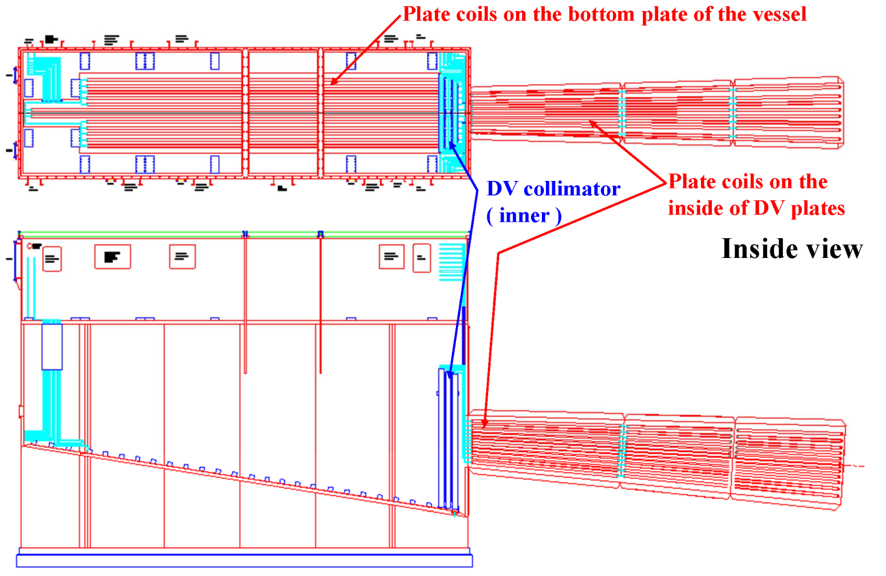}
        \caption{\small Water cooling channels for the helium vessel and the Decay Volume.}
        \label{fig:helium_vessel}
\end{figure}

\begin{figure}
        \centering
        \includegraphics[width=0.8\linewidth]{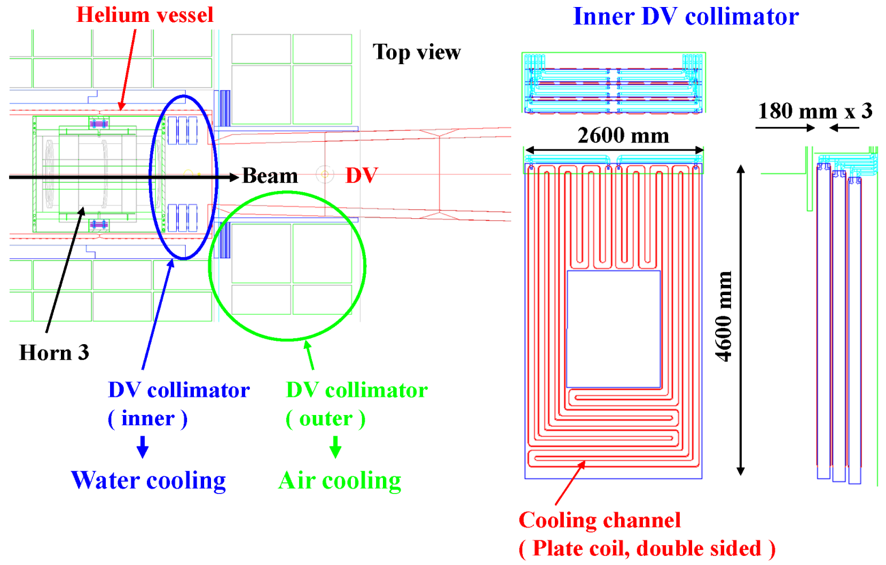}
        \caption{\small Water cooling channels for the collimator for the Decay Volume.}
        \label{fig:DV_collimator}
\end{figure}

\begin{figure}
        \centering
        \includegraphics[width=0.8\linewidth]{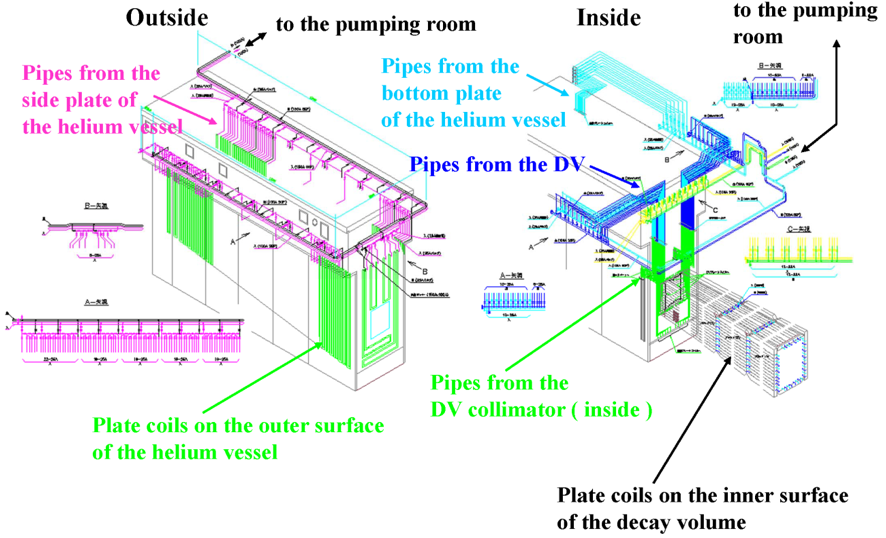}
        \caption{\small Water cooling lines in the Target Station.}
        \label{fig:TS_cooling_line}
\end{figure}

\begin{figure}
        \centering
        \includegraphics[width=0.8\linewidth]{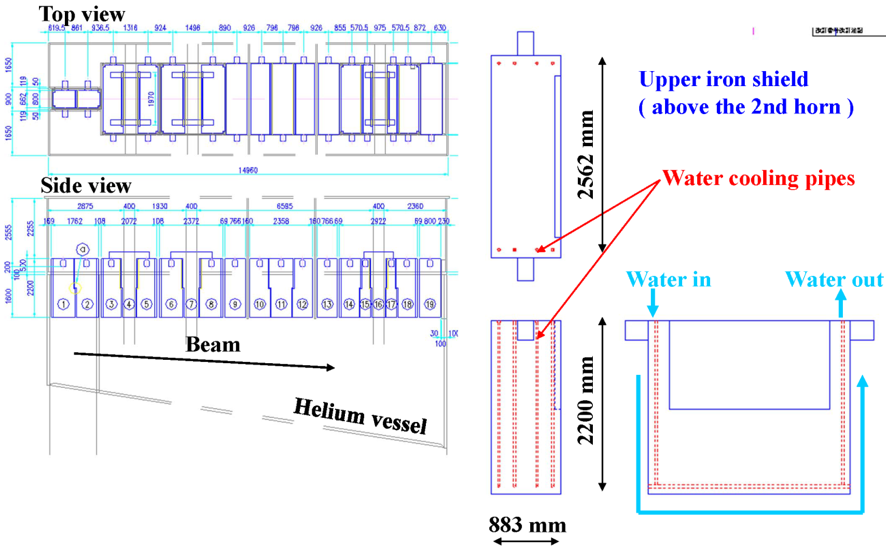}
        \caption{\small Water cooling pipes for the iron shields above the magnetic horns.}
        \label{fig:iron_shield}
\end{figure}

The helium vessel, the decay volume, the collimator for the decay volume, the iron shields 
in the beam dump, and the vessel for the beam dump are cooled by cooling water through 
the iron channels called “plate coils”, as shown in Figs.~\ref{fig:helium_vessel},
\ref{fig:DV_collimator}, and \ref{fig:TS_cooling_line}. 
The water cooling pipes made of the carbon steel are cast into the iron shields above the horns,
as shown in Fig.~\ref{fig:iron_shield}. 
The beam dump core is made of graphite and cooled with the attached cooling modules. 
The cooling module is made of aluminum alloy, and the water cooling pipes made of 
the carbon steel are cast into the module.

\color{\MODCOLOR}
A water leak from the plate coil can be detected by decrease of cooling water and
observation of the drain. Because there are the plate coils more than the necessary number,
when the water leaks from a cooling channel, we can continue the operation with closing that channel.
\color{black}
The plate coils and the cooling water pipes have several watercourses, and it is possible to 
change the watercourse by switching the valves. Several channels are connected in series at 
present to suppress the total flow rate, but it is designed in order to remove heat load at 
4 MW beam operation by connecting in parallel. Present connection is for 750 kW beam operation. 
It is necessary to increase the total flow rate in parallel connection.
Cooling water pipes in the beam line area are designed in order to accommodate 4 MW beam operation, 
but pipes in the machine room are designed for 750 kW beam operation. The flow 
velocity of the cooling water in the pipes on the machine room side rises with increase of the total 
flow at the 1.3 MW beam operation, but it is not necessary to replace the pipes.
It is necessary to replace the circulation pumps, the heat exchangers, chillers, and cooling towers 
to more large capacity ones, 
because they were designed based on 750 kW beam operation. 

The capacity of the pumps and the heat exchanger in the primary coolant for the target and the horns 
is enough, but it is necessary to increase the heat removal in the secondary coolant by increasing 
flow rate or lowering the coolant temperature. It is possible to lower the temperature of the secondary 
coolant to 12~$^{\circ}$C from the present operation temperature of 
25~$^{\circ}$C at present, 
because there is a margin to lower 
the temperature sufficiently. However, when lowering the temperature, it is necessary to install heat 
insulation for dew condensation prevention to the pipes. In case of increasing the flow rate of the secondary 
coolant, it is necessary to replace the pump to a larger capacity one.

\begin{figure}
        \centering
        \includegraphics[width=0.8\linewidth]{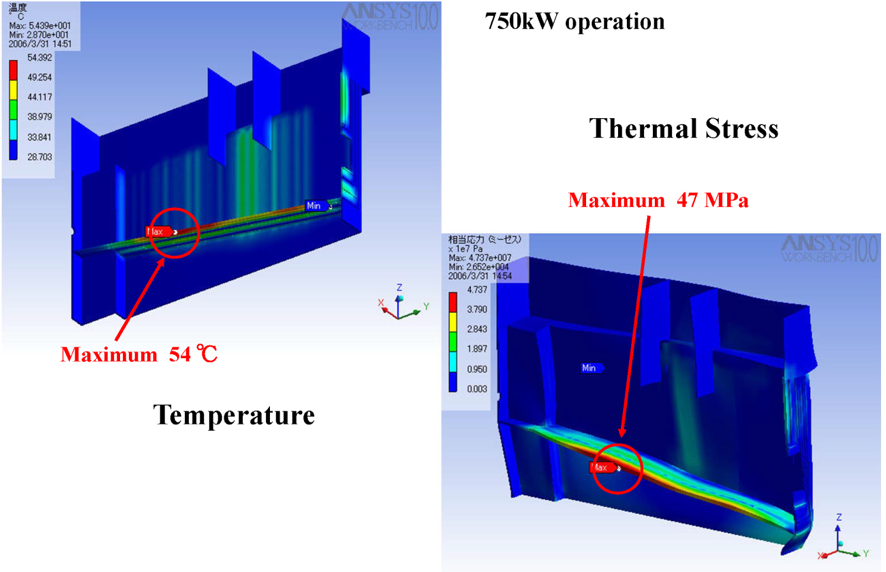}
        \includegraphics[width=0.8\linewidth]{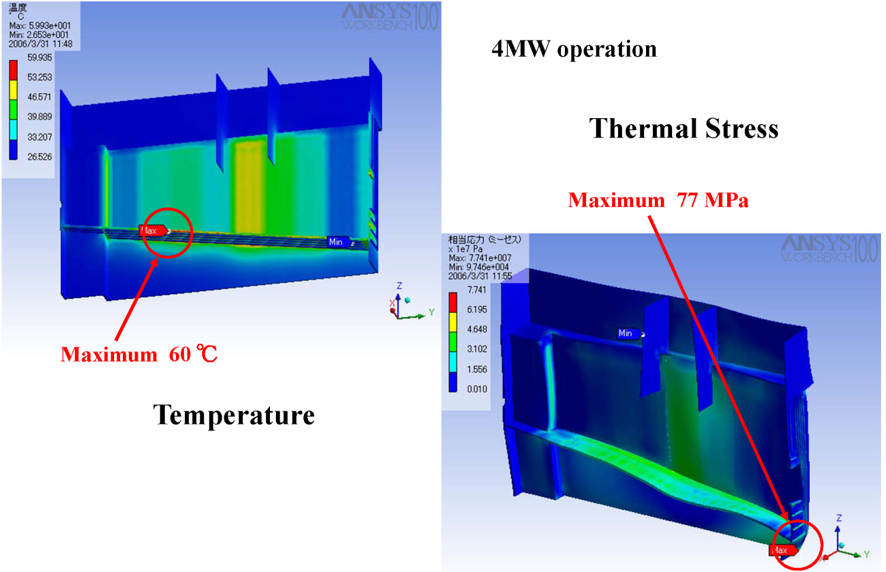}
        \caption{\small A sample of ANSYS heat simulation (for helium vessel).}
        \label{fig:TS_thermal_stress}
\end{figure}

Tabs.~\ref{tab:cooling_temp_iron_nu} and \ref{tab:cooling_temp_iron_nubar} show measured maximum 
temperature rises of “iron line” devices at 410~kW neutrino beam and 450~kW antineutrino beam
operations with calculated values, respectively.
\begin{table}
        \centering
        \small
        \caption{\small Measured maximum temperature rise ($\nu$ beam operation).}
        \begin{tabular}{lccc}
        \hline\hline
                                  & Present          & Extrapolation  & Calculated value        \\
                                  & (410~kW, $\nu$)  & (750~kW, $\nu$)& (750~kW, $\nu$)         \\\hline
        Helium vessel (side wall) & 7.1~$^{\circ}$C  & 13~$^{\circ}$C & 22~$^{\circ}$C          \\
        Iron shields above horns  & 8.7~$^{\circ}$C  & 16~$^{\circ}$C & 22~$^{\circ}$C          \\
        Decay volume              & 7.3~$^{\circ}$C  & 13~$^{\circ}$C & 34~$^{\circ}$C          \\
        DV collimator             & 11.0~$^{\circ}$C & 20~$^{\circ}$C & 38~$^{\circ}$C          \\
        Beam dump core            & 36.9~$^{\circ}$C & 68~$^{\circ}$C & -                       \\
        Vessel for BD core        & 7.5~$^{\circ}$C  & 14~$^{\circ}$C & -                       \\
        Iron shields behind BD    & 20.1~$^{\circ}$C & 37~$^{\circ}$C & -                       \\
        \hline\hline
        \end{tabular}
        \label{tab:cooling_temp_iron_nu}
\end{table}
\begin{table}
        \centering
        \small
        \caption{\small Measured maximum temperature rise (anti-$\nu$ beam operation).}
        \begin{tabular}{lccc}
        \hline\hline
                        & Present               & Extrapolation         & Calculated value      \\
                        & (450~kW, anti-$\nu$)  & (750~kW, anti-$\nu$)  & (750~kW, $\nu$)       \\\hline
        Helium vessel (side wall) & 7.8~$^{\circ}$C & 13~$^{\circ}$C & 22~$^{\circ}$C           \\
        Iron shields above horns  & 7.5~$^{\circ}$C & 12~$^{\circ}$C & 22~$^{\circ}$C           \\
        Decay volume              & 11.9~$^{\circ}$C & 20~$^{\circ}$C & 34~$^{\circ}$C          \\
        DV collimator             & 13.3~$^{\circ}$C & 22~$^{\circ}$C & 38~$^{\circ}$C          \\
        Bema dump core            & 20.5~$^{\circ}$C & 34~$^{\circ}$C & -                       \\
        Vessel for BD core        &  6.0~$^{\circ}$C & 10~$^{\circ}$C & -                       \\
        Iron shiedls behind BD    & 18.0~$^{\circ}$C & 30~$^{\circ}$C & -                       \\
        \hline\hline
        \end{tabular}
        \label{tab:cooling_temp_iron_nubar}
\end{table}
The calculated value is by ANSYS heat simulation, supposing heat transfer coefficients on 
the plate coils and pipes and inputting heat loads calculated using a 2-dimmensional mode by MARS simulation.
Fig.~\ref{fig:TS_thermal_stress} shows a sample of ANSYS heat simulation (for helium vessel). 
The inputted heat transfer coefficient is conservative in this simulation.
The temperatures of “iron line” devices are kept below 60 degree (temperature rises are kept below 
30 degree). This condition is important in particular of the side wall of the helium vessel, 
because the side wall supports the magnetic horns. When the temperature of all parts of the side 
wall rises 30 degree, the position of the magnetic horn becomes 1mm higher.

Tabs.~\ref{tab:cooling_heat_removal_nu} and \ref{tab:cooling_heat_removal_nubar} show 
heat removals of “iron line” devices at 410~kW neutrino beam and at 450~kW anti-neutrino beam 
operation with design values, respectively.
\begin{table}
        \centering
        \small
        \caption{\small Measured heat removals ($\nu$ beam operation).}
        \begin{tabular}{lccc}
        \hline\hline
                        & Present          & Extrapolation    & Calculated value      \\
                        & (410~kW, $\nu$)  & (750~kW, $\nu$)  & (750~kW, $\nu$)       \\\hline
        Helium vessel   & 40~kW            & 73~kW            & 158~kW                \\
        Iron shields    & 11~kW            & 20~kW            & 29~kW                 \\
        Decay volume    & 42~kW            & 77~kW            & 85~kW                 \\
        DV collimator   & 40~kW            & 73~kW            & 105~kW                \\
        TS Total        & 133~kW           & 243~kW           & 377~kW                \\
        NU3 Total       & 129~kW           & 236~kW           & 332~kW                \\
        \hline\hline
        \end{tabular}
        \label{tab:cooling_heat_removal_nu}
\end{table}
\begin{table}
        \centering
        \small
        \caption{\small Measured heat removals (anti-$\nu$ beam operation).}
        \begin{tabular}{lccc}
        \hline\hline
                        & Present               & Extrapolation         & Calculated value      \\
                        & (450~kW, anti-$\nu$)  & (750~kW, anti-$\nu$)  & (750~kW, $\nu$)       \\\hline
        Helium vessel (side wall) & 54~kW       & 90~kW                 & 158~kW                \\
        Iron shields              & 19~kW       & 32~kW                 & 29~kW                 \\
        Decay volume              & 52~kW       & 87~kW                 & 85~kW                 \\
        DV collimator             & 55~kW       & 92~kW                 & 105~kW                \\
        TS Total                  & 179~kW      & \color{\MODCOLORB}298~kW\color{black}                & 377~kW                \\
        NU3 Total                 & 101~kW      & 168~kW                & 332~kW                \\
        \hline\hline
        \end{tabular}
        \label{tab:cooling_heat_removal_nubar}
\end{table}
Measured values are calculated from measured values of temperature rises and flow rates of the coolant. 
Design values were determined by heat loads from 2D-MARS simulation.
Heat load is higher at anti-neutrino beam operation than at neutrino one in the Target Station,
on the other hand, it is higher at neutrino beam operation in the NU3.
The measured total heat removal in the Target Station at 450kW anti-neutrino beam operation is 179kW, 
and it becomes 289kW when it is 
extrapolated at 750kW. It is \color{\MODCOLOR}1.6\color{black} times as small as design value 377kW. 
\color{\MODCOLOR}The heat input was overestimated much 1.4 times by using 2D-MARS in this simulation.\color{black}

When this extrapolation is right, 
the cooling system becomes effective to beam power of 980kW.
The measured total heat removal in NU3 at 410 kW neutrino beam operation is 129 kW, and it becomes
236 kW when it is extrapolated at 750 kW. It is 1.4 times as small as design value, 332 kW.
When this extrapolation is right, the cooling system becomes effective to beam power of 1060 kW.
\color{\MODCOLOR}We will do the FEM simulation at 1.3MW beam operation with operational feedback.\color{black}

The oxygen density of the cooling water is suppressed low for pipes not to rust,
using deoxidation device with deaeration films, 
because the material of pipes is carbon steel in “iron line”.
The deoxidation devices are installed in the target station and NU3.
Because the black layer of steel inside the pipes is left intentionally in order to prevent making rust,
the deaeration films are filled with the black powder from the black layer.
Therefore the deoxidation devices has filters, but it is necessary to exchange the filters periodically.
And it is also  necessary to exchange the deaeration films periodically.
The amount of the black powder which collects on deoxidation device increases according to
the beam power empirically, so when operating at beam power beyond the current state, 
it is necessary to add the deoxidation devices and the filters.
\color{\MODCOLOR}And we will do a chemical analysis of cooling water to research the character 
and the amount of the black powder quantitatively, and estimate future soundness of plate coils and pipes.
\color{black}

Summarizing, except the deoxidation devices, the present water cooling system of 
the secondary beam line is capable up to 750kW beam operation, 
but it is necessary to do the following upgrade in case of the operation beyond 750kW.
\begin{itemize}
\item Lowering the secondary coolant temperature in the “target-horn line”
\item Replacement of the circulation pumps and the heat exchangers to larger capacity ones in the “iron line”
\item Adding chillers and cooling towers in the “iron line”
\end{itemize}
For the deoxidation devices, it is necessary to add the deoxidation devices and the filters as soon 
as possible.

\graphicspath{{figures/main_secondary}}

\subsection{Radiation shielding}

\begin{figure}
        \centering
        \includegraphics[width=0.9\linewidth]{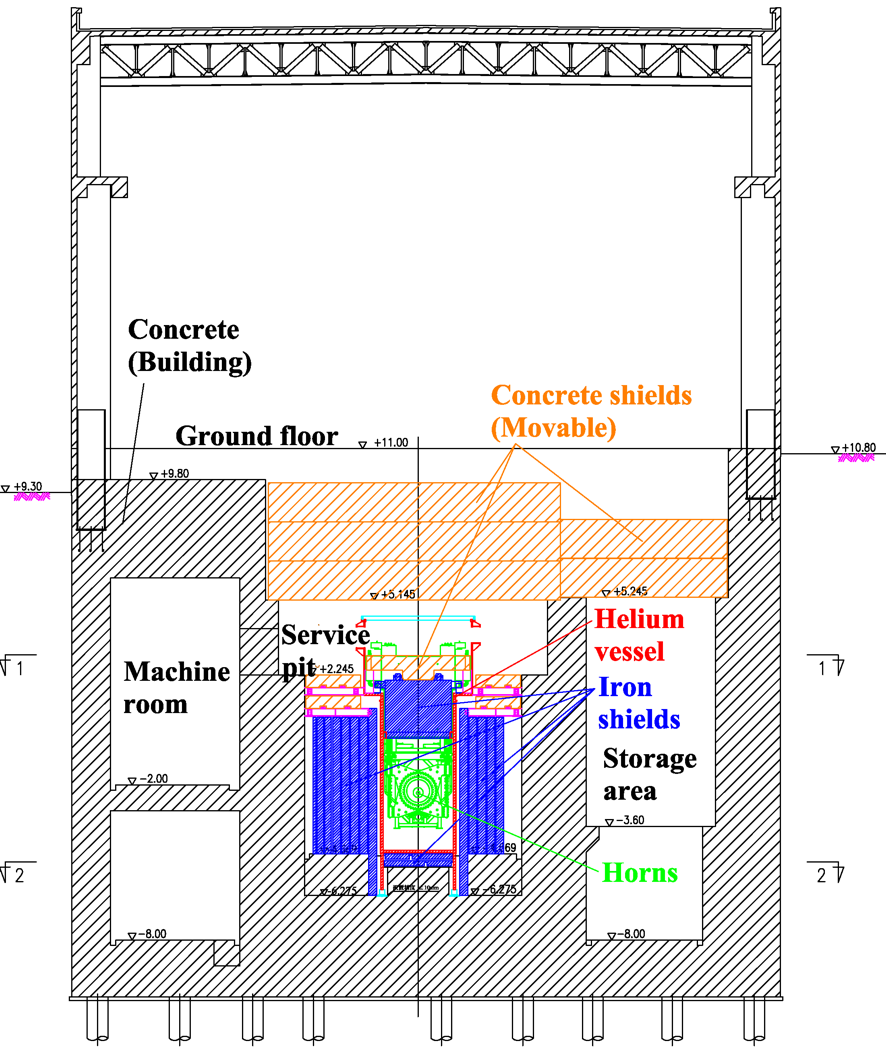}
        \caption{\small Cross section of the Target Station.}
        \label{fig:TS_shield}
\end{figure}

Figure~\ref{fig:TS_shield} shows the cross section of the target station.
Target station is the facility for operation and maintenance of the target and the magnetic horns. 
The iron shields surround the helium vessel in which the target and the magnetic horns are installed, 
and the concrete shields surround the iron shields.
\color{\MODCOLOR}
The iron shields above the magnetic horns (inside the helium vessel) are cooled by water and 
the other iron shields are cooled by air. There is the air conditioning system to cool the iron shields.
\color{black}
At the side and bottom part, concrete serves both as a shield and a skeleton structure of the target 
station building and iron shields are fixed to the building. At the upper part, all the shields are movable. 
There are 19 iron shield blocks and 10 concrete shield blocks inside the helium vessel, 
and 147 concrete shield blocks above the helium vessel.
Tab.~\ref{tab:shield_thickness} shows total thickness of shields.
\color{\MODCOLOR}
The maximum temperature rise of the air cooling iron shields is 10 degree at 450kW beam operation. 
The extrapolation value for 1.3MW is 29 degree. It is low enough.
\color{black}
\begin{table}
        \centering
        \small
        \caption{\small Thickness of shields.}
        \begin{tabular}{lcc}
        \hline
                                        & Iron shields  & Concrete shields      \\\hline
        Top (outside helium vessel)     & -             & 4,500~mm              \\
                                        &               & Movable               \\\hline
        Top (inside helium vessel)      & 2,250~mm      & 940~mm                \\
                                        & Movable       & Movable               \\\hline
        Side (machine room side)        & 2,200~mm      & 4,000~mm              \\\hline
        Side (storage area side)        & 1,600~mm      & 5,000~mm              \\\hline
        Bottom                          & 500~mm        & 4,770$\sim$7,190~mm   \\\hline
        \end{tabular}
        \label{tab:shield_thickness}
\end{table}

The radiation level has to be below the legal limit of the off-limits area ($<$~25~$\mu$Sv/h) for a person 
to be able to enter the ground floor of the target station even during beam operation.
The thickness of the upper shield was determined by calculation as the maximum radiation level on
 the ground floor is below the half of the legal limit (12.5~$\mu$Sv/h) at 750kW beam operation. 
 The thicknesses of the side shield and the bottom shield were determined by calculation 
 as the maximum radiation level is below 5mSv/h at the boundary with soil. Cord MCNP was used for calculation.

Tab.~\ref{tab:radiation_level} shows the measured radiation level on the ground floor (above the beam line) of 
the target station at 460kW beam operation and the extrapolation values (750kW and 1.3MW).
\begin{table}
        \centering
        \small
        \caption{\small Measured radiation level on the ground floor of the Target Station.}
        \begin{tabular}{cccc}
        \hline
                                & Present       & Extrapolation & Extrapolation \\
                                & (460~kW)      & (750~kW)      & (1.3~MW)      \\\hline
        Neutrino operation      &               &               &               \\
        $\gamma$                & 0.7~$\mu$Sv/h & 1.1~$\mu$Sv/h & 2.0~$\mu$Sv/h \\
        neutron                 & 1.3~$\mu$Sv/h & 2.1~$\mu$Sv/h & 3.7~$\mu$Sv/h \\
        total                   & 2.0~$\mu$Sv/h & 3.3~$\mu$Sv/h & 5.7~$\mu$Sv/h \\\hline
        Anti-neutrino operation &               &               &               \\
        $\gamma$                & 0.7~$\mu$Sv/h & 1.1~$\mu$Sv/h & 2.0~$\mu$Sv/h \\
        neutron                 & 1.2~$\mu$Sv/h & 2.0~$\mu$Sv/h & 3.4~$\mu$Sv/h \\
        total                   & 1.9~$\mu$Sv/h & 3.1~$\mu$Sv/h & 5.4~$\mu$Sv/h \\\hline
        \end{tabular}
        \label{tab:radiation_level}
\end{table}
The extrapolation value indicates that the radiation level is below the specified value (12.5~$\mu$Sv/h) 
even at 1.3MW beam operation, in other words that it is not necessary to add more shields. 
\color{\MODCOLOR}
But the neutron counter used these measurements have no response for neutrons above 
20 MeV and counting loss due to the short pulsed beam operation. 
So we will measure the radiation level with 
\color{\MODCOLORB}
sensors sensitive for neutrons (e.g., a CR-39 track detector and/or a thermolu-minescent dosimeter) 
\color{black}
above the concrete shields in the target station.
\color{black}
However, we have to determine whether we add more shields based on the calculation, 
because the application to Nuclear Regulation Agency is based on calculation. 
So we have to redo radiation calculation at 1.3MW beam operation.

\graphicspath{{figures/main_secondary/}}

\subsection{Disposal of radioactive water}

\subsubsection{Overview}

To cool the beamline instruments, cooling water systems are
equipped in the neutrino beamline. Three cooling water system are operated
as listed in Tab.~\ref{table:cwsystem}.
The cooling water circulates between objects for cooling and heat exchangers.
Neutrons are produced from the proton beam at the target.
They hit oxygen atoms of cooling water and many radioisotopes are produced.
Since one oxygen atom is made of 8 protons and 8 neutrons,
number of both protons and neutrons in one radioisotope
are less than 8. 
In the cooling water, some other materials are resolved in water.
Most of them are metals from water circulation system.
They are irons from steal pipes and aluminum from the magnetic horns.
Radioisotopes are also produced from breaks of those metals.

Among them, all radioisotopes except $^{3}$H can be removed by
ion-exchangers. Although management of the ion-exchangers are quite
hard work, we do not have any essential difficulties.

\begin{table}[b!]
\small
\caption{\small
Cooling water system in the neutrino beamline. 
Name of the system, volume of cooling water, objects for cooling, drainage system for cooling water are listed. Total $^{3}$H produced by 750kW $\times$ 10$^{7}$ seconds and by 1.3MW $\times$ 10$^{7}$ seconds
are also shown. 
}
\begin{center}
\begin{tabular}{lclccc}
\hline
\hline
Cooling  & Total volume & Objects for  &  drainage & Total $^{3}$H (GBq)  & Total $^{3}$H (GBq)  \\
system   & of water  (m$^{3}$)  & cooling     &         & 750kW x 10$^{7}$s   & 1.3MW x 10$^{7}$s  \\
\hline
Horn & 2.7 & 3 magnetic horns  &  NU2 & 126 & 218\\ 
TS32$^{\circ}$C & 7.8 &  Helium vessel & NU2 & 150 & 260\\
  &  &  decay volume (upstream) & \\ 
NU3-32$^{\circ}$C & 10.0  &  decay volume (downstream) & NU3 & 60 &104\\
  &  &   beam dump &  \\ 
\hline
\hline
\end{tabular}
\label{table:cwsystem}
\end{center}
\end{table}

$^{3}$H accumulated in the neutrino beam line are disposed by two methods.
One is the drainage of the radioactive water, and the other is disposal
using tank truck.
For drainage, radioactive water are moved to disposal tanks and are diluted by industrial water. 
The effective volume of the NU2 and NU3 disposal tank are 84m$^{3}$ and
17m$^{3}$ respectively. From a regulation, concentration of $^{3}$H in the
disposal water must be less than 60Bq/cc. Because of the safety reason,
the radiation safety section requires that the concentration should not exceed
42Bq/cc. Accordingly, total $^{3}$H disposed from NU2 and NU3 disposal tank
in one drainage cycle are 3.53 GBq  and 0.72 GBq, respectively.
At present, one cycle of drainage from the disposal tank takes every three business days.
This schedule is arranged from following constraints;
\begin{itemize}
\item Local government request that the drainage should be done in business day morning
\item Overnight operation is needed to measure the concentration of $^{3}$H
\item One business day is needed for paper work related to permission of drainage
\end{itemize}

The back-end section of JAEA provides a service to take over radioactive water by a tank truck.
The takeover by the tank truck in NU3 started, successfully.
This can be also done in NU2 after a change of the water circuit.
In one-day tank truck service, about 10GBq of $^{3}$H can be taken over.
The maximum frequency of the tank truck service is once every month, or 10 times
per year. (They have about 2 months of maintenance period per year.) 
If present agreement between neutrino group and JAEA is considered,
radioactive water containing 100GBq can be taken out by the tank truck service.

\begin{figure}[t]
\centerline{\includegraphics[width=16cm]{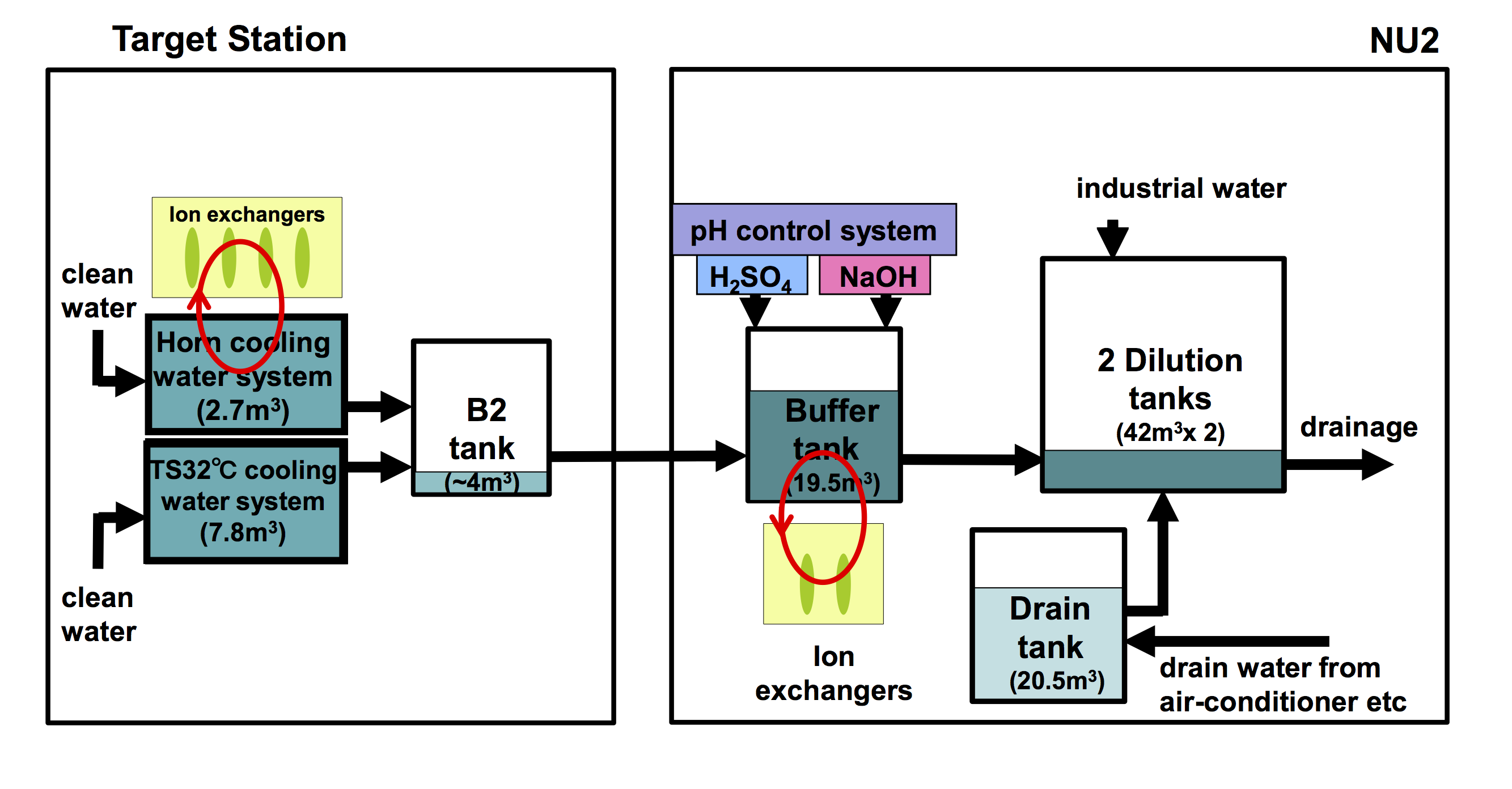}}
\caption{\small
Schematic view of the drainage for horn and TS32$^{\circ}$C radioactive water.
}
\label{fig:drainage-schematic}
\end{figure}

\begin{figure}[t]
\centerline{\includegraphics[width=16cm]{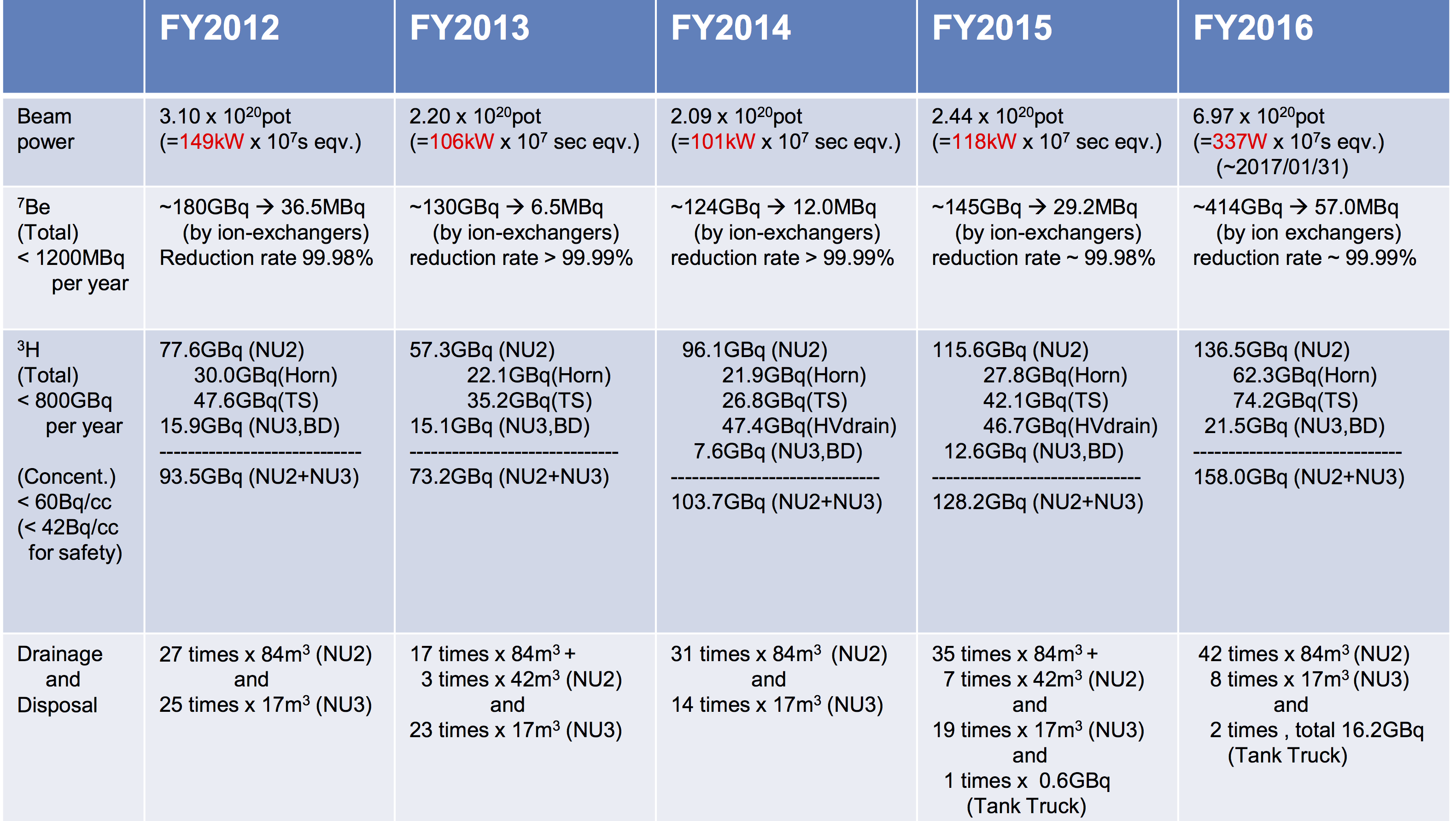}}
\caption{\small
Annual summary of drainage of radioactive water generated in the neutrino beam line.
}
\label{fig:haisui-summary}
\end{figure}

\subsubsection{Operation status}

The procedure of the radioactive water drainage is schematically
shown in Fig.~\ref{fig:drainage-schematic}.
During beam operation, cooling water in the target station building
become highly radio-activated. Nominal contaminations of $^{3}$H are
$\sim$6000Bq/cc for horn cooling water and $\sim$2500Bq/cc for TS32 cooling water
after 470 kW x 1 month beam operation.
An access to the cooling water system is done during periodical maintenance period.
Fresh water is supplied to the system, and radioactive
water is sent to the buffer tank which is placed in another
building, namely NU2. The effective volume of the buffer tank is 18.2m$^{3}$.
After the replacement dilution, beam operation can be resumed.

The NU2 building is not off-limit area even during the beam period.
First, radioactivities except $^{3}$H are removed by the ion-exchangers.
It is known that radioactivities can be removed effectively for acid water.
The sulfuric acid and sodium hydroxide are occasionally used for the pH control.
After radioactivities except $^{3}$H are removed, dilution/disposal process start.
In an usual case, $^{3}$H concentration of radioactive water in the buffer tank exceed 1000Bq/cc.
Drainage of radioactive water is permitted if the $^{3}$H concentration is less than
42Bq/cc. Small volume (a few m$^{3}$) of radioactive water is sent to the dilution tank.
Industrial water are added, and radioactive water is diluted to be less than 42Bq/cc.
After the measurement of the radioactivities and paper works, the water is disposed.
It takes 3 business days for one cycle of the drainage. By $\sim$10 times of
the dilution/drainage procedure, radioactive water in the buffer tank can be
completely disposed. It will take more than one month.

Annual status of radioactive water disposal is summarized in  
Fig.~\ref{fig:haisui-summary}.
\color{\MODCOLORB}
In term of the annual drainage, we usually drained not only the radioactive water 
generated in the latest beam operation but also the water remained from the previous operation 
due to the maintenance schedule.
\color{black}
Although maximum beam power exceeded 450kW, accumulated proton on
target in a year was still less than 400kW x 10$^{7}$ second equivalent.
In FY2016, radioactive water from NU2 building is disposed from the
disposal tank by dilution. In total, 136.5GBq of $^{3}$H was disposed by 42 times of
drainage cycle. On the other hand,  total $^{3}$H from NU3 was 21.5 GBq,
and the radioactive water was disposed mainly by the tank truck.

\color{\MODCOLOR}
\subsubsection{Prospect and issues for the operation beyond 750kW beam power}
\color{black}

Total $^{3}$H produced by 750kW $\times$ 10$^{7}$ seconds and by 1.3MW $\times$ 10$^{7}$ seconds
are shown in Tab.~\ref{table:cwsystem}. 
\color{\MODCOLORB}
We estimated the total Tritium production based on the measured increase of the Tritium with the accumulated POT. 
\color{black}

\color{\MODCOLOR}
In 750kW  x 10$^{7}$ second operation, 
total $^{3}$H production from NU3-32$^{\circ}$C cooling water system
will be 60GBq.
They can be disposed by 6 times of tank truck disposal.
On the other hand, 276GBq of $^{3}$H must be disposed from NU2.
It corresponds to 78 times of dilution/drainage cycle from the disposal tank.
Considering the present disposal frequency, 78 times per year is not realistic. 
The tank truck disposal from NU2 facility is also under planning. 
However, total number of the tank truck disposal is 10 times per year, 
and only 4 times remains for NU2. The tank truck disposal will be
slightly contribute, but it does not drastically change the situation.

In 1.3MW x 10$^{7}$ second operation, 104GBq of $^{3}$H are produced in
the NU3 building. Most of them can be disposed by the tank truck.
However, the tank truck quota for the neutrino facility is almost full
for the NU3, and no quota is left for NU2.
In NU2, 478GBq of $^{3}$H must be disposed. It corresponds to 135 times of
dilution/drainage. Because the quota of tank truck disposal will be fully used
by NU3, the $^{3}$H from NU2 must be disposed by the dilution/drainage
from the disposal tank. 

\subsubsection{Upgrade plan and schedule}

\begin{figure}[hbt]
\centering
\includegraphics[width=0.7\linewidth]{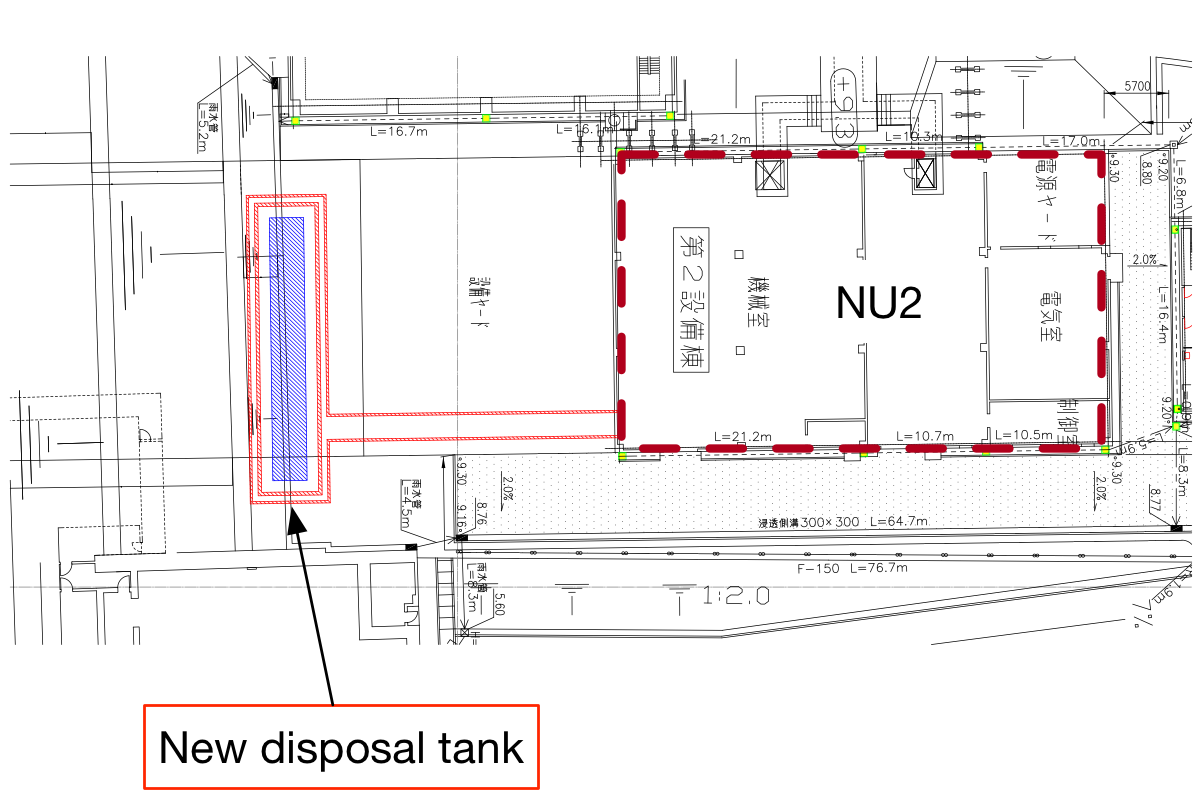}
\caption{\small
\color{\MODCOLOR}A ground plan of the new disposal tank. 
The feasibility of the construction and the cost are under consideration.\color{black}
}
\label{fig:nu4}
\end{figure}

\color{\MODCOLOR}
For the operation beyond 750kW beam power, larger disposal tanks are certainly necessary.
Construction of $\sim$400 m$^3$ disposal tank was proposed and the budget request
was submitted to KEK.
The conceptual design of the NU2 extension  for the new disposal tank is in progress.
Figure~\ref{fig:nu4} shows a ground plan of the new disposal tank. 
It is planned to complete the conceptual and the execution design of the new building  
and the new disposal tank by the end of JFY2019 and 
to make ready to start the construction in JFY2020. 
\color{black}

\graphicspath{{figures/main_secondary/}}

\color{\MODCOLOR}       
\subsection{Secondary beamline alignment}

\begin{figure}
        \centering
        \includegraphics[width=0.65\linewidth]{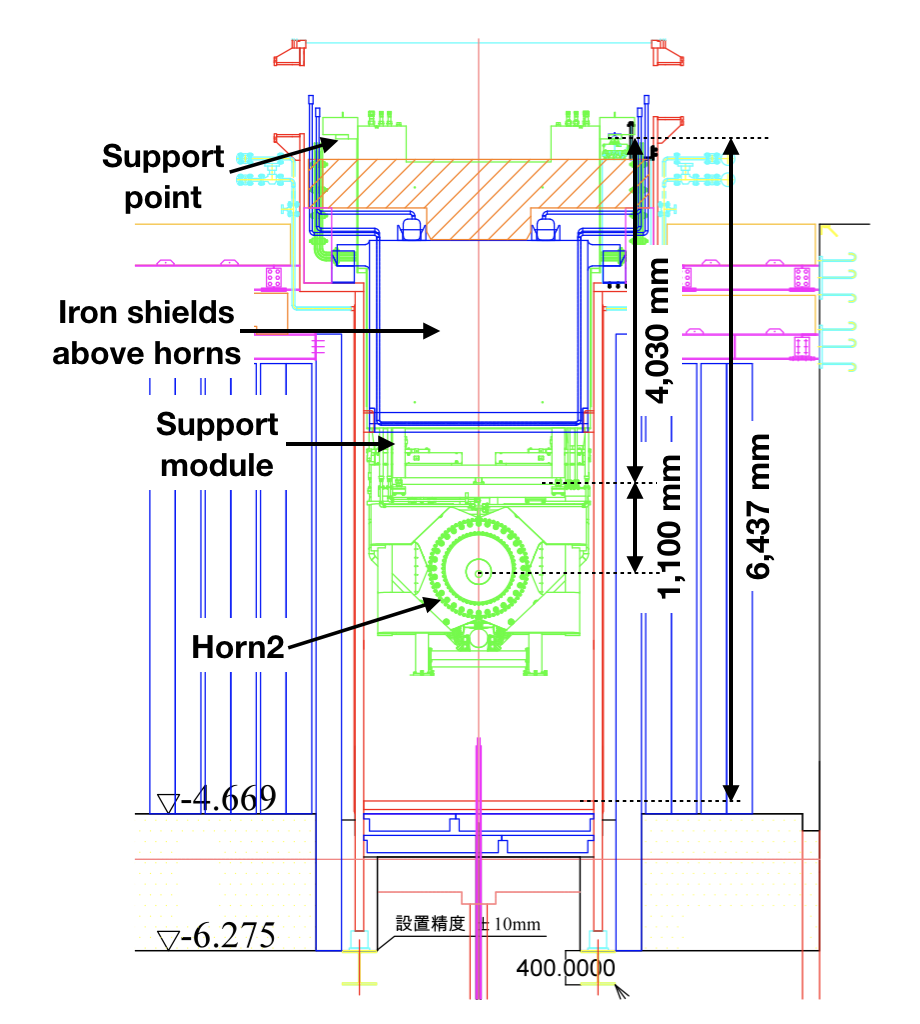}
        \caption{\small Cross section of helium vessel at horn2.}
        \label{fig:alignment}
\end{figure}

The magnetic horns and baffle are suspended by the support modules that are supported at the top of the helium vessel
iron structure, as shown in Fig.~\ref{fig:alignment}. 
The temperature rise of each component inside the helium vessel causes thermal expansion which can introduce
mis-alignment of beamline equipments, especially in vertical direction. 
The iron structure of the helium vessel is cooled by water and the requirement of allowed temperature rise is 30~$^{\circ}$C
to limit the thermal expansion within 1~mm. The aluminum vertical frames of the horns also have cooling channels inside them to avoid
thermal expansion. The horn support modules also have water cooling loops in their structure, which were not adopted to the first-generation
horns and newly adopted to the second-generation ones. Their water circulation line is not prepared so far, thus they should be added 
for the higher power operation. Table~\ref{tab:alignment} summarizes the expected thermal expansion of each component.
\begin{table}
\centering
\small
\caption{\small Expected thermal expansion of secondary beamline component. The value of the measured temperature rise of
the support module is taken from that of the iron shields above horns.}
\label{tab:alignment}
\begin{tabular}{lrrr}
\hline\hline
Parameters & Horn2 & Support module & Helium vessel \\ 
\hline\hline
Material       & Aluminum & Iron & Iron \\
Thermal Expansion Coefficient ($10^{-6}/^{\circ}$C) & 23.5 & 11.7 & 11.7 \\
Length (mm) & 1,100 & 4,030 & 6,437 \\
Measured temperature rise ($^{\circ}$C) & 8.3 (@480~kW) & 7.5$^{\ast}$ (@450~kW) & 7.8 (@450~kW) \\
Expected thermal expansion @1.3~MW (mm) &  0.580 downward & 1.020 downward & 1.700 upward \\
\hline\hline
\end{tabular}
\end{table}
The maximum heat load at the helium vessel iron structure is deposited around side wall of the horn2.
The values of the thermal expansion in Tab.~\ref{tab:alignment} are for the horn2 or neighboring helium vessel iron wall.
Because the water circulation to the support module cooling loops is not prepared yet, the measured temperature rise of
the support module is taken from that of the iron shields above horns, which is located inside the support modules and cooled by water.
Based on this estimation, the center of the horn2 will be moved upward by 0.1~mm, since the horn + support module and
the helium vessel expand in opposite direction and they are mostly cancelled out. 
The horn1 alignment compared to the upstream beamline components can be checked by extrapolation of fitted beam orbit by
beam monitor measurements and the actual measurement of OTR which is placed in front of target.
No significant beam orbit drift was observed within measurement errors during warming-up process by beam operation.
This implies that the alignment of the horn1 was not changed largely. In summary, thermal expansion effect on the secondary beamline
alignment is not so significant and can be well controlled by measuring the temperature of each component.

\color{black}

\clearpage

\graphicspath{{figures/main_ctrldaq}}
\section{Beamline DAQ and control} \label{sec:ctrldaq}
\subsection{Introduction}
\begin{figure}[h] 
\centering 
\includegraphics[width=0.8\textwidth] {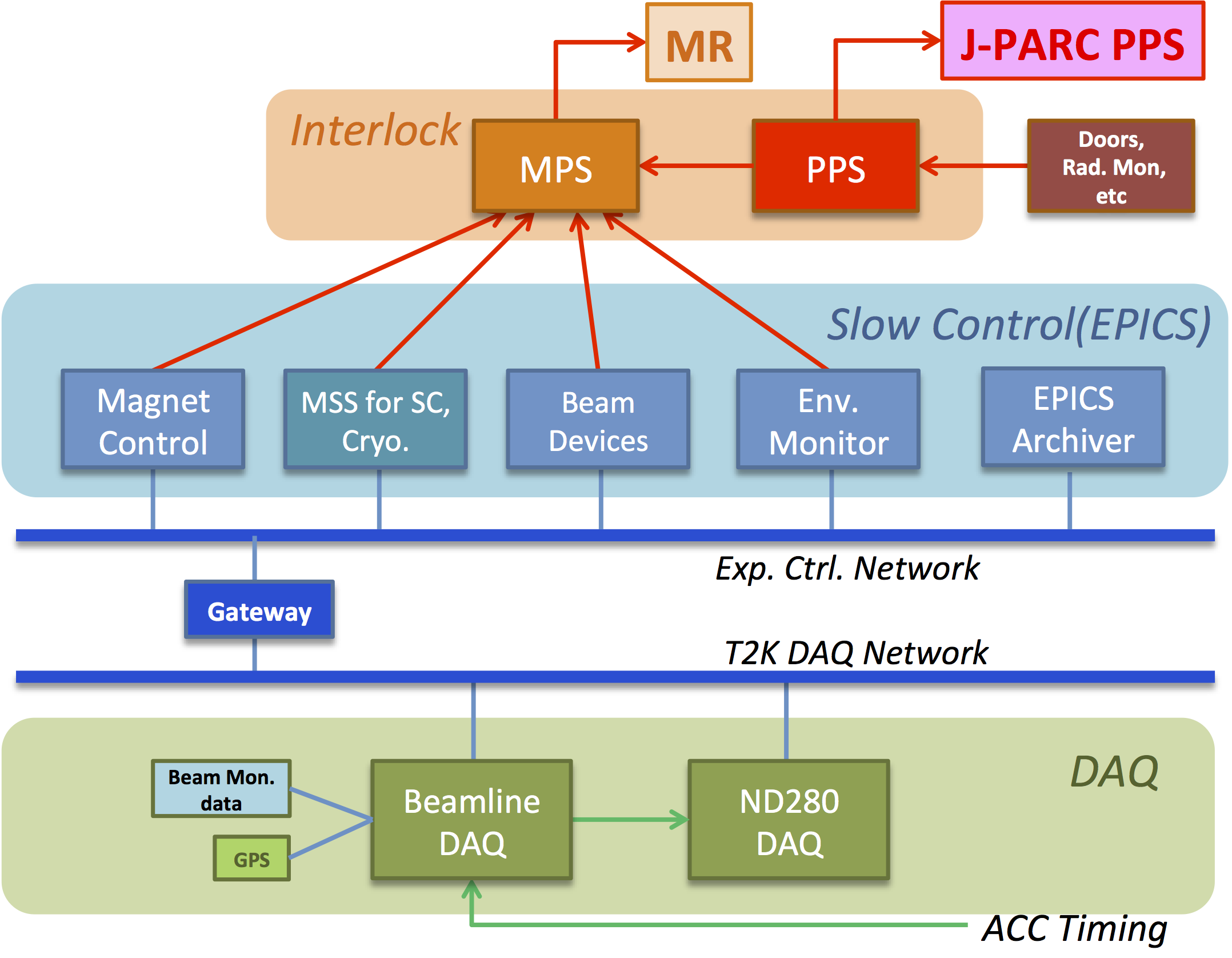}
\caption{Overview of the beamline control and DAQ system.}
\label{fig:daqoverview}
\end{figure}

The beamline DAQ and control system
collects all necessary informations from
the beam monitors and the beamline equipments
and is utilized to control those beamline equipments.
Figure~\ref{fig:daqoverview} shows an overview of the
beamline DAQ and control system.
It consists of 
a data aquisition for the beam monitor
including the GPS and beam timing distribution,
a slow control and monitor sytem with EPICS framework,
and an interlock system (Machine Protection System, MPS) which stops beam operation
immediately when some failures are happened in the
beamline equipments. 

The beamline DAQ and control system
has been stably operated so far. 
Toward 1.3~MW operation, 
the upgrade of beamline DAQ system is necessary
in order to operate with faster repetition time
than the present 2.48~sec repetition. 
The GPS system is essential to identity the
T2K beam signal at far detector.
The upgrade of GPS sytem is planned to improve
its stability. 
Moreover, improvement of the beam interlock is
desired to realize further safe operation with the
high intensity beam. 
The details of the beamline DAQ upgrade,
GPS upgrade and beam interlock improvement are
discussed in the following sections.

\subsection{Beamline DAQ upgrade}
\begin{figure}[h] 
\centering 
\includegraphics[width=0.8\textwidth] {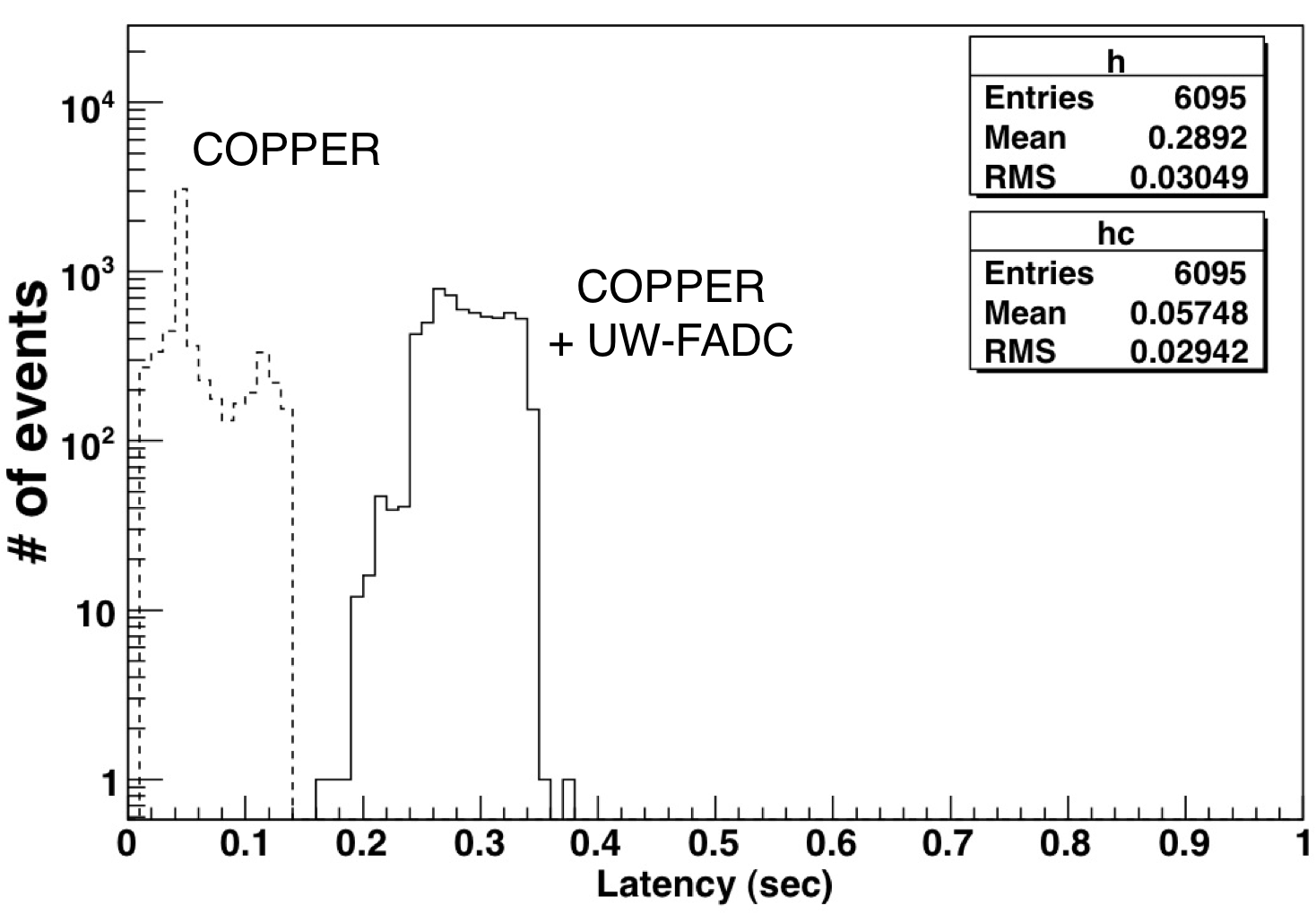}
\caption{Histogram of the latency of the beamline DAQ for each readout event.}
\label{fig:latency}
\end{figure}

\begin{figure}[h] 
\centering 
\includegraphics[width=0.4\textwidth] {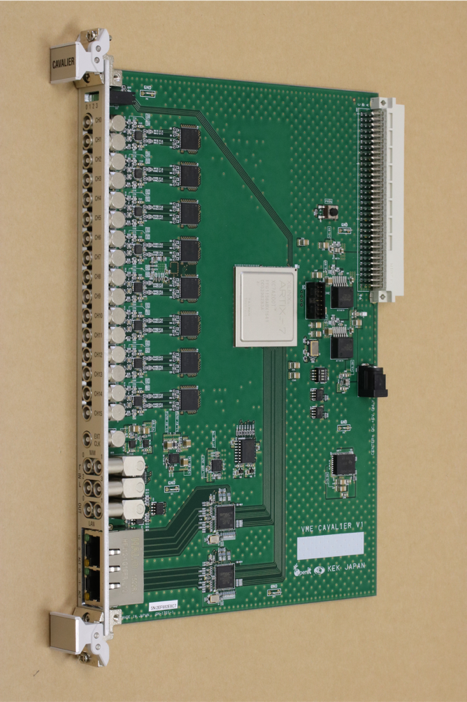}
\caption{A photograph of the prototype ADC module.}
\label{fig:cavalier}
\end{figure}

Improvement of readout latency is ncessary
toward the 1~Hz operation.
Table~\ref{tab:readout} summarizes the number of readout channels and
the type of readout electornics.
A custom-made ADC module (so-called UW-FADC) is used to
readout signal from CT and ESM.
The UW-FADC is read out via VME bus with a single access
readout mode. The measured readout latency of the UW-FADC
during a DAQ run with the usual operation mode in the past is
around 0.2$\sim$0.4 seconds and it's longer than one of
other readout electronics, COPPER, as shown in Fig.~\ref{fig:latency}.
Since the collected beam monitor data is utilized
to calculate the beam intensity, position and width 
in the online analysis computers in order to quickly check
if all the beam quantities satisfy the requirement of
normal beam condition, it is desired to collect data from
all the beam monitors in a short period.
In order to improve the readout latency of the UW-FADC
as similar to one of the COPPER, a new readout electronics
module for CT and ESM is being developed recently. 
Figure~\ref{fig:cavalier} shows a photograph of the prototype ADC module.
New ADC module consists of 16~channel inputs which is digitized 
with maximul 250~MHz sampling and 12~bits ADC.
The data is transfered from this module to DAQ computer
using the embedded SiTCP technology\cite{Uchida:2008fha} which enables a network readout.
This new module is under development. The basic performance such as
the linearity of the digitized data to the input analogue signal,
the cross talk among the input channels, the singal-to-noise ratio,
has been evaluated and confirmed to satisfy the requirements
while the readout latency will be confirmed in 2018.
After the vilidation, it is planned to install 
into T2K beamline DAQ system during SK-Gd upgrade work in 2018.

\begin{table}
  \caption{Summary of readout electroncis and the number of channels
    for each beam monitors and GPS. a:160~MHz FADC, b:COPPER 65~MHz
    FADC, b':COPPER 65~MHz FADC with 200~kHz sampling mode, c:a dedicated
    electronics.}
\label{tab:readout}
\vspace{1ex}
\centering
\begin{tabular}{lllc}
\hline\hline
Monitor & \# of monitors & \# of ch. & Readout type\\\hline
CT & 5 & 5 & a\\\hline
BLM & 50 & 50 & b\\\hline
SSEM & 19 & 912 & b\\\hline
ESM & 21 & 84 & a \\\hline
OTR & 1 & Image Data & c \\\hline
Mumon & Si:49 + IC:49 & 98 & b \\\hline
Horn Current & 14 & 14 & b' \\\hline
GPS/LTC & 1 & - & c \\
\hline\hline
\end{tabular}
\end{table}

\begin{figure}[h] 
\centering 
\includegraphics[width=0.8\textwidth] {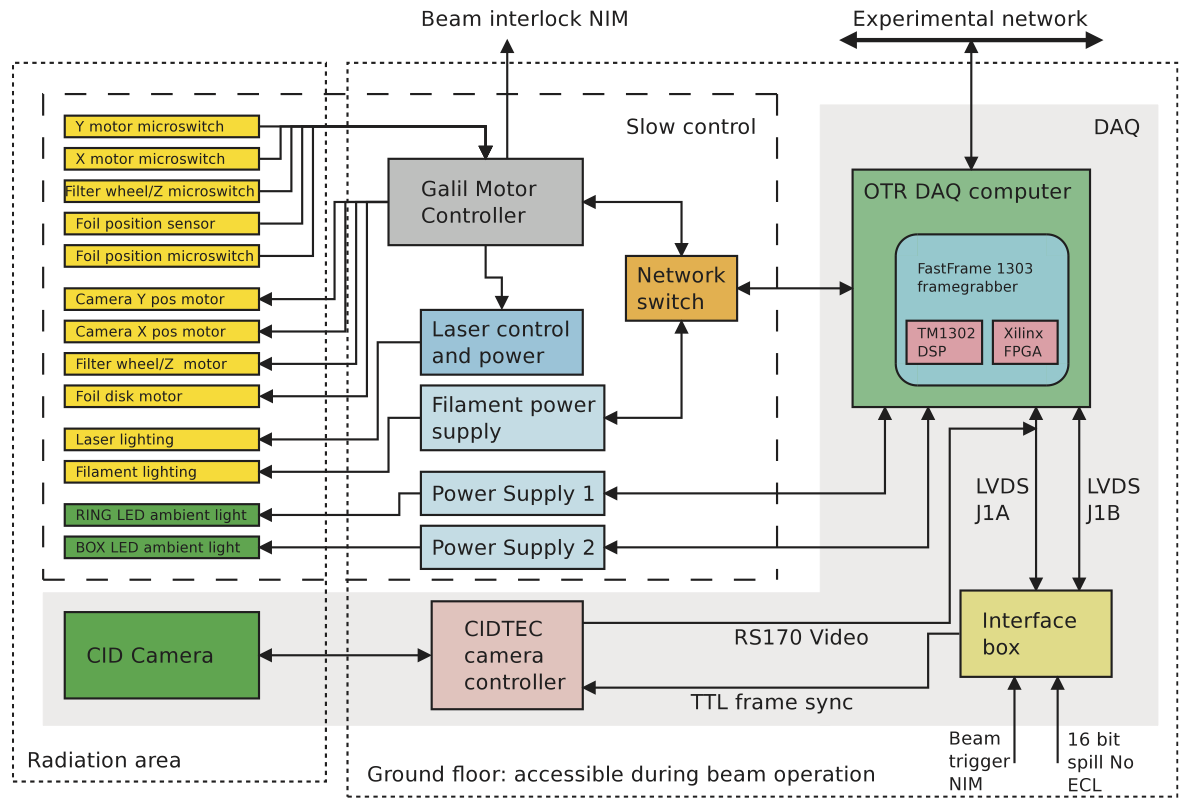}
\caption{The present OTR image readout DAQ\cite{Bhadra:2012st}.}
\label{fig:otr}
\end{figure}

Moreover, the OTR image readout DAQ will be also upgraded
in order to improve the readout speed as well as to improve
the stability and robustness of the system.
Figure~\ref{fig:otr} shows the present OTR image readout DAQ.
The present system uses a Windonws PC to readout and control
the radiation tolerant camera, Thermo Fischer CID87325,
and its interface board on the PC, FastFrame 1303 PCI board.
Every beam spill, three OTR images are read out
for one pedestal image and two beam profile images
in the spill and after the spill.  
The total processing time to collect all the three images
takes around 800~ms.
\color{\MODCOLORB}
Toward the 1 Hz operation, improvement of the processing time is planned in order to collect 
the OTR images and perform an online monitoring of the beam profile at the target within a
single spill cycle. One of ideas is that the readout PC will be replaced with a faster PC 
in order to improve the processing speed because most of the processing time spends on reading 
the image data from the FastFrame 1303 PCI board and performing the PNG compression. 
Another idea is to stop storing the third image in order to speed up the data transfer and reduce the file size.
The new readout PC is planned to use a Linux-based system and a custom-made dedicated digitizer board 
with a ethernet interface to eliminate the requirement of the FastFrame 1303 PCI board whose drivers are 
only available for Windows and difficult to upgrade and maintain.
\color{black}

\subsection{GPS upgrade plan}
GPS system is utilized to synchronize 
event timing at the far detector with J-PARC beam timing. 
There are almost identical systems at both J-PARC and SK site and 
each system consists of two independent GPS receivers,
a rubidium atomic clock and a custom local time clock (LTC) board
which generates time stamps when a trigger arrives.
The 1 pulse per second (1~PPS) signal
from the GPS receivers and 100~MHz clock from the rubidium atomic clock
are used to generate the time stamps every 10~ns.
When a beam trigger signal is provided to the LTC, 
a time stamp is recorded to each trigger.
This time stamp information is sent to the SK DAQ system via
a virtual private network (VPN) in realtime and utilized to select 
T2K beam events in SK detector.

The GPS system plays important role at T2K. On the other hand, 
there are recently several troubles happened on both J-PARC and SK GPS system
due to antiquated equipment. It is also realized that one of GPS receivers 
is no longer available in the market and it is difficult to keep maintaining them.

Upgrade plan has been discussed in not only T2K but also SK collaboration.
The following step-by-step approach is performed to improve the known issues.
\begin{enumerate}
\item Upgrade the primary GPS receiver with a recent product
\item Upgrade the secondary GPS receiver and new LTC module
\end{enumerate}
The first step will be performed in 2018 during no beam period
for the SK-Gd upgrade. A recycled equipments from T2K TOF study\cite{Abe:2015gna}
are used as much as possible to reduce cost.
The second step will be performed during the J-PARC MR upgrade.
Development of new LTC module has been started.

\subsection{Beam interlock improvement}
Toward 1.3~MW beam power, 
improvements of the interlock system 
for further safe operation is important. 
If such high intensity beam continuously hits off-centered at the target,
it is expected from the stress analysis that the target is damaged.
Moreover, the high intensity beam can cause a serious damage on the beamline
when a beam hits the beamline equipment and produces a huge thermal shock. 
In order to avoid such the cases, interlock system will be improved. 

\begin{figure}[htb]
\begin{center}
\begin{minipage}{7.5cm}
 \includegraphics[width=7.5cm]{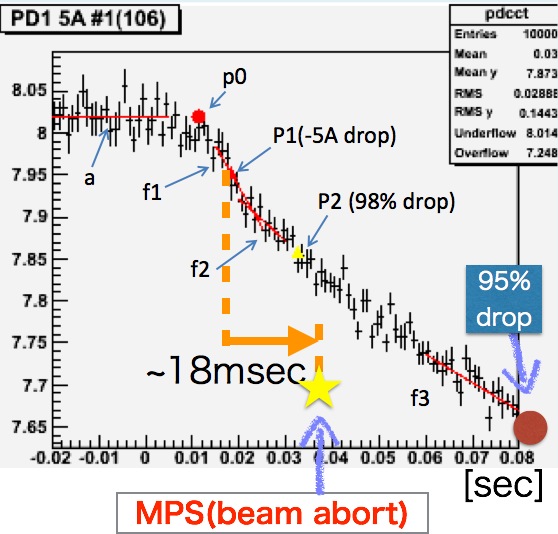}
 \caption{Latency measurement of the magnet MPS.}
  \label{fig:magmps}%
\end{minipage}
\hfill
\begin{minipage}{7cm}
  \includegraphics*[width=7.5cm]{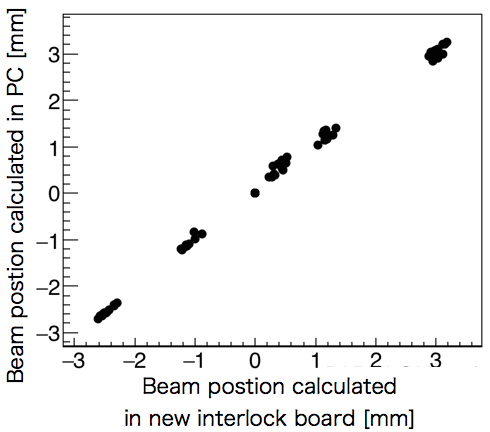}
   \caption{Comparison between the beam position calculated by the FPGA and 
the online computer}
   \label{fig:papillon_result}
\end{minipage}
\end{center}
\end{figure}

One of plans is improvement of the interlock for magnet current deviations.
At present, a comparator module detects a deviation of magnet current and
issues an interlock signal to stop beam immediately\cite{Nakayoshi:2015tbt}. 
There is a certain latency and therefore the magnetic field reaches the 95$\sim$98~\% 
of the nominal value based on the measurement (Fig.~\ref{fig:magmps}).
It can avoid the case which the beam hit any magnets and/or ducts but beam
can hit the off-centered position on the target. 
Further improvement is under consideration where a shorter latency and
enable to issue the interlock signal before the magnet field reaches
the 98~\% of the nominal value.

One of other plans is to calculate the beam position and profile every spills 
and issue an interlock to stop beam immediately.
In the present beamline control system, a similar scheme is already
implemented. Signals from one of the beam profile monitor,
SSEM~19 which is placed at most downstream 
of the primary proton beamline and 3.6~m upstream of the target,
are digitized and those data is sent to an online computer.
The beam center position and profile is then calculated at the online computer. 
An interlock signal is issued if the calculated
beam position exceeds the allow range, but it takes 
over 1~second after the beam injection and therefore next beam cycle is
already started. 
New electronics module is recently developed in order to improve the
latency of the calculation. The signals from SSEM~19 are sent to the new
module and the beam position and profile is calculated in a FPGA which
placed on the module in a real-time way. 
A prototype module has been developed. 
\color{\MODCOLOR}
A verification test with actual beam was also performed so far. 
In the verification test, SSEM06, which is placed in the upstream section of 
the primary beamline, was utilized. 
The proton beam position at SSEM06 was intentionally changed by changing the upstream 
dipole and horizontal steering magnets while the beam power was set to 35 kW. 
The beam position was also calculated in the online computer using a detailed off-line analysis method. 
Figure \ref{fig:papillon_result} shows the comparison between the beam position calculated in the FPGA and 
the online computer. 
It is confirmed that the beam position calculation by the prototype board was consistent with the one calculated by  
the online computer. 
The measured latency of this module was 9$\mu$s 
which is fast enough to stop the next beam injection to the MR\footnote{The expected time 
between the beam extraction to the neutrino primary line and the next beam injection to the MR 
is a few hundred millisecond.}. 
In addition to the beam position calculation, the beam profile calculation will be also implemented in the FPGA. 
It is planned to finish R\&D including the further verification test and start its operation before the MR starts 
operation with the shortened repetition time. 

\color{\MODCOLORB}
We will also consider further developments of an interlock system for 
off-center beam at the target based on an FPGA module with utilizing 
information from other beam position monitors including OTR and MUMON. 
\color{black}

For the safety operation with 1.3MW beam, 
it is also necessary to keep maintaining the present MPS as well as the its upgrades discussed above. 
All the important interlock items were tested to find potential loopholes before the integration to the MPS. 
In the same manner, we will test new interlock devices in order to avoid any loopholes in the MPS.

\subsection{Radiation effects on the electronics}
The possible radiation effects on the electronics used at the neutrino beamline is considered 
because there is a potential risk of the dangerous beam operation 
if any malfunction of the interlock devices due to the radiation effects is happened. 
The total dose effect(TID) and the single event effect (SEE) are 
considered as the possible radiation effects on the electronics. 

All the electronics devices for the MPS are placed at the ground floor of the neutrino facility buildings. 
In the case of the target station (TS) and the 3rd neutrino utility building (NU3), 
the electronics devices are placed at the 1st class radiation control area.  
Although the areas where the electronics is placed are the 1st class radiation control area, 
the people can enter to there during the beam operation.  
These areas are controlled under the J-PARC radiation safety regulation and 
the radiation level should be less than 12.5$\mu$Sv/h. 
The actual radiation level is periodically measured during the beam operation. 
The measured value during 480kW of the beam power was 0.7 $\mu$Sv/h (1.4 $\mu$Sv/h) for the 
gamma (neutron) at TS. It was also 1.5 $\mu$Sv/h (undetected) for the 
gamma (neutron) at NU3. 

Assuming the 12.5$\mu$Sv/h of the gamma is continuously exposed on the electronics devices, 
the expected accumulated dose is approximately 0.1Gy/year. 
Since the TID effects on the semiconductor devices can be typically appeared 
when the total accumulated dose exceeds 100~Gy\cite{Gonella:2007zz}, 
we expect that the TID effect is small for the electronics at the neutrino beamline. 

On the other hand, there is a possibility to occur the SEE and cause 
any malfunction of the interlock devices even though the radiation level of 
the neutron is $\sim 1\mu$Sv/h level. 
In the case of the neutrino beamline, a conceivable case is that 
the CPU unit of the programable controller logic (PLC) in the interlock system 
is stopped due to the SEE. 
When this case is happened, we can reset the PLC by going to the PLC location. 
However, a certain countermeasure is necessary to avoid any malfunction of the interlock. 
All the PLC logic used in the neutrino beamline will be improved to stop the beam operation
when the CPU unit of the PLC is stopped.  
\color{black}

\clearpage

\section{Neutrino Flux Prediction} \label{sec:flux}
Improvement of the systematic uncertainty is essential
to achieve the goal of T2K-II CP violation study
as well as the precise study of neutrino interaction. 
The present systematic uncertainty on the number of
predicted events at the far detector is $\sim$6~\% in total.
The neutrino flux uncertainty is contributed in the
product of flux and neutrino cross section after the ND280
constraints. 
If those errors are not improved, the significant reduction of
the CP violation sensitivity is expected with the T2K-II statistics.

The neutrino flux prediction uncertainty is currently
around 9~\% at the neutrino energy peak on the absolute flux at
both the near and far detector. It is dominated by 
uncertainties on the hadron production in the target and surrounding
materials in the neutrino beamline and by the proton beam orbit
measurement\cite{Abe:2012av}. 
Although the far-to-near ratio uncertainty is 0.2~\% is achieved
with a tuning based on the NA61/SHINE thin target data\cite{Abgrall:2014xwa, Abgrall:2015hmv}, 
the uncertainty on the absolute flux is relevant for neutrino cross section measurements and
it results 3~\% error of the flux and cross section after the near detector 
constraints. 
While the extrapolation uncertainty is already small, the oscillation
analysis may benefit from further reduction of flux uncertainty
since the neutrino cross section model can be more strongly
constrained with smaller flux uncertainties.

\begin{figure}[h] 
\centering 
\includegraphics[width=0.9\textwidth] {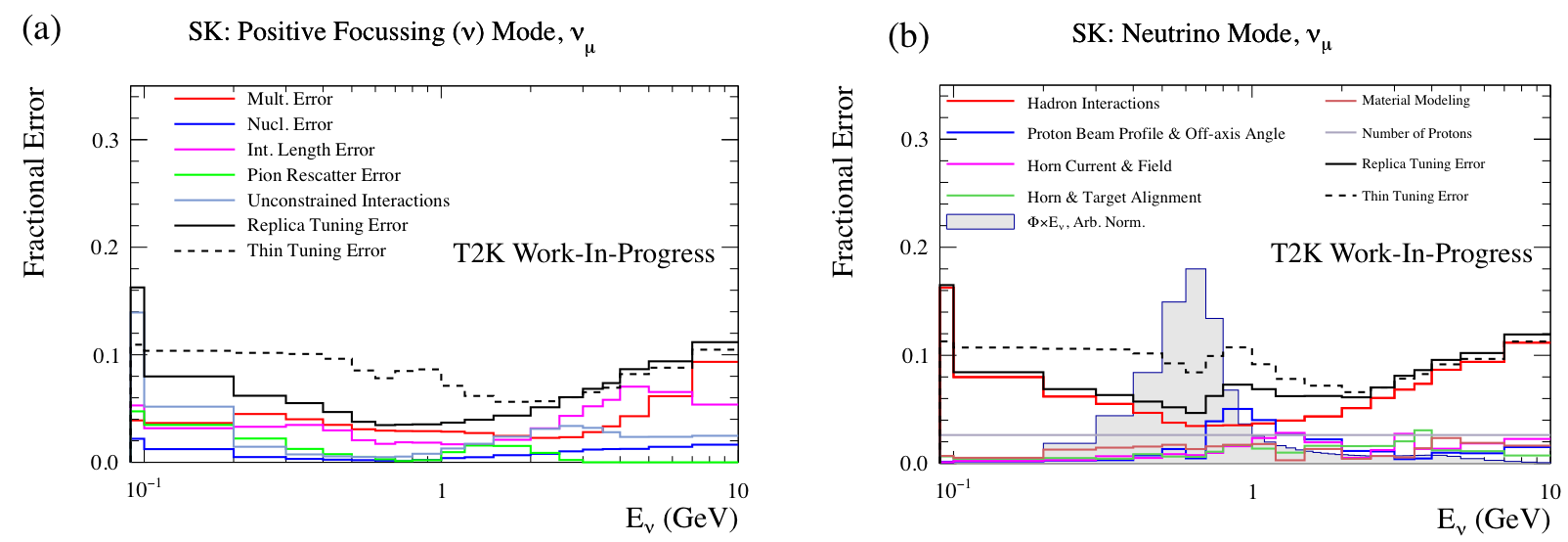}
\caption{(a) The flux uncertainties at the far detector due to
  the hadronic interaction modeling. The black
  solid (dashed) line shows the uncertainties evaluated with the NA61/SHINE replica
  (thin) target data. (b) The total flux uncertainties at the far detector.  }
\label{fig:replicaerror}
\end{figure}

NA61/SHINE measurements of the hadron production from a
replica T2K target can reduce the absolute flux uncertainty.
The charged pion production with the replica target is already measured
and released from the NA61/SHINE\cite{Abgrall:2016jif,Berns:2018tap} and has been achieved $\sim$8~\%
precision. The use of these measurements in the T2K flux calculation
can reduce the hadron production error down to the half of current
size of uncertainty.
Figure~\ref{fig:replicaerror} shows the preliminary results of
flux uncertainty evaluation with the NA61/SHINE replica target data.
Additional improvements on the flux is expected 
using the higher statistics replica target data taken in 2010, 
in particular kaon and proton production
yields are also measured in addition to the charged pions.
Although the flux prediction with the replica target data,
the major uncertainty is still hadron production. 
Further measurements of the hadron production can be expected\cite{NA61+}. 

Another large source of the uncertainty is the beam direction
uncertainty due to the alignment of beamline components and the
position of the proton beam on the upstream of the T2K target.
This uncertainty will be reduced by implementing to use the
INGRID beam direction measurement to constrain the neutrino beam
direction.

\clearpage

\section{Summary}
\label{sec:summary}

Toward a discovery of CP violation in neutrino sector, T2K Collaboration is planning to upgrade
the neutrino beamline, the near detectors, and the analysis methods with improved statistics of
2$\times 10^{22}$ POT. The J-PARC accelerator group is aiming to achieve beam power
improvement for 1.3~MW operation. The neutrino beamline will also be upgraded to accommodate
the 1.3~MW beam. In this document, technical details of the neutrino beamline upgrades were
described. The beamline DAQ and horn electrical system will be upgraded for the higher repetetion
operation at 1.3 s cycle. Beam monitors will be upgraded for the operation at the higher beam intensity
of 3.2$\times 10^{14}$ protons/pulse for 1.3~MW operation. Most of the secondary beamline components
need improvement of their cooling capacity for 1.3~MW beam. Radioactive water disposal system
should also be upgraded to sufficiently treat the large amount of radioactivity from the higher power
operation. 
The required cost for these upgrades was evaluated.
All the necessary upgrades are to be completed by JFY2021.

\clearpage
\appendix 
\section{Possible further neutrino beam-line upgrade}
\label{sec:possibleupgrade}

\graphicspath{{figures/main_primary/beamMonitors}}

\subsection{Beam Induced Fluorescence Monitor (BIF)} \label{sec:BIF}

A Beam Induced Fluorescence (BIF) monitor, which can continuously and
non-destructively measure the proton beam profile, is under development \cite{Friend:2016BIF}.

\begin{figure}[h] 
\centering 
\includegraphics[width=6cm] {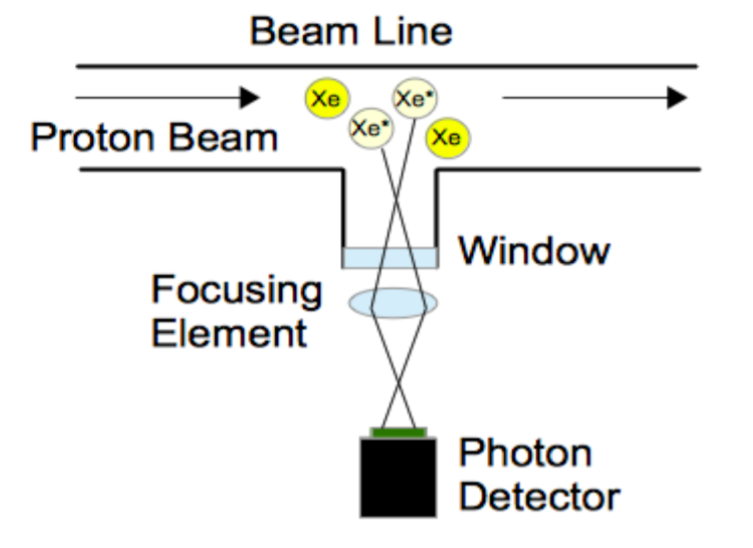}
\caption{Simple schematic of BIF monitor.}
\label{fig:BIFschematic} 
\end{figure}

In a BIF monitor, the beam profile is measured when the passing beam ionizes
\textcolor{\MODCOLOR}{or excites} some of the gas particles in the beamline. The particles then isotropically fluoresce when
returning to the ground state, and the transverse profile of this fluorescence
light will match the transverse profile of the proton beam.  This light could
then be observed from the bottom of the beampipe (to measure the beam X profile) 
and from the side (to measure the beam Y profile).  A simple BIF monitor
schematic is shown in Fig. \ref{fig:BIFschematic}.

\begin{figure}[h] 
\centering 
\includegraphics[width=11cm]{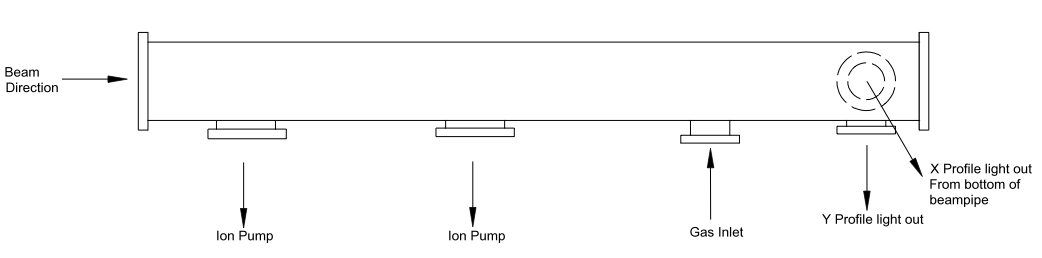}
\caption{Simple diagram of BIF beamline design.}
\label{fig:BIFbeampipe} 
\end{figure}

The neutrino primary beamline has a \textcolor{\MODCOLOR}{3.894}-m-long empty duct in the final focusing (FF) section 
between FH1 and FV2.  We plan to install a BIF profile monitor at the downstream end of
this duct.  The current monitor design consists of (from upstream to downstream) : 
\begin{itemize}
\item Two additional 500~L/s ion pumps
\item A series of valves to inject pulsed N\(_2\) gas
\item Two composite quartz viewports (one at the bottom of the beampipe and one
  on the pathway side of the beampipe) to allow BIF light to exit the beampipe
\item An optical system for transporting and detecting the light, as described below
\end{itemize}
A simple diagram of the current beamline design is shown in Fig.~\ref{fig:BIFbeampipe}.

\paragraph{Space Charge Effects} 

The transverse field from the charge of the proton beam itself can
reach as high as \(4\times10^6\)~V/m for a 2 mm width bunch containing \(1.5 \times 10^{13}\) protons.
Acceleration of ions or electrons in this field would distort the measured beam
profile, where a highly impractical magnet of \(>\)1 T would be required to
counteract this effect in an Ionization Profile Monitor.  Therefore a BIF
monitor is deemed more suitable.

Space charge effects can also cause distortion of the BIF profile if particles
ionized by the beam passage move in the beam field before fluorescing. In this
case, choosing a gas with a short fluorescence lifetime or a large mass can help
to mitigate this distortion. Another option is to use fast photosensors with
ns-timescale readout or gated timing such that only early fluorescence light is
measured.  Because pumping N\(_2\) from our beamline is much simpler than
pumping a heavier gas, the second option is our current baseline proposal.
Relevant fluorescence parameters of N\(_2\) can be found in
Ref.~\cite{Plum:2002gas}.  

\paragraph{Light Detection} 

Two light detection options are currently under development in parallel;
both should have similar light collection acceptance and efficiency.

Both light detection options will collect fluorescence light from the
beampipe through a composite quartz beam window.  Light will be focused
point-to-parallel by a plano-convex lens and
transported to a second plano-convex lens, which can be placed as far from the beampipe as necessary
to reduce radiation damage to other optical elements.  The second lens
will shrink the image to a suitable size for detection and focus the 
light onto a detection system.  A 45\degree~mirror will be placed between the 
2 lenses to bend the light by 90\degree~to the floor for the Y profile measurement arm.
Work to optimize the lens size and focal length is underway : a larger lens
should be able to collect more light (without being too large to be practically
usable), while the final image size must also be reasonable.  Lenses must also 
be robust to radiation damage.

\begin{figure}[h] 
\centering 
\includegraphics[width=6cm]{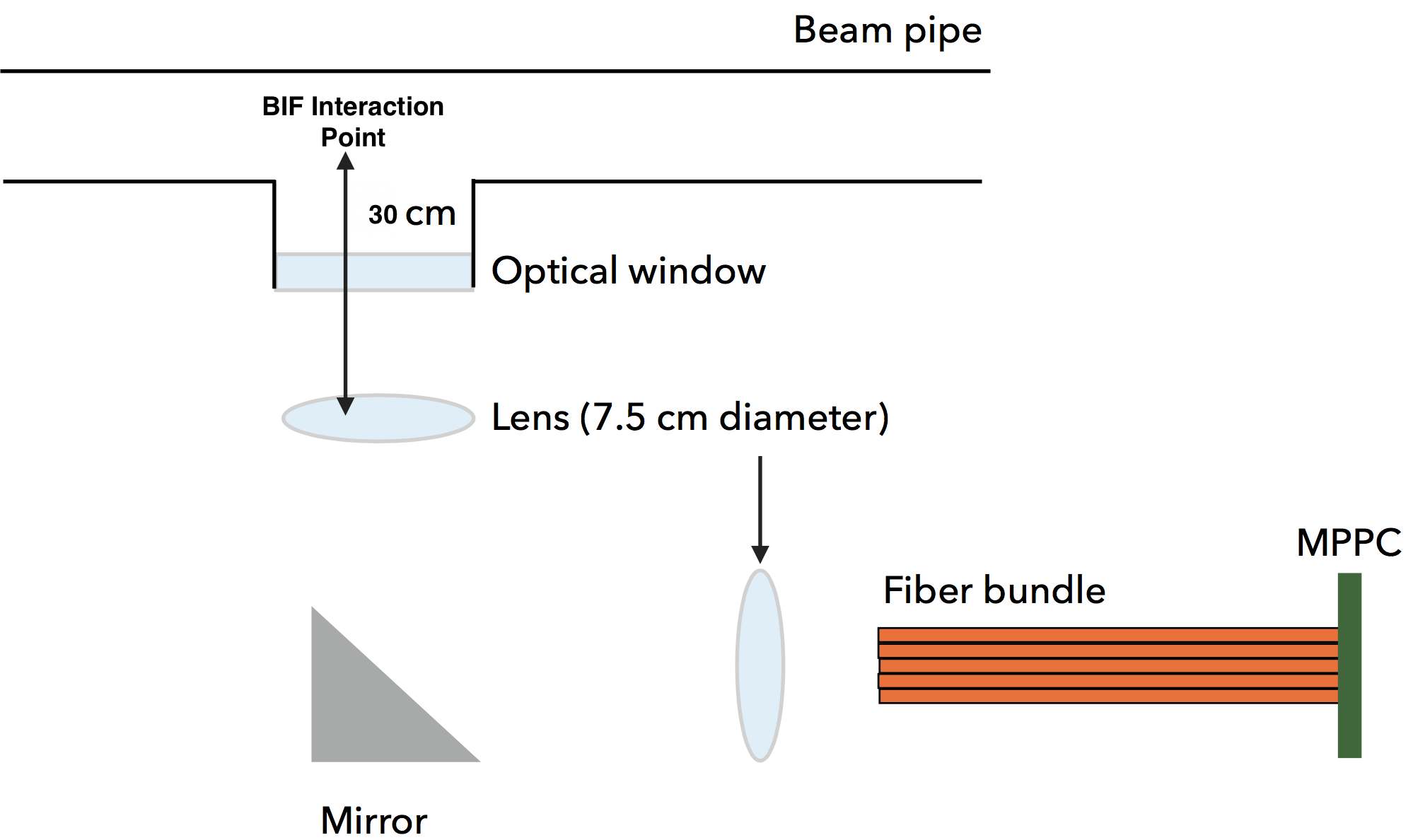}
\caption{Schematic of BIF optical fiber/MPPC optical system option.} 
\label{fig:BIFMPPCopt} 
\end{figure}

The first light detection option uses a bundle optical 
fibers which will transport the light \(\sim\)30~m into subtunnel D-E, where
light detection will be done by an array of Multi-Pixel Photon Counters (MPPCs), 
as shown in Fig.~\ref{fig:BIFMPPCopt}.  The fibers at the beamline end will be 
bundled into an array held by a custom-made fiber holder. 
Currently, 30-m-long, silica core (800\(\mu\)m core diameter) optical fibers
from Thorlabs coupled to a 4\(\times\)4 array of 3 mm \(\times\) 3 mm sensor S13361 series MPPCs from Hamamatsu 
are under consideration.  

Based on our studies, MPPCs are not radiation hard, and therefore must be kept in a
relatively low-radiation area.  Beam induced backgrounds and degradation of MPPCs in the
neutrino primary beamline tunnel have been extensively studied, and it has been
found that subtunnel D-E will be a suitable location to keep both MPPCs and the
amplifier boards used for signal amplification and readout.  Signals from the
amplifier board will be transported to the surface level (NU2) by BNC cables
(type NH-5D-FB) and read out by
FADCs in the NU2 DAQ.  An input bias voltage (\(\sim\)60V) will be supplied from NU2, while a
\(\pm\)5 or 12~V power signal may be supplied from NU2 or by a small radiation-hard
power supply in the subtunnel. 

MPPCs have the advantage that they can detect light down to single pe levels and have
a fast readout.  

Studies of optical fibers are currently underway, and, although there is some
beam-induced background observed when optical fibers are placed near the
beamline (with an approximately equal fast component and slow component)
corresponding to \(\sim\)\textcolor{\MODCOLOR}{150} photo-electrons per beam spill, work
is ongoing to reduce these backgrounds by improved shielding, etc.  

\begin{figure}[h] 
\centering 
\includegraphics[width=6cm]{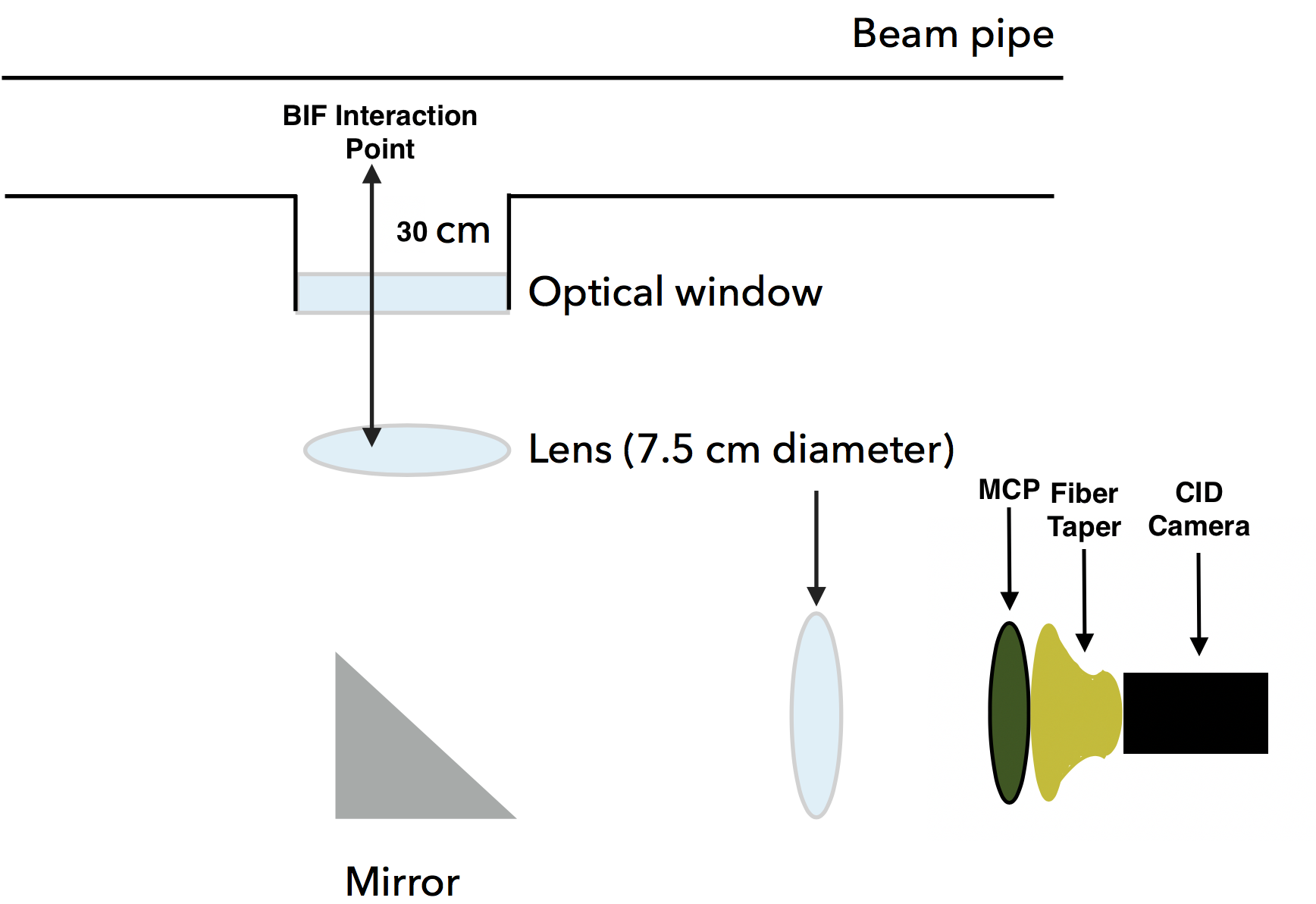}
\caption{Schematic of BIF MCP/CID camera optical system option.} 
\label{fig:BIFMCPopt} 
\end{figure}

The second light detection option is to collect light onto a gatable image
intensifier with a Micro Channel Plate (MCP) coupled to a radiation-hard 
CID camera by a fiber taper, as shown in Fig.~\ref{fig:BIFMCPopt}.  
A dual-stage gatable image intensifier can have a gain of up to \textcolor{\MODCOLOR}{\(\sim\)\(1\times10^6\)} 
and gate timing down to \textcolor{\MODCOLOR}{\(\sim\)10ns}.
The radiation-hard CID camera is functionally identical to the one used in the 
OTR monitor, and therefore its properties are well known and a readout scheme 
already exists. 

The number of detected photons, \(N_\gamma^{det}\), should be \(\gtrsim\)1000 to 
reconstruct a profile with reasonable error \(1/\sqrt{N} = < 3\%\).   \(N_\gamma^{det}\) 
depends on the number of protons \(N_p\), the number of photons produced per incident 
proton \(N^p_\gamma\), and the light collection acceptance and efficiency~\(\epsilon\):
\begin{equation}
  \label{eq:bifgammadet}
N_\gamma^{det} = N_pN^p_\gamma\epsilon. \end{equation}

The energy loss and number of photons per deposited energy 
have been experimentally measured as \(dE/dx =0.2 keVm^2/g\) 
and \(\alpha_\gamma= 0.278keV^{-1}\) for N\(_2\) respectively \cite{Plum:2002gas}.
The number of gas molecules per area (integrated distance \(\times\) gas density) 
is \(\Delta z \times \rho\) for gas density \(\rho\).  
Over \(\Delta z = 0.02\)~m, assuming \(\rho=PM/RT=0.011P\) for N\(_2\) 
(by the ideal gas law for pressure \(P\), molar mass \(M\), ideal gas constant \(R\) 
and temperature \(T\)), gives a number of produced photons, \(N^p_\gamma\)
\begin{equation}
  \label{eq:bifnpgamma}
N^p_\gamma=  dE/dx \times \alpha_\gamma \times \Delta z\times  \rho =1.2\times10^{-5}P \end{equation}
at room temperature.

For the optical fiber/MPPC option discussed above, we approximate the light collection 
acceptance and efficiency to be
\begin{equation}
  \label{eq:bifeff}
  \epsilon=
  \epsilon_{opt}\epsilon_{opt-fiber}\epsilon_{fiber}\epsilon_{fiber-mppc}\epsilon_{mppc} = 
  4.1 \times10^{-5}\sim 7.1\times10^{-5}\end{equation}
  for:
\begin{itemize}
\item Lens/mirror acceptance + efficiency \(\epsilon_{opt}=0.99\times10^{-3}\), 
\item Fiber collection efficiency \(\epsilon_{opt-fiber}= 0.22 \sim 0.38\),
\item Fiber transmission efficiency \(\epsilon_{fiber}= 0.66\),
\item Fiber to MPPC coupling efficiency \(\epsilon_{fiber-mppc}=  0.95\),
\item MPPC photon detection efficiency \(\epsilon_{mppc} = 0.3\).
\end{itemize}
Work to increase this acceptance and efficiency by improvements to the optical
system and MPPC detection efficiency is ongoing.

Equations \ref{eq:bifgammadet}, \ref{eq:bifnpgamma}, and \ref{eq:bifeff} give the requirement of \(P=5\sim8\times10^{-3}\)~Pa to detect \(\gtrsim\)1000 photons per beam spill
assuming \(N_p=2.5\times10^{14}\) (approximately the current number of protons-per-pulse).
Given these parameters, we hope to make a spill-by-spill beam profile
measurement with 0.2~mm width precision.

\paragraph{Gas Injection} 

The FF beamline vacuum is kept at \(\sim1\times10^{-5}\)~Pa, while the 
vacuum level in the Super Conducting (SC) section is \(\sim1\times10^{-7}\)~Pa.
The average vacuum level in the FF section should be kept \(<1\times10^{-4}\) 
Pa in order to maintain a beamline ion pump lifetime of 3\(\sim\)4 years.
The average vacuum level at the exit to the SC section of the beamline must 
be strictly kept to \(<5\times10^{-6}\)~Pa to prevent a quench of the SC magnets.  
General beamline components are required to have a gas flow-out rate of \(< 10^{-7}\) 
Pam\(^3\)/s, although some leeway will be given to the BIF monitor if the other 
vacuum requirements can be met.

A differential pressure of 3\(\sim\)4 orders of magnitude is challenging in the case of 
continuous gas injection.  Generally, differential pumping on either side of a ``pinhole'' 
in the beampipe is used, but this is impossible to achieve in the NU FF section 
without either using an impractically large beampipe, or having the BIF pinhole become 
a limiting aperture on the beampipe size.  
Instead, pulsed gas injection, where a short pulse of N\(_2\) gas will be injected before each beam
spill, triggered by the accelerator NU beam timing, will be used.  

Extensive steady-state simulations of the FF beamline by COMSOL's Molecular 
Flow modeling package have been carried out.  These simulation results, which have 
been benchmarked by smaller test bench studies using a custom 16\(\sim\)20~L 
vacuum chamber, are also planned to be confirmed in 2018 by tests in the 
actual FF beamline.

Our COMSOL simulation models the full FF section beamline.
Four 500~L/s ion pumps are already installed along the neutrino primary beamline
FF section \textcolor{\MODCOLOR}{6.0, 8.0, 17.7, and 18.7}~m downstream of the SC section exit.
An additional two ``BIF'' vacuum pumps will also be added to the FF beamline upstream of
the BIF gas injection point, approximately \textcolor{\MODCOLOR}{8.5 and 9.5} m downstream of the SC section exit.  
These pumps will be identical to the 500~L/s, ICF203 ion pumps 
already in use elsewhere in the neutrino primary beamline.
The SC section exit is modeled as a 1000~L/s pump in this simulation.  The
downstream end of the beamline is modeled as a 100~L/s vacuum pump, which
accounts for a turbo-molecular pump at the top of the monitor stack. 

The COMSOL simulation result is shown in Table \ref{tab:BIFpressure}: by injecting
\(2\times10^{-7}\)~kg/s N\(_2\) with a duty factor of 0.5\%, a pressure at the BIF
interaction point of \(\sim10^{-2}\)~Pa during the beam spill can be
achieved, while still maintaining a reasonable average pressure at the nearby ion
pumps and SC section exit.  

\begin{table}
  \centering
  \caption{Expected average pressure and incident molar flux at vacuum pumps, SC
    section exit, and BIF viewport for \(2\times10^{-7}\)~kg/s N\(_2\) gas injection 
    based on COMSOL Molecular Flow model of the FF beamline.
  \label{tab:BIFpressure}}
  \begin{tabular}{l l l } 
    \hline\hline
    \multicolumn{3}{c}{Pulsed 0.5\% Duty Factor} \\
  \hline
    & Pressure (Pa) & Molar flux (mol/m\(^2\)/s) \\
    Upstream (New) Pump & 6.5e-5 & 6.3e-6 \\
    Downstream (Orig.) Pump & 1.6e-5 & 1.6e-6 \\
    SC Sec. & 4.6e-6 & 4.5e-7 \\
  \hline
  \multicolumn{3}{c}{Steady State (During \(\sim\)600ns Beam Spill)} \\
  \hline
    BIF Interaction Region & 9.7e-3 & 9.4e-4 \\
    \hline\hline
  \end{tabular}
\end{table}

The current gas injection design will be to inject a short pulse of N\(_2\) gas
before each beam spill using two Parker Hannifin (PH) Series 9 pulse valves connected 
in series with a small (\(\sim\)0.009~L) chamber between them.  PH 
pulse valves ship with PTFE poppets, but custom PEEK poppets may be required for
use in the radiation environment near the beamline.  The PH standard Pulse Valve controller
(Iota One) takes a standard TTL input pulse (which will be triggered by the beam trigger) 
and generates a HV DC pulse suitable
to open the pulse valve for between 150\(\mu\)s and 10s of minutes, and is rated 
to use a \(\sim\)4-m-long cable.  However,  R\&D is ongoing to
either design a suitable cable which is 30-m-long (to install the controller in the
subtunnel D-E) or 100-m-long (to install the pulse valve in NU2).  

\paragraph{BIF Schedule} \label{sec:BIFschedule}

The schedule for the BIF testing and installation is as follows :
\begin{itemize}
  \item \(\sim\)JFY2015 
\begin{itemize}
  \item Started tests using test vacuum chamber
  \item Simulated light transport system
  \item Tested parts of light detection system
\end{itemize}
\item JFY2016
\begin{itemize}
  \item Characterized gas uniformity using simulation + test vacuum
    chamber; designed, built prototype pulsed gas system
  \item  Designed, built, tested light transport system
  \item  Continued testing parts of light detection system 
\end{itemize}
\item JFY2017
\begin{itemize}
  \item Build, test complete gas inlet, vacuum system
  \item Build, test complete light transport and light detection system
  \item Try beam test (observe BIF) by temporarily causing poor vacuum in NU beamline 
  \item Procure various vacuum parts
\end{itemize}
\item JFY2018
\begin{itemize}
  \item \textcolor{\MODCOLOR}{Install most parts of prototype monitor in NU beamline FF section
    for gas injection tests and first BIF signal observation}
\end{itemize}
\item JFY2019
\begin{itemize}
  \item \textcolor{\MODCOLOR}{Install final parts of complete monitor}
\end{itemize}
\end{itemize}

\clearpage
\bibliographystyle{unsrt}

\end{document}